\documentclass[12pt, a4paper, twoside]{report}
\usepackage{amsmath}
\usepackage{amssymb}
\usepackage{epsfig}
\usepackage{calc}
\usepackage[ansinew]{inputenc}
\usepackage{color}
\usepackage{fancyhdr}

\def\opone{\leavevmode\hbox{\small1\kern-3.8pt\normalsize1}}
\newcommand{\ket}[1]{\left | \, #1 \right \rangle}

\def\be{\begin{equation}}
\def\ee{\end{equation}}

\def\w{\wedge}
\def\d{{\rm d}}
\def\k{\kappa}
\def\r{\rho}
\def\a{\alpha}

\def\b{\beta}

\def\g{\gamma}
\def\G{\Gamma}
\def\dd{\delta}
\def\D{\Delta}
\def\e{\epsilon}

\def\m{\mu}
\def\n{\nu}

\def\l{\lambda}
\def\L{\Lambda}
\def\s{\sigma}
\def\S{\Sigma}

\def\Ct{{\widetilde C}}
\def\Gt{{\widetilde G}}

\def\et{{\widetilde\epsilon}}

\def\St{{\widetilde S}}

\def\t{\theta}

\def\tr{\,{\rm tr}\,}
\def\Tr{\,{\rm Tr}\,}

\def\d{{\rm d}}

\def\k{\kappa}
\def\s{\sigma}
\def\o{\omega}
\def\f{\phi}

\def\O{\Omega}

\def\ks{{k \kern-.5em /}}
\def\es{{\e \kern-.4em /}}
\def\ds{{\partial \kern-.5em /}}
\def\Ds{{D \kern-.7em /}}

\def\gh{{\hat g}}

\def\R{{\cal R}}

\def\hN{\hat{N}}
\def\hg{\hat{g}}
\def\zb{\bar{z}}
\def\zt{\tilde{z}}
\def\ztb{\bar{\tilde{z}}}
\def\wb{\bar{w}}
\def\wt{\widetilde}
\def\eb{{\overline e}}
\def\st{{}^*}
\def\la{\langle}
\def\ra{\rangle}
\def\mN{\mathcal{N}}
\def\mF{\mathcal{F}}
\def\mD{\mathcal{D}}
\def\mO{\mathcal{O}}
\def\mQ{\mathcal{Q}}
\def\sty{\, {}^{*_Y}\kern-.2em}
\def\rh{\hat r}

\newtheorem{proposition}{Proposition}[chapter]
\newtheorem{theorem}[proposition]{Theorem}
\newtheorem{definition}[proposition]{Definition}

\begin{document}
\setcounter{footnote}{0}
\setcounter{page}{0} 
\pagenumbering{roman}

\begin{titlepage}

\begin{center}{\bf \Large SUPERSYMMETRIC GAUGE THEORIES\vspace{2mm}}\\
{\bf \Large FROM STRING THEORY}

\vskip 1.5cm PhD-thesis of

\vskip 1.5cm {\bf  STEFFEN METZGER}

\vskip.3cm  Laboratoire de Physique Th\'eorique,
\'Ecole Normale Sup\'erieure\\
24 rue Lhomond, 75231 Paris Cedex 05, France

\vskip.3cm and

\vskip.3cm  Arnold-Sommerfeld-Center for Theoretical
Physics,\\
Department f\"ur Physik, Ludwig-Maximilians-Universit\"at\\
Theresienstr. 37, 80333 Munich, Germany\\

\vskip.3cm {\small e-mail: {\tt
steffen.metzger@physik.uni-muenchen.de}}

\vskip 1cm December 2005
\end{center}

\end{titlepage}

\cleardoublepage

\vspace*{5cm} {\hspace{4.5cm}
\begin{tabular}{l}
{\it Wenn ich in den Grübeleien eines langen Lebens}\\
{\it eines gelernt habe, so ist es dies,}\\
{\it dass wir von
einer tieferen Einsicht}\\
{\it in die elementaren Vorgänge viel weiter entfernt sind}\\
{\it als die meisten unserer Zeitgenossen glauben.}\\
\\
A. Einstein (1955)
\end{tabular}}

\cleardoublepage

\vspace*{5cm} {\hspace{7.5cm} F\"ur Constanze}

\cleardoublepage

\begin{center}
{\bf \LARGE Abstract}\\
\end{center}
\vspace{5mm} The subject of this thesis are various ways to
construct four-dimensional quantum field theories from string
theory.

In a first part we study the generation of a supersymmetric
Yang-Mills theory, coupled to an adjoint chiral superfield, from
type IIB string theory on non-compact Calabi-Yau manifolds, with
D-branes wrapping certain subcycles. Properties of the gauge
theory are then mapped to the geometric structure of the
Calabi-Yau space. In particular, the low energy effective
superpotential, governing the vacuum structure of the gauge
theory, can in principle be calculated from the open (topological)
string theory. Unfortunately, in practice this is not feasible.
Quite interestingly, however, it turns out that the low energy
dynamics of the gauge theory is captured by the geometry of
another non-compact Calabi-Yau manifold, which is related to the
original Calabi-Yau by a geometric transition. Type IIB string
theory on this second Calabi-Yau manifold, with additional
background fluxes switched on, then generates a four-dimensional
gauge theory, which is nothing but the low energy effective theory
of the original gauge theory. As to derive the low energy
effective superpotential one then only has to evaluate certain
integrals on the second Calabi-Yau geometry. This can be done, at
least perturbatively, and we find that the notoriously difficult
task of studying the low energy dynamics of a non-Abelian gauge
theory has been mapped to calculating integrals in a well-known
geometry. It turns out, that these integrals are intimately
related to quantities in holomorphic matrix models, and therefore
the effective superpotential can be rewritten in terms of matrix
model expressions. Even if the Calabi-Yau geometry is too
complicated to evaluate the geometric integrals explicitly, one
can then always use matrix model perturbation theory to calculate
the effective superpotential.

This intriguing picture has been worked out by a number of authors
over the last years. The original results of this thesis comprise
the precise form of the special geometry relations on local
Calabi-Yau manifolds. We analyse in detail the cut-off dependence
of these geometric integrals, as well as their relation to the
matrix model free energy. In particular, on local Calabi-Yau
manifolds we propose a pairing between forms and cycles, which
removes all divergences apart from the logarithmic one. The
detailed analysis of the holomorphic matrix model leads to a
clarification of several points related to its saddle point
expansion. In particular, we show that requiring the planar
spectral density to be real leads to a restriction of the shape of
Riemann surfaces, that appears in the planar limit of the matrix
model. This in turns constrains the form of the contour along
which the eigenvalues have to be integrated. All these results are
used to exactly calculate the planar free energy of a matrix model
with
cubic potential.\\

The second part of this work covers the generation of
four-dimensional supersymmetric gauge theories, carrying several
important characteristic features of the standard model, from
compactifications of eleven-dimensional supergravity on
$G_2$-manifolds. If the latter contain conical singularities,
chiral fermions are present in the four-dimensional gauge theory,
which potentially lead to anomalies. We show that, locally at each
singularity, these anomalies are cancelled by the non-invariance
of the classical action through a mechanism called ``anomaly
inflow". Unfortunately, no explicit metric of a compact
$G_2$-manifold is known. Here we construct families of metrics on
compact weak $G_2$-manifolds, which contain two conical
singularities. Weak $G_2$-manifolds have properties that are
similar to the ones of proper $G_2$-manifolds, and hence the
explicit examples might be useful to better understand the generic
situation. Finally, we reconsider the relation between
eleven-dimensional supergravity and the $E_8\times E_8$-heterotic
string. This is done by carefully studying the anomalies that
appear if the supergravity theory is formulated on a ten-manifold
times the interval. Again we find that the anomalies cancel
locally at the boundaries of the interval through anomaly inflow,
provided one suitably modifies the classical action.

\cleardoublepage

\tableofcontents

\cleardoublepage

\pagenumbering{arabic}

\chapter{Setting the Stage}
{\it Wer sich nicht mehr wundern,\\
nicht mehr staunen kann, \\
der ist sozusagen tot \\
und sein Auge erloschen.\\
\\
A. Einstein}
\\
\\
When exactly one hundred years ago Albert Einstein published his
famous articles on the theory of special relativity
\cite{EinsteinSR}, Brownian motion \cite{EinsteinBM} and the
photoelectric effect \cite{EinsteinPE}, their tremendous impact on
theoretical physics could not yet be foreseen. Indeed, the
development of the two pillars of modern theoretical physics, the
theory of general relativity and the quantum theory of fields, was
strongly influenced by these publications. If we look back today,
it is amazing to see how much we have learned during the course of
the last one hundred years. Thanks to Einstein's general
relativity we have a much better understanding of the concepts of
space, time and the gravitational force. Quantum mechanics and
quantum field theory, on the other hand, provide us with a set of
physical laws which describe the dynamics of elementary particles
at a subatomic scale. These theories have been tested many times,
and always perfect agreement with experiment has been found
\cite{PDG}. Unfortunately, their unification into a quantum theory
of gravity has turned out to be extremely difficult. Although it
is common conviction that a unified mathematical framework
describing both gravity and quantum phenomena should exist, it
seems to be still out of reach. One reason for the difficulties is
the fundamentally different nature of the physical concepts
involved. Whereas relativity is a theory of space-time, which does
not tell us much about matter, quantum field theory is formulated
in a fixed background space-time and deals with the nature and
interactions of elementary particles. In most of the current
approaches to a theory of quantum gravity one starts from quantum
field theory and then tries to extend and generalise the concepts
to make them applicable to gravity. A notable exception is the
field of loop quantum gravity\footnote{For a recent review from a
string theory perspective and an extensive list of references see
\cite{NPZ05}.}, where the starting point is general relativity.
However, given the conceptual differences it seems quite likely
that one has to leave the familiar grounds of either quantum field
theory or relativity and try to think of something fundamentally
new.

Although many of its concepts are very similar to the ones
appearing in quantum field theory, string theory \cite{GSW87},
\cite{Pol} is a branch of modern high energy physics that
understands itself as being in the tradition of both quantum field
theory and relativity. Also, it is quite certainly the by far most
radical proposition for a unified theory. Although its basic ideas
seem to be harmless - one simply assumes that elementary particles
do not have a point-like but rather a string-like structure - the
consequences are dramatic. Probably the most unusual prediction of
string theory is the existence of ten space-time dimensions.

As to understand the tradition on which string theory is based and
the intuition that is being used, it might be helpful to comment
on some of the major developments in theoretical physics during
the last one hundred years. Quite generally, this might be done by
thinking of a physical process as being decomposable into the
scene on which it takes place and the actors which participate in
the play. Thinking about the scene means thinking about the
fundamental nature of space and time, described by general
relativity. The actors are elementary particles that interact
according to a set of rules given by quantum field theory. It is
the task of a unified theory to think of scene and actors as of
two interdependent parts of the successful play of nature.

\bigskip
{\bf High energy physics in a nutshell}\\
Special relativity tells us how we should properly think of space
and time. From a modern point of view it can be understood as the
insight that our world is $(\mathbb{R}^4,\eta)$, a topologically
trivial four-dimensional space carrying a flat metric, i.e. one
for which all components of the Riemann tensor vanish, with
signature $(-,+,+,+)$. Physical laws should then be formulated as
tensor equations on this four-dimensional space. In this
formulation it becomes manifest that a physical process is
independent of the coordinate system in which it is described.

General relativity \cite{EinsteinGR} then extends these ideas to
cases where one allows for a metric with non-vanishing Riemann
tensor. On curved manifolds the directional derivatives of a
tensor along a vector in general will not be tensors, but one has
to introduce connections and covariant derivatives to be able to
write down tensor equations. Interestingly, given a metric a very
natural connection and covariant derivative can be constructed.
Another complication that appears in general relativity is the
fact that the metric itself is a dynamical field, and hence there
should be a corresponding tensor field equation which describes
its dynamics. This equation is know as the Einstein equation and
it belongs to the most important equations of physics. It is quite
interesting to note that the nature of space-time changes
drastically when going from special to general relativity. Whereas
in the former theory space-time is a rigid spectator on which
physical theories can be formulated, this is no longer true if we
allow for general metrics. The metric is both a dynamical field
and it describes the space in which dynamical processes are
formulated. It is this double role that makes the theory of
gravity so intriguing and complicated.

Einstein's explanation of the photoelectric effect made use of the
quantum nature of electro-magnetic waves, which had been the main
ingredient to derive Planck's formula for black body radiation,
and therefore was one major step towards the development of
quantum mechanics. As is well known, this theory of atomic and
subatomic phenomena was developed during the first decades of the
twentieth century by Sommerfeld, Bohr, Heisenberg, Schödinger,
Dirac, Pauli and many others.\footnote{A list of many original
references can be found in \cite{CDL}.} It describes a physical
system (in the Schrödinger picture) as a time-dependent state in
some Hilbert space with a unitary time evolution that is
determined by the Hamilton operator, which is specific to the
system. Observables are represented by operators acting on the
Hilbert space, and the measurable quantities are the eigenvalues
of these operators. The probability (density) of measuring an
eigenvalue is given by the modulus square of the system state
vector projected onto the eigenspace corresponding to the
eigenvalue. Shortly after the measurement the physical state is
described by an eigenvector of the operator. This phenomenon is
known as the collapse of the wave function and here unitarity
seems to be lost. However, it is probably fair to say that the
measurement process has not yet been fully understood.

In the late nineteen forties one realised that the way to combine
the concepts of special relativity and quantum mechanics was in
terms of a quantum theory of fields\footnote{See \cite{Wb00} for a
beautiful introduction and references to original work.}. The
dynamics of such a theory is encoded in an action $S$, and the
generating functional of correlation functions is given as the
path integral of $e^{{i\over\hbar} S}$ integrated over all the
fields appearing in the action. In the beginning these theories
had been plagued by infinities, and only after these had been
understood and the concept of renormalisation had been introduced
did it turn into a powerful calculational tool. Scattering cross
sections and decay rates of particles could then be predicted and
compared to experimental data. The structure of this theoretical
framework was further explored in the nineteen fifties and sixties
and, together with experimental results, which had been collected
in more and more powerful accelerators and detectors, culminated
in the formulation of the standard model of particle physics. This
theory elegantly combines the strong, the electro-magnetic and the
weak force and accounts for all the particles that have been
observed so far. The Higgs boson, a particle that is responsible
for the mass of some of the other constituents of the model, is
the only building block of the standard model that has not yet
been discovered. One of the major objectives of the Large Hadron
Collider (LHC), which is currently being built at CERN in Geneva,
is to find it and determine its properties.

For many years all experimental results in particle physics could
be explained from the standard model. However, very recently a
phenomenon, known as {\it neutrino oscillation}, has been observed
that seems to be inexplicable within this framework. The standard
model contains three types of neutrinos, which do not carry charge
or mass, and they only interact via the weak force. In particular,
neutrinos cannot transform into each other. However, observations
of neutrinos produced in the sun and the upper part of the
atmosphere seem to indicate that transitions between the different
types of neutrinos do take place in nature, which is only possible
if neutrinos carry mass. These experiments are not only
interesting because the standard model has to be extended to
account for these phenomena, but also because the mass of the
neutrinos could be relevant for open questions in cosmology.
Neutrinos are abundant in the universe and therefore, although
their mass is tiny, they might contribute in a non-negligible way
to the dark matter which is known to exist in our universe.
Although interesting, since beyond the standard model, neutrino
oscillation can be described by only slightly modifying the
standard model action. Therefore, it does not seem to guide us in
formulating a theory of quantum gravity.

There are also some theoretical facts which indicate that after
all the standard model has to be modified. For some time it was
believed that quantum field theories of the standard model type
form the most general setting in which particle physics can be
formulated. The reason is that combining some weak assumptions
with the basic principles of relativity and quantum mechanics
leads to no-go theorems which constrain the possible symmetries of
the field theory. However, in the middle of the nineteen seventies
Wess and Zumino discovered that quantum field theories might carry
an additional symmetry that relates bosonic and fermionic
particles and which is known as {\it supersymmetry} \cite{WZ74a},
\cite{WZ74b}. It was realised very soon that the no-go theorems
had been to restrictive since they required the symmetry
generators to form a Lie algebra. This can be generalised to
generators forming a {\it graded} Lie algebra and it was then
shown by Haag, Lopuszanski and Sohnius that the most general
graded symmetry algebra consistent with the concepts of quantum
field theory is the supersymmetry algebra. This provides a strong
motivation to look for supersymmetry in our world since it would
in some sense be amazing if nature had chosen not to use all the
freedom that it has. On the classical level supersymmetry can also
be applied to theories containing the metric which are then known
as supergravity theories. Of course one might ask whether it was
simply the lack of this additional supersymmetry that made the
problem of quantising gravity so hard. However, unfortunately it
turns out that these problems persist in supergravity theories. In
order to find a consistent quantum theory of gravity which
contains the standard model one therefore has to proceed even
further.

Another important fact, which has to be explained by a unified
theory, is the difference between the Planck scale of $10^{19}$
GeV (or the GUT scale at $10^{16}$ GeV) and the preferred scale of
electro-weak theory, which lies at about $10^2$ GeV. The lack of
understanding of this huge difference of scales is known as the
{\it hierarchy problem}. Finally it is interesting to see what
happens if one tries to estimate the value of the cosmological
constant from quantum field theory. The result is by 120 orders of
magnitude off the measured value! This {\it cosmological constant
problem} is another challenge for a consistent quantum theory of
gravity.

\bigskip
{\bf Mathematical rigour and experimental data}\\
Physics is a science that tries to formulate abstract mathematical
laws from observing natural phenomena. The experimental setup and
its theoretical description are highly interdependent. However,
whereas it had been the experiments that guided theoretical
insight for centuries, the situation is quite different today.
Clearly, Einstein's 1905 papers were still motivated by
experiments - Brownian motion and the photoelectric effect had
been directly observed, and, although the Michelson-Morley
experiment was much more indirect, it finally excluded the ether
hypotheses thus giving way for Einstein's ground breaking theory.
Similarly, quantum mechanics was developed as very many data of
atomic and subatomic phenomena became available and had to be
described in terms of a mathematical theory. The explanation of
the energy levels in the hydrogen atom from quantum mechanics is
among the most beautiful pieces of physics. The questions that are
answered by quantum field theory are already more abstract. The
theory is an ideal tool to calculate the results of collisions in
a particle accelerator. However, modern  accelerators are
expensive and tremendously complicated technical devices. It takes
years, sometimes decades to plan and build them. Therefore, it is
very important to perform theoretical calculations beforehand and
to try to predict interesting phenomena from the mathematical
consistency of the theory. Based on these calculations one then
has to decide which machine one should build and which phenomena
one should study, in order to extract as much information on the
structure of nature as possible.

An even more radical step had already been taken at the beginning
of the twentieth century with the development of general
relativity. There were virtually no experimental results, but
relativity was developed from weak physical assumptions together
with a stringent logic and an ingenious mathematical formalism. It
is probably fair to say that Einstein was the first theoretical
physicist in a modern sense, since his reasoning used mathematical
rigour rather than experimental data.

Today physics is confronted with the strange situation that
virtually any experiment can be explained from known theories, but
these theories are themselves known to be incomplete. Since the
gravitational force is so weak, it is very difficult to enter the
regime where (classical) general relativity is expected to break
down. On the other hand, the energy scales where one expects new
phenomena to occur in particle physics are so high that they can
only be observed in huge and expensive accelerators. The scales
proposed by string theory are even way beyond energy scales that
can be reached using standard machines. Therefore, today physics
is forced to proceed more or less along the lines of Einstein,
using pure thought and mathematical consistency. This path is
undoubtedly difficult and dangerous. As we know from special and
general relativity, a correct result can be extremely
counterintuitive, so a seemingly unphysical theory, like string
theory with its extra dimensions, should not be easily discarded.
On the other hand physics has to be aware of the fact that finally
its purpose is to explain experimental data and to quantitatively
predict new phenomena. The importance of experiments cannot be
overemphasised and much effort has to be spent in setting up
ingenious experiments which might tell us something new about the
structure of nature. This is a natural point where one could delve
into philosophical considerations. For example one might muse
about how Heisenberg's positivism has been turned on the top of
its head, but I will refrain from this and rather turn to the
development of string theory.

\vspace{2cm}
{\bf The development of string theory}\\
String theory was originally developed as a model to understand
the nature of the strong force in the late sixties. However, when
Quantum Chromodynamics (QCD) came up it was quickly abandoned,
with only very few people still working on strings. One of the
first important developments was the insight of Jo\"{e}l Scherk
and John Schwarz that the massless spin two particle that appears
in string theory can be interpreted as the graviton, and that
string theory might actually be a quantum theory of gravitation
\cite{SS74}. However, it was only in 1984 when string theory
started to attract the attention of a wider group of theoretical
physicists. At that time it had become clear that a symmetry of a
classical field theory does not necessarily translate to the
quantum level. If the symmetry is lost one speaks of an {\it
anomaly}. Since local gauge and gravitational symmetries are
necessary for the consistency of the theory, the requirement of
anomaly freedom of a quantum field theory became a crucial issue.
Building up on the seminal paper on gravitational anomalies
\cite{AGW84} by Alvarez-Gaumé and Witten, Green and Schwarz showed
in their 1984 publication \cite{GS84} that $\mathcal{N}=1$
supergravity coupled to super Yang-Mills theory in ten dimensions
is free of anomalies, provided the gauge group of the Yang-Mills
theory is either $SO(32)$ or $E_8\times E_8$. The anomaly freedom
of the action was ensured by adding a local counterterm, now known
as the {\it Green-Schwarz term}. Quite remarkably, both these
supergravity theories can be understood as low energy effective
theories of (ten-dimensional) superstring theories, namely the
Type I string with gauge group $SO(32)$ and the heterotic string
with gauge group $SO(32)$ or $E_8\times E_8$. The heterotic string
was constructed shortly after the appearance of the Green-Schwarz
paper in \cite{GHMR85}. This discovery triggered what is know as
the first string revolution. In the years to follow string theory
was analysed in great detail, and it was shown that effective
theories in four dimensions with $\mathcal{N}=1$ supersymmetry can
be obtained by compactifying type I or heterotic string theories
on Calabi-Yau manifolds. These are compact K\"ahler manifolds that
carry a Ricci-flat metric and therefore have $SU(3)$ as holonomy
group. Four-dimensional effective actions with $\mathcal{N}=1$
supersymmetry are interesting since they provide a framework
within which the above mentioned hierarchy problem can be
resolved. Calabi-Yau manifolds were studied intensely, because
many of their properties influence the structure of the effective
four-dimensional field theory. A major discovery of mathematical
interest was the fact that for a Calabi-Yau manifold $X$ with
Hodge numbers $h^{1,1}(X)$ and $h^{2,1}(X)$ there exists a {\it
mirror manifold} $Y$ with $h^{1,1}(Y)=h^{2,1}(X)$ and
$h^{2,1}(Y)=h^{1,1}(X)$. Furthermore, it turned out that the
compactification of yet another string theory, known as Type IIA,
on $X$ leads to the same effective theory in four dimensions as a
fifth string theory, Type IIB, on $Y$. This fact is extremely
useful, since quantities related to moduli of the complex
structure on a Calabi-Yau manifold can be calculated from
integrals in the Calabi-Yau geometry. Quantities related to the
K\"ahler moduli on the other hand obtain corrections from
world-sheet instantons and are therefore very hard to compute.
Mirror symmetry then tells us that the K\"ahler quantities of $X$
can be obtained from geometric integrals on $Y$. For a detailed
exposition of mirror symmetry containing many references see
\cite{Horietal}.

Equally important for string theory was the discovery that string
compactifications on singular Calabi-Yau manifold make sense and
that there are smooth paths in the space of string
compactifications along which the topology of the internal
manifold changes \cite{GMS95}. All these observations indicate
that stringy geometry is quite different from the point particle
geometry we are used to.

Another crucial development in string theory was Polchinski's
discovery of {\it D-branes} \cite{Po95}. These are extended
objects on which open strings can end. Their existence can be
inferred by exploiting a very nice symmetry in string theory,
known as T-duality. In the simplest case of the bosonic string it
states that string theory compactified on a circle of radius $R$
is isomorphic to the same theory on a circle of radius $1/R$,
provided the momentum quantum numbers and the winding numbers are
exchanged. Note that here once again the different notion of
geometry in string theory becomes apparent. One of the many
reasons why D-branes are useful is that they can be used to
understand black holes in string theory and to calculate their
entropy.

There also has been progress in the development of quantum field
theory. For example Seiberg and Witten in 1994 exactly solved the
four-dimensional low-energy effective $\mathcal{N}=2$ theory with
gauge group $SU(2)$ \cite{SW94}. The corresponding action is
governed by a holomorphic function $\mathcal{F}$, which they
calculated from some auxiliary geometry. Interestingly, one can
understand this geometry as part of a Calabi-Yau compactification
and the Seiberg-Witten solution can be embedded into string theory
in a beautiful way.

Originally, five different consistent string theories had been
constructed: Type I with gauge group $SO(32)$, Type IIA, Type IIB
and the heterotic string with gauge groups $SO(32)$ and $E_8\times
E_8$. In the middle of the nineties is became clear, however, that
these theories, together with eleven-dimensional supergravity, are
all related by dualities and therefore are part of one more
fundamental theory, that was dubbed M-theory \cite{Wi95b}.
Although the elementary degrees of freedom of this theory still
remain to be understood, a lot of evidence for its existence has
been accumulated. The discovery of these dualities triggered
renewed interest in string theory, which is known today as the
second string revolution.

Another extremely interesting duality, discovered by Maldacena and
known as the AdS/CFT correspondence \cite{Ma97}, \cite{GKP98},
\cite{Wi98}, relates Type IIB theory on the space $AdS_5\times
S^5$ and a four-dimensional $\mathcal{N}=4$ supersymmetric
conformal field theory on four-dimensional Minkowski space.
Intuitively this duality can be understood from the fact that IIB
supergravity has a brane solution which interpolates between
ten-dimensional Minkowski space and $AdS_5\times S^5$. This brane
solution is thought to be the supergravity description of a
D3-brane. Consider a stack of D3-branes in ten-dimensional
Minkowski space. This system can be described in various ways. One
can either consider the effective theory on the world-volume of
the branes which is indeed an $\mathcal{N}=4$ SCFT or one might
want to know how the space backreacts on the presence of the
branes. The backreaction is described by the brane solution which,
close to the location of the brane, is $AdS_5\times S^5$.

\bigskip
{\bf Recent developments in string theory}\\
String theory is a vast field and very many interesting aspects
have been studied in this context. In the following a quick
overview of the subjects that are going to be covered in this
thesis will be given.

After it had become clear that string theories are related to
eleven-dimensional supergravity, it was natural to analyse the
seven-dimensional space on which one needs to compactify to obtain
an interesting four-dimensional theory with the right amount of
supersymmetry. It is generally expected that the four-dimensional
effective field theory should live on Minkowski space and carry
$\mathcal{N}=1$ supersymmetry. Compactification to Minkowski space
requires the internal manifold to carry a Ricci-flat metric. From
a careful analysis of the supersymmetry transformations one finds
that the vacuum is invariant under four supercharges if and only
if the internal manifold carries one covariantly constant spinor.
Ricci-flat seven-dimensional manifolds carrying one covariantly
constant spinor are called $G_2$-manifolds. Indeed, one can show
that their holonomy group is the exceptional group $G_2$. Like
Calabi-Yau compactifications in the eighties,
$G_2$-compactifications have been analysed in much detail
recently. An interesting question is of course, whether one can
construct standard model type theories from compactifications of
eleven-dimensional supergravity on $G_2$-manifolds. Characteristic
features of the standard model are the existence of non-Abelian
gauge groups and of chiral fermions. Both these properties turn
out to be difficult to obtain from $G_2$-compactifications. In
order to generate them, one has to introduce singularities in the
$G_2$-manifolds. Another important question, which arises once a
chiral theory is constructed, is whether it is free of anomalies.
Indeed, the anomaly freedom of field theories arising in string
theory is a crucial issue and gives important consistency
constraints. The anomalies in the context of $G_2$-manifolds have
been analysed, and the theories have been found to be anomaly
free, if one introduces an extra term into the effective action of
eleven-dimensional supergravity. Interestingly, this term reduces
to the standard Green-Schwarz term when compactified on a circle.
Of course, this extra term is not specific to
$G_2$-compactifications, but it has to be understood as a first
quantum correction of the classical action of eleven-dimensional
supergravity. In fact, it was first discovered in the context of
anomaly cancellation for the M5-brane \cite{DLM95}, \cite{Wi96}.

\bigskip
Another important development that took place over the last years
is the construction of realistic field theories from Type II
string theory. In general, if one compactifies Type II on a
Calabi-Yau manifold one obtains an $\mathcal{N}=2$ effective field
theory. However, if $D$-branes wrap certain cycles in the internal
manifold they break half of the supersymmetry. The same is true
for suitably adjusted fluxes, and so one has new possibilities to
construct $\mathcal{N}=1$ theories. Very interestingly,
compactifications with fluxes and compactifications with branes
turn out to be related by what is known as geometric transition.
As to understand this phenomenon recall that a singularity of
complex codimension three in a complex three-dimensional manifold
can be smoothed out in two different ways. In mathematical
language these are know as the small resolution and the
deformation of the singularity. In the former case the singular
point is replaced by a two-sphere of finite volume, whereas in the
latter case it is replaced by a three-sphere. If the volume of
either the two- or the three-sphere shrinks to zero one obtains
the singular space. The term ``geometric transition" now describes
the process in which one goes from one smooth space through the
singularity to the other one. It is now interesting to see what
happens if we compactify Type IIB string theory on two Calabi-Yau
manifolds that are related by such a transition. Since one is
interested in $\mathcal{N}=1$ effective theories it is suitable to
add either fluxes or branes in order to further break
supersymmetry. It is then very natural to introduce D5-branes
wrapping the two-spheres in the case of the small resolution of
the singularity. The manifold with a deformed singularity has no
suitable cycles around which D-branes might wrap, so we are forced
to switch on flux in order to break supersymmetry. In fact, we
can, very similar in spirit to the AdS/CFT correspondence,
consider the deformed manifold as the way the geometry backreacts
on the presence of the branes. This relation between string theory
with flux or branes on topologically different manifolds is by
itself already very exciting. However, the story gets even more
interesting if we consider the effective theories generated from
these compactifications. In the brane setup with $N$ D5-branes
wrapping the two-cycle we find an $\mN=1$ theory with gauge group
$U(N)$ in four dimensions. At low energies this theory is believed
to confine and the suitable description then is in terms of a
chiral superfield $S$, which contains the gaugino bilinear. Quite
interestingly, it has been shown that it is precisely this low
energy effective action which is generated by the compactification
on the deformed manifold. In a sense, the geometric transition and
the low energy description are equivalent. In particular, the
effective superpotential of the low-energy theory can be
calculated from geometric integrals on the deformed manifold.

For a specific choice of manifolds the structure gets even richer.
In three influential publications, Dijkgraaf and Vafa showed that
IIB on the resolved manifold is related to a holomorphic matrix
model. Furthermore, from the planar limit of the model one can
calculate terms in the low energy effective action of the $U(N)$
$\mN=1$ gauge theory. More precisely, the integrals in the
deformed geometry are mapped to integrals in the matrix model,
where they are shown to be related to the planar free energy.
Since this free energy can be calculated from matrix model Feynman
diagrams, one can use the matrix model to calculate the effective
superpotential.\\

{\bf Plan of this thesis}\\
My dissertation is organised in two parts. In the first part I
explain the intriguing connection between four-dimensional
supersymmetric gauge theories, type II string theories and matrix
models. As discussed above, the main idea is that gauge theories
can be ``geometrically engineered" from type II string theories
which are formulated on the direct product of a four-dimensional
Minkowski space and a six-dimensional non-compact Calabi-Yau
manifold. I start by reviewing some background material on
effective actions in chapter \ref{effac}. Chapter \ref{RSCY} lists
some important properties of Riemann surfaces and Calabi-Yau
manifolds. Furthermore, I provide a detailed description of local
Calabi-Yau manifolds, which are the spaces that appear in the
context of the geometric transition. In particular, the fact that
integrals of the holomorphic three-form on the local Calabi-Yau
map to integrals of a meromorphic one-form on a corresponding
Riemann surface is reviewed. In chapter \ref{holMM} I study the
holomorphic matrix model in some detail. I show how the planar
limit and the saddle point approximation have to be understood in
this setup, and how special geometry relations arise. Quite
interestingly, the Riemann surface that appeared when integrating
the holomorphic form on a local Calabi-Yau is the same as the one
appearing in the planar limit of a suitably chosen matrix model.
In chapter \ref{TSMM} I explain why the matrix model can be used
to calculate integrals on a local Calabi-Yau manifold. The reason
is that there is a relation between the open B-type topological
string on the Calabi-Yau and the holomorphic matrix model. All
these pieces are then put together in chapter \ref{SSGTMM}, where
it is shown that the low energy effective action of a class of
gauge theories can be obtained from integrals in the geometry of a
certain non-compact Calabi-Yau manifold. After a specific choice
of cycles, only one of these integral is divergent. Since the
integrals appear in the formula for the effective superpotential,
this divergence has to be studied in detail. In fact, the integral
contains a logarithmically divergent part, together with a
polynomial divergence, and I show that the latter can be removed
by adding an exact term to the holomorphic three-form. The
logarithmic divergence cancels against a divergence in the
coupling constant, leading to a finite superpotential. Finally, I
review how the matrix model can be used to calculate the effective
superpotential.

\bigskip
In the second part I present work done during the first half of my
PhD about M-theory on $G_2$-manifolds and (local) anomaly
cancellation. I start with a short exposition of the main
properties of $G_2$-manifolds, eleven-dimensional supergravity and
anomalies. An important consistency check for chiral theories is
the absence of anomalies. Since singular $G_2$-manifolds can be
used to generate standard model like chiral theories, anomaly
freedom is an important issue. In chapter \ref{G2anom} it is shown
that M-theory on singular $G_2$-manifolds is indeed anomaly free.
In this context it will be useful to discuss the concepts of
global versus local anomaly cancellation. Then I explain the
concept of weak $G_2$-holonomy in chapter \ref{weakG2}, and
provide examples of explicit metrics on compact singular manifolds
with weak $G_2$ holonomy. Finally, in chapter \ref{HW} I study
M-theory on $M_{10}\times I$, with $I$ an interval, which is known
to be related to the $E_8\times E_8$ heterotic string. This setup
is particularly fascinating, since new degrees of freedom living
on the boundary of the space have to be introduced for the theory
to be consistent. Once again a careful analysis using the concepts
of local anomaly cancellation leads to new results. These
considerations will be brief, since some of them have been
explained rather extensively in the following review article,
which is a very much extended version of my diploma thesis:

\bigskip
\noindent[P4] S. Metzger, {\it M-theory compactifications,
$G_2$-manifolds and anomalies},\\ \indent\ \ {\tt hep-th/0308085}

\bigskip
In the appendices some background material is presented, which is
necessary to understand the full picture. I start by explaining
the notation that is being used throughout this thesis. Then I
turn to some results in mathematics. The definition of divisors on
Riemann surfaces is presented and the notion of relative
(co-)homology is discussed. Both concepts will appear naturally in
our discussion. Furthermore, I quickly explain the Atiyah-Singer
index theorem, which is important in the context of anomalies. One
of the themes that seems to be omnipresent in the discussion is
the concept of special geometry and of special K\"ahler manifolds.
A detailed definition of special K\"ahler manifolds is given and
their properties are worked out. Another central building block is
the B-type topological string, and therefore I quickly review its
construction. Finally, the concept of an anomaly is explained, and
some of their properties are discussed.

\newpage\noindent {\bf\large Acknowledgements}

\vskip 3.mm

\noindent I gratefully acknowledge support by the Gottlieb
Daimler- und Karl Benz-Stiftung, as well as the Studienstiftung
des deutschen Volkes.

\bigskip
It is a pleasure to thank Adel Bilal for the fantastic
collaboration over the last years. He was an excellent teacher of
string theory, taking a lot of time answering my questions in a
clean and pedagogical way. Doing research with him was always
motivating and enjoyable. Not only do I want to thank him for his
scientific guidance but also for his support in all sorts of
administrative problems.

Furthermore, I would like to thank Julius Wess, my supervisor in
Munich, for supporting and encouraging me during the time of my
PhD. Without him this binational PhD-thesis would not have been
possible. I am particularly grateful that he came to Paris to
participate in the commission in front of which my thesis was
defended.

It was an honour to defend my thesis in front of the jury
consisting of L. Alvarez-Gaumé, A. Bilal, R. Minasian, D. L\"ust,
J. Wess and J.B. Zuber.

Of course, it is a great pleasure to thank the Laboratoire de
Physique Théorique at École Normale Supérieure in Paris for the
warm hospitality during the last years, and its director Eugène
Cremmer, who always helped in resolving administrative problems of
all kinds. The support from the secretaries at LPTENS, Véronique
Brisset Fontana, Cristelle Login and especially Nicole Ribet
greatly facilitated my life. Thank you very much! I also thank the
members of LPTENS and in particular Costas Bachas, Volodya
Kazakov, Boris Pioline and Jan Troost for many discussions and for
patiently answering many of my questions. Furthermore, I very much
enjoyed to share a room with Christoph Deroulers, Eytan Katzav,
Sebastien Ray and Guilhem Semerjian. It was nice to discuss
cultural gaps, not only between Germany, France and Israel, but
also between statistical and high energy physics. Studying in
Paris would not have been half as enjoyable as it was, had I not
met many people with whom it was a pleasure to collaborate, to
discuss physics and other things, to go climbing and travelling. I
thank Yacine Dolivet, Gerhard G\"otz, Dan Israel, Kazuhiro Sakai
and Cristina Toninelli for all the fun during the last years.

I am also grateful that I had the chance to be a visitor in the
group of Dieter L\"ust at the Arnold Sommerfeld Center for
Theoretical Physics at Munich during the summer term 2005. It was
a great experience to get to know the atmosphere of the large and
active Munich group. I thank Peter Mayr and Stefan Stieberger for
discussions.

There are many physicists to which I owe intellectual debt. I
thank Alex Altmeyer, Marco Baumgartl, Andreas Biebersdorf, Boris
Breidenbach, Daniel Burgarth, Davide Cassani, Alex Colsmann,
Sebastian Guttenberg, Stefan Kowarik and Thomas Klose for so many
deep and shallow conversations.

Mein besonderer Dank gilt meinen Eltern. Herzlichen Dank für Eure
Unterstützung und Geduld!

Schließlich danke ich Dir, Constanze.

\part{Gauge Theories, Matrix Models and Geometric Transitions}

\chapter{Introduction and Overview}
The structure of the strong nuclear interactions is well know to
be captured by Quantum Chromodynamics, a non-Abelian gauge theory
with gauge group $SU(3)$, which is embedded into the more general
context of the standard model of particle physics. Although the
underlying theory of the strong interactions is known, it is
actually very hard to perform explicit calculations in the
low-energy regime of this theory. The reason is, of course, the
behaviour of the effective coupling constant of QCD, which goes to
zero at high energies, an effect called {\it asymptotic freedom},
but becomes of order 1 at energies of about 200MeV. At this energy
scale the perturbative expansion in the coupling constant breaks
down, and it becomes much harder to extract information about the
structure of the field theory. It is believed that the theory will
show a property known as {\it confinement}, in which the quarks
form colour neutral bound states, which are the particles one
observes in experiments. Most of the information we currently have
about this energy regime comes from numerical calculations in
lattice QCD. Pure Yang-Mills theory is also asymptotically free
and it is expected to behave similarly to QCD. In particular, at
low energies the massless gluons combine to colour neutral bound
states, known as {\it glueball fields}, which are massive.
Therefore, at low energies the microscopic degrees of freedom are
irrelevant for a description of the theory, but it is the vacuum
expectation values of composite fields which are physically
interesting. These vacuum expectation values can be described by
an effective potential that depends on the relevant low energy
degrees of freedom. The expectation values can then be found from
minimising the potential.

Understanding the low energy dynamics of QCD is a formidable
problem. On the other hand, it is known that $\mN=1$
supersymmetric non-Abelian gauge theories share many properties
with QCD. However, because of the higher symmetry, calculations
simplify considerably, and some exact results can be deduced for
supersymmetric theories. They might therefore be considered as a
tractable toy model for QCD. In addition, we mentioned already
that indications exist that the action governing physics in our
four-dimensional world might actually be supersymmetric. Studying
supersymmetric field theories can therefore not only teach us
something about QCD but it may after all be the correct
description of nature.

In order to make the basic ideas somewhat more concrete let us
quickly consider the simplest example, namely $\mN=1$ Yang-Mills
theory with gauge group $SU(N)$. The relevant degrees of freedom
at low energy are captured by a chiral superfield $S$ which
contains the gaugino bilinear. The effective superpotential, first
written down by Veneziano and Yankielowicz, reads
\begin{equation}
W_{eff}(S)=S\left[\log\left(\L^{3N}\over S^N\right)+N\right]\ ,
\end{equation}
where $\L$ is the dynamical scale of the theory. The minima of the
corresponding effective potential can be found by determining the
critical points of the effective superpotential. Indeed, from
extremising the superpotential we find
\begin{equation}
\langle S\rangle^N=\L^{3N}\ ,
\end{equation}
which is the correct result. All this will be explained in more
detail in the main text.

Over the last decades an intimate relation between supersymmetric
field theories and string theory has been unveiled.
Ten-dimensional supersymmetric theories appear as low energy
limits of string theories and four-dimensional ones can be
generated from string compactifications. The structure of a
supersymmetric field theory can then often be understood from the
geometric properties of the manifolds appearing in the string
context. This opens up the intriguing possibility that one might
actually be able to learn something about the vacuum structure of
four-dimensional field theories by studying geometric properties
of certain string compactifications. It is this idea that will be
at the heart of the first part of this thesis.

\begin{figure}[h]
\centering
\includegraphics[width=0.5\textwidth]{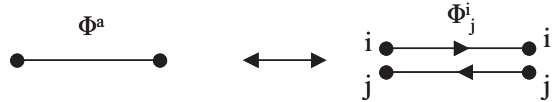}\\
\caption[]{The propagator of an adjoint field $\Phi^a$ can be
represented as a fatgraph. The indices $i,j$ then run from 1 to
$N$.}\label{fatgraph1}
\end{figure}

\begin{figure}[h]
\centering
\includegraphics[width=0.6\textwidth]{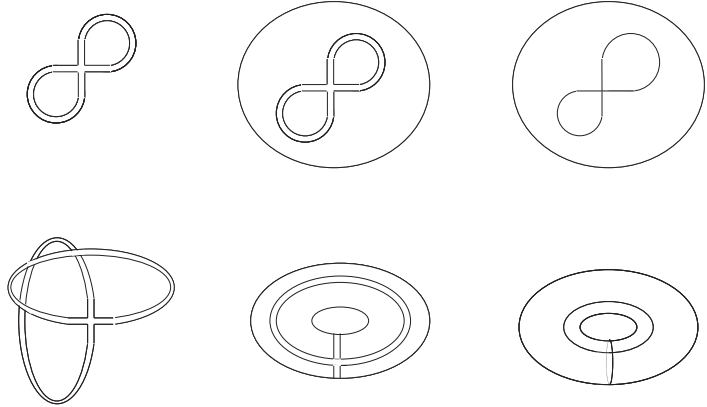}\\
\caption[]{The appearance of a Riemann surface for two fatgraphs.
The first graph can be drawn on a sphere and, after having shrunk
the fatgraph to a standard graph, it can be understood as its
triangulation with three faces, two edges and one vertex. The
second graph, on the other hand, can be drawn on a torus. The
corresponding triangulation has two edges, one vertex and only one
face.}\label{triangulation}
\end{figure}

\bigskip
{\bf Gauge theory - string theory duality}\\
In fact, there is yet another intriguing relation between
non-Abelian gauge theories and string theories that goes back to
't Hooft \cite{tH74}. Let us consider the free energy of a
non-Abelian gauge theory, which is known to be generated from
connected Feynman diagrams. 't Hooft's idea was to introduce
fatgraphs by representing an $U(N)$ adjoint field as the direct
product of a fundamental and an anti-fundamental representation,
see Fig. \ref{fatgraph1}.
The free energy can then be calculated by summing over all
connected vacuum amplitudes, which are given in terms of fatgraph
Feynman diagrams. By rescaling the fields one can rewrite the
Lagrangian of the gauge theory in such a way that it is multiplied
by an overall factor of $1\over g_{YM}^2$. This means that every
vertex comes with a factor $1\over g_{YM}^2$, whereas the
propagators are multiplied by $g_{YM}^2$. The gauge invariance of
the Lagrangian manifests itself in the fact that all index lines
form closed (index) loops. For each index loop one has to sum over
all possible indices which gives a factor of $N$. If $E$ denotes
the number of propagators in a given graph, $V$ its vertices and
$F$ the number of index loops we find therefore that a given graph
is multiplied by
\begin{equation}
\left({1\over g_{YM}^2}\right)^V\left(g_{YM}^2\right)^E
N^F=(g_{YM}^2)^{E-V-F}t^F=(g_{YM}^2)^{-\chi}\
t^F=(g_{YM}^2)^{2\gh-2}\ t^F\ .
\end{equation}
Here we defined what is know as the {\it 't Hooft coupling},
\begin{equation}
t:=g_{YM}^2N\ .
\end{equation}
Furthermore, the Feynman diagram can be understood as the
triangulation of some two-dimensional surface with $F$ the number
of faces, $E$ the number of edges and $V$ the number of vertices
of the triangulation, see Fig. \ref{triangulation}.
Then the result follows immediately from $V-E+F=\chi=2-2\gh$,
where $\gh$ is the genus and $\chi$ the Euler characteristic of
the surface. Summing over these graphs gives the following
expansion of the free energy
\begin{equation}
F^{gauge}\left(g_{YM}^2,t\right)=\sum_{\gh=0}^\infty\left(g_{YM}^2\right)^{2\gh-2}
\sum_{h=1}^\infty F_{\gh,h} t^h +\mbox{non-perturbative}\
.\label{GTexpansion}
\end{equation}
Here we slightly changed notation, using $h$ instead of $F$. This
is useful, since an open string theory has an expansion of
precisely the same form, where now $h$ is the number of holes in
the world-sheet Riemann surface. Next we define
\begin{equation}
F_\gh^{gauge}(t):=\sum_{h=1}^\infty F_{\gh,h} t^h\ ,
\end{equation}
which leads to
\begin{equation}\label{expansion}
F^{gauge}\left(g_{YM}^2,t\right)=\sum_{\gh=0}^\infty\left(g_{YM}^2\right)^{2\gh-2}
F_{\gh}^{gauge}(t)+\mbox{non-perturbative}\ .
\end{equation}
In the 't Hooft limit $g_{YM}^2$ is small but $t$ is fixed, so $N$
has to be large. The result can now be compared to the well-known
expansion of the free energy of a closed string theory, namely
\begin{equation}
F^{string}(g_s)=\sum_{\gh=0}^\infty g_s^{2\gh-2}
F_{\gh}^{string}+\mbox{non-perturbative}\ .
\end{equation}
This leads us to the obvious question whether there exists a
closed string theory which, when expanded in its coupling constant
$g_s$, calculates the free energy of our gauge theory, provided we
identify
\begin{equation}
g_s\sim g_{YM}^2\ .
\end{equation}
In other words, is there a closed string theory such that
$F_\gh^{gauge}=F_\gh^{string}$? Note however, that
$F_\gh^{gauge}=F_\gh^{gauge}(t)$, so we can only find a reasonable
mapping if the closed string theory depends on a parameter $t$.

For some simple gauge theories the corresponding closed string
theory has indeed been found. The most spectacular example of this
phenomenon is the AdS/CFT correspondence. Here the gauge theory is
four-dimensional $\mN=4$ superconformal gauge theory with gauge
group $U(N)$ and the corresponding string theory is Type IIB on
$AdS_5\times S^5$. However, in the following we want to
concentrate on simpler examples of the gauge theory - string
theory correspondence. One example is Chern-Simons theory on $S^3$
which is known to be dual to the A-type topological string on the
resolved conifold. This duality will not be explained in detail
but we will quickly review the main results at the end of this
introduction. The second example is particularly simple, since the
fields are independent of space and time and the gauge theory is a
matrix model. To be more precise, we are interested in a
holomorphic matrix model, which can be shown to be dual to the
B-type closed topological string on some non-compact Calabi-Yau
manifold. It will be part of our task to study these relations in
more detail and to see how we can use them to extract even more
information about the vacuum structure of the supersymmetric gauge
theory.

\begin{figure}[h]
\centering
\includegraphics[width=\textwidth]{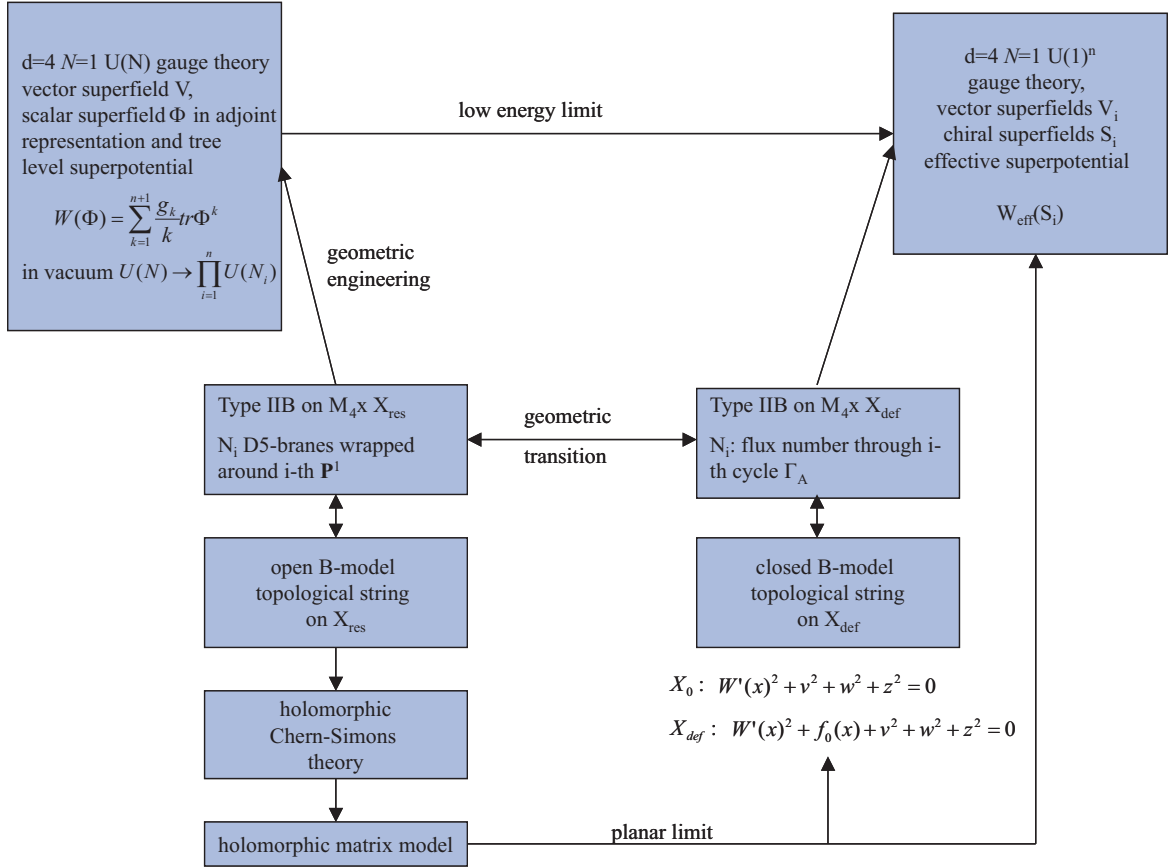}\\
\caption[]{A sketch of the relation of supersymmetric Yang-Mills
theory with Type IIB string theory on non-compact Calabi-Yau
manifolds and with the holomorphic matrix model.}
\label{bigpictureIIB}
\end{figure}

\bigskip
{\bf Gauge theories, the geometric transition and matrix models}\\
After these general preliminary remarks let us turn to the
concrete model we want to study. Background material and many of
the fine points are going to be be analysed in the main part of
this thesis. Here we try to present the general picture and the
relations between the various theories, see Fig.
\ref{bigpictureIIB}.

To be specific, we want to analyse an $\mN=1$ supersymmetric gauge
theory with gauge group $U(N)$. Its field content is given by a
vector superfield $V$ and an adjoint chiral superfield $\Phi$. The
dynamics of the latter is governed by the tree-level
superpotential
\begin{equation}
W(\Phi):=\sum_{k=1}^{n+1}{g_k\over k}\tr\Phi^k+g_0\
,\label{suppot}
\end{equation}
with complex coefficients $g_k$. Here, once again, we used the
equivalence of the adjoint representation of $U(N)$ and the direct
product of the fundamental and anti-fundamental representation,
writing $\Phi=\Phi_{ij}$. Note that the degrees of freedom are the
same as those in an $\mN=2$ vector multiplet. In fact, we can
understand the theory as an $\mN=2$ theory that has been broken to
$\mN=1$ by switching on the superpotential (\ref{suppot}).

It turns out that, in order to make contact with string theory, we
have to expand the theory around one of its classical vacua. These
vacua are obtained from distributing the eigenvalues of $\Phi$ at
the critical points\footnote{The critical points of $W$ are always
taken to be non-degenerate in this thesis, i.e. if $W'(p)=0$ then
$W''(p)\neq0$.} of the superpotential, where $W'(x)=0$. Then a
vacuum is specified by the numbers $N_i$ of eigenvalues of $\Phi$
sitting at the $i$-th critical point of $W$. Note that we
distribute eigenvalues at all critical points of the
superpotential. Of course, we have the constraint that
$\sum_{i=1}^nN_i=N$. In such a vacuum the gauge group is broken
from $U(N)$ to $\prod_{i=1}^nU(N_i)$.

It is this gauge theory, in this particular vacuum, that can be
generated from string theory in a process known as {\it geometric
engineering}. To construct it one starts from Type IIB string
theory on some non-compact Calabi-Yau manifold $X_{res}$, which is
given by the small resolutions of the singular points
of\footnote{The space $X_{res}$ is going to be explained in detail
in section \ref{propCY}, for a concise exposition of singularity
theory see \cite{AGLV}.}
\begin{equation}
W'(x)^2+v^2+w^2+z^2=0 \ ,\label{sing}
\end{equation}
where $W(x)=\sum_{k=1}^{n+1}{g_k\over k}x^k+g_0$. This space
contains precisely $n$ two-spheres from the resolution of the $n$
singular points. One can now generate the gauge group and break
$\mN=2$ supersymmetry by introducing $D5$-branes wrapping these
two-spheres. More precisely, we generate the theory in the
specific vacuum with gauge group $\prod_iU(N_i)$, by wrapping
$N_i$ D5-branes around the $i$-th two-sphere. The scalar fields in
$\Phi$ can then be understood as describing the position of the
various branes and the superpotential is natural since D-branes
have tension, i.e. they tend to wrap the minimal cycles in the
non-compact Calabi-Yau manifold. The fact that, once pulled away
from the minimal cycle, they want to minimise their energy by
minimising their world-volume is expressed in terms of the
superpotential on the gauge theory side.

Mathematically the singularity (\ref{sing}) can be smoothed out in
yet another way, namely by what is know as deformation. The
resulting space $X_{def}$ can be described as an equation in
$\mathbb{C}^4$,
\begin{equation}
W'(x)^2+f_0(x)+v^2+w^2+z^2=0 \ ,\label{defsing}
\end{equation}
where $f_0(x)$ is a polynomial of degree $n-1$. In (\ref{defsing})
the $n$ two-spheres of (\ref{sing}) have been replaced by $n$
three-spheres. The transformation of the resolution of a
singularity into its deformation is know as {\it geometric
transition}.

Our central physical task is to learn something about the low
energy limit of the four-dimensional $U(N)$ gauge theory. Since
$X_{res}$ and $X_{def}$ are intimately related one might want to
study Type IIB on $X_{def}$. However, the resulting effective
action has $\mN=2$ supersymmetry and therefore cannot be related
to our original $\mN=1$ theory. There is, however, a heuristic but
beautiful argument that leads us on the right track. The geometric
transition is a local phenomenon, in which one only changes the
space close to the singularity. Far from the singularity an
observer should not even realise that the transition takes place.
We know on the other hand, that D-branes act as sources for flux
and an observer far from the brane can still measure the flux
generated by the brane. If the branes disappear during the
geometric transition we are therefore forced to switch on
background flux on $X_{def}$, which our observer far from the
brane can measure. We are therefore led to analyse the effective
field theory generated by Type IIB on $X_{def}$ in the presence of
background flux.

The four-dimensional theory is $\mN=1$ supersymmetric and has
gauge group $U(1)^n$, i.e. it contains $n$ Abelian vector
superfields. In addition there are also $n$ chiral superfields
denoted by\footnote{Later on we will introduce fields $S_i,\ \bar
S_i,\ \tilde S_i$ with slightly different definitions. Since we
are only interested in a sketch of the main arguments, we do not
distinguish between these fields right now. Also, $S_i$ sometimes
denotes the full chiral multiplet, and sometimes only its scalar
component. It should always be clear from the context, which of
the two is meant.} $S_i$. Their scalar components describe the
volumes of the $n$ three-spheres $\G_{A_i}$, which arose from
deforming the singularity. Since the holomorphic $(3,0)$-form
$\O$, which comes with every Calabi-Yau manifold, is a
calibration\footnote{A precise definition of calibrations and
calibrated submanifolds can be found in \cite{Jo00}.} (i.e. it
reduces to the volume form on suitable submanifolds such as
$\G_{A_i}$) the volume can be calculated from
\begin{equation}
S_i=\int_{\G_{A_i}}\O\ .
\end{equation}
Type IIB string theory is known to generate a four-dimensional
superpotential in the presence of three-form flux $G_3$, which is
given by the Gukov-Vafa-Witten formula \cite{GVW99}
\begin{equation}
W_{eff}(S_i)\sim\sum_i\left(\int_{\G_{A^i}}G_3\int_{\G_{B_i}}
\O-\int_{\G_{B_i}}G_3\int_{\G_{A^i}} \O\right)\ . \label{GVW}
\end{equation}
Here $\G_{B_i}$ is the three-cycle dual to $\G_{A^i}$.

Now we are in the position to formulate an amazing conjecture,
first written down by Cachazo, Intriligator and Vafa \cite{CIV01}.
It simply states that {\it the theory generated from $X_{def}$ in
the presence of fluxes is nothing but the effective low energy
description of the theory generated from $X_{res}$ in the presence
of D-branes}. Indeed, in the low energy limit one expects the
$SU(N_i)$ part of the $U(N_i)$ gauge groups to confine. The theory
should then be described by $n$ chiral multiplets which contain
the corresponding gaugino bilinears. The vacuum structure can be
encoded in an effective superpotential. The claim is now that the
$n$ chiral superfields are nothing but the $S_i$ and that the
effective superpotential is given by (\ref{GVW}). In their
original publication Cachazo, Intriligator and Vafa calculated the
effective superpotential directly for the $U(N)$ theory using
field theory methods. On the other hand, the geometric integrals
of (\ref{GVW}) can be evaluated explicitly, at least for simple
cases, and perfect agreement with the field theory results has
been found.

\bigskip
These insights are very profound since a difficult problem in
quantum  field theory has been rephrased in a beautiful geometric
way in terms of a string theory. It turns out that one can extract
even more information about the field theory by making use of the
relation between Type II string theory compactified on a
Calabi-Yau manifold and the topological string on this Calabi-Yau.
It has been known for a long time that the topological string
calculates terms in the effective action of the Calabi-Yau
compactification \cite{AGNT93}, \cite{BCOV93b}. For example, if we
consider the Type IIB string we know that the vector multiplet
part of the four-dimensional effective action is determined from
the prepotential of the moduli space of complex structures of the
Calabi-Yau manifold. But this function is nothing else than the
genus zero free energy of the corresponding B-type topological
string. Calculating the topological string free energy therefore
gives information about the effective field theory. In the case we
are interested in, with Type IIB compactified on $X_{res}$ with
additional D5-branes, one has to study the open B-type topological
string with topological branes wrapping the two-cycles of
$X_{res}$. It can be shown \cite{Wi95a} that in this case the
corresponding string field theory reduces to holomorphic
Chern-Simons theory and, for the particular case of $X_{res}$,
this was shown by Dijkgraaf and Vafa \cite{DV02a} to simplify to a
holomorphic matrix model with partition function
\begin{equation}
Z=C_{\hN}\int\d M\exp\left(-{1\over g_s}\tr W(M)\right)\ ,
\end{equation}
where the potential $W(x)$ is given by the same function as the
superpotential above. Here $g_s$ is a coupling constant, $\hN$ is
the size of the matrices and $C_{\hN}$ is some normalisation
constant. Clearly, this is a particularly simple and tractable
theory and one might ask whether one can use it to calculate
interesting physical quantities. The holomorphic matrix model had
not been studied until very recently \cite{La03}, and in our work
\cite{BM05} some more of its subtleties have been unveiled.
Similarly to the case of a Hermitean matrix model one can study
the planar limit in which the size $\hN$ of the matrices goes to
infinity, the coupling $g_s$ goes to zero and the product
$t:=g_s\hN$ is taken to be fixed. In this limit there appears a
Riemann surface of the form
\begin{equation}
y^2=W'(x)^2+f_0(x)\ ,\label{Riemann}
\end{equation}
which clearly is intimately related to $X_{def}$. Indeed, as we
will see below, the integrals in the geometry of $X_{def}$, which
appear in (\ref{GVW}), can be mapped to integrals on the Riemann
surface (\ref{Riemann}). These integrals in turn can be related to
the free energy of the matrix model at genus zero,
$\mathcal{F}_0$. After this series of steps one is left with an
explicit formula for the effective superpotential,
\begin{equation}
W_{eff}(S)\sim\sum_{i=1}^n\left(N_i{\partial\mathcal{F}_0(
S)\over\partial S_i}-S_i\log\L_i^{2N_i}-2\pi i S_i\tau\right)\ ,
\end{equation}
where dependence on $S$ means dependence on all the $S_i$. The
constants $\L_i$ and $\tau$ will be explained below. The free
energy can be decomposed into a perturbative and a
non-perturbative part (c.f. Eq. (\ref{expansion})). Using
monodromy arguments one can show that
${\partial\mF_0^{np}\over\partial S_i}\sim S_i\log S_i$, and
therefore
\begin{equation}
W_{eff}(S)\sim\sum_{i=1}^n\left(N_i{\partial\mathcal{F}_0^{p}(
S)\over\partial S_i}+ N_iS_i\log\left(
S_i\over\L_i^{2}\right)-2\pi i S_i\tau\right)\ ,
\end{equation}
where now $\mathcal{F}_0^{p}$ is the perturbative part of the free
energy at genus zero, i.e. we can calculate it by summing over all
the planar matrix model vacuum amplitudes. This gives a
perturbative expansion of $W_{eff}$, which upon extremisation
gives the vacuum gluino condensate $\langle S\rangle$. Thus using
a long chain of dualities in type IIB string theory we arrive at
the beautiful result that the low energy dynamics and vacuum
structure of a non-Abelian gauge theory can be obtained from
perturbative calculations in a matrix model.\footnote{In fact,
this result can also be proven without making use of string theory
and the geometric transition, see \cite{DGLVZ02}, \cite{CDSW02}.}

\begin{figure}[h]
\centering
\includegraphics[width=\textwidth]{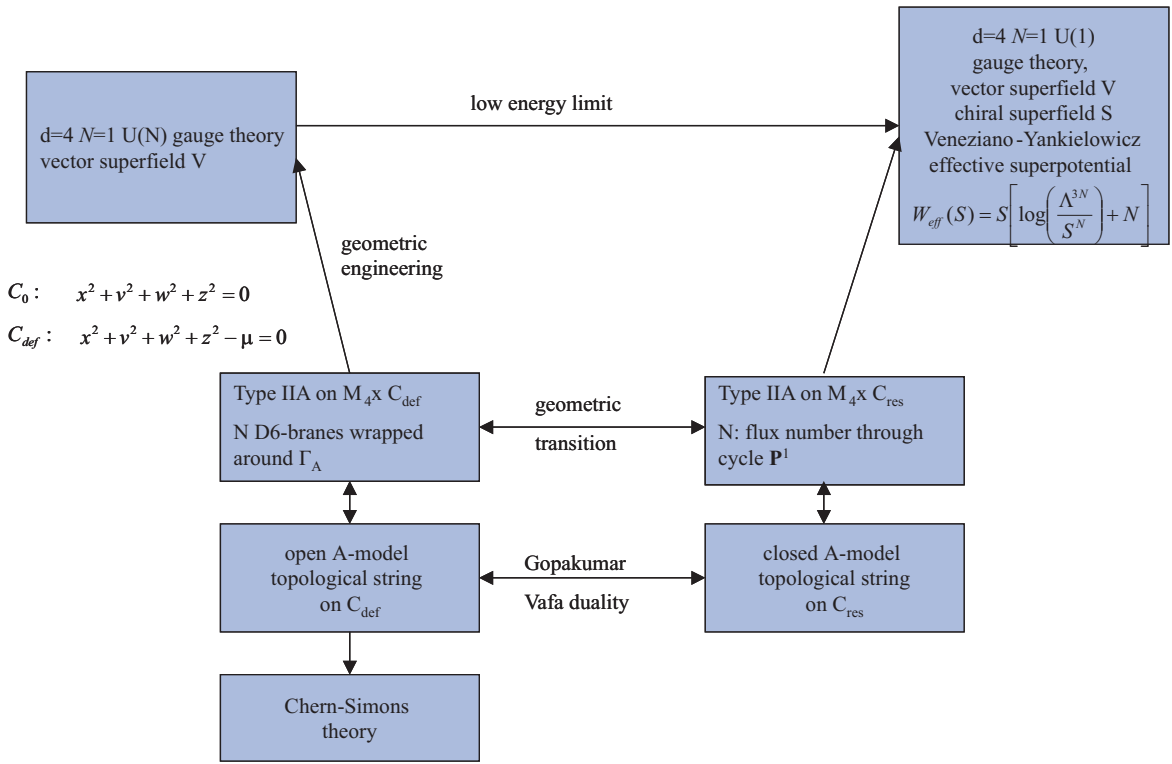}\\
\caption[]{A sketch of the relation of supersymmetric Yang-Mills
theory with Type IIA string theory on the conifold and with
Chern-Simons theory.} \label{bigpictureIIA}
\end{figure}

\bigskip
{\bf Chern-Simons theory and the Gopakumar-Vafa transition}\\
In the following chapters many of the points sketched so far are
going to be made more precise by looking at the technical details
and the precise calculations. However, before starting this
endeavour, it might be useful to give a quick overview of what
happens in the case of IIA string theory instead. As a matter of
fact, in this context very many highly interesting results have
been uncovered over the last years. The detailed exposition of
these developments would certainly take us too far afield, but we
consider it nevertheless useful to provide a quick overview of the
most important results. For an excellent review including many
references to original work see \cite{M04a}. In fact, the results
in the context of IIB string theory on which we want to report
have been discovered only after ground breaking work in IIA string
theory. The general picture is quite similar to what happens in
Type IIB string theory, and it is sketched in Fig.
\ref{bigpictureIIA}.

The starting point here is to consider the open A-model
topological string on $T^*S^3$ with topological branes wrapped
around the three-cycle at the center. Witten's string field theory
then does not reduce to holomorphic Chern-Simons theory, as is the
case on the B-side, but to ordinary Chern-Simons theory
\cite{Wi95a}, \cite{Wi89} on $S^3$,
\begin{equation}
S={k\over 4\pi}\int_{S^3}\left(A\w\d A+{2\over3}A\w A\w A\right).
\end{equation}
To be more precise, Witten showed that the $F_{\gh,h}^{CS}$ of the
expansion (\ref{GTexpansion}) for Chern-Simons theory on $S^3$
equals the free energy of the open A-model topological string on
$T^*S^3$ at genus $\gh$ and $h$ holes, $F_{\gh,h}^{A-tst}$. The
details of this procedure can be found in \cite{Wi95a}.

Of course, the space $T^*S^3$ is isomorphic to the deformed
conifold, a space we want to call $C_{def}$, which is given by
\begin{equation}
x^2+v^2+w^2+z^2=\m\ .
\end{equation}
Clearly, this is the deformation of the singularity
\begin{equation}
x^2+v^2+w^2+z^2=0\ .
\end{equation}
As we have seen above, these singularities can be smoothed out in
yet another way, namely by means of a small resolution. The
resulting space is known as the resolved conifold and will be
denoted by $C_{res}$. Both $C_{def}$ and $C_{res}$ are going to be
studied in detail in \ref{propCY}. Motivated by the AdS/CFT
correspondence, in which a stack of branes in one space has a dual
description in some other space without branes (but with fluxes),
in \cite{GV98a}, \cite{GV98b}, \cite{GV99} Gopakumar and Vafa
studied whether there exists a dual closed string description of
the open topological string on $C_{def}$, and hence of
Chern-Simons theory. This turns out to be the case and the dual
theory is given by the closed topological A-model on $C_{res}$. To
be somewhat more precise, the Gopakumar-Vafa conjecture states
that Chern-Simons gauge theory on $S^3$ with gauge group $SU(N)$
and level $k$ is equivalent to the closed topological string of
type A on the resolved conifold, provided we identify
\begin{equation}
g_s={2\pi\over k+N}=g_{CS}^2\ \ \ ,\ \ \ \k={2\pi i N\over{k+N}}\
,\label{gslambda}
\end{equation}
where $\k$ is the K\"ahler modulus of the two-sphere appearing in
the small resolution. Note that $\k=it$ where $t=g_{CS}^2N$ is the
't Hooft coupling.

On the level of the partition function this conjecture was tested
in \cite{GV99}. Here we only sketch the main arguments, following
\cite{M04a}. The partition function
$Z^{CS}(S^3)=\exp\left(-F^{CS}\right)$ of $SU(N)$ Chern-Simons
theory on $S^3$ is known, including non-perturbative terms
\cite{Wi89}. The free energy splits into a perturbative $F^{CS,p}$
and a non-perturbative piece $F^{CS,np}$, where the latter can be
shown to be
\begin{equation}
F^{CS,np}=\log{(2\pi g_s)^{{1\over2}N^2}\over{\rm
vol}\left(U(N)\right)}\ .
\end{equation}
We find that the non-perturbative part of the free energy comes
from the volume of the gauge group in the measure. As discussed
above, the perturbative part has an expansion
\begin{equation}
F^{CS,p}=\sum_{\hg=0}^\infty g_s^{2\gh-2}\sum_{h=1}^\infty
F_{\gh,h}t^h\ .
\end{equation}
The sum over $h$ can actually be performed and gives
$F_\gh^{CS,p}$. The non-perturbative part can also be expanded in
the string coupling and, for $g\geq 2$, the sum of both pieces
leads to (see e.g. \cite{M04a})
\begin{equation}
F_\gh^{CS}=F_\gh^{CS,p}+F_\gh^{CS,np}={(-1)^\gh|B_{2\gh}B_{2\gh-2}|\over2\gh(2\gh-2)(2\gh-2)!}
+{|B_{2\gh}|\over2\gh(2\gh-2)!}{\rm
Li}_{3-2\gh}\left(e^{-\k}\right)\ ,\label{CSFg}
\end{equation}
where $B_n$ are the Bernoulli numbers and ${\rm
Li}_j(x):=\sum_{n=1}^\infty{x^n\over n^j}$ is the polylogarithm of
index $j$.

This result can now be compared to the free energy of the
topological A-model on the resolved conifold. Quite generally,
from results of \cite{COGP91}, \cite{BCOV93a}, \cite{BCOV93b},
\cite{GV98a}, \cite{GV98b} and \cite{FP98} it can be shown that
the genus $\gh$ contribution to the free energy of the topological
A-model on a Calabi-Yau manifold $X$ reads \cite{M04a}
\begin{eqnarray}
F_\gh^{A-tst}&=&{(-1)^\gh\chi(X)|B_{2\gh}B_{2\gh-2}|\over4\gh(2\gh-2)(2\gh-2)!}\nonumber\\
&&+ \sum_{\b}\left({|B_{2\gh}|n^0_\b\over2\gh(2\gh-2)!}+
{2(-1)^\gh
n^2_\b\over(2\gh-2)!}\pm\ldots-{\gh-2\over12}n_\b^{\gh-1}+
n^\gh_\b\right){\rm Li}_{3-2\gh}\left(Q^\b \right)\
.\nonumber\\\label{ATSTFg}
\end{eqnarray}
Here $\b=\sum_i n_i[\g^{(2)}_i]$ is a homology class, where the
$[\g_i^{(2)}]$ form a basis of $H_2(X)$. In general there are more
than one K\"ahler parameters $\k_i$ and $Q^\b$ has to be
understood as $\prod_iQ_i^{n_i}$ with $Q_i:=e^{-\k_i}$.
Furthermore, the $n_\b^\gh$, known as {\it Gopakumar-Vafa
invariants}, are integer numbers.

We see that precise agreement can be found between (\ref{CSFg})
and (\ref{ATSTFg}), provided we set $\chi(X)=2$ and $n_1^0=1$,
with all other Gopakumar-Vafa invariants vanishing. This is indeed
the correct set of geometric data for the resolved conifold, and
we therefore have shown that the conjecture holds, at least at the
level of the partition function. In order to have a full duality
between two theories, however, one should not only compare the
free energy but also the observables, which in Chern- Simons
theory are given by Wilson loops. In \cite{OV99} the corresponding
quantities were constructed in the A-model string, thus providing
further evidence for the conjecture. Finally, a nice and intuitive
proof of the duality from a world-sheet perspective has been given
by the same authors in \cite{OV02}. Quite interestingly this
duality can be lifted to non-compact $G_2$-manifolds, where the
transition is a flop \cite{A00}, \cite{AMV01}.

After having established the Gopakumar-Vafa duality we can now
proceed similarly to the above discussion on the IIB side. Indeed,
in \cite{V01} the duality was embedded into the context of full
string theory. There the statement is that IIA string theory on
the direct product of four-dimensional Minkowski space and the
deformed conifold with $N$ D6-branes wrapping around the $S^3$ in
$T^*S^3$ is dual to the IIA string on Minkowski times the resolved
conifold, where one now has to switch on flux with flux number $N$
through the $S^2$. As above one can also study the
four-dimensional effective field theories generated by these
compactifications. The $N$ D6-branes clearly lead to pure
supersymmetric Yang-Mills theory with gauge group $U(N)$, whereas
the dual theory on $C_{res}$ with fluxes switched on, leads to an
effective $U(1)$ theory in four dimensions. In \cite{V01} it is
also shown that the effective superpotential generated from IIA on
the resolved conifold is nothing but the Veneziano-Yankielowicz
potential. Thus, like in the IIB case sketched above, the
geometric transition is once again equivalent to the low energy
description.

\bigskip
Both the resolved and the deformed conifold can be described in
the language of toric varieties. In fact, one can study the A-type
topological string on more general toric varieties. The geometry
of these spaces can be encoded in terms of toric diagrams and the
geometric transition then has a nice diagrammatic representation.
Quite interestingly, it was shown in \cite{AKMV03} that, at least
in principle, one can compute the partition function of the A-type
topological string on any toric variety. This is done by
understanding the toric diagram as some sort of ``Feynman
diagram", in the sense that to every building block of the diagram
one assigns a mathematical object and the partition function
corresponding to a toric variety is then computed by putting these
mathematical objects together, following a simple and clear cut
set of rules. Many more results have been derived in the context
of the A-type topological string on toric varieties, including the
relation to integrable models \cite{ADMV03}. These developments
are, however, outside the scope of my thesis.

\chapter{Effective Actions}\label{effac}
In what follows many of the details of the intriguing picture
sketched in the introduction will be explained. Since the full
picture consists of very many related but different theories we
will not be able to study all of them in full detail. However, we
are going to provide references wherever a precise explanation
will not be possible. Here we start by an exposition of various
notions of effective actions that exist in quantum field theory.
We quickly review the definitions of the generating functional of
the one-particle-irreducible correlation functions and explain how
it can be used to study vacua of field theories. The Wilsonian
effective action is defined somewhat differently, and it turns out
that it is particularly useful in the context of supersymmetric
gauge theories. Finally a third type of effective action is
presented. It is defined in such a way that it captures the
symmetries and the vacuum structure of the theory, and in some but
not in all cases it coincides with the Wilsonian action.

\section{The 1PI effective action and the background field
method}
We start from the Lagrangian density of a field theory
\begin{equation}
\mathcal{L}=\mathcal{L}(\Phi)
\end{equation}
and couple the fields $\Phi(x)$ to a set of classical currents
$J(x)$,
\begin{equation}
Z[J]=\exp(iF[J])=\int D\Phi\ \exp\left(i\int\d^4x\
\mathcal{L}(\Phi)+i\int\d^4x\ J(x)\Phi(x) \right)\ .
\end{equation}
The quantity $iF[J]$ is the sum of all connected vacuum-vacuum
amplitudes. Define
\begin{equation}\label{expPhi}
\Phi_J(x):={\delta\over\delta J(x)}F[J]=\langle\Phi(x)\rangle_J\ .
\end{equation}
This equation can also be used to define a current $J_{\Phi_0}(x)$
for a given classical field $\Phi_0(x)$, s.t.
$\Phi_J(x)=\Phi_0(x)$ if $J(x)=J_{\Phi_0}(x)$. Then one defines
the {\it quantum effective action} $\G[\Phi_0]$ as
\begin{equation}
\G[\Phi_0]:=F[J_{\Phi_0}]-\int\d^4x\ \Phi_0(x)J_{\Phi_0}(x)\ .
\end{equation}
The functional $\G[\Phi_0]$ is an effective action in the sense
that $iF[J]$ can be calculated as a sum of connected {\it tree}
graphs for the vacuum-vacuum amplitude, with vertices calculated
as if $\G[\Phi_0]$ and not $S[\Phi]$ was the action. But this
implies immediately that $i\G[\Phi_0]$ is the sum of all
one-particle-irreducible (1PI) connected graphs with arbitrary
number of external lines, each external line corresponding to a
factor of $\Phi_0$. Another way to put it is (see for example
\cite{Wb00} or \cite{PS} for the details)
\begin{equation}\label{Gamma1PI}
\exp(i\G[\Phi_0])=\int_{1PI} D\phi\ \exp(iS[\Phi_0+\phi])\ .
\end{equation}
Furthermore, varying $\G$ gives
\begin{equation}
{\delta\G[\Phi_0]\over\delta \Phi_0(x)}=-J_{\Phi_0}(x)\ ,
\end{equation}
and in the absence of external currents
\begin{equation}
{\delta\G[\Phi_0]\over\delta \Phi_0(x)}=0\ .
\end{equation}
This can be regarded as the equation of motion for the field
$\Phi_0$, where quantum corrections have been taken into account.
In other words, it determines the stationary configurations of the
background field $\Phi_0$.

The {\it effective potential} of a quantum field theory is defined
as the non-derivative terms of its effective Lagrangian. We are
only interested in translation invariant vacua, for which
$\Phi_0(x)=\Phi_0$ is constant and one has
\begin{equation}
\G[\Phi_0]=-V_4\ V_{eff}(\Phi_0)\ ,
\end{equation}
where $V_4$ is the four-dimensional volume of the space-time in
which the theory is formulated.

To one loop order the 1PI generating function can be calculated
from
\begin{equation}
\exp(i\G[\Phi_0])\approx\int D\phi\ \exp\left(i\int\d^4x\
\mathcal{L}^q(\Phi_0,\phi)\right)\ ,
\end{equation}
$\mathcal{L}^q(\Phi_0,\phi)$ contains all those terms of
$\mathcal{L}(\Phi_0+\phi)$ that are at most quadratic in $\phi$.
One writes
$\mathcal{L}^q(\Phi_0,\phi)=\mathcal{L}(\Phi_0)+\tilde{\mathcal{L}}^q(\Phi_0,\phi)$
and performs the Gaussian integral over $\phi$, which
schematically leads to
\begin{equation}
\G[\Phi_0]\approx V_4\mathcal{L}(\Phi_0)-{1\over2}\log
\det(A(\Phi_0))\ ,
\end{equation}
where $A$ is the matrix of second functional derivatives of
$\mathcal{L}$ with respect to the fields $\phi$.\footnote{Since we
have to include only 1PI diagrams to calculate $\G$ (c.f. Eq.
\ref{Gamma1PI}) the terms linear in the fluctuations $\phi$ do not
contribute to $\G$. See \cite{Ab81} for a nice discussion.} For
some concrete examples the logarithm of this determinant can be
calculated and one can read off the effective action and the
effective potential to one loop order. Minimising this potential
then gives the stable values for the background fields, at least
to one loop order. The crucial question is, of course, whether the
structure of the potential persists if higher loop corrections are
included.

\bigskip
{\bf Example: Yang-Mills theory}\\
In order to make contact with some of the points mentioned in the
introduction, we consider the case of Yang-Mills theory with gauge
group $SU(N)$ and Lagrangian
\begin{equation}
\mathcal{L}=-{1\over4g^2}F_{\m\n}^aF^{\m\n a}\ ,
\end{equation}
and try to determine its effective action using the procedure
described above. Since the details of the calculation are quite
complicated we only list the most important results. One start by
substituting
\begin{equation}
A_\m^a(x)\rightarrow A_\m^a(x)+a_\m^a(x)\
\end{equation}
into the action and chooses a gauge fixing condition. This
condition is imposed by adding gauge fixing and ghost terms to the
action. Then, the effective action can be evaluated to one loop
order from
\begin{equation}
\exp(i\G[A])\approx\int Da\ Dc \ D\bar c \ \exp\left(i\int\d^4x\
\mathcal{L}^q(A,a,c,\bar c)\right)\ ,
\end{equation}
where $\mathcal{L}^q(A,a,c,\bar c)$ only contains those terms of
$\mathcal{L}(A+a)+\mathcal{L}^{gf}+\mathcal{L}^{ghost}$ that are
at most quadratic in the fields $a,c,\bar c$. Here
$\mathcal{L}^{gf}$ is the gauge fixing and $\mathcal{L}^{ghost}$
the ghost Lagrangian. The Gaussian integrals can then be evaluated
and, at least for small $N$, one can work out the structure of the
determinant (see for example chapter 17.5. of \cite{Wb00}). The
form of the potential is similar in shape to the famous Mexican
hat potential, which implies that the perturbative vacuum where
one considers fluctuations around the zero-field background is an
unstable field configuration. The Yang-Mills vacuum lowers its
energy by spontaneously generating a non-zero ground state.

However, this one-loop calculation can only be trusted as long as
the effective coupling constant is small. On the other hand, from
the explicit form of the effective action one can also derive the
one-loop $\b$-function. It reads
\begin{equation}
\b(g)=-{{11\over3}Ng^3\over16\pi^2}+\ldots\ ,
\end{equation}
where the dots stand for higher loop contributions. The
renormalisation group equation
\begin{equation}
\m{\partial\over\partial\m}g(\m)=\b(g)
\end{equation}
is solved by
\begin{equation}
{1\over
g(\m)^2}=-{{11\over3}N\over8\pi^2}\log\left({|\L|\over\m}\right)\
.
\end{equation}
Therefore, for energies lower or of order $|\L|$ we cannot trust
the one-loop approximation. Nevertheless, computer calculations in
lattice gauge theories seem to indicate that even for small
energies the qualitative picture remains true, and the vacuum of
Yang-Mills theory is associated to a non-trivial background field
configuration, which gives rise to confinement and massive
glueball fields. However, the low energy physics of non-Abelian
gauge theories is a regime which has not yet been understood.

\section{Wilsonian effective actions of supersymmetric theories}
For a given Lagrangian one can also introduce what is known as the
{\it Wilsonian effective action} \cite{Wi71}, \cite{Wi75}. Take
$\l$ to be some energy scale and define the Wilsonian effective
Lagrangian $\mathcal{L}_\l$ as the local Lagrangian that, with
$\l$ imposed as an ultraviolet cut-off, reproduces precisely the
same results for S-matrix elements of processes at momenta below
$\l$ as the original Lagrangian $\mathcal{L}$. In general, masses
and coupling constants in the Wilsonian action will depend on $\l$
and usually there are infinitely many terms in the Lagrangian.
Therefore, the Wilsonian action might not seem very attractive.
However, it can be shown that its form is quite simple in the case
of supersymmetric theories.

Supersymmetric field theories are amazingly rich and beautiful.
Independently on whether they turn out to be the correct
description of nature, they certainly are useful to understand the
structure of quantum field theory. This is the case since they
often possess many properties and characteristic features of
non-supersymmetric field theories, but the calculations are much
more tractable, because of the higher symmetry. For an
introduction to supersymmetry and some background material see
\cite{B00}, \cite{Wb00}, \cite{WB92}. Here we explain how one can
calculate the Wilsonian effective superpotential in the case of
$\mN=1$ supersymmetric theories.

The $\mN=1$ supersymmetric action of a vector superfield $V$
coupled to a chiral superfield $\Phi$ transforming under some
representation of the gauge group is given by\footnote{We follow
the notation of \cite{Wb00}, in particular $\left[\tr W^\tau \e
W\right]_F=\left[\e_{\a\b}\tr W_\a W_\b\right]_F={1\over2}\tr
F_{\m\n}F^{\m\n}-{i\over4}\e_{\m\n\r\s}\tr F^{\m\n}F^{\r\s}+\tr
\bar\l\ds(1-\g_5)\l-\tr D^2$. Here $F_{\m\n}$ etc. are to be
understood as $F_{\m\n}^at^a$, where $t^a$ are the Hermitean
generators of the gauge group, which satisfy $\tr
t^at^b=\delta^{ab}$.}
\begin{equation}
S=\int\d^4x\ [\Phi^\dagger e^{-V}\Phi]_D-\int\d^4x\
\left[\left({\tau\over 16\pi i}\tr W^\tau\e
W\right)_F+c.c.\right]+\int\d^4x\ \left[(W(\Phi))_F+c.c.\right]\ ,
\end{equation}
where $W(\Phi)$ is known as the (tree-level) {\it superpotential}.
The subscripts $F$ and $D$ extract the $F$- respectively
$D$-component of the superfield in the bracket. Renormalisability
forces $W$ to be at most cubic in $\Phi$, but since we are often
interested in theories which can be understood as effective
theories of some string theory, the condition of renormalisability
will often be relaxed.\footnote{Of course, for non-renormalisable
theories the first two terms can have a more general structure as
well. The first term, for instance, in general reads
$K(\Phi,\Phi^\dagger e^{-V})$ where $K$ is known as the K\"ahler
potential. However, these terms presently are not very important
for us. See for example \cite{WB92}, \cite{Wb00} for the details.}
The constant $\tau$ is given in terms of the bare coupling $g$ and
the $\Theta$-angle,
\begin{equation}\label{tau}
\tau={4\pi i\over g^2}+{\Theta\over2\pi}\ .
\end{equation}
The ordinary bosonic potential of the theory reads
\begin{equation}
V(\phi)=\sum_n\left|{\partial W\over\partial
\phi_n}\right|^2+{g^2\over2}\sum_a\left(\sum_{mn}\phi_n^*\phi_m(t^a)_{mn}\right)^2\
,
\end{equation}
where $\phi$ is the lowest component of the superfield $\Phi$ and
$t^a$ are the Hermitean generators of the gauge group. A
supersymmetric vacuum $\phi_0$ of the theory is a field
configuration for which $V$ vanishes \cite{Wb00}, so we have the
so called F-flatness condition
\begin{equation}
\left.{\partial \over\partial \phi}W(\phi)\right|_{\phi_0}=0\ ,
\end{equation}
as well as the D-flatness condition
\begin{equation}\label{Dflatness}
\left.\sum_{mn}\phi_n^*\phi_m(t^a)_{mn}\right|_{\phi_0}=0\ .
\end{equation}
The space of solutions to the D-flatness condition is known as the
{\it classical moduli space} and it can be shown that is can
always be parameterised in terms of a set of independent
holomorphic gauge invariants $X_k(\phi)$.

The task is now to determine the effective potential of this
theory in order to learn something about its quantum vacuum
structure. Clearly, one possibility is to calculate the 1PI
effective action, however, for supersymmetric gauge theories there
exist non-renormalisation theorems which state that the Wilsonian
effective actions of these theories is particularly simple.

\begin{proposition}Perturbative non-renormalisation theorem\\
If the cut-off $\l$ appearing in the Wilsonian effective action
preserves supersymmetry and gauge invariance, then the Wilsonian
effective action to all orders in perturbation theory has the form
\begin{equation}
S_\l=\int\d^4x\ [(W(\Phi))_F+c.c.]-\int\d^4x\
\left[\left({\tau_\l\over 16\pi i}\tr W^\tau\e
W\right)_F+c.c.\right]+\mbox{D-terms}\ ,
\end{equation}
where
\begin{equation}
\tau_\l={4\pi i\over g^2_\l}+{\Theta\over2\pi}\ ,
\end{equation}
and $g_\l$ is the {\it one-loop} effective coupling.
\end{proposition}
Note in particular that the superpotential remains unchanged in
perturbation theory, and that the gauge kinetic term is
renormalised only at one loop. The theorem was proved in
\cite{GSR79} using supergraph techniques, and in \cite{Se93} using
symmetry arguments and analyticity.

Although the superpotential is not renormalised to any finite
order in perturbation theory it does in fact get corrected on the
non-perturbative level, i.e. one has
\begin{equation}
W_{eff}=W_{tree}+W_{non-pert}\ .
\end{equation}
The non-perturbative contributions were thoroughly studied in a
series of papers by Affleck, Davis, Dine and Seiberg \cite{DDS83},
\cite{ADS83}, using dimensional analysis and symmetry
considerations. For some theories these arguments suffice to
exactly determine $W_{non-pert}$. An excellent review can be found
in \cite{Wb00}. This effective Wilsonian superpotential can now be
used to study the quantum vacua of the gauge theory, which have to
be critical points of the effective superpotential.

\bigskip
So far we defined two effective actions, the generating functional
of 1PI amplitudes and the Wilsonian action. Clearly, it is
important to understand the relation between the two. In fact, for
the supersymmetric theories studied above one can also evaluate
the 1PI effective action. It turns out that this functional
receives contributions to all loop orders in perturbation theory,
corresponding to Feynman diagrams in the background fields with
arbitrarily many internal loops. Therefore, we find that the
difference between the two effective actions seems to be quite
dramatic. One of them is corrected only at one-loop and the other
one obtains corrections to all loop orders. The crucial point is
that one integrates over all momenta down to zero to obtain the
1PI effective action, but one only integrates down to the scale
$\l$ to calculate the Wilsonian action. In other words, whereas
one has to use tree-diagrams only if one is working with the 1PI
effective action, one has to include loops in the Feynman diagrams
if one uses the Wilsonian action. However, the momentum in these
loops has an ultraviolet cutoff $\l$. Taking this $\l$ down to
zero then gives back the 1PI generating functional. Therefore, the
difference between the two has to come from the momentum domain
between 0 and $\l$. Indeed, as was shown by Shifman and Vainshtein
in \cite{SV86}, in supersymmetric theories the two-loop and higher
contributions to the 1PI effective action are infrared effects.
They only enter the Wilsonian effective action as the scale $\l$
is taken to zero. For finite $\l$ the terms in the Wilsonian
effective action arise only from the tree-level and one-loop
contributions, together with non-perturbative corrections.

Furthermore, it turns out that the fields and coupling constants
that appear in the Wilsonian effective action are {\it not} the
physical quantities one would measure in experiment. For example,
the non-renormalisation theorem states that the coupling constant
$g$ is renormalised only at one loop. However, from explicit
calculations one finds that the 1PI $g$ is renormalised at all
loops. This immediately implies that there are two different
coupling constants, the Wilsonian one and the 1PI coupling. The
two are related in a non-holomorphic way and again the difference
can be shown to come from infrared effects. It is an important
fact that the Wilsonian effective superpotential does depend
holomorphically on both the fields and the (Wilsonian) coupling
constants, whereas the 1PI effective action is non-holomorphic in
the (1PI) coupling constants. The relation between the two
quantities has been pointed out in \cite{SV91}, \cite{DS94}. In
fact, one can be brought into the other by a non-holomorphic
change of variables. Therefore, for supersymmetric theories we can
confidently use the Wilsonian effective superpotential to study
the theory. If the non-perturbative corrections to the
superpotential are calculable (which can often be done using
symmetries and holomorphy) then one can obtain the exact effective
superpotential and therefore exact results about the vacuum
structure of the theory. However, the price one has to pay is that
this beautiful description is in terms of unphysical Wilsonian
variables. The implications for the true physical quantities can
only be found after undoing the complicated change of variables.

\section{Symmetries and effective potentials}\label{effactionsym}
There is yet another way (see \cite{IS95} for a review and
references), to calculate an effective superpotential, which uses
Seiberg's idea \cite{Se93} to interpret the coupling constants as
chiral superfields. Let
\begin{equation}\label{Wtree}
W(\Phi)=\sum_{k}g_kX_k(\Phi)
\end{equation}
be the tree-level superpotential, where the $X_k$ are gauge
invariant polynomials in the matter chiral superfield $\Phi$. In
other words, the $X_k$ are themselves chiral superfields. One can
now regard the coupling constants $g_k$ as the vacuum expectation
value of the lowest component of another chiral superfield $G_k$,
and interpret this field as a source \cite{Se93}. I.e. instead of
(\ref{Wtree}) we add the term $W(G,\Phi)=\sum_kG_kX_k$ to the
action. Integrating over $\Phi$ then gives the partition function
$Z[G]=\exp(iF[G])$. If we assume that supersymmetry is unbroken
$F$ has to be a supersymmetric action of the chiral superfields
$G$, and therefore it can be written as
\begin{equation}
F[G]=\int\d^4x \ \left[(W_{low}(G))_F+c.c.\right]+\ldots\ ,
\end{equation}
with some function $W_{low}(G)$. As we will see, this function can
often be determined from symmetry arguments. For standard fields
(i.e. not superfields) we have the relation (\ref{expPhi}). In a
supersymmetric theory this reads
\begin{equation}
\langle X_k\rangle_G={\delta\over\delta
G_{k}}F[G]={\partial\over\partial G_k}W_{low}(G)
\end{equation}
where we used that $W_{low}$ is holomorphic in the fields $G_k$.
On the other hand we can use this equation to define $\la G_k\ra$
as the solution of
\begin{equation}
X_k^0=\left.{\partial\over\partial G_k}W_{low}(G)\right|_{\langle
G_k\rangle}\ .\label{legendre}
\end{equation}
Then we define the Legendre transform of $W_{low}$
\begin{equation}
W_{dyn}(X_k^0):=W_{low}(\langle G_k\rangle)-\sum_k\langle
G_k\rangle X_k^0\ ,\label{Wdyn}
\end{equation}
where $\langle G_k\rangle$ solves (\ref{legendre}), and finally we
set
\begin{equation}
W_{eff}(X_k,g_k):=W_{dyn}(X_k)+\sum_kg_kX_k\ .\label{effsup}
\end{equation}
This effective potential has the important property that the
equations of motion for the fields $X_k$ derived from it determine
their expectation values. Note that (\ref{effsup}) is nothing but
the tree-level superpotential corrected by the term $W_{dyn}$.
This looks similar to the Wilsonian superpotential, of which we
know that it is uncorrected perturbatively but it obtains
non-perturbative corrections. Indeed, for some cases the Wilsonian
superpotential coincides with (\ref{effsup}), however, in general
this is not the case (see \cite{IS95} for a discussion of these
issues). Furthermore, since $W_{dyn}$ does not depend on the
couplings $g_k$, the effective potential depends linearly on
$g_k$. This is sometimes known as the {\it linearity principle},
and it has some interesting consequences. For instance one might
want to integrate out the field $X_i$ by solving
\begin{equation}
{\partial W_{eff}\over\partial X_i}=0\ ,
\end{equation}
which can be rewritten as
\begin{equation}
g_i=-{\partial W_{dyn}\over\partial X_i}\ .
\end{equation}
If one solves this equation for $X_i$ in terms of $g_i$ and the
other variables and plugs the result back in $W_{eff}$, the
$g_i$-dependence will be complicated. In particular, integrating
out all the $X_i$ gives back the superpotential $W_{low}(g)$.
However, during this process one does actually not loose any
information, since this procedure of integrating out $X_i$ can
actually be inverted by {\it integrating in} $X_i$. This is
obvious from the fact that, because of the linearity in $g_k$,
integrating out $X_i$ is nothing but performing an (invertible)
Legendre transformation.

\bigskip
{\bf Super Yang-Mills theory and the Veneziano-Yankielowicz potential}\\
In order to see how the above recipe is applied in practice, we
study the example of $\mN=1$ Super-Yang-Mills theory.  Its action
reads
\begin{equation}
S_{SYM}=-\int\d^4x\ \left[\left({\tau\over 16\pi i}\tr\ W^\tau\e
W\right)_F+c.c.\right]\ ,\label{SYMclassical}
\end{equation}
which, if one defines the chiral superfield
\begin{equation}
S:={1\over32\pi^2}\tr\ W^\tau\e W\ ,
\end{equation}
can be rewritten as
\begin{equation}
S_{SYM}=\int\d^4x\ \left[(2\pi i\tau S)_F+c.c.\right]\
.\label{tauSaction}
\end{equation}
$S$ is known as the {\it gaugino bilinear superfield}, whose
lowest component is proportional to $\l\l\equiv\tr\l^\tau\e \l$.
Note that both $S$ and $\tau$ are complex.

The classical action (\ref{SYMclassical}) is invariant under a
chiral $U(1)$ R-symmetry that acts as $W_\a(x,\t)\rightarrow
e^{i\varphi}W_\a(x,e^{-i\varphi}\t)$, which implies in particular
that $\l\rightarrow e^{i\varphi}\l$. The quantum theory, however,
is not invariant under this symmetry, which can be understood from
the fact that the measure of the path integral is not invariant.
The phenomenon in which a symmetry of the classical action does
not persist at the quantum level is known as an anomaly. For a
detailed analysis of anomalies and some applications in string and
M-theory see the review article \cite{Me03}. The most important
results on anomalies are listed in appendix \ref{anomalies}. The
precise transformation of the measure for a general transformation
$\l\rightarrow e^{i\e(x)}\l$ can be evaluated \cite{Wb00} and
reads
\begin{equation}
D\l D\bar\l\rightarrow D\l'D\bar\l'=\exp\left(i\int\d^4x\ \e(x)
G[x;A]\right)D\l D\bar\l
\end{equation}
where
\begin{equation}
G[x;A]=-{N\over32\pi^2}\e_{\m\n\r\s}F_a^{\m\n}F_a^{\r\s}\ .
\end{equation}
For the global R-symmetry $\e(x)=\varphi$ is constant, and
${1\over64\pi^2}\int\d^4x\ \e_{\m\n\r\s}F^{\m\n}F^{\r\s}=\n$ is an
integer, and we see that the symmetry is broken by instantons.
Note that, because of the anomaly, the chiral rotation
$\l\rightarrow e^{i\varphi}\l$ is equivalent to
$\Theta\rightarrow\Theta-2N\varphi$ (c.f. Eq. (\ref{tau})). Thus,
the chiral rotation is a symmetry only if $\varphi={k\pi\over N}$
with $k=0,\ldots 2N-1$, and the $U(1)$ symmetry is broken to
$\mathbb{Z}_{2N}$.

The objective is to study the vacua of $\mN=1$ super Yang-Mills
theory by probing for gaugino condensates, to which we associate
the composite field $S$ that includes the gaugino bilinear. This
means we are interested in the effective superpotential
$W_{eff}(S)$, which describes the symmetries and anomalies of the
theory. In particular, upon extremising $W_{eff}(S)$ the value of
the gaugino condensate in a vacuum of $\mN=1$ Yang-Mills is
determined.

The $\b$-function of $\mN=1$ super Yang-Mills theory reads at one
loop
\begin{equation}\label{betaSYM}
\b(g)=-{3Ng^3\over16\pi^2}
\end{equation}
and the solution of the renormalisation group equation is given by
\begin{equation}
{1\over g^2(\m)}=-{3N\over 8\pi^2}\log{|\L|\over\m}\ .
\end{equation}
Then
\begin{eqnarray}
\tau_{1-loop}&=&{4\pi i\over g^2(\m)}+{\Theta\over2\pi}
={1\over2\pi
i}\log\left({|\L|e^{i\Theta/3N}\over\m}\right)^{3N}=:{1\over2\pi
i}\log\left({\L\over\m}\right)^{3N}\label{tauoneloop}
\end{eqnarray}
enters the one-loop action
\begin{equation}
S_{1-loop}=\int\d^4x\ \left[(2\pi i\tau_{1-loop}
S)_F+c.c.\right]=\int\d^4x\
\left[\left(3N\log\left(\L\over\m\right) S\right)_F+c.c.\right]\
.\label{action}
\end{equation}

In order to determine the superpotential $W_{low}(\tau)$ one can
use Seiberg's method \cite{Se93}, and one interprets $\tau$ as a
background chiral superfield. This is useful, since the effective
superpotential is known to depend {\it holomorphically} on all the
fields, and therefore it has to depend holomorphically on $\tau$.
Furthermore, once $\tau$ is interpreted as a field, spurious
symmetries occur. In the given case one has the spurious
R-symmetry transformation\\
\parbox{14cm}{
\begin{eqnarray}
W(x,\t)&\rightarrow& e^{i\varphi}W(x,e^{-i\varphi}\t)\ ,\nonumber\\
\tau&\rightarrow& \tau+{N\varphi\over\pi}\ ,\nonumber
\end{eqnarray}}\hfill\parbox{8mm}{\begin{eqnarray}\end{eqnarray}}\\
and the low energy potential has to respect this symmetry. This
requirement, together with dimensional analysis, constrains the
superpotential $W_{low}$ uniquely to
\begin{equation}
W_{low}=N\m^3\exp\left({2\pi i\over N}\tau\right)=N\L^3\
,\label{Wlow}
\end{equation}
where $\m$ has dimension one. Indeed, this $W_{low}(\tau)$
transforms as $W_{low}(\tau)\rightarrow
e^{2i\varphi}W_{low}(\tau)$, and therefore the action is
invariant.

Before deriving the effective action let us first show that a
non-vanishing gaugino condensate exists. One starts from
(\ref{SYMclassical}) and now one treats $\tau$ as a background
field. Then, since $W_{\a a}=\l_{\a a}+\ldots$, the $F$-component
of $\tau$, denoted by $\tau_F$, acts as a source for $\l\l$. One
has
\begin{eqnarray}
\langle\l\l\rangle&=&{1\over Z}\int D\Phi\ e^{iS}\l\l={1\over
Z}\int D\Phi\ \exp\left[-i\int d^4x\left[\left({\tau\over16\pi i}
W^\tau\e W\right)_F+c.c.\right]\right]\l\l\nonumber\\
&=&-16\pi{1\over Z}\int D\Phi\ {\delta \over\delta
\tau_F}e^{iS}=-{16\pi\over Z}{\delta\over \delta
\tau_F}Z=-16\pi{\delta\over \delta \tau_F}\log Z\nonumber\\
&=&-16\pi i{\delta\over \delta \tau_F}\int\d^4x\
\left[(W_{low})_F+c.c.\right]+\ldots=-16\pi i{\partial\over
\partial \tau}W_{low}(\tau)\nonumber\\
&=&32\pi^2 \m^3\exp\left({2\pi i\tau\over N}\right)\ ,
\end{eqnarray}
where $D\Phi$ stands for the path integral over all the fields.
$\tau$ is renormalised only at one loop and non-perturbatively,
and it has the general form
\begin{equation}\label{taunonpert}
\tau={3N\over2\pi i}\log\left({\L\over\m}\right)+\sum_{n=1}^\infty
a_n\left({\L\over\m}\right)^{3Nn}\ .
\end{equation}
Therefore, the non-perturbative terms of $\tau$ only contribute to
a phase of the gaugino condensate, and it is sufficient to plug in
the one-loop expression for $\tau$. The result is a non-vanishing
gaugino condensate,
\begin{equation}
\langle\l\l\rangle=32\pi^2\L^3\ .\label{gauginocondensate}
\end{equation}
The presence of this condensate means that the vacuum does not
satisfy the $\mathbb{Z}_{2N}$ symmetry, since
$\langle\l\l\rangle\rightarrow e^{2i\varphi}\langle\l\l\rangle$
and only a $\mathbb{Z}_2$-symmetry survives. The remaining
$|{\mathbb{Z}_{2N}\over\mathbb{Z}_2}|-1=N-1$ transformations, i.e.
those with $\varphi={k\pi\over N}$ with $k=1,\ldots N-1$ transform
one vacuum into another one, and we conclude that there are $N$
distinct vacua.

Next we turn to the computation of the effective action. From
(\ref{tauSaction}) we infer that $2\pi iS$ and $\tau$ are
conjugate variables. We apply the recipe of the last section,
starting from $W_{low}(\tau)$, as given in (\ref{Wlow}). From
\begin{equation}
2\pi iS=\left.{\partial\over\partial
\tau}W_{low}(\tau)\right|_{\langle \tau\rangle}=2\pi
i\m^3\exp\left({2\pi i\over N}\langle\tau\rangle\right)
\end{equation}
we infer that $\langle \tau\rangle={N\over2\pi i}\log\left({S\over
\m^3}\right)$. According to (\ref{Wdyn}) one finds
\begin{equation}
W_{dyn}(S)=NS-NS\log \left({S\over\m^3}\right)\ .
\end{equation}
Finally, using the one-loop expression for $\tau$ and identifying
the $\m$ appearing in (\ref{Wlow}) with the one in
(\ref{tauoneloop}), one obtains the Veneziano-Yankielowicz
potential \cite{VY82}
\begin{equation}
W_{eff}(\L,S)=W_{VY}(\L,S)=S\left[N+\log\left(\L^{3N}\over
S^N\right)\right]\ .
\end{equation}
In order to see in what sense this is the correct effective
potential, one can check whether it gives the correct expectation
values. Indeed, $\left.{\partial W_{eff}(\L,S)\over\partial
S}\right|_{\langle S\rangle}=0$ gives the $N$ vacua of $\mN=1$
super Yang-Mills theory,
\begin{equation}
\langle S\rangle=\L^3e^{2\pi ik\over N}\ ,\ \ \ k=0,\ldots N-1\ .
\end{equation}
Note that this agrees with (\ref{gauginocondensate}), since $S$ is
defined as ${1\over32\pi^2}$ times $\tr W_aW^\a$, which accounts
for the difference in the prefactors. Furthermore, the
Veneziano-Yankielowicz potential correctly captures the symmetries
of the theory. Clearly, under R-symmetry $S$ transforms as
$S\rightarrow \tilde S=e^{2i\varphi}S$ and therefore the effective
Veneziano-Yankielowicz action $\mathcal{S}_{VY}$ transforms as
\begin{eqnarray}
\mathcal{S}_{VY}\rightarrow\tilde{\mathcal{S}}_{VY}&=&\int\d^4x\ \left[\left(\tilde W_{VY}\right)_{\tilde F}+c.c.\right]+\ldots\nonumber\\
&=&\int\d^4x\ \left[e^{2i\varphi}\left( SN+S\log\left(\L^{3N}\over
S^N\right)-2iN\varphi
S\right)_{\tilde F}+c.c.\right]\nonumber\\
&=&\mathcal{S}_{VY}-\int\d^4x\
\left[\left({iN\varphi\over16\pi^2}\tr W^\tau\e
W\right)_F+c.c.\right]\ ,
\end{eqnarray}
which reproduces the anomaly. However, for the effective theory to
have the $\mathbb{Z}_{2N}$ symmetry, one has to take into account
that the logarithm is a multi-valued function. For the $n$-th
branch one must define
\begin{equation}
W_{VY}^{(n)}(\L,S)=S\left[N+\log\left(\L^{3N}\over S^N\right)+2\pi
in\right]\ .
\end{equation}
Then the discrete symmetry shifts
$\mathcal{L}_n\rightarrow\mathcal{L}_{n-k}$ for $\varphi={\pi
k\over N}$. The theory is invariant under $\mathbb{Z}_{2N}$ if we
define
\begin{equation}
Z=\sum_{n=-\infty}^\infty\int D S \exp\left(i\int\d^4x\
\left[W_{VY}^{(n)}+c.c.\right]_F+\ldots\right)\ .
\end{equation}
Thus, although the Veneziano-Yankielowicz effective action is not
the Wilsonian effective action, it contains all symmetries,
anomalies and the vacuum structure of the theory.

\bigskip
{\bf Super Yang-Mills coupled to matter}\\
The main objective of the next chapters is to find an effective
superpotential in the Veneziano-Yankielowicz sense, i.e. one that
is not necessarily related to the 1PI or Wilsonian effective
action, but that can be used to find the vacuum structure of the
theory, for $\mN=1$ Yang-Mills theory coupled to a chiral
superfield $\Phi$ in the adjoint representation. The tree-level
superpotential in this case is given by
\begin{equation}
W(\Phi)=\sum_{k=1}^{n+1}{g_k\over k}\tr \Phi^k+g_0\ ,\
g_k\in\mathbb{C}
\end{equation}
where without loss of generality $g_{n+1}=1$. Furthermore, the
critical points of $W$ are taken to be non-degenerate, i.e. if
$W'(p)=0$ then $W''(p)\neq0$. As we mentioned above, the
corresponding effective superpotential can be evaluated
perturbatively from a holomorphic matrix model \cite{DV02a},
\cite{DV02b}, \cite{DV02c}. This can either be shown using string
theory arguments based on the results of \cite{V01} and
\cite{CIV01}, or from an analysis in field theory \cite{DGLVZ02},
\cite{CDSW02}. The field theory itself has been studied in
\cite{CSW03a}, \cite{CSW03b}, \cite{Fe02}. However, before we turn
to explaining these development we need to present background
material on the manifolds appearing in this context.

\chapter{Riemann Surfaces and Calabi-Yau Manifolds}\label{RSCY}
In this section we explain some elementary properties of Riemann
surfaces and Calabi-Yau manifolds. Of course, both types of
manifolds are ubiquitous in string theory and studying them is of
general interest. Here we will concentrate on those aspects that
are relevant for our setup. As we mentioned in the introduction,
the theory we are interested in can be geometrically engineered by
``compactifying" Type II string theory on non-compact Calabi-Yau
manifolds, the structure of which will be presented in detail. The
superpotential can be calculated from geometric integrals of a
three-form over a basis of three-cycles in these manifolds. Quite
interestingly, it turns out that the non-compact Calabi-Yau
manifolds are intimately related to Riemann surfaces and that the
integrals on the Calabi-Yau can be mapped to integrals on the
Riemann surface. As we will see in the next chapter, it is
precisely this surface which also appears in the large $\hN$ limit
of a holomorphic matrix model.

Our main reference for Riemann surfaces is \cite{FK}. An excellent
review of both Riemann surfaces and Calabi-Yau manifolds, as well
as their physical applications can be found in \cite{Horietal}.
The moduli space of Calabi-Yau manifolds was first studied in
\cite{CdlO91}.

\section{Properties of Riemann surfaces}\label{propRS}
\begin{definition}
{\em A} Riemann surface {\em is a complex one-dimensional
connected analytic manifold.}
\end{definition}

There are many different description of Riemann surfaces. We are
interested in the so called hyperelliptic Riemann surfaces of
genus $\hg$,
\begin{equation}
y^2=\prod_{i=1}^{\hg+1}(x-a_i^+)(x-a_i^-)\
,\label{hyperellipticRS}
\end{equation}
with $x,y,a_i^\pm\in\mathbb{C}$ and all $a_i^\pm$ different. These
can be understood as two complex sheets glued together along cuts
running between the branch points $a_i^-$ and $a_i^+$, together
with the two points at infinity of the two sheets, denoted by $Q$
on the upper and $Q'$ on the lower sheet. On these surfaces the
set $\left\{{\d x\over y},{x\d x\over y},\ldots,{x^{\hg-1}\d
x\over y}\right\}$ forms a basis of holomorphic differentials.
This fact can be understood by looking at the theory of divisors
on Riemann surfaces, presented in appendix \ref{div}, see also
\cite{FK}. The divisors capture the zeros and divergences of
functions on the Riemann surface. Let $P_1,\ldots P_{2\gh+2}$
denote those points on the Riemann surface which correspond to the
zeros of $y$ (i.e. to the $a_i^\pm$). Close to $a_i^\pm$ the good
coordinates are $z_i^\pm=\sqrt{x-a_i^\pm}$. Then the divisor of
$y$ is given by ${P_1\ldots P_{2\gh+2}\over Q^{\gh+1}Q'^{\gh+1}}$,
since $y$ has simple zeros at the $P_i$ in the good coordinates
$z_i$, and poles of order $\gh+1$ at the points $Q,Q'$. If we let
$R,R'$ denote those points on the Riemann surface which correspond
to zero on the upper, respectively lower sheet then it is clear
that the divisor of $x$ is given by ${RR'\over Q Q'}$. Finally,
close to $a_i^\pm$ we have $\d x\sim z_i^\pm\d z^\pm_i$, and
obviously $\d x$ has double poles at $Q$ and $Q'$, which leads to
a divisor ${P_1\ldots P_{2\gh+2}\over Q^2Q'^2}$. In order to
determine the zeros and poles of more complicated objects like
${x^k\d x\over y}$ we can now simply multiply the divisors of the
individual components of this object. In particular, the divisor
of ${\d x\over y}$ is $Q^{\gh-1}Q'^{\gh-1}$, and for $\gh\geq 1$
it has no poles. Similarly, we find that ${x^k\d x\over y}$ has no
poles if $k\leq\gh-1$. Quite generally, for any compact Riemann
surface $\S$ of genus $\hg$ one has,
\begin{equation}
{\rm dim}\  \textsf{Hol}^1_{\gh}(\S)=\hg\ ,
\end{equation}
where $\textsf{Hol}^1_{\gh}(\S)$ is the first holomorphic de Rham
cohomology group on $\S$ with genus $\gh$. On the other hand,
later on we will be interested in integrals of the form $y\d x$,
with divisor ${P_1^2\ldots P_{2\gh+2}^2\over
Q^{\gh+3}Q'^{\gh+3}}$, showing that $y\d x$ has poles of order
$\gh+3,\gh+2,\ldots 1$ at $Q$ and $Q'$. To allow for such forms
with poles one has to mark points on the surface. This marking
amounts to pinching a hole into the surface. Then one can allow
forms to diverge at this point, since it no longer is part of the
surface. The dimension of the first holomorphic de Rham cohomology
group on $\S$ of genus $\gh$ with $n$ marked points,
$\textsf{Hol}^1_{\gh,n}(\S)$, is given by \cite{FK}
\begin{equation}
{\rm dim}\  \textsf{Hol}^1_{\gh,n}(\S)=2\hg+n-1\ .
\end{equation}

\begin{figure}[h]
\centering
\includegraphics[width=\textwidth]{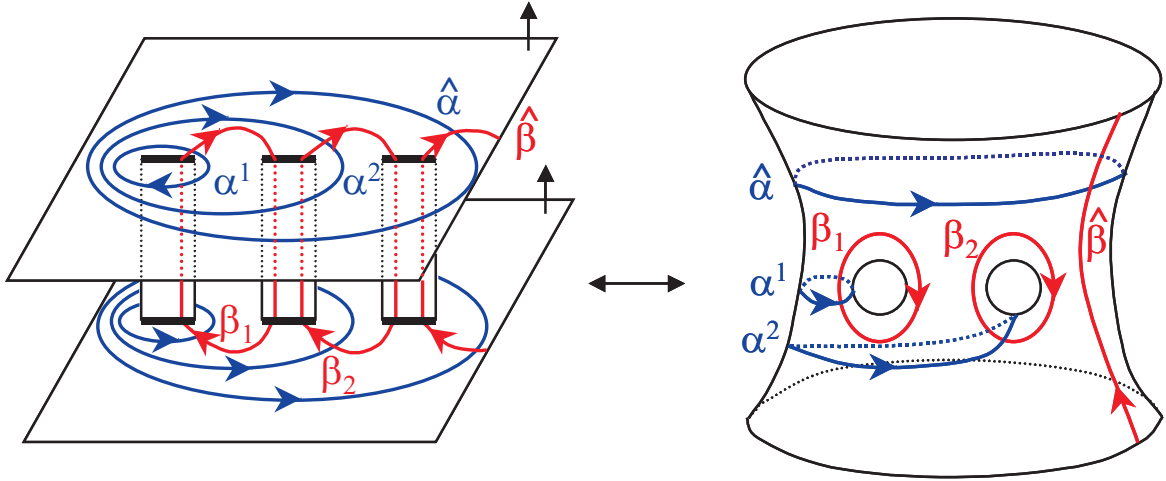}\\
\caption[]{The hyperelliptic Riemann surface
(\ref{hyperellipticRS}) can be understood as two complex sheets
glued together along cuts running between $a_i^-$ and $a_i^+$.
Here we indicate a symplectic set of cycles for $\gh=2$. It
consists of $\gh$ compact cycles $\a^i$, surrounding $i$ of the
cuts, and their compact duals $\b_i$, running from cut $i$ to
$i+1$ on the upper sheet and from cut $i+1$ to cut $i$ on the
lower one. We also indicated the relative cycles $\hat\a,\hat\b$,
which together with $\a^i,\b_i$ form a basis of the relative
homology group $H_1(\S,\{Q,Q'\})$. Note that the orientation of
the two planes on the left-hand side is chosen such that both
normal vectors point to the top. This is why the orientation of
the $\a$-cycles is different on the two planes. To go from the
representation of the Riemann surface on the left to the one on
the right one has to flip the upper plane.} \label{albecycles}
\end{figure}

The first homology group $H_1(\S;\mathbb{Z})$ of the Riemann
surface $\S$ has $2\hg$ generators $\a^i,\b_i,\
i\in\{1,\ldots,\hg\}$, with intersection matrix\\
\parbox{14cm}{
\begin{eqnarray}
&&\a^i\cap \a^j=\emptyset\ \ \ ,\ \ \ \b_i\cap \b_j=\emptyset\ ,\nonumber\\
&&\a^i\cap \b_j=-\b_j\cap \a^i=\delta_j^i\ .\nonumber
\end{eqnarray}}\hfill\parbox{8mm}{\begin{eqnarray}\label{sympbasis}\end{eqnarray}}\\
Note that this basis is defined only up to a\footnote{There are
different conventions for the definition of the symplectic group
in the literature. We adopt the following:
\begin{equation}
Sp(2m,\mathbb{K}):=\{S\in GL(2m,\mathbb{K}):S^\tau\mho S=\mho\},
\end{equation}
where
\begin{equation}
\mho:=\left(\begin{array}{cc}0&\opone\\
-\opone&0\end{array}\right)
\end{equation}
and $\opone$ is the $m\times m$ unit matrix. $\mathbb{K}$ stands
for any field.} $Sp(2\hg,\mathbb{Z})$ transformation. To see this
formally, consider the vector $v:=\left(\a^i\ \b_j\right)^\tau$
that satisfies $v\cap v^\tau=\mho$. But for $S\in
Sp(2\hg,\mathbb{Z})$ we have for $v':=Sv$ that $v'\cap
v'^\tau=S\mho S^\tau=\mho$. A possible choice of the cycles
$\a^i,\b_i$ for the hyperelliptic Riemann surface
(\ref{hyperellipticRS}) is given in Fig. \ref{albecycles}.
As we mentioned already, later on we will be interested in
integrals of $y\d x$, which diverges at $Q,Q'$. Therefore, we are
led to consider a Riemann surface with these two points excised.
On such a surface there exists a very natural homology group,
namely the relative homology $H_1(\S,\{Q,Q'\})$. For a detailed
exposition of relative (co-)homology see appendix \ref{relhom}.
$H_1(\S,\{Q,Q'\})$ not only contains the closed cycles
$\a^i,\b_i$, but also a cycle $\hat\b$, stretching from $Q$ to
$Q'$, together with its dual $\hat\a$. As an example one might
look at the simple Riemann surface
\begin{equation}
y^2=x^2-\m=(x-\sqrt\m)(x-\sqrt\m)\ ,
\end{equation}
with only one cut between $-\sqrt\m$ and $\sqrt\m$ surrounded by a
cycle $\hat\a$. The dual cycle $\hat\b$ simply runs from $Q'$
through the cut to $Q$.

\begin{figure}[h]
\centering
\includegraphics[width=\textwidth]{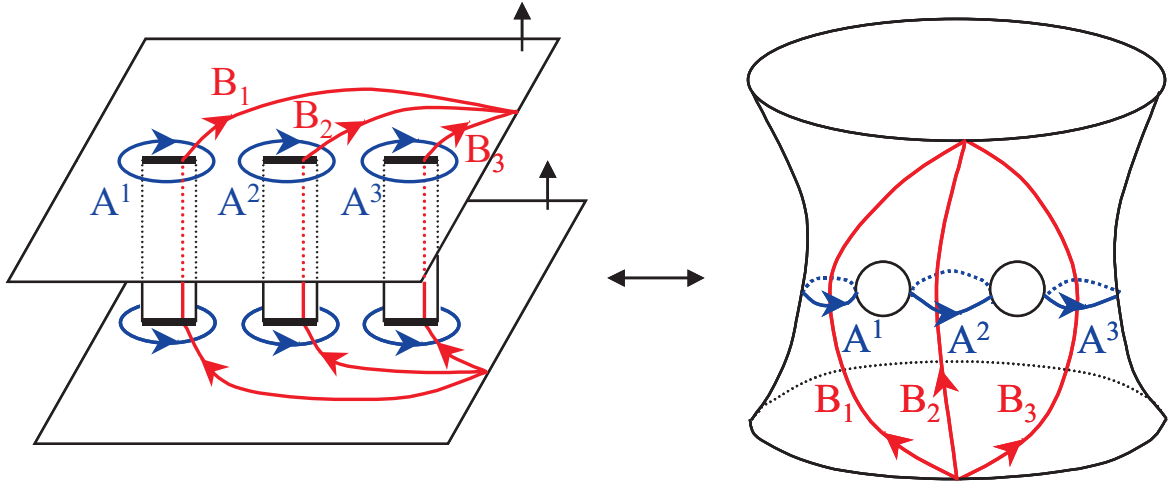}\\
\caption[]{Another choice of basis for $H_1(\S,\{Q,Q'\})$
containing compact A-cycles and non-compact B-cycles.}
\label{ABcycles}
\end{figure}

There are various symplectic bases of $H_1(\S,\{Q,Q'\})$. Next to
the one just presented, another set of cycles often appears in the
literature. It contains $\gh+1$ compact cycles $A^i$, each
surrounding one cut only, and their duals $B_i$, which are all
non-compact, see Fig. \ref{ABcycles}.
Although string theory considerations often lead to this basis, it
turns out to be less convenient, basically because of the
non-compactness of the $B$-cycles.

Next we collect a couple of properties which hold for any
(compact) Riemann surface $\S$. Let $\o$ be any one-form on $\S$,
then
\begin{equation}
\left(\begin{array}{c}\int_{\a^i}\o\\\int_{\b_j}\o\end{array}\right)
\end{equation}
is called the {\it period vector} of $\o$. For any pair of closed
one-forms $\o,\chi$ on $\S$ one has
\begin{equation}\label{intformriemann}
\int_{\S}\o\wedge\chi=\sum_{i=1}^{\hg}\left(\int_{\a^i}\o\int_{\b_i}\chi-\int_{\b_i}\o\int_{\a^i}\chi\right)\
.
\end{equation}
This is the {\it Riemann bilinear} relation for Riemann surfaces.

Denote the $\hg$ linearly independent holomorphic one-forms on
$\S$ by $\{\l_k\},\ k\in\{1,\ldots,\hg\}$. Define
\begin{equation}
e^i_k:=\int_{\a^i}\l_k\ \ \ ,\ \ \ h_{ik}:=\int_{\b_i}\l_k\ ,
\end{equation}
and from these the {\it period matrix}
\begin{equation}
\Pi_{ij}:=h_{ik}(e^{-1})^k_j\ .
\end{equation}
Inserting two holomorphic forms $\l_i,\l_j$ for $\o,\chi$ in
(\ref{intformriemann}) the left-hand side vanishes, which tells us
that the period matrix is symmetric. This is known as {\it
Riemann's first relation}. Furthermore, using
(\ref{intformriemann}) with $\l_k,\bar \l_{\bar l}$ one finds {\it
Riemann's second relation},
\begin{equation}
{\rm Im}(\Pi_{ij})>0.
\end{equation}
Riemann's relations are invariant under a symplectic change of the
homology basis.

\bigskip
{\bf The moduli space of Riemann surfaces}\\
Let $\mathcal{M}_{\hg}$ denote the moduli space of complex
structures on a genus $\hg$ Riemann surface. As is reviewed in
appendix \ref{CG}, infinitesimal changes of the complex structure
of a manifold $X$ are described by $H^1_{\bar\partial}(TX)$ and
therefore this vector space is the tangent space to
$\mathcal{M}_{\hg}$ at the point corresponding to $X$. This is
interesting because the dimension of $\mathcal{M}_{\hg}$ coincides
with the dimension of its tangent space and the latter can be
computed explicitly (using the Grothendieck-Riemann-Roch formula,
see \cite{Horietal}). The result is
\begin{eqnarray}
\mathcal{M}_0&=&\{{\rm point}\}\ ,\nonumber\\
\rm{dim}_\mathbb{C}\mathcal{M}_1&=&1\ ,\\
\rm{dim}_\mathbb{C}\mathcal{M}_{\hg}&=&3\hg-3\ \ \ \mbox{for} \
\hg\geq2\ .\nonumber
\end{eqnarray}
One might consider the case in which one has additional marked
points on the Riemann surface. The corresponding moduli space is
denoted by $\mathcal{M}_{\hg,n}$ and its dimension is given by
\begin{eqnarray}
\rm{dim}_\mathbb{C}\mathcal{M}_{0,n}&=&n-3\ \ \ \mbox{for}\ n\geq3\ ,\nonumber\\
\rm{dim}_\mathbb{C}\mathcal{M}_{1,n}&=&n\ ,\\
\rm{dim}_\mathbb{C}\mathcal{M}_{\hg}&=&3\hg-3+n\ \ \ \mbox{for} \
\hg\geq2\ .\nonumber
\end{eqnarray}

\section{Properties of (local) Calabi-Yau manifolds}\label{propCY}

\subsection{Aspects of compact Calabi-Yau manifolds}\label{compCY}
Our definition of a Calabi-Yau manifold is similar to the one of
\cite{Jo01}.

\begin{definition}\label{CY}
{\em Let $X$ be a compact complex manifold of complex dimension
$m$ and $J$ the complex structure on $X$. A} Calabi-Yau manifold
{\em is a triple $(X,J,g)$, s.t. $g$ is a Kähler metric on $(X,J)$
with holonomy group $Hol(g)=SU(m)$.}
\end{definition}
A Calabi-Yau manifold of dimension $m$ admits a nowhere vanishing,
covariantly constant holomorphic $(m,0)$-form $\O$ on $X$ that is
unique up to multiplication by a non-zero complex number.

\begin{proposition}
Let $(X,J,g)$ be a Calabi-Yau manifold, then $g$ is Ricci-flat.
Conversely, if $(X,J,g)$ of dimension $m$ is simply connected with
a Ricci-flat K\"ahler metric, then its holonomy group is contained
in $SU(m)$.
\end{proposition}
Note that Ricci-flatness implies $c_1(TX)=0$. The converse follows
from Calabi's conjecture:
\begin{proposition}
Let $(X,J)$ be a compact complex manifold with $c_1(X)=[0]\in
H^2(X;\mathbb{R})$. Then every K\"ahler class $[\o]$ on $X$
contains a unique Ricci-flat K\"ahler metric $g$.
\end{proposition}
In our definition of a Calabi-Yau manifold we require the holonomy
group to be precisely $SU(m)$ and not a proper subgroup. It can be
shown that the first Betti-number of these manifolds vanishes,
$b_1=b^1=0$.

We are mainly interested in Calabi-Yau three-folds. The Hodge
numbers of these can be shown to form the following {\it Hodge
diamond}:
\begin{equation*}
\begin{array}{ccccccc}
&&&h^{0,0}&&&\\
&&h^{1,0}&&h^{0,1}&&\\
&h^{2,0}&&h^{1,1}&&h^{0,2}&\\
h^{3,0}&&h^{2,1}&&h^{1,2}&&h^{0,3}\\
&h^{3,1}&&h^{2,2}&&h^{1,3}&\\
&&h^{3,2}&&h^{2,3}&&\\
&&&h^{3,3}&&&
\end{array}=
\begin{array}{ccccccc}
&&&1&&&\\
&&0&&0&&\\
&0&&h^{1,1}&&0&\\
1&&h^{2,1}&&h^{2,1}&&1\\
&0&&h^{1,1}&&0&\\
&&0&&0&&\\
&&&1&&&
\end{array}
\end{equation*}
The dimension of the homology group $H_3(X;\mathbb{Z})$ is
$2h^{2,1}+2$ and one can always choose a symplectic
basis\footnote{We use the letters $(\a^i,\b_j)$, $(A^i, B_j)$,
$(a^i,b_j)$, $\ldots$ to denote symplectic bases on Riemann
surfaces and $(\G_{\a^I},\G_{\b_J})$, $(\G_{A^I},\G_{B_J})$,
$(\G_{a^I},\G_{b_J})$, $\ldots$ for symplectic bases of
three-cycles on Calabi-Yau three-folds. Also, the index $i$ runs
from 1 to $\gh$, whereas $I$ runs from 0 to $h^{2,1}$.}
$\G_{\a^I}, \G_{\b_J}$, $I,J\in\{0,\ldots,h^{2,1}\}$ with
intersection matrix similar to the one in (\ref{sympbasis}).

The period vector of the holomorphic form $\O$ is defined as
\begin{equation}
\Pi(z):=\left(\begin{array}{c}
\int_{\G_{\a^I}}\O\\\int_{\G_{\b_J}}\O
\end{array}\right)\ .
\end{equation}

Similar to the bilinear relation on Riemann surfaces one has for
two closed three-forms $\S,\Xi$ on a (compact) Calabi-Yau
manifold,
\begin{equation}
\langle
\S,\Xi\rangle:=\int_X\S\w\Xi=\sum_I\left(\int_{\G_{\a^I}}\S\int_{\G_{\b_I}}\Xi
-\int_{\G_{\b_I}}\S\int_{\G_{\a^I}}\Xi\right)\ .\label{RiemannCY}
\end{equation}

\bigskip
{\bf The moduli space of (compact) Calabi-Yau three-folds}\\
This and the next subsection follow mainly the classic paper
\cite{CdlO91}. Let $(X,J,\O,g)$ be a Calabi-Yau manifold of
complex dimension $m=3$. We are interested in the moduli space
$\mathcal{M}$, which we take to be the space of all Ricci-flat
K\"ahler metrics on $X$. Note that in this definition of the
moduli space it is implicit that the topology of the Calabi-Yau
space is kept fixed. In particular the numbers $b_2=h_{1,1}$ for
the two-cycles and $b_3=2(h_{2,1}+1)$ for the three-cycles are
fixed once and for all.\footnote{In fact this condition can be
relaxed if one allows for singularities. Then moduli spaces of
Calabi-Yau manifolds of different topology can be glued together
consistently. See \cite{CdlO90} and \cite{Gr97} for an
illuminating discussion.} Following \cite{CdlO91} we start from
the condition of Ricci-flatness, $\mathcal{R}_{AB}(g)=0$,
satisfied on every Calabi-Yau manifold. Here $A,B,\ldots$ label
real coordinates on the Calabi-Yau $X$. In order to explore the
space of metrics we simply deform the original metric and require
Ricci-flatness to be maintained,
\begin{equation}
\mathcal{R}_{AB}(g+\delta g)=0\ . \label{Ricciflatdef}
\end{equation}
Of course, starting from one ``background" metric and deforming it
only explores the moduli space in a neighbourhood of the original
metric and we only find a local description of $\mathcal{M}$. Its
global structure is in general very hard to describe. After some
algebra (\ref{Ricciflatdef}) turns into the Lichnerowicz equation
\begin{equation}
\nabla^C\nabla_C\delta g_{AB}+2R_{A\ B}^{\ D\ E}\delta g_{DE}=0\
.\label{Lich}
\end{equation}
Next we introduce complex coordinates $x^\m$ on $X$, with
$\m,\n,\ldots=1,2,3$. Then there are two possible deformations of
the metric, namely $\delta g_{\mu\nu}$ or $\delta g_{\mu\bar
\nu}$. Plugging these into (\ref{Lich}) leads to two independent
equations, one for $\delta g_{\m\n}$ and one for $\delta
g_{\m\bar\n}$ and therefore the two types of deformations can be
studied independently. To each variation of the metric of mixed
type one can associate the real $(1,1)$-form $i\delta g_{\mu\bar
\nu}\d x^\mu\w\d \bar x^{\bar \nu}$, which can be shown to be
harmonic if and only if $\delta g_{\mu\bar \nu}$ satisfies the
Lichnerowicz equation. A variation of pure type can be associated
to the (2,1)-form $\O_{\k\l}^{\ \ \bar \n}\delta g_{\bar \m\bar
\n}\d x^\k\w\d x^\l\w\d \bar x^{\bar \m}$, which also is harmonic
if and only if $\delta g_{\bar \m\bar \n}$ satisfies (\ref{Lich}).
This tells us that the allowed transformations of the metric are
in one-to-one correspondence with $H^{(1,1)}(X)$ and
$H^{(2,1)}(X)$. The interpretation of the mixed deformations
$\delta g_{\m\bar \n}$ is rather straightforward as they lead to a
new K\"ahler form,
\begin{eqnarray}
\tilde K&=&i\tilde g_{\m\bar \n}\d x^\m\w\d\bar x^{\bar
\n}=i(g_{\m\bar
\n}+\delta g_{\m\bar \n})\d x^\m\w\d \bar x^{\bar \n}\nonumber\\
&=&K+i\delta g_{\m\bar \n}\d x^\m\w\d\bar x^{\bar \n}=K+\delta K\
.
\end{eqnarray}
The variation $\delta g_{\m\n}$ on the other hand is related to a
variation of the complex structure. To see this note that $\tilde
g_{AB}=g_{AB}+\delta g_{AB}$ is a Kähler metric close to the
original one. Then there must exist a coordinate system in which
the pure components of the metric $\tilde g_{AB}$ vanish. Under a
change of coordinates $x^A\rightarrow x'^A:=x^A+f^A(x)=h^A(x)$ we
have
\begin{eqnarray}
\tilde g_{AB}'&=&\left(\partial h\over\partial
x\right)^{-1C}_A\left(\partial h\over\partial
x\right)^{-1D}_B\tilde g_{CD}\nonumber\\
&=&g_{AB}+\delta
g_{AB}-(\partial_Af^C)g_{CB}-(\partial_Bf^D)g_{AD}\ .
\end{eqnarray}
We start from a mixed metric $g_{\m\bar \n}$ and add a pure
deformation $\delta g_{\bar \m\bar \n}$. The resulting metric can
be written as a mixed metric in some coordinate system, we only
have to chose $f$ s.t.
\begin{equation}
\delta g_{\bar \m\bar \n}-(\partial_{\bar \m}f^\r) g_{\r\bar
\n}-(\partial_{\bar \n}f^\r) g_{\r\bar \m}=0\ .
\end{equation}
But this means that $f$ cannot be chosen to be holomorphic and
thus we change the complex structure. Note that the fact that the
deformations of complex structure are characterised by
$H^{2,1}(X)$ is consistent with the discussion in appendix
\ref{CG} since $H^{2,1}(X)\cong H^1_{\bar\partial}(TX)$.

Next let us define a metric on the space of all Ricci-flat
K\"ahler metrics,
\begin{equation}
\d s^2={1\over 4 \rm{vol}(X)}\int_Xg^{AC}g^{BD}(\delta
g_{AB}\delta g_{CD})\sqrt{g}\ \d^6x\ .
\end{equation}
In complex coordinates one finds
\begin{eqnarray}
\d s^2={1\over 2\rm{vol}(X)}\int_X g^{\m\bar \k}g^{\n\bar
\l}\left[\delta g_{\m\n}\delta g_{\bar \k\bar \l}+\delta g_{\m\bar
\l}\delta g_{\n\bar \k}\right]\sqrt{g}\ \d^6x\ .
\end{eqnarray}
Interestingly, this metric is block-diagonal with separate blocks
corresponding to variations of the complex and Kähler structure.

\bigskip
{\bf Complex structure moduli}\\
Starting from one point in the space of all Ricci-flat metrics on
a Calabi-Yau manifold $X$, we now want to study the space of those
metrics that can be reached from that point by deforming the
complex structure of the manifold, while keeping the K\"ahler form
fixed. The space of these metrics is the moduli space of complex
structures and it is denoted $\mathcal{M}_{cs}$. Set
\begin{equation}
\chi_i:={1\over2}\chi_{i\m\n\bar \l}\d x^\m\w\d x^\n\w\d \bar
x^{\bar \l}\ \ \ \mbox{with}\ \ \ \chi_{i\m\n\bar
\l}:=-{1\over2}\O_{\m\n}^{\ \ \bar \r}{\partial g_{\bar \l\bar
\r}\over\partial z^i}\ ,\label{chi}
\end{equation}
where the $z^i$ for $i\in\{1,\ldots,h_{2,1}\}$ are the parameters
for the complex structure deformation, i.e. they are coordinates
on $\mathcal{M}_{cs}$. Clearly, $\chi_i$ is a $(2,1)$-form
$\forall i$. One finds
\begin{equation}
\bar{\O}_{\bar \r}^{\ \m\n}\chi_{i\m\n\bar \l}=-||\O||^2{\partial
g_{\bar \r\bar \l}\over \partial z^i}\ ,
\end{equation}
where we used that $\O_{\m\n\l}=\O_{123}\tilde\e_{\m\n\l}$ with
$\tilde \e$ a tensor density, and
$||\O||^2:=\sqrt{g}^{-1}\O_{123}\bar\O_{123}={1\over3!}\O_{\m\n\r}\bar\O^{\m\n\r}$.
This gives
\begin{equation}
\delta g_{\bar \r\bar \l}=-{1\over||\O||^2}\bar \O_{\bar \r}^{\ \
\m\n}\chi_{i\m\n\bar \l}\delta z^i\label{delg}\ .
\end{equation}
We saw that the metric on moduli space can be written in block
diagonal form. At the moment we are interested in the complex
structure only and we write a metric on $\mathcal{M}_{cs}$
\begin{equation}
2G_{i\bar j}^{(cs)}\delta z^i\delta \bar z^{\bar j}:={1\over
2\rm{vol}(X)}\int_X g^{\k\bar \n}g^{\m\bar \l}\delta
g_{\k\m}\delta g_{\bar \l\bar \n}\sqrt{g}\ \d ^6x\
.\label{CSmodmetric}
\end{equation}
Using (\ref{delg}) we find
\begin{equation}
G_{i\bar j}^{(cs)}=-{\int_X\chi_i\w\bar\chi_{\bar
j}\over\int_X\O\w\bar \O}\ ,
\end{equation}
where we used that $||\O||^2$ is a constant on $X$, which follows
from the fact that $\O$ is covariantly constant, and
$\int_X\O\w\bar\O=||\O||^2\rm{vol}(X)$. The factor of 2
multiplying $G_{i\bar j}^{(cs)}$ was chosen to make this formula
simple. To proceed we need the important formula
\begin{equation}
{\partial \O\over \partial z^i}=k_i\O+\chi_i\
,\label{partialOmega}
\end{equation}
where $k_i$ may depend on the $z^j$ but not on the coordinates of
$X$. See for example \cite{CdlO91} for a proof. Using
(\ref{partialOmega}) it is easy to show that
\begin{equation}
G_{i\bar j}^{(cs)}=-{\partial\over\partial
z^i}{\partial\over\partial \bar z^{\bar
j}}\log\left(i\int_X\O\w\bar\O\right)\ ,
\end{equation}
which tells us that the metric (\ref{CSmodmetric}) on the moduli
space of complex structures $\mathcal{M}_{cs}$ is K\"ahler with
K\"ahler potential
\begin{equation}
K=-\log\left(i\int_X\O\w\bar\O \right)\ .
\end{equation}
If we differentiate this equation with respect to $z^i$ we can use
(\ref{partialOmega}) to find that $k_i=-{\partial\over\partial
z^i}K$.

Next we consider the {\it Hodge bundle} $\mathcal{H}$, which is
nothing but the cohomology bundle over $\mathcal{M}_{cs}$ s.t. at
a given point $z\in\mathcal{M}_{cs}$ the fiber is given by
$H^3(X_z;\mathbb{C})$, where $X_z$ is the manifold $X$ equipped
with the complex structure $J(z)$ determined by the point
$z\in\mathcal{M}_{cs}$. This bundle comes with a natural flat
connection which is known as the {\it Gauss-Manin connection}. Let
us explain this in more detail. One defines a Hermitian metric on
the Hodge bundle as
\begin{equation}
(\eta,\t):=i\int_X\eta\w\t\ \ \ \mbox{for}\ \eta,\ \t\in
H^3(X;\mathbb{C})\ .
\end{equation}
This allows us to define a symplectic basis of real integer
three-forms $\o_I,\eta^J\in H^3(X;\mathbb{C})$,
$I,J\in\{0,1,\ldots h^{2,1}\}$ with the property
\begin{equation}
(\o_I,\eta^J)=-(\eta^J,\o_I)=\delta_I^J\ ,
\end{equation}
which is unique up to a symplectic transformation. Dual to the
basis of $H^3(X;\mathbb{Z})\subset H^3(X;\mathbb{C})$ there is a
symplectic basis of three-cycles $\{\G_{\a^I},\G_{\b_J}\}\in
H_3(X;\mathbb{Z})$, s.t.
\begin{equation}
\int_{\G_{\a^I}}\o_J=\delta_J^I\ \ \ ,\ \ \
\int_{\G_{\b_I}}\eta^J=\delta_I^J\ ,
\end{equation}
and all other combinations vanish. Clearly, the corresponding
intersection matrix is\\
\parbox{14cm}{
\begin{eqnarray}
&&\G_{\a^I}\cap \G_{\a^j}=\emptyset\ \ \ ,\ \ \ \G_{\b_J}\cap \G_{\b_J}=\emptyset\ ,\nonumber\\
&&\G_{\a^I}\cap \G_{\b_J}=-\G_{\b_j}\cap \G_{\a^i}=\delta_j^i\
,\nonumber
\end{eqnarray}}\hfill\parbox{8mm}{\begin{eqnarray}\end{eqnarray}}\\
see for example \cite{GH} for a detailed treatment of these
issues. Note that the Hermitian metric is defined on every fibre
of the Hodge bundle, so we actually find three-forms and
three-cycles at every point, $\o_I(z),\ \eta^J(z),\ \G_{\a^I}(z),\
\G_{\b^J}(z)$. Since the topology of $X$ does not change if we
move in the moduli space one can identify the set of basis cycles
at one point $p_1$ in moduli space with the set of basis cycles at
another point $p_2$. To do so one takes a path connecting $p_1$
and $p_2$ and identifies a cycle at $p_1$ with the cycle at $p_2$
which arises from the cycle at $p_1$ by following the chosen
path.\footnote{This might sound somewhat complicated but it is in
fact very easy. Consider as an example a torus $T^2$ with its
standard cycles $\a,\b$. If we now change the size (or the shape)
of the torus we will find a new set of cycles $\a',\b'$. These can
in principle be chosen arbitrarily. However, there is a natural
choice which we want to identify with $\a,\b$, namely those that
arise from $\a,\b$ by performing the scaling.} For a detailed
explanation in the context of singularity theory see \cite{AGLV}.
Note that this identification is unique if the space is simply
connected. The corresponding connection is the Gauss-Manin
connection. Clearly, this connection is flat, since on a simply
connected domain of $\mathcal{M}_{cs}$ the identification
procedure does not depend on the chosen path. If the domain is not
simply connected going around a non-contractible closed path in
moduli space will lead to a monodromy transformation of the
cycles. Since we found that there is a natural way to identify
basis elements of $H_3(X_z;\mathbb{Z})$ at different points
$z\in\mathcal{M}_{cs}$ we can also identify the corresponding dual
elements $\o_I(z),\eta^J(z)$. Then, by definition, a section $\s$
of the cohomology bundle is covariantly constant with respect to
the Gauss-Manin connection if, when expressed in terms of the
basis elements $\o_I(z),\eta^J(z)$, its coefficients do not change
if we move around in $\mathcal{M}_{cs}$,
\begin{equation}
\s(z)=\sum_{I=0}^{h^{2,1}(X)}c^I\o_I(z)+\sum_{J=0}^{h^{2,1}(X)}d_J\eta^J(z)\
\ \ \forall\ z\in\mathcal{M}_{cs}\ .
\end{equation}
In particular, the basis elements $\o_I,\eta^J$ are covariantly
constant. A holomorphic section $\r$ of the cohomology bundle is
given by
\begin{equation}
\rho(z)=\sum_{I=0}^{h^{2,1}(X)}f^I(z)\o_I(z)+\sum_{J=0}^{h^{2,1}(X)}g_J(z)\eta^J(z)\
,
\end{equation}
where $f^I(z), g_J(z)$ are holomorphic functions on
$\mathcal{M}_{cs}$.

If we move in the base space $\mathcal{M}_{cs}$ of the Hodge
bundle $\mathcal{H}$ we change the complex structure on $X$ and
therefore we change the Hodge-decomposition of the fibre
$H^3(X_z;\mathbb{C})=\oplus_{k=0}^3H^{(3-k,k)}(X_z)$. Thus
studying the moduli space of Calabi-Yau manifolds amounts to
studying the variation of the Hodge structure of the Hodge bundle.
Consider the holomorphic $(3,0)$-form, defined on every fibre
$X_z$. The set of all these forms defines a holomorphic section of
the Hodge bundle. On a given fibre $X_z$ the form $\O$ is only
defined up to multiplication by a non-zero constant. The section
$\O$ of the Hodge bundle is then defined only up to a
multiplication of a nowhere vanishing holomorphic function
$e^{f(z)}$ on $\mathcal{M}_{cs}$ which amounts to saying that one
can multiply $\O$ by different constants on different fibres, as
long as they vary holomorphically in $z$. It is interesting to see
in which way this fact is related to the properties of the
K\"ahler metric on $\mathcal{M}_{cs}$. The property that $\O$ is
defined only up to multiplication of a holomorphic function,
\begin{equation}
\O\rightarrow\O':=e^{f(z)}\O\ ,\label{transOmega}
\end{equation}
implies
\begin{eqnarray}
K&\rightarrow& \tilde K=K+f(z)+\bar f(\bar z)\ ,\\
G_{i\bar j}^{(CS)}&\rightarrow& \tilde G_{i\bar j}^{(CS)}=G_{i\bar
j}^{(CS)}\ .
\end{eqnarray}
So a change of $\O(z)$ can be understood as a K\"ahler
transformation which leaves the metric on moduli space unchanged.

Next we express the holomorphic section $\O$ of the Hodge bundle
as\footnote{Since the $\o_I,\eta_J$ on different fibres are
identified we omit their $z$-dependence.}
\begin{equation}
\O(z)=X^I(z)\o_I+\mathcal{F}_J(z)\eta^J\ ,
\end{equation}
where we have
\begin{equation}
X^I(z)=\int_{\G_{\a^I}}\O(z)\ \ ,\ \
\mathcal{F}_J(z)=\int_{\G_{\b_J}}\O(z)\ .
\end{equation}
Since $\O(z)$ is holomorphic both $X^I$ and $\mathcal{F}_J$ are
holomorphic functions of the coordinates $z$ on moduli space. By
definition for two points in moduli space, say $z,z'$, the
corresponding three-forms $\O(z),\O(z')$ are different. Since the
bases $\o^I(z),\eta_j(z)$ and $\o^I(z'),\eta_j(z')$ at the two
points are identified and used to compare forms at different
points in moduli space it is the coefficients $X^I,\mathcal{F}_J$
that must change if we go from $z$ to $z'$. In fact, we can take a
subset of these, say the $X^I$ to form coordinates on moduli
space. Since the dimension of $\mathcal{M}_{cs}$ is $h^{2,1}$ but
we have $h^{2,1}+1$ functions $X^I$ these have to be homogeneous
coordinates. Then the $\mathcal{F}_J$ can be expressed in terms of
the $X^I$. Let us then take the $X^I$ to be homogeneous
coordinates on $\mathcal{M}_{cs}$ and apply the Riemann bilinear
relation (\ref{RiemannCY}) to
\begin{equation}
\int \O\w{\partial \O\over\partial X^I}=0\ .
\end{equation}
This gives\footnote{To be more precise one should have defined
$\tilde{\mathcal{F}}_J(z):=\int_{\G_{\b_J}}\O(z)$ and
$\mathcal{F}_J(X):=\tilde{\mathcal{F}}_J(z(X))$.}
\begin{equation}
\mathcal{F}_I=X^J{\partial\over\partial
X^I}\mathcal{F}_J={1\over2}\partial_J(X^I\mathcal{F}_I)={\partial\over\partial
X^I}\mathcal{F}\ ,
\end{equation}
where
\begin{equation}
\mathcal{F}:={1\over2}X^I\mathcal{F}_I\ .
\end{equation}
So the $\mathcal{F}_I$ are derivatives of a function
$\mathcal{F}(X)$ which is homogeneous of degree 2. $\mathcal{F}$
is called the {\it prepotential}. This nomenclature comes from the
fact that the K\"ahler potential can itself be expressed in terms
of $\mathcal{F}$,
\begin{equation}
K=-\log\left(
i\int\O\w\bar\O\right)=-\log\left(i\sum_{I=0}^{h^{2,1}}\left(X^I\bar{\mathcal{F}}_I-\bar
X^I\mathcal{F}_I\right)\right)\ .
\end{equation}
We will have to say much more about this structure below.

To summarise, we found that the moduli space of complex structures
of a Calabi-Yau manifold carries a K\"ahler metric with a K\"ahler
potential that can be calculated from the geometry of the
Calabi-Yau. A K\"ahler transformation can be understood as an
irrelevant multiplication of the section $\O(z)$ by a nowhere
vanishing holomorphic function. The coordinates of the moduli
space can be obtained from integrals of $\O(z)$ over the
$\G_{\a^I}$-cycles and the integrals over the corresponding
$\G_{\b_J}$-cycles can then be shown to be derivatives of a
holomorphic function $\mathcal{F}$ in the coordinates $X$, in
terms of which the K\"ahler potential can be expressed. As
explained in appendix \ref{SGCY} these properties determine
$\mathcal{M}_{cs}$ to be a special K\"ahler manifold.

\bigskip
Obviously the next step would be to analyse the structure of the
moduli space of K\"ahler structures, $\mathcal{M}_{KS}$. Indeed,
this has been studied in \cite{CdlO91} and the result is that
$\mathcal{M}_{KS}$ also is a special K\"ahler manifold, with a
K\"ahler potential calculable from some prepotential. However, the
prepotential now is no longer a simple integral in the geometry,
but it receives instanton correction. Hence, in general it is very
hard to calculate it explicitly. This is where results from mirror
symmetry come in useful. Mirror symmetry states that Calabi-Yau
manifolds come in pairs and that the K\"ahler structure
prepotential can be calculated by evaluating the geometric
integral on the mirror manifold and using what is known as the
mirror map. Unfortunately, we cannot delve any further into this
fascinating subject, but must refer the reader to \cite{Gr97} or
\cite{Horietal}.

\subsection{Local Calabi-Yau manifolds}\label{locCY}
After an exposition of the properties of (compact) Calabi-Yau
spaces we now turn to the spaces which are used to geometrically
engineer the gauge theories that we want to study.

\begin{definition}{\em A} local Calabi-Yau manifold {\em is a
non-compact K\"ahler manifold with vanishing first Chern-class.}
\end{definition}

Next we give a series of definitions which will ultimately lead us
to an explicit local Calabi-Yau manifold. For completeness we
start from the definition of $\mathbb{CP}^1$.

\begin{definition}
{\em The} complex projective space
$\mathbb{CP}^1\equiv\mathbb{P}^1$ {\em is defined as
\begin{equation}
\mathbb{C}^2\backslash \{0\}\slash \sim\ .
\end{equation}
For $(z_1,z_2)\in\mathbb{C}^2\backslash \{0\}$ the equivalence
relation is defined as
\begin{equation}
(z_1,z_2)\sim(\l z_1,\l z_2)=\l(z_1,z_2)
\end{equation}
for $\l\in\mathbb{C}\backslash\{0\}$. Note that this implies that
$\mathbb{CP}^1$ is the space of lines through 0 in $\mathbb{C}^2$.
We can introduce patches on $\mathbb{CP}^1$ as follows\\
\parbox{14cm}{
\begin{eqnarray}
U_1^{\mathbb{P}^1}&:=&\{(z_1,z_2)\in\mathbb{C}^2\backslash\{0\}:z_1\neq 0,(z_1,z_2)\sim\l(z_1,z_2)\}\ ,\nonumber\\
U_2^{\mathbb{P}^1}&:=&\{(z_1,z_2)\in\mathbb{C}^2\backslash\{0\}:z_2\neq
0,(z_1,z_2)\sim\l(z_1,z_2)\}\ ,\nonumber
\end{eqnarray}}\hfill\parbox{8mm}{\begin{eqnarray}\end{eqnarray}}\\
and on these patches we can introduce coordinates
\begin{equation}
\xi_{1}:={z_2\over z_1}\ \ \ ,\ \ \ \xi_{2}:={z_1\over z_2}\ .
\end{equation}
On the overlap $U_1\cap U_2$ we have
\begin{equation}
\xi_{2}={1\over\xi_{1}}\ ,
\end{equation}
and we find that $\mathbb{CP}^1$ is isomorphic to a Riemann sphere
$S^2$.}
\end{definition}

\begin{definition}
{\em The space $\mathcal{O}(n)\rightarrow \mathbb{CP}^1$ is a line
bundle over $\mathbb{CP}^1$. We can define it in terms of charts\\
\parbox{14cm}{
\begin{eqnarray}
U_1&:=&\{(\xi_1,\Phi):\xi_1\in
U_1^{\mathbb{P}^1}\cong\mathbb{C},\Phi\in\mathbb{C}\}\ ,\nonumber\\
U_2&:=&\{(\xi_2,\Phi'):\xi_2\in
U_2^{\mathbb{P}^1}\cong\mathbb{C},\Phi'\in\mathbb{C}\}\ ,\nonumber
\end{eqnarray}}\hfill\parbox{8mm}{\begin{eqnarray}\end{eqnarray}}\\
with
\begin{equation}
\xi_2={1\over\xi_1}\ \ \ , \ \ \ \Phi'=\xi_1^{-n}\Phi
\end{equation}
on $U_1\cap U_2$.}
\end{definition}

\begin{definition}\label{OmOn}
{\em Very similarly $\mathcal{O}(m)\oplus\mathcal{O}(n)\rightarrow
\mathbb{CP}^1$ is a fibre bundle over $\mathbb{CP}^1$ where the
fibre is a direct sum of two complex planes. We define it via
coordinate charts and transition functions\\
\parbox{14cm}{
\begin{eqnarray}
U_1&:=&\{(\xi_1,\Phi_0,\Phi_1):\xi_1\in
U_1^{\mathbb{P}^1},\Phi_0\in\mathbb{C},\Phi_1\in\mathbb{C}\}\ ,\nonumber\\
U_2&:=&\{(\xi_2,\Phi_0',\Phi_1'):\xi_2\in
U_2^{\mathbb{P}^1},\Phi_0'\in\mathbb{C},\Phi_1'\in\mathbb{C}\}\
,\nonumber
\end{eqnarray}}\hfill\parbox{8mm}{\begin{eqnarray}\end{eqnarray}}\\
with
\begin{eqnarray}
\xi_2={1\over\xi_1}\ \ \ , \ \ \ \Phi_0'=\xi_1^{-m}\Phi_0\ \ \ ,\
\ \ \Phi_1'=\xi_1^{-n}\Phi_1\ \ \ \mbox{on}\ U_1\cap U_2\ .
\end{eqnarray}}
\end{definition}
These manifolds are interesting because of the following
proposition, which is explained in \cite{M04b} and \cite{GH}.
\begin{proposition}
The first Chern class of
$\mathcal{O}(m)\oplus\mathcal{O}(n)\rightarrow \mathbb{CP}^1$
vanishes if $m+n=-2$.
\end{proposition}

\begin{figure}[h]
\centering
\includegraphics[width=0.8\textwidth]{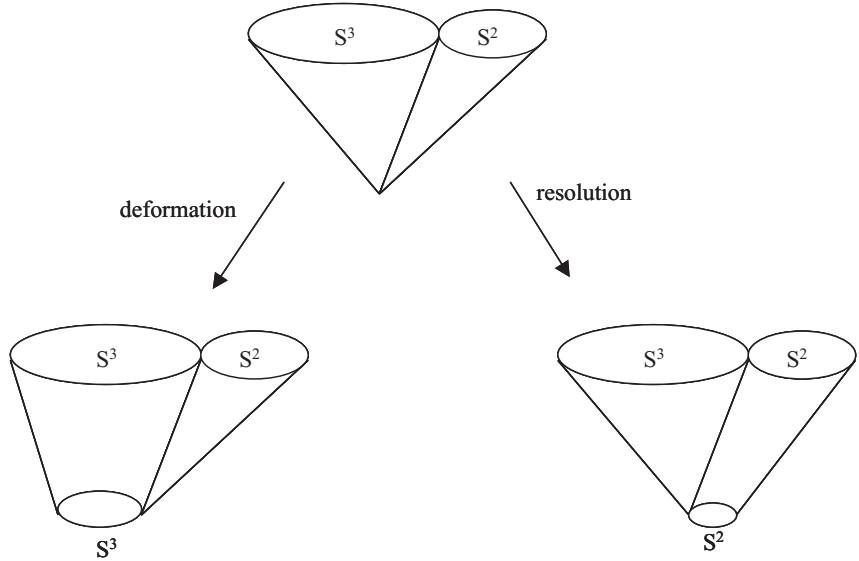}\\
\caption[]{The conifold is a cone over $S^3\times S^2$ with a
conical singularity at its tip that can be smoothed out by a
deformation or a small resolution.} \label{conifold}
\end{figure}

\bigskip
{\bf The conifold}
\begin{definition}{\em
The} conifold $C_0$ {\em is defined as $f^{-1}(0)$ with $f$ given by}\\
\parbox{14cm}{
\begin{eqnarray}
f:\mathbb{C}^4&\rightarrow&\mathbb{C}\nonumber\\
(w_1,w_2,w_3,w_4)&\mapsto
&f(w_1,w_2,w_4,w_4):=w_1^2+w_2^2+w_3^2+w_4^2\ .\nonumber
\end{eqnarray}}\hfill\parbox{8mm}{\begin{eqnarray}\label{conifoldsing}\end{eqnarray}}\\
\end{definition}
In other words
\begin{equation}
C_0:=\{\vec w\in\mathbb{C}^4:w_1^2+w_2^2+w_3^2+w_4^2=0\}\ .
\end{equation}
Setting $\Phi_0=w_1+iw_2$, $\Phi_1=iw_3-w_4$, $\Phi_0'=iw_3+w_4$
and $\Phi_1'=w_1-iw_2$ this reads
\begin{equation}
C_0:=\{(\Phi_0,\Phi_0',\Phi_1,\Phi_1')^\tau\in\mathbb{C}^4:\Phi_0\Phi_1'-\Phi_0'\Phi_1=0\}\
.
\end{equation}
Clearly, $f$ has a singularity at zero with singular value zero
and, therefore $C_0$ is a singular manifold. To study the
structure of $C_0$ in more detail we set $w_i=x_i+iy_i$. Then
$f=0$ reads
\begin{equation}
{\vec x}^2-{\vec y}^2=0\ \ \ ,\ \ \ \vec x\cdot\vec y=0\ .
\end{equation}
The first equation is ${\vec x}^2={1\over2}r^2$ if $r^2:=\vec
x^2+\vec y^2$, so $\vec x$ lives on an $S^3$. $\vec y$ on the
other hand is perpendicular to $\vec x$. For given $r$ and $x$ we
have an $S^2$ of possible $\vec y\,$s and so for given $r$ we have
a fibre bundle of $S^2$ over $S^3$. However, there is no
nontrivial fibration of $S^2$ over $S^3$ and we conclude that
$f=0$ is a cone over $S^3\times S^2$. Fig. \ref{conifold} gives an
intuitive picture of the conifold, together with the two possible
ways to resolve the singularity, namely its deformation and its
small resolution, to which we will turn presently.
\\

\bigskip
{\bf The deformed conifold}
\begin{definition}
{\em The} deformed conifold $C_{def}$ {\em is the set $f^{-1}(\m)$
with $\m\in\mathbb{R}_+$ and  $f$ as in (\ref{conifoldsing}).}
\end{definition}
In other words the deformed conifold is given by
\begin{eqnarray}
C_{def}:=\{\vec w\in\mathbb{C}^4:w_1^2+w_2^2+w_3^2+w_4^2=\m\}
\end{eqnarray}
or
\begin{equation}
C_{def}:=\{(\Phi_0,\Phi_0',\Phi_1,\Phi_1')^\tau\in\mathbb{C}^4:\Phi_0\Phi_1'-\Phi_0'\Phi_1=\m\}\
.\label{defcon}
\end{equation}
The above analysis of the structure of the conifold remains valid
for fixed $r$, where again we have $S^3\times S^2$. However, now
we have an $S^3$ of minimal radius $\sqrt{\m}$ that occurs for
$\vec y=0$. If we define $\vec q:={1\over\sqrt{\m+\vec y^2}}\vec
x$ we find
\begin{equation}
{\vec q}^{\ 2}=1\ \ \ ,\ \ \ \vec q\cdot\vec y=0\ .
\end{equation}
This shows that the deformed conifold is isomorphic to $T^*S^3$.
Both the deformed conifold and the conifold are local Calabi-Yau.
Interestingly, for these spaces this fact can be proven by writing
down an explicit Ricci-flat K\"ahler metric \cite{CdlO90}.

\bigskip
{\bf The resolved conifold}\\
The resolved conifold is given by the small resolution (see Def.
\ref{smallresolution}) of the set $f=0$. We saw that the singular
space can be characterised by
\begin{equation}
C:=\{(\Phi_0,\Phi_0',\Phi_1,\Phi_1')\in\mathbb{C}^4:\Phi_0\Phi_1'=\Phi_0'\Phi_1\}\
.
\end{equation}
The small resolution of this space at $\vec 0\in
C\subset\mathbb{C}^4$ is in fact the space $\tilde
C:=C_{res}:=\mathcal{O}(-1)\oplus\mathcal{O}(-1)\rightarrow\mathbb{CP}^1$.
To see this we have to construct a map $\pi:\tilde C\rightarrow C$
such that $\pi:\tilde C\backslash\pi^{-1}(\vec 0)\rightarrow
C\backslash\vec0$ is an isomorphism and $\pi^{-1}(\vec
0)\cong\mathbb{CP}^1$. On the two patches of $\tilde C$ it is
given by\\
\parbox{14cm}{
\begin{eqnarray}
(\xi_1,\Phi_0,\Phi_1)&\stackrel{\pi}{\mapsto}&(\Phi_0,\xi_1\Phi_0,\Phi_1,\xi_1\Phi_1)\ ,\nonumber\\
(\xi_2,\Phi_0',\Phi_1')&\stackrel\pi\mapsto&(\xi_2\Phi_0',\Phi_0',\xi_2\Phi_1',\Phi_1')\
.\nonumber
\end{eqnarray}}\hfill\parbox{8mm}{\begin{eqnarray}\end{eqnarray}}\\
A point in the overlap of the two charts in $\tilde C$ has to be
mapped to the same point in $C$, which is indeed the case, since
on the overlap we have $\xi_2={1\over\xi_1}$. Note also that as
long as $\Phi_0,\Phi_1$ do not vanish simultaneously this map is
one to one. However, $(U_1^{\mathbb{P}^1},0,0)$ and
$(U_2^{\mathbb{P}^1},0,0)$ are mapped to $(0,0,0,0)$, s.t.
$\pi^{-1}((0,0,0,0))\cong \mathbb{P}^1$. this proves that
$\mathcal{O}(-1)\oplus\mathcal{O}(-1)\rightarrow \mathbb{CP}^1$ is
indeed the small resolution of the conifold.

As in the case of the deformed conifold one can write down a
Ricci-flat K\"ahler metric for the resolved conifold, see
\cite{CdlO90}.

\bigskip
{\bf The resolved conifold and toric geometry}\\
There is another very important description of the resolved
conifold which appears in the context of linear sigma models and
makes contact with toric geometry (see for example \cite{Wi93},
\cite{HV00}, \cite{Horietal}). Let $\vec
z=(z_1,z_2,z_3,z_4)^\tau\in\mathbb{C}^4$ and consider the space
\begin{equation}
C_{toric}:=\{\vec z\in
\mathbb{C}^4:|z_1|^2+|z_2|^2-|z_3|^2-|z_4|^2=t\}\slash \sim
\end{equation}
where the equivalence relation is generated by a $U(1)$ group that
acts as
\begin{equation}
(z_1,z_2,z_3,z_4)\mapsto(e^{i\theta}z_1,e^{i\theta}z_2,e^{-i\theta}z_3,e^{-i\theta}z_4)\
.
\end{equation}
This description appears naturally in the linear sigma model . In
order to see that this is indeed isomorphic to $C_{res}$ note that
for $z_3=z_4=0$ the space is isomorphic to $\mathbb{CP}^1$.
Consider then the sets $U_i:=\{\vec z\in C_{toric}:z_i\neq0\}$,
$i=1,2$. On $U_1$ we define the $U(1)$ invariant coordinates
\begin{equation}
\xi_1:={z_2\over z_1}\ ,\ \Phi_0:=z_1z_3\ ,\ \Phi_1:=z_1z_4\ ,
\end{equation}
and similarly for $U_2$,
\begin{equation}
\xi_2:={z_1\over z_2}\ ,\ \Phi_0':=z_2z_3\ ,\ \Phi_1':=z_2z_4\ ,
\end{equation}
Clearly, the $\xi_i$ are the inhomogeneous coordinate on
$U_i^{\mathbb{P}^1}$. On the overlap $U_1\cap U_2$ we have
\begin{equation}
\xi_2=\xi_1^{-1}\ ,\ \Phi_0'=\xi_1\Phi_0\ ,\ \Phi_1'=\xi_1\Phi_1\
,
\end{equation}
which are the defining equations of $C_{res}$. Therefore, indeed,
$C_{res}\cong C_{toric}$.

\begin{figure}[h]
\centering
\includegraphics[width=0.6\textwidth]{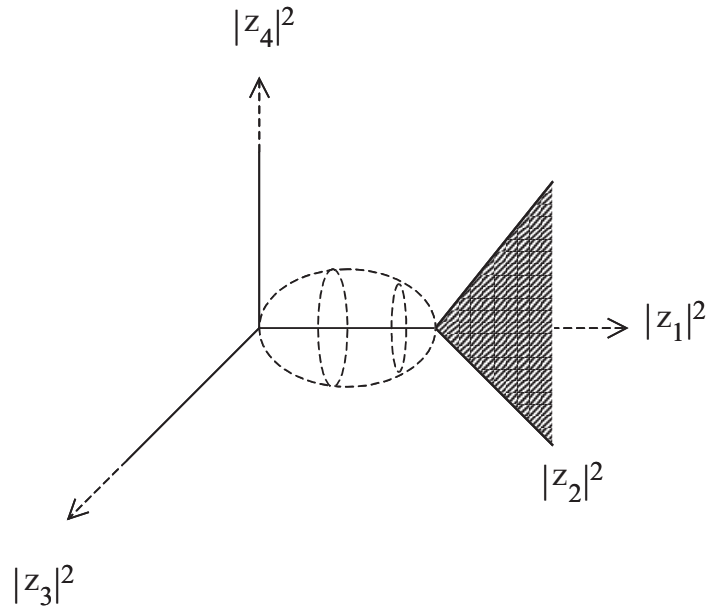}\\
\caption[]{The part of $\mathbb{R}^3_+$ that is the base of the
$T^3$ fibration of the resolved conifold is bounded by the
surfaces $|z_1|^2=0,\ |z_3|^2=0,\ |z_4|^2=0$ and $|z_2|^2=0$. If
$k$ of these equations are satisfied simultaneously, $k$ of the
$S^1$s in $T^3$ shrinks to zero size. In particular, at
$|z_3|^2=|z_4|^2=0$ one has a single $S^1$ in the fibre that
shrinks at $|z_1|^2=0$ and $|z_1|^2=t$. The set of these $S^1$
form the $\mathbb{P}^1$ in $C_{res}$.} \label{rescon}
\end{figure}

From this description of the resolved conifold we can now
understand it as a $T^3$ fibration over (part of) $\mathbb{R}^3_+$
parameterised by $|z_1|^2,\ |z_3|^2,\ |z_4|^2$. Because of the
defining equation $|z_2|^2=t-|z_1|^2+|z_3|^2+|z_4|^2$ the base
cannot consist of the entire $\mathbb{R}^3_+$. For example for
$z_3=z_4=0$ and $|z_1|^2>t$ this equation has no solution. In
fact, the boundary of the base is given by the hypersurfaces
$|z_1|^2=0,\ |z_3|^2=0,\ |z_4|^2=0$ and $|z_2|^2=0$, see Fig.
\ref{rescon}.
The $T^3$ of the fiber is given by the phases of all the $z_i$
modulo the $U(1)$ transformation. The singularity locus of this
fibration is then easily determined. In fact on every hypersurface
$|z_i|^2=0$ the corresponding $S^1$ shrinks to zero size and the
fibre consists of a $T^2$ only. At the loci where two of these
surfaces intersect two $S^1$s shrink and we are left with an
$S^1$. Finally, there are two points where three hypersurfaces
intersect and the fibre degenerates to a point. This happens at
$|z_3|^2=|z_4|^2=0$ and $|z_1|^2=0$ or $|z_1|^2=t$. If we follow
the $|z_1|^2$-axis from 0 to $t$ an $S^1$ opens up and shrinks
again to zero. The set of these cycles form a sphere
$S^2\cong\mathbb{P}^1$, which is the $\mathbb{P}^1$ in $C_{res}$.
For a detailed description of these circle fibrations see for
example \cite{LV97}.

\bigskip
{\bf More general local Calabi-Yau manifolds}\\
There is a set of more general local Calabi-Yau manifolds that was
first constructed in \cite{Fr86} and which appeared in the physics
literature in \cite{KKLM99} and \cite{CIV01}. One starts from the
bundle $\mathcal{O}(-2)\oplus\mathcal{O}(0)\rightarrow
\mathbb{P}^1$ which is local Calabi-Yau. To make the discussion
clear we once again write down the charts and the transition
functions,
\begin{eqnarray}
U_1:=\{(\xi_1,\Phi_0,\Phi_1):\xi_1\in
U_1^{\mathbb{P}^1},\Phi_0\in\mathbb{C},\Phi_1\in\mathbb{C}\}\nonumber\\
U_2:=\{(\xi_2,\Phi_0',\Phi_1'):\xi_1\in
U_2^{\mathbb{P}^1},\Phi_0'\in\mathbb{C},\Phi_1'\in\mathbb{C}\}\nonumber\\
\mbox{with}\ \ \ \xi_2={1\over\xi_1}\ \ \ , \ \ \
\Phi_0'=\xi_1^{2}\Phi_0\ \ \ ,\ \ \ \Phi_1'=\Phi_1\ \ \ \mbox{on}\
\  U_1\cap U_2\ .
\end{eqnarray}
To get an intuitive picture of the structure of this space, we
take $\Phi_0=\Phi_0'=0$ and fix $\Phi_1=\Phi_1'$ arbitrarily. Then
we can ``walk around" in the $\xi_i$ direction ``consistently",
i.e. we can change $\xi_i$ without having to change the fixed
values of $\Phi_0,\Phi_1$.

Next we consider a space $X_{res}$ with coordinate patches
$U_1,U_2$ as above but with transition functions
\begin{equation}\label{Xres}
\xi_2={1\over\xi_1}\ \ \ , \ \ \
\Phi_0'=\xi_1^{2}\Phi_0+W'(\Phi_1)\xi_1\ \ \ ,\ \ \
\Phi_1'=\Phi_1\ \ \ \mbox{on}\ U_1\cap U_2\ .
\end{equation}
Here $W$ is a polynomial of degree $n+1$. To see that the
structure of this space is very different from the one of
$\mathcal{O}(-2)\oplus\mathcal{O}(0)\rightarrow\mathbb{CP}^1$ we
set $\Phi_0=\Phi_0'=0$ and fix $\Phi_1$ arbitrarily, as before.
Note that now $\xi_1$ is fixed and changing $\xi_1$ amounts to
changing $\Phi_0'$. Only for those specific values of $\Phi_1$ for
which $W'(\Phi_1)=0$ can we consistently ``walk" in the
$\xi_1$-direction.
Mathematically this means that the space that is deformed by a
polynomial $W$ contains $n$ different $\mathbb{CP}^1$s.

We have seen that the ``blow-down" of
$\mathcal{O}(-1)\oplus\mathcal{O}(-1)\rightarrow \mathbb{CP}^1$ is
given by the conifold $C$. Now we are interested in the blow-down
geometry of our deformed space, i.e. the geometry where the size
of the $\mathbb{CP}^1$s is taken to zero. We claim that $X_{res}$
can be obtained from a small resolution of all the singularities
of the space
\begin{equation}
X:=\{(\Phi_0,\Phi_0',\Phi_1,z)\in\mathbb{C}^4:
4\Phi_0\Phi_0'+z^2+W'(\Phi_1)^2=0\}\ .\label{X}
\end{equation}
As to prove that $X_{res}$ is the small resolution of $X$ we have
to find a map $\pi:X_{res}\rightarrow X$. It is given by
\begin{eqnarray}
(\xi_1,\Phi_0,\Phi_1)&\stackrel{\pi}{\mapsto}&(\Phi_0,\xi_1^2\Phi_0+W'(\Phi_1)\xi_1,\Phi_1,i(2\xi_1\Phi_0+W'(\Phi_1)))\ ,\nonumber\\
(\xi_2,\Phi_0',\Phi_1')&\stackrel{\pi}{\mapsto}&(\xi_2^2\Phi_0'-\xi_2W'(\Phi_1'),\Phi_0',\Phi_1',i(2\xi_2\Phi_0'-W'(\Phi_1')))\
.\nonumber
\end{eqnarray}
It is easy to check that\\
$\bullet$ $\pi(U_i)\subset X$ ,\\
$\bullet$ on $U_1\cap U_2$ the two maps map to the same point ,\\
$\bullet$ the map is one to one as long as $W'(\Phi_1)\neq0$ ,\\
$\bullet$ for $\Phi_1=\Phi_1'$ s.t. $W'(\Phi_1)=0$ one finds that
$(U_1^{\mathbb{P}^1},0,\Phi_1)$ and
$(U_2^{\mathbb{P}^1},0,\Phi_1')$ are mapped to $(0,0,\Phi_1,0)$,
i.e. $\pi^{-1}((0,0,\Phi_1,0))\cong \mathbb{CP}^1$ $\forall
\Phi_1$ s.t. $W'(\Phi_1)=0$. This shows that $X_{res}$ can indeed
be understood as the small resolution of the singularities in $X$.

Changing coordinates $x=\Phi_1,\ v=\Phi_0+\Phi_0',\
w=i(\Phi_0'-\Phi_0)$ the expression (\ref{X}) for $X$ can be
rewritten as
\begin{equation}
X=\{(v,w,x,z)\in\mathbb{C}^4:W'(x)^2+v^2+w^2+z^2=0\}\
.\label{Xsing}
\end{equation}

\bigskip
Finally, we note that it is easy to find the deformation of the
singularities of (\ref{Xsing}). If $f_0(x)$ is a polynomial of
degree $n-1$ then\\
\parbox{14cm}{
\begin{eqnarray}
X_{def}&:=&\{(v,w,x,z)\in\mathbb{C}^4:F(x,v,w,z)=0\}\label{Xdef}\nonumber\\
\mbox{with}\ \ \ F(x,v,w,z)&:=&W'(x)^2+f_0(x)+v^2+w^2+z^2\nonumber
\end{eqnarray}}\hfill\parbox{8mm}{\begin{eqnarray}\end{eqnarray}}\\
is the space where (for generic coefficients of $f_0$) all the
singularities are deformed. The spaces $X_{res},\ X$ and $X_{def}$
are the local Calabi-Yau manifolds that will be used in order to
geometrically engineer the gauge theories we are interested in.
Going from the resolved space $X_{res}$ through the singular space
to the deformed one $X_{def}$ is known as a {\it geometric
transition}.

\subsection{Period integrals on local Calabi-Yau manifolds and
Riemann surfaces}\label{periodint}
We mentioned already in the introduction that one building block
that is necessary to obtain the effective superpotential
(\ref{GVW}) of gauge theories is given by the integrals of the
holomorphic $(3,0)$-form $\O$, which exists on any (local)
Calabi-Yau manifold, over all the (relative) three-cycles in the
manifold. Here we will analyse these integrals on the space
$X_{def}$, and review how they map to integrals on a Riemann
surface, which is closely related to the local Calabi-Yau manifold
we are considering.

Let us first concentrate on the definition of $\O$. $X_{def}$ is
given by a (non-singular) hypersurface in $\mathbb{C}^4$. Clearly,
on $\mathbb{C}^4$ there is a preferred holomorphic $(4,0)$-form,
namely $\d x\w\d v\w\d w\w\d z$. Since $X_{def}$ is defined by
$F=0$, where $F$ is a holomorphic function in the $x,v,w,z$, the
$(4,0)$-form on $\mathbb{C}^4$ induces a natural holomorphic
$(3,0)$-form on $X_{def}$. To see this note that $\d F$ is
perpendicular to the hypersurface $F=0$. Then there is a unique
holomorphic $(3,0)$-form on $F=0$, such that $\d x\w\d v\w\d w\w\d
z=\O\w\d F$. If $z\neq0$ it can be written as
\begin{equation}\label{Omega}
\O={\d x\w\d v\w\d w\w\d z\over\d F}={\d x\w\d v\w\d w\over2z}\ ,
\end{equation}
where $z$ is a solution of $F=0$. Turning to the three-cycles we
note that, because of the simple dependence of the surface
(\ref{Xdef}) on $v,w$ and $z$ every three-cycle of $X_{def}$ can
be understood as a fibration of a two-sphere over a line segment
in the hyperelliptic Riemann surface $\S$,
\begin{equation}
y^2=W'(x)^2+f_0(x)=:\prod_{i=1}^n(x-a_i^+)(x-a_i^-)\
,\label{heRiemann}
\end{equation}
of genus $\gh=n-1$. This was first realised in \cite{KLMVW96} in a
slightly different context, see \cite{Le96} for a review. As
explained above, this surface can be understood as two complex
planes glued together along cuts running between $a_i^-$ and
$a_i^+$. Following the conventions of \cite{BM05} $y_0$, which is
the branch of the Riemann surface with $y_0\sim W'(x)$ for
$|x|\rightarrow\infty$, is defined on the upper sheet and
$y_1=-y_0$ on the lower one. For compact three-cycles the line
segment connects two of the branch points of the curve and the
volume of the $S^2$-fibre depends on the position on the base line
segment. At the end points of the segment one has $y^2=0$ and the
volume of the sphere shrinks to zero size. Non-compact
three-cycles on the other hand are fibrations of $S^2$ over a
half-line that runs from one of the branch points to infinity on
the Riemann surface. Integration over the fibre is elementary and
gives
\begin{equation}\label{fibration}
\int_{S^2}\O=\pm 2\pi i\ y(x)\d x\ ,
\end{equation}
(the sign ambiguity will be fixed momentarily) and thus the
integral of the holomorphic $\O$ over a three-cycle is reduced to
an integral of $\pm 2\pi i y\d x$ over a line segment in $\S$.
Clearly, the integrals over line segments that connect two branch
points can be rewritten in terms of integrals over compact cycles
on the Riemann surface, whereas the integrals over non-compact
three-cycles can be expressed as integrals over a line that links
the two infinities on the two complex sheets. In fact, the
one-form
\begin{equation}
\zeta:=y\d x
\end{equation}
is meromorphic and diverges at infinity (poles of order $n+2$) on
the two sheets and therefore it is well-defined only on the
Riemann surface with the two infinities $Q$ and $Q'$ removed.
Then, we are naturally led to consider the relative homology
$H_1(\S,\{Q,Q'\})$, which we encountered already when we discussed
Riemann surfaces in section \ref{propRS}. To summarise, one ends
up with a one-to-one correspondence between the (compact and
non-compact) three-cycles in (\ref{Xdef}) and $H_1(\S,\{Q,Q'\})$.
Referring to our choice of bases $\{A^i,B_j\}$ respectively
$\{\a^i,\b_j\}$ for $H_1(\S,\{Q,Q'\})$, as defined in Figs.
\ref{ABcycles} and \ref{albecycles}, we define $\G_{A^i},\G_{B_j}$
to be the $S^2$-fibrations over $A^i,B_j$, and
$\G_{\a^i},\G_{\b_j}$ are $S^2$-fibrations over $\a^i,\b_j$. So
the problem effectively reduces to calculating the
integrals\footnote{The sign ambiguity of (\ref{fibration}) has now
been fixed, since we have made specific choices for the
orientation of the cycles. Furthermore, we use the (standard)
convention that the cut of $\sqrt{x}$ is along the negative real
axis of the complex $x$-plane. Also, on the right-hand side we
used that the integral of $\zeta$ over the line segment is $1\over
2$ times the integral over a closed cycle $\g$.}
\begin{equation}
\int_{\G_{\g}}\O=-i\pi\int_{\g}\zeta\ \ \ \mbox{for}\ \ \
\g\in\{\a^i,\b_j,\hat\a,\hat\b\}\ .\label{integrals}
\end{equation}
As we will see in the next chapter, these integrals can actually
be calculated from a holomorphic matrix model.

As mentioned already, one expects new features to be contained in
the integral $\int_{\hat\b}\zeta$, where $\hat\b$ runs from $Q'$
on the lower sheet to $Q$ on the upper one. Indeed, it is easy to
see that this integral is divergent. It will be part of our task
to understand and properly treat this divergence. As usual, the
integral will be regulated and one has to make sure that physical
quantities do not depend on the regulator and remain finite once
the regulator is removed. In the literature this is achieved by
simply discarding the divergent part. Here we want to give a more
intrinsic geometric prescription that will be similar to standard
procedures in relative cohomology. To render the integral finite
we simply cut out two ``small'' discs around the points $Q,Q'$. If
$x,x'$ are coordinates on the upper and lower sheet respectively,
one only considers the domains $|x|\leq \L_0$, $|x'|\leq\L_0$,
$\L_0\in\mathbb{R}$. Furthermore, we take the cycle $\hat\b$ to
run from the point $\L_0'$ on the real axis of the lower sheet to
$\L_0$ on the real axis of the upper sheet. (Actually one could
take $\L_0$ and $\L_0'$ to be complex. We will come back to this
point later on.)

\chapter{Holomorphic Matrix Models and Special Geometry}\label{holMM}
After having collected some relevant background material let us
now come back to the main line of our arguments. Our principal
goal is to determine the effective superpotential (in the
Veneziano-Yankielowicz sense, see section \ref{effactionsym}) of
super Yang-Mills theory coupled to a chiral superfield in the
adjoint representation with tree-level superpotential
\begin{equation}
W(\Phi)=\sum_{k=1}^{n+1}{g_k\over k}\tr \Phi^k+g_0\
.\label{potential}
\end{equation}
We mentioned in the introduction that this theory can be
geometrically engineered from type IIB string theory on the local
Calabi-Yau manifolds $X_{res}$ studied in section \ref{locCY}.
Furthermore, as we will review below, Cachazo, Intriligator and
Vafa claim that the effective superpotential of this theory can be
calculated from integrals of $\O$ over all the three-cycles in the
local Calabi-Yau $X_{def}$, which is obtained from $X_{res}$
through a geometric transition.

We reviewed the structure of the moduli space of a {\it compact}
Calabi-Yau manifold $X$, and we found the special geometry
relations\\
\parbox{14cm}{
\begin{eqnarray}
X^I&=&\int_{\G_{A^I}}\O,\nonumber\\
\mF_I\equiv{\partial\mF\over\partial X^I}&=&\int_{\G_{B_I}}\O\
,\nonumber
\end{eqnarray}}\hfill\parbox{8mm}{\begin{eqnarray}\label{SG}\end{eqnarray}}\\
where $\O$ is the unique holomorphic $(3,0)$-form on $X$, and
$\{\G_{A^I},\G_{B_J}\}$ is a symplectic basis of $H_3(X)$. On
Riemann surfaces similar relations hold.

An obvious and important question to ask is whether we can find
special geometry relations on the {\it non-compact} manifolds
$X_{def}$, which would then be relevant for the computation of the
effective superpotential. We already started to calculate the
integrals of $\O$ over the three-cycles in section \ref{periodint}
and we found that they map to integrals of a meromorphic form on a
Riemann surface. It is immediately clear that the naive special
geometry relations have to be modified, since we have at least one
integral over a non-compact cycle $\G_{\hat \b}$, which is
divergent. This can be remedied by introducing a cut-off $\L_0$,
but then the integral over the regulated cycle depends on the
cut-off. The question we want to address in this chapter is how to
evaluate the integrals of $\zeta=y\d x$ of Eq. (\ref{integrals})
on the hyperelliptic Riemann surface (\ref{heRiemann}).
Furthermore, we derive a set of equations for these integrals on
the Riemann surface which is similar to the special geometry
relations (\ref{SG}), but which contain the cut-off $\L_0$.
Finally, a clear cut interpretation of the function $\mF$ that
appears in these relations is given. It turns out to be nothing
but the free energy of a holomorphic matrix model at genus zero.
For this reason we will spend some time explaining the holomorphic
matrix model.

\section{The holomorphic matrix model}
The fact that the holomorphic matric model is relevant in this
context was first discovered by Dijkgraaf and Vafa in
\cite{DV02a}, who noticed that the open topological B-model on
$X_{res}$ is related to a holomorphic matrix model with $W$ as its
potential. Then, in \cite{DV02c} they explored how the matrix
model can be used to evaluate the effective superpotential of a
quantum field theory. A general reference for matrix models is
\cite{DFGZJ93}, particularly important for us are the results of
\cite{BIPZ78}. Although similar to the Hermitean matrix model, the
holomorphic matrix model has been studied only recently. In
\cite{La03} Lazaroiu described many of its intriguing features,
see also \cite{KLLR03}. The subtleties of the saddle point
expansion in this model, as well as some aspects of the special
geometry relations were first studied in our work \cite{BM05}.

\subsection{The partition function and convergence properties}
We begin by defining the partition function of the {\it
holomorphic} one-matrix model following \cite{La03}. In order to
do so, one chooses a smooth path
$\g:\mathbb{R}\rightarrow\mathbb{C}$ without self-intersection,
such that $\dot{\g}(u)\neq 0\ \forall u\in\mathbb{R}$ and
$|\g(u)|\rightarrow \infty$ for $u\rightarrow\pm\infty$. Consider
the ensemble $\G(\g)$ of\footnote{We reserve the letter $N$ for
the number of colours in a $U(N)$ gauge theory. It is important to
distinguish between $N$ in the gauge theory and $\hat{N}$ in the
matrix model.} $\hN\times\hN$ complex matrices $M$ with spectrum
spec$(M)=\{\l_1,\ldots\l_{\hN}\}$ in\footnote{Here and in the
following we will write $\g$ for both the function and its image.}
$\g$ and distinct eigenvalues,
\begin{equation}
\Gamma(\gamma):=\{M\in\mathbb{C}^{\hN\times\hN}:\mbox{spec}(M)\subset\gamma,\
\mbox{all}\ \l_m \ \mbox{distinct}\}\ .
\end{equation}
The holomorphic measure on $\mathbb{C}^{\hat{N}\times \hat{N}}$ is
just $\d M\equiv\wedge_{p,q}\d M_{pq}$ with some appropriate sign
convention \cite{La03}. The potential is given by the tree-level
superpotential of Eq. (\ref{potential})
\begin{equation}
W(x):=g_0+\sum_{k=1}^{n+1} {g_k\over k} x^k,\ \ \ g_{n+1}=1\ .
\end{equation}
Without loss of generality we have chosen $g_{n+1}=1$. The only
restriction for the other complex parameters $\{g_k\}_{k=0,\ldots
n}$, collectively denoted by $g$, comes from the fact that the $n$
critical points $\m_i$ of $W$ should not be degenerate, i.e.
$W''(\m_i)\neq0$ if $W'(x)=\prod_{i=1}^n(x-\m_i)$. Then the
partition function of the holomorphic one-matrix model is
\begin{equation}
Z(\Gamma(\g),g,g_s,\hN):=C_{\hN}\int_{\Gamma(\g)}\d M \
\exp\left(-{1\over g_s} \tr W(M)\right),
\end{equation}
where $g_s$ is a positive coupling constant and $C_{\hN}$ is some
normalisation factor. To avoid cluttering the notation we will
omit the dependence on $\g$ and $g$ and write
$Z(g_s,\hN):=Z(\G(\g),g,g_s,\hN)$. As usual \cite{DFGZJ93} one
diagonalises $M$ and performs the integral over the diagonalising
matrices. The constant $C_{\hN}$ is chosen in such a way that one
arrives at
\begin{equation}\label{partmm}
Z(g_s,\hN)={1\over{\hat{N}!}}\int_\gamma\d\lambda_1\ldots\int_\gamma\d\lambda_{\hat{N}}
\exp\left({-\hN^2S(g_s,\hN;\l_m)}\right)=:e^{-F(g_s,\hN)}\ ,
\end{equation}
where
\begin{equation}
S(g_s,\hN;\l_m)={1\over\hN^2g_s}\sum_{m=1}^{\hN}W(\l_m)-{1\over\hN^2}\sum_{p\neq
q}\ln(\l_p-\l_q)\ .
\end{equation}
See \cite{La03} for more details.

\begin{figure}[h]
\centering
\includegraphics{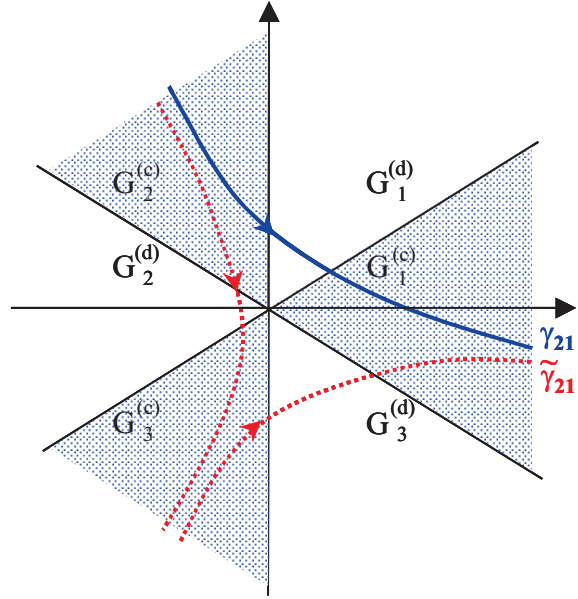}\\
\caption[]{Example of convergence and divergence domains for $n=2$
and a possible choice of $\g_{21}$. Because of holomorphicity the
path can be deformed without changing the partition function, for
instance one could use the path $\tilde\g_{21}$ instead.}
\label{convergence1}
\end{figure}

The convergence of the $\l_m$ integrals depends on the polynomial
$W$ and the choice of the path $\g$. For instance, it is clear
that once we take $W$ to be odd, $\g$ cannot coincide with the
real axis but has to be chosen differently. For given $W$ the
asymptotic part of the complex plane ($|x|$ large) can be divided
into convergence domains $G_l^{(c)}$ and divergence domains
$G_l^{(d)}$, $l=1,\ldots n+1$, where $e^{-{1\over g_s}W(x)}$
converges, respectively diverges as $|x|\rightarrow\infty$. To see
this in more detail take $x=re^{i\theta}$ and
$g_k=r_ke^{-i\theta_k}$, with $r,r_k\geq0$, $\theta,\theta_k\in
\left[0,2\pi\right.)$ for $k=1,\ldots n$ and $r_{n+1}=1$,
$\t_{n+1}=0$, and plug it into the potential,
\begin{equation}
W(re^{i\t})=g_0+\sum_{k=1}^{n+1} {r_k r^k\over
k}\cos(k\theta-\theta_k)+i\sum_{k=1}^{n+1}{r_k r^k\over
k}\sin(k\theta-\theta_k)\ .
\end{equation}
The basic requirement is that $\left|e^{-{1\over
g_s}W(re^{i\t})}\right|=e^{-{1\over g_s} Re W(re^{i\t})}$ should
vanish for $r\rightarrow \infty$. If we fix $\theta$, s.t.
$\cos((n+1)\theta)\neq0$, then $e^{-{1\over g_s}W(re^{i\t})}$
decreases exponentially for $r\rightarrow \infty$ if and only if
$\cos((n+1)\theta)>0$ which gives
\begin{equation}
\theta={{\alpha }\over {n+1}}+\pi {2(l-1)\over {n+1}}\ \ \
\mbox{with} \ l=1,2\ldots,n+1\ \mbox{and}\
\alpha\in(-\pi/2,\pi/2)\ .
\end{equation}
This defines $n+1$ open wedges in $\mathbb{C}$ with apex at the
origin, which we denote by $G_l^{(c)},\ l=1\ldots,n+1$. The
complementary sectors $G_l^{(d)}$ are regions where
$\cos((n+1)\theta)<0$, i.e.
\begin{equation}
\theta={\alpha \over {n+1}}+\pi {{2l-1}\over {n+1}}\ \ \
\mbox{with}\  l=1,2\ldots,n+1\ \mbox{and}\
\alpha\in(-\pi/2,\pi/2)\ .
\end{equation}
The path $\g$ has to be chosen \cite{La03} to go from some
convergence domain $G_k^{(c)}$ to some other $G_l^{(c)}$, with
$k\neq l$; call such a path $\g_{kl}$, see Fig.
\ref{convergence1}.
Note that, for the case of $n+1$ even, the convergence sectors
$G_k^{(c)}$ come in pairs, symmetric with respect to the inversion
$x\rightarrow -x$, so that $G_1^{(c)}$ and $G_{(n+3)/2}^{(c)}$ lie
opposite each other and cover the real axis. Then we can choose
$\g$ to coincide with the real axis. In this case the holomorphic
matrix model reduces to the eigenvalue representation of the
Hermitian matrix model. In the case of odd $n+1$, the image of
$G_k^{(c)}$ under $x\rightarrow -x$ is $G_{k+[(n+1)/2]}^{(d)}$,
and the contour cannot chosen to be the real line. The value of
the partition function depends only on the pair $(k,l)$ and,
because of holomorphicity, is not sensitive to deformations of
$\g_{kl}$. In particular, instead of $\g_{kl}$ we can make the
equivalent choice \cite{BM05}
\begin{equation}
\tilde\g_{kl}=\g_{p_1p_2}\cup\g_{p_2p_3}\cup\ldots\cup\g_{p_{n-1}p_{n}}\cup\g_{p_{n}p_{n+1}}\
\ \ \mbox{with}\ p_1=k,\ p_{n+1}=l ,\label{tildegamma}
\end{equation}
as shown in Fig. \ref{convergence1}. Here we split the path into
$n$ components, each component running from one convergence sector
to another. Again, due to holomorphicity we can choose the
decomposition in such a way that every component $\g_{p_ip_{i+1}}$
runs through one of the $n$ critical points of $W$ in
$\mathbb{C}$, or at least comes close to it. This choice of
$\tilde\g_{kl}$ will turn out to be very useful to understand the
saddle point approximation discussed below. Hence, the partition
function and the free energy depend on the pair $(k,l),g,g_s$ and
$\hN$. Of course, one can always relate the partition function for
arbitrary $(k,l)$ to one with $(k',1)$, $k'=k-l+1$ mod $n$, and
redefined coupling constants $g_1,\ldots g_{n+1}$.

\subsection{Perturbation theory and fatgraphs}\label{fatgraphs}
\begin{figure}[h]
\centering
\includegraphics[width=0.6\textwidth]{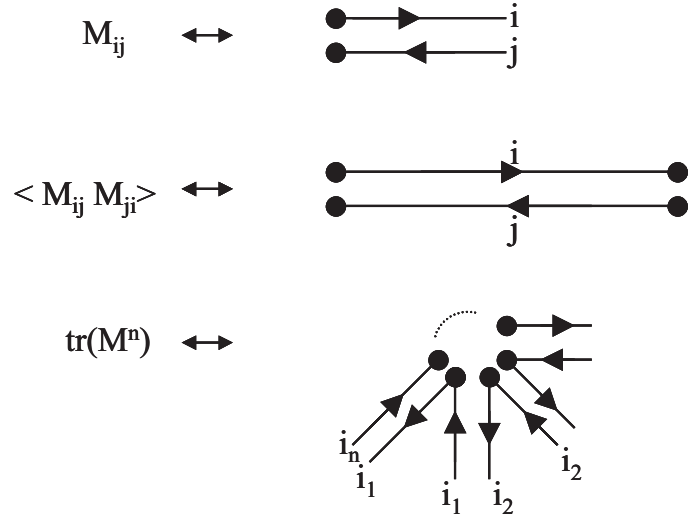}\\
\caption[]{The Feynman diagrammatic representation of the matrix
$M_{ij}$, the propagator $\langle M_{ij}M_{ji}\rangle_G$ and an
$n$-valent vertex $\tr(M^n)$.}\label{Feynman}
\end{figure}

Later we will discuss a method how one can calculate (at least
part of) the free energy non-perturbatively. There is, however,
also a Feynman diagram technique that can be used to evaluate the
partition function. Here we follow the exposition of \cite{DF04},
where more details can be found. Define a Gaussian expectation
value,
\begin{equation}
\langle f(M)\rangle_{G}:={{\int \d M\ f(M)\exp\left(-{1\over g_s}
{m\over2}\tr M^2\right)}\over{\int \d M\ \exp\left(-{1\over
g_s}{m\over2}\tr M^2\right)}}\ ,
\end{equation}
and let us for simplicity work with a cubic superpotential in this
subsection, $W(x)={m\over2}x^2+{g\over3}x^3$, with $m,g$ real and
positive. (Note that we take $g_{n+1}=g\neq1$ in this section,
since we want to do perturbation theory in $g$.) We set
$Z^G:=C_{\hN}\int \d M\ \exp\left(-{1\over g_s} {m\over2}\tr
M^2\right)$ and expand the interaction term,
\begin{eqnarray}
Z(g_s,\hN)&=&C_{\hN}\int_{\G}\d M\ \exp\left(-{1\over
g_s}{m\over2}\tr
M^2\right)\sum_{V=0}^{\infty}{1\over V!}\left(-{1\over g_s}{g\over3}\tr M^3\right)^V\nonumber\\
&=&\sum_{V=0}^{\infty}{1\over V!}Z^G\left\langle \left(-{1\over
g_s}{g\over3}\tr M^3\right)^V\right\rangle_{G}\ .
\end{eqnarray}
The standard way to calculate $\langle (-{1\over g_s}{g\over3}\tr
M^3)^V\rangle_{G}$ is, of course, to introduce sources $J$ with
\begin{equation}
\left\langle \exp\left(\tr
JM\right)\right\rangle_G=\exp\left({g_s\over2m}\tr J^2\right)\ ,
\end{equation}
such that one obtains the propagator
\begin{equation}
\langle M_{ij}M_{kl}\rangle_{G}=\left.{\partial\over\partial
J_{ji}}{\partial\over\partial J_{lk}}\left\langle \exp\left(\tr
JM\right)\right\rangle_G\right|_{J=0}={g_s\over
m}\delta_{il}\delta_{jk}\ .
\end{equation}
More generally one deduces \cite{DF04} the {\it matrix Wick
theorem}
\begin{equation}
\left\langle\prod_{(i,j)\in I} M_{ij}\right\rangle_{G}=\sum_{{\rm
parings}\ P}\prod_{((i,j),(k,l))\in P}\langle
M_{ij}M_{kl}\rangle_{G}\ ,\label{Wick}
\end{equation}
where $I$ is an index family containing pairs $(i,j)$, and the sum
runs over the set of possibilities to group the index pairs
$(i,j)$ again into pairs. Of course the propagator relation tells
us that most of these pairings give zero. The remaining ones can
be captured by Feynman graphs if we establish the diagrammatic
rules of Fig. \ref{Feynman}.
For obvious reasons these Feynman graphs are called {\it
fatgraphs}. We are then left with the relation
\begin{equation}\label{corrs}
\langle(\tr
M^k)^V\rangle_G=\sum_{\begin{array}{c}\mbox{\scriptsize{ fatgraphs
$\G$ with }}\\\mbox{\scriptsize{$V$ k-valent
vertices}}\end{array}}\mN_\G\left({g_s\over
m}\right)^{E(\G)}\hN^{F(\G)}\ ,
\end{equation}
where $F(\G)$ is the number of index loops in the fatgraph $\G$,
$E(\G)$ is the number of propagators and $\mN_\G$ is the number of
different ways the propagators can be glued together to build the
fatgraph $\G$. The partition function for our cubic example can
then be expressed as
\begin{eqnarray}
Z(g_s,\hN)&=&\sum_{V=0}^{\infty}{1\over V!}Z^G\left(-{1\over
g_s}{g\over3}\right)^V\sum_{\begin{array}{c}\mbox{\scriptsize{
fatgraphs $\G$ with}}\\\mbox{\scriptsize{ $V$ 3-valent vertices}}\end{array}}\mN_\G\left({g_s\over m}\right)^{E(\G)}\hN^{F(\G)}\nonumber\\
&=&Z^G\sum_{\mbox{\scriptsize{fatgraphs}}\
\G}\left(-g\right)^{V(\G)}m^{-E(\G)}{1\over|Aut(\G)|}t^Fg_s^{2\gh-2}\
,
\end{eqnarray}
where $|Aut(\G)|$ is the symmetry group of $\G$, $\gh$ is the
genus of the Riemann surface on which the fatgraph $\G$ can be
drawn (c.f. our discussion in the introduction), and
\begin{equation}
t:=g_s\hN
\end{equation}
is the (matrix model) 't Hooft coupling. Similarly, for the free
energy we find
\begin{equation}
F(g_s,t)=\sum_{\mbox{\scriptsize{connected\ fatgraphs}}\
\G}-\left(-g\right)^{V(\G)}m^{-E(\G)}{1\over|Aut(\G)|}t^Fg_s^{2\gh-2}-\log(Z^G)\
.
\end{equation}
Note that this has precisely the structure
\begin{equation}
F(g_s,t)=\sum_{\gh=0}^\infty\sum_{h=1}^\infty
F_{\gh,h}t^hg_s^{2\gh-2}+\mbox{non-perturbative}\ ,
\end{equation}
that we encountered already in the introduction. This tells us
that we can not determine the entire free energy from a Feynman
diagram expansion, since we also have to take care of the
non-perturbative piece. Furthermore, note that in the particular
limit in which $\hN\rightarrow\infty$ with fixed $t$ only planar
diagrams, i.e. those with $\hg=0$, contribute to the perturbative
expansion.

The free energy can then be calculated perturbatively by
following a set of Feynman rules:\\
$\bullet$ To calculate $F$ up to order $g^k$ draw all possible
fatgraph diagrams that contain up to $k$ vertices.\\
$\bullet$ Assign a factor $-{g\over g_s}$ to every vertex.\\
$\bullet$ Assign a factor ${g_s\over m}$ to every fatgraph
propagator.\\
$\bullet$ Assign a factor $\hN$ to each closed index line.\\
$\bullet$ Multiply the contribution of a given diagram by
$|Aut(\G)|^{-1}$, where $Aut(\G)$ is the automorphism group of the
diagram.\\
$\bullet$ Sum all the contributions and multiply the result by an
overall minus sign, which comes from the fact, that
$F(g_s,t)=-\log Z(g_s,t)$.

\begin{figure}[h]
\centering
\includegraphics[width=0.6\textwidth]{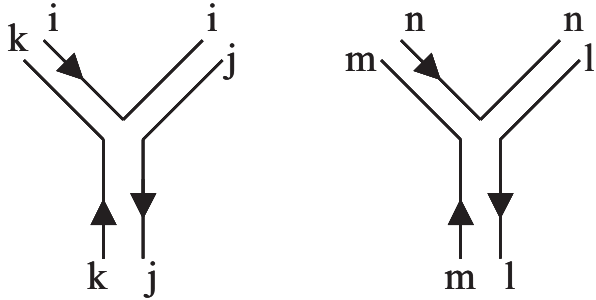}\\
\caption[]{To calculate $\left\langle(\tr M^3)^2\right\rangle$ on
has to sum over all the possibilities to connect two trivalent
vertices.} \label{vertices}
\end{figure}

\begin{figure}[h]
\centering
\includegraphics[width=0.6\textwidth]{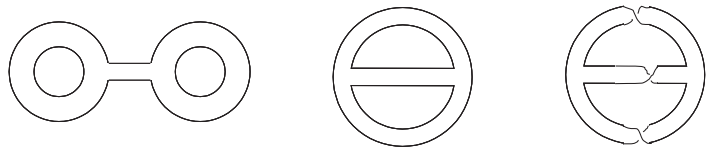}\\
\caption[]{From two three-valent vertices one can draw three
Feynman diagrams, two of which can be drawn on the sphere, whereas
the last one is ``nonplanar".} \label{fatgraphs2}
\end{figure}

\begin{figure}[h]
\centering
\includegraphics[width=0.8\textwidth]{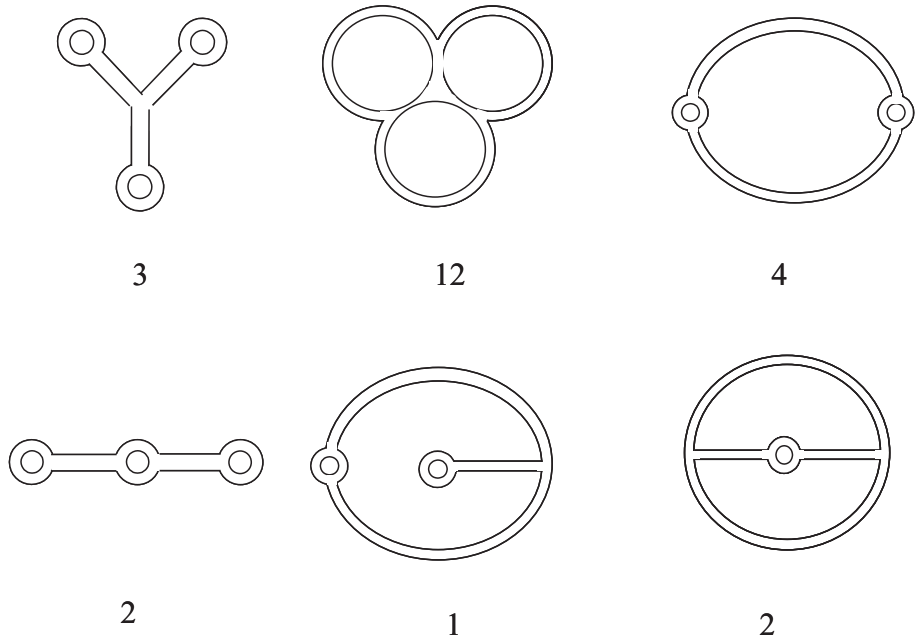}\\
\caption[]{All possible planar diagrams containing four vertices,
together with their symmetry factors $|Aut(\G)|$.}
\label{fourvertices}
\end{figure}

As an example let us calculate the planar free energy up to order
$g^4$. Clearly, we cannot build a vacuum diagram from an odd
number of vertices, which is, of course, consistent with
$\langle(\tr M^3)^V\rangle_G=0$ for odd $V$. The first non-trivial
contribution to the free energy comes from $\langle(\tr
M^3)^2\rangle_G$, which is given by the sum of all possibilities
to connect two trivalent vertices, see Fig. \ref{vertices}.
The three diagrams that contribute are sketched in Fig.
\ref{fatgraphs2}.
According to our rules the first diagram gives ${1\over
g_s^2}{t^3g^2\over m^3}{1\over 2}$, the second one contributes
${1\over g_s^2}{t^3g^2\over m^3}{1\over 6}$ and the last one
${tg^2\over m^3}{1\over 6}$. Then we have\footnote{Alternatively
we can use the explicit formula for $F$. From (\ref{corrs}) one
obtains
\begin{eqnarray}
F(g_s,t)&=&-\log\left(\sum_{V=0}^\infty{1\over V!}\left(-{1\over
g_s}{g\over3}\right)^V\left\langle\left(\tr
M^3\right)^V\right\rangle_G\right)-\log\left(Z^G\right)\nonumber\\
&=&-\log\left(1+{1\over2}{1\over
g_s^2}{g^2\over3^2}(9+3)N^3\left(g_s\over
m\right)^3+{1\over2}{1\over g_s^2}{g^2\over3^2}3N\left(g_s\over
m\right)^3+\ldots\right)-\log\left(Z^G\right)\nonumber
\end{eqnarray}In the second line we used the fact that when writing down all
possible pairings one obtains nine times the first diagram of Fig.
\ref{fatgraphs2}, three times the second and three times the last
one. Of course, the result coincides with the one obtained using
the Feynman rules.}
\begin{eqnarray}
F(g_s,t)&=&-{1\over g_s^2}{2\over3}{g^2\over
m^3}t^3-{1\over6}{g^2\over m^3}t+\ldots-\log\left(Z^G\right)\
.\label{Fexpansion}
\end{eqnarray}
The terms of order $g^4$ in $F_0(t)$ can be obtained by writing
down all fatgraphs with four vertices that can be drawn on a
sphere. They are sketched, together with their symmetry factor
$|Aut(\G)|$, in Fig. \ref{fourvertices}.
Adding all these contributions to the result of order $g^2$ gives
an expression for the planar free energy up to order $g^4$:
\begin{equation}\label{F0expansion}
F_0(t)=-{2\over3}{g^2t^3\over m^3}-{8\over3}{g^4t^4\over
m^6}+\ldots+F_0^{np}(t)\ .
\end{equation}

\subsection{Matrix model technology}
Instead of doing perturbation theory one can evaluate the matrix
model partition function using some specific matrix model
technology. Next, we will recall some of this standard technology
adapted to the holomorphic matrix model. Let us first assume that
the path $\g$ consists of a single connected piece. The case
(\ref{tildegamma}) will be discussed later on. Let $s$ be the
length coordinate of the path $\g$, centered at some point on
$\g$, and let $\l(s)$ denote the parameterisation of $\g$ with
respect to this coordinate. Then, for an eigenvalue $\l_m$ on
$\g$, one has $\l_m=\l(s_m)$ and the partition function
(\ref{partmm}) can be rewritten as
\begin{equation}\label{part(s)}
Z(g_s,\hN)={1\over{\hN!}}\int_{\mathbb{R}}\d
s_1\ldots\int_{\mathbb{R}}\d
s_{\hat{N}}\prod_{l=1}^{\hat{N}}\dot{\l}(s_l)
\exp\left(-\hN^2S(g_s,\hN;\l(s_m))\right)\ .
\end{equation}
The spectral density is defined as
\begin{equation}\label{rozero}
\r(s,s_m):={1\over \hat{N}}\sum_{m=1}^{\hat{N}}\delta(s-s_m)\ ,
\end{equation}
so that $\r$ is normalised to one,
$\int_{-\infty}^\infty\r(s,s_m)\d s=1$. The normalised trace of
the resolvent of the matrix $M$ is given by
\begin{equation}\label{omrho}
\o (x,s_m):={1\over \hat{N}}\tr {1\over {x-M}}={1\over
\hat{N}}\sum_{m=1}^{\hat{N}}{1\over {x-\l(s_m)}}=\int\d
s{\r(s,s_m)\over {x-\l(s)}}\ ,
\end{equation}
for $x\in\mathbb{C}$. Following \cite{La03} we decompose the
complex plane into domains $D_i$, $i\in\{1,\ldots, n\}$, with
mutually disjoint interior, $( \cup_i\overline{D}_i=\mathbb{C},\ \
D_i\cap D_j=\emptyset\ \ \mbox{for}\ i\neq j)$. These domains are
chosen in such a way that $\g$ intersects each $\overline{D}_i$
along a single line segment $\D_i$, and $\cup_i\D_i=\g$.
Furthermore, $\m_i$, the $i$-th critical point of $W$, should lie
in the interior of $D_i$. One defines
\begin{equation}
\chi_i(M):=\int_{\partial D_i} {\d x\over 2\pi i}{1\over{x-M}}\ ,
\end{equation}
(which projects on the space spanned by the eigenvectors of $M$
whose eigenvalues lie in $D_i$), and the filling fractions
$\tilde\s_i(\l_m):={1\over\hN}\tr\chi_i(M)$ and
\begin{equation}
\s_i(s_m):=\tilde\s_i(\l(s_m))=\int\d s\
\r(s,s_m)\chi_i(\l(s))=\int_{\partial D_i}{\d x\over 2\pi i}\
\o(x,s_m)\ ,\label{fillingfraction}
\end{equation}
(which count the eigenvalues in the domain $D_i$, times $1/\hN$).
Obviously
\begin{equation}\label{sumnu}
\sum_{i=1}^{n}\s_i(s_m)=1\ .
\end{equation}

{\bf Loop equations}\\
Next we apply the methods of \cite{K99} to derive the loop
equations of the holomorphic matrix model. We define the
expectation value
\begin{equation}
\langle h(\l_m)\rangle:={1\over Z(g_s,\hN)}\cdot{1\over
\hN!}\int_{\g}\d\l_1\ldots\int_{\g}\d\l_{\hat{N}}\ h(\l_m)
\exp\left(-\hN^2S(g_s,\hN;\l_m)\right)\ .
\end{equation}
From the translational invariance of the measure one finds the
identity \cite{K99}, \cite{La03}
\begin{equation}
\int_{\g}\d\l_1\ldots\int_{\g}\d\l_{\hat{N}}\sum_{m=1}^{\hat{N}}{\partial\over\partial\l_m}\left[\prod_{k\neq
l}(\l_k-\l_l) e^{-{1\over
g_s}\sum_{j=1}^{\hat{N}}W(\l_j)}{1\over{x-\l_m}}\right]=0\ .
\end{equation}
Evaluating the derivative gives
\begin{equation}
\left\langle\sum_{m=1}^{\hat{N}}{1\over{(x-\l_m)^2}}-{1\over
g_s}\sum_{m=1}^{\hat{N}}{W'(\l_m)\over{x-\l_m}}+2\sum_{m=1}^{\hN}\sum_{\stackrel{l=1}{l\neq
m}}^{\hN} {1\over{(\l_m-\l_l)(x-\l_m)}}\right\rangle=0\ .
\end{equation}
Using
\begin{equation}\label{fractions}
{1\over{(x-\a)(x-\b)}}={1\over{\a-\b}}\left[{1\over{x-\a}}-{1\over{x-\b}}\right]
\end{equation}
we find
\begin{equation}
\sum_{m=1}^{\hat{N}}{1\over{(x-\l_m)^2}}+2\sum_{m=1}^{\hN}\sum_{\stackrel{l=1}{l\neq
m}}^{\hN} {1\over{(\l_m-\l_l)(x-\l_m)}}=\sum_{l,m=1}^{\hN}{1\over
x-\l_m}{1\over x-\l_l}
\end{equation}
and therefore
\begin{equation}
\left\langle\o(x;s_m)^2-{1\over{t\hat{N}}}\sum_{m=1}^{\hat{N}}{W'(\l(s_m))\over{x-\l(s_m)}}\right\rangle=0\
.
\end{equation}
If we define the polynomial
\begin{equation}
f(x;s_m):=-{4t\over{\hN}}\sum_{m=1}^{\hat{N}}{{W'(x)-W'(\l(s_m))}\over{x-\l(s_m)}}=-4t\int
\d s\ \r(s;s_m){{W'(x)-W'(\l(s))}\over{x-\l(s)}}\ ,
\end{equation}
we obtain the {\it loop equations}
\begin{equation}\label{loop}
\langle \o(x;s_m)^2\rangle-{1\over
t}W'(x)\langle\o(x;s_m)\rangle-{1\over 4t^2}\langle
f(x;s_m)\rangle=0\ .
\end{equation}

{\bf Equations of motion}\\
It will be useful to define an effective action as
\begin{eqnarray}\label{effaction}
&&S_{eff}(g_s,\hN;s_m):=S(g_s,\hN;\l(s_m))-{1\over\hN^2}\sum_{m=1}^{\hN}\ln(\dot\l(s_m))\nonumber\\
&&=\int\d s\ \r(s;s_m)\left({1\over
t}W(\l(s))-{1\over{\hN}}\ln(\dot{\l}(s))-\mathcal{P}\int\d s' \
\r(s';s_p)\ln(\l(s)-\l(s'))\right)\ \ \ \ \ \ \ \ \ \
\end{eqnarray}
so that
\begin{equation}
Z(g_s,\hN)={1\over {\hN!}}\int\d s_1\ldots\int\d s_{\hat{N}}\
\exp\left(-\hN^2 S_{eff}(g_s,\hN;s_m)\right)\ .
\end{equation}
Note that the principal value is defined as
\begin{equation}\label{Pvalue}
\mathcal{P}\
\ln\left(\l(s)-\l(s')\right)={1\over2}\lim_{\e\rightarrow0}\left[\ln\left(\l(s)-\l(s')+
i\e\dot\l(s)\right)+\ln\left(\l(s)-\l(s')-i\e\dot\l(s)\right)\right]\
.
\end{equation}
The equations of motion corresponding to this effective action,
${\delta S_{eff}\over\delta s_m}=0$, read
\begin{equation}\label{eommatrix}
{1\over t}W'(\l(s_m))={2\over{\hat{N}}}\sum_{p=1,\ p\neq
m}^{\hN}{1\over{\l(s_m)-\l(s_p)}}+{1\over{\hat{N}}}{\ddot{\l}(s_m)\over\dot{\l}(s_m)^2}\
.
\end{equation}
Using these equations of motion one can show that
\begin{eqnarray}\label{eomloop}
&&\o(x,s_m)^2-{1\over t}W'(x)\o(x,s_m)-{1\over 4 t^2}f(x,s_m)+\nonumber\\
&&+{1\over\hN}{\d\over\d x}\o(x,s_m)+{1\over\hN^2}
\sum_{m=1}^{\hN}{\ddot\l(s_m)\over\dot\l(s_m)^2}{1\over{x-\l(s_m)}}=0\
.
\end{eqnarray}

\begin{figure}[h]
\centering
\includegraphics{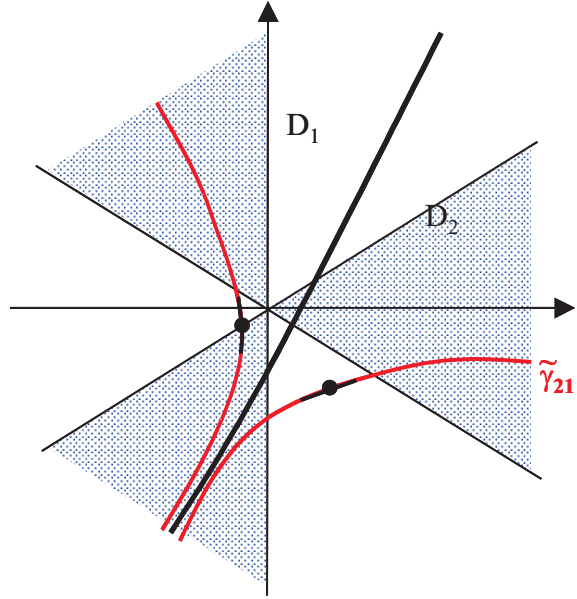}\\
\caption[]{For the cubic potential of Fig. \ref{convergence1} we
show the choice of the domains $D_1$ and $D_2$ (to the left and
right of the bold line) and of the path $\tilde\g_{21}$ with
respect to the two critical points, as well as the cuts that form
around these points.} \label{curvesinD}
\end{figure}

\bigskip
{\bf Solutions of the equations of motion}\\
Note that in general the effective action is a complex function of
the real $s_m$. Hence, in general, i.e. for a generic path
$\g_{kl}$ with parameterisation $\l(s)$, there will be no solution
to (\ref{eommatrix}). One clearly expects that the existence of
solutions must constrain the path $\l(s)$ appropriately. Let us
study this in more detail.\\
Recall that we defined the domains $D_i$ in such a way that
$\m_i\subset D_i$. Let $\hN_i$ be the number of eigenvalues
$\l(s_m)$ which lie in the domain $D_i$, so that
$\sum_{i=1}^n\hN_i=\hN$, and denote them by $\l(s_a^{(i)})$,
$a\in\{1,\ldots\hN_i\}$.\\
Solving the equations of motion in general is a formidable
problem. To get a good idea, however, recall the picture of
$\hN_i$ fermions filled into the $i$-th ``minimum" of ${1\over
t}W$ \cite{K91}. For small $t$ the potential is deep and the
fermions are located not too far from the minimum, in other words
all the eigenvalues are close to $\m_i$. To be more precise
consider (\ref{eommatrix}) and drop the last term, an
approximation that will be justified momentarily. Let us take $t$
to be small and look for solutions\footnote{One might try the
general ansatz $\l(s_a^{(i)})=\m_i+\e\delta\l_a^{(i)}$ but it
turns out that a solution can be found only if $\e\sim\sqrt{t}$.}
$\l(s_a^{(i)})=\m_i+\sqrt{t}\delta\l_a^{(i)}$, where
$\delta\l_a^{(i)}$ is of order one. So, we assume that the
eigenvalues $\l(s_a^{(i)})$ are not too far from the critical
point $\m_i$. Then the equation reads
\begin{equation}
W''(\m_i)\delta\l_a^{(i)}={2\over\hN}\sum_{b=1,\ b\neq
a}^{\hN_i}{1\over{\delta\l^{(i)}_a-\delta\l_b^{(i)}}}+o(\sqrt{t})\
,
\end{equation}
so we effectively reduced the problem to finding the solution for
$n$ distinct quadratic potentials. If we set $z_a:=\sqrt{\hN
W''(\m_i)\over2}\delta\l_a^{(i)}$ and neglect the
$o(\sqrt{t})$-terms this gives
\begin{equation}
z_a=\sum_{b=1,\ b\neq a}^{\hN_i}{1\over {z_a-z_b}}\ ,
\end{equation}
which can be solved explicitly for small $\hN_i$. It is obvious
that $\sum_{a=1}^{\hN_i}z_a=0$, and one finds that there is a
unique solution (up to permutations) with the $z_a$ symmetrically
distributed around 0 on the real axis. This justifies a posteriori
that we really can neglect the term proportional to the second
derivative of $\l(s)$, at least to leading order. Furthermore,
setting $W''(\m_i)=|W''(\m_i)|e^{i\phi_i}$ one finds that the
$\l(s_a^{i})$ sit on a tilted line segment around $\m_i$ where the
angle of the tilt is given by $-\phi_i/2$. This means for example
that for a potential with $W'(x)=x(x-1)(x+1)$ the eigenvalues are
distributed on the real axis around $\pm1$ and on the imaginary
axis around 0. Note further that, in general, the reality of $z_a$
implies that
${W''(\m_i)\over2}\left({\delta\l^{(i)}_a}\right)^2>0$ which tells
us that, close to $\m_i$, $W(\l(s))-W(\m_i)$ is real with a {\it
minimum} at $\l(s)=\m_i$.\\
So we have found that the path $\g_{kl}$ has to go through the
critical points $\m_i$ with a tangent direction fixed by the phase
of the second derivative of $W$. On the other hand, we know that
the partition function does not depend on the form of the path
$\g_{kl}$. Of course, there is no contradiction: if one wants to
compute the partition function from a saddle point expansion, as
we will do below, and as is implicit in the planar limit, one has
to make sure that one expands around solutions of
(\ref{eommatrix}) and the existence of these solutions imposes
conditions on how to choose the path $\g_{kl}$. From now on we
will assume that the path is chosen in such a way that it
satisfies all these constraints. Furthermore, for later purposes
it will be useful to use the path $\tilde\g_{kl}$ of
(\ref{tildegamma}) chosen such that its part $\g_{p_ip_{i+1}}$
goes through all $\hN_i$ solutions $\l_a^{(i)}$,
$a=1,\ldots\hN_i$, and lies entirely in $D_i$, see Fig.
\ref{curvesinD}.
\\
It is natural to assume that these properties together with the
uniqueness of the solution (up to permutations) extend to higher
numbers of $\hN_i$ as well. Of course once one goes beyond the
leading order in $\sqrt{t}$ the eigenvalues are no longer
distributed on a straight line, but on a line segment that is
bent in general and that might or might not pass through $\m_i$.\\

{\bf The large $\hN$ limit}\\
We are interested in the large $\hN$ limit of the matrix model
free energy. It is well known that the expectation values of the
relevant quantities like $\r$ or $\o$ have expansions of the form
\begin{equation}
\langle\r(s,s_m)\rangle=\sum_{I=0}^{\infty}\r_I(s)\hN^{-I}\ \ \ ,\
\ \ \langle\o(x,s_m)\rangle=\sum_{I=0}^{\infty}\o_I(x)\hN^{-I}\
.\label{omexp}
\end{equation}
Clearly, $\o_0(x)$ is related to $\r_0(s)$ by the large $\hN$
limit of (\ref{omrho}), namely
\begin{equation}\label{omegaofrho}
\o_0(x)=\int\d s{\r_0(s)\over{x-\l(s)}}\ .
\end{equation}
We saw already that an eigenvalue ensemble that solves the
equations of motion is distributed along line segments around the
critical points $\m_i$. In the limit $\hN\rightarrow \infty$ this
will turn into a continuous distribution on the segments
$\mathcal{C}_i$, through or close to the critical points of $W$.
Then $\r_0(s)$ has support only on these $\mathcal{C}_i$ and
$\o_0(x)$ is analytic in $\mathbb{C}$ with cuts $\mathcal{C}_i$.
Conversely, $\r_0(s)$ is given by the discontinuity of $\o_0(x)$
across its cuts:
\begin{equation}\label{solution}
\r_0(s):=\dot{\l}(s)\lim_{\e\rightarrow0}{1\over2\pi i}
[\o_0(\l(s)-i\e\dot{\l}(s))-\o_0(\l(s)+i\e\dot{\l}(s))]\ .
\end{equation}

The planar limit we are interested in is $\hN\rightarrow\infty$,
$g_s\rightarrow0$ with $t=g_s\hN$ held fixed. Hence we rewrite all
$\hN$ dependence as a $g_s$ dependence and consider the limit
$g_s\rightarrow 0$. Then, the equation of motion (\ref{eommatrix})
reduces to
\begin{equation}\label{eomplanar}
{1\over t}W'(\l(s))=2\mathcal{P}\int \d s'\
{\r_0(s')\over{\l(s)-\l(s')}}\ .
\end{equation}
Note that this equation is only valid for those $s$ where
eigenvalues exist, i.e. where $\r_0(s)\neq 0$. In principle one
can use this equation to compute the planar eigenvalue
distribution $\r_0(s)$ for given $W'$.\\

The leading terms in the expansions (\ref{omexp}) for
$\langle\r(s,s_m)\rangle$ or $\langle\o(x,s_m)\rangle$, i.e.
$\r_0(s)$ or $\o_0(x)$, can be calculated from a saddle point
approximation, where the $\{s_m\}$ are given by a solution
$\{s_m^*\}$ of (\ref{eommatrix}): $\r_0(s)=\r(s;s_m^*)$, or
explicitly from eq. (\ref{rozero})
\begin{equation}\label{rozeroplanar}
\r_0(s)={1\over \hN} \sum_{m=1}^{\hat{N}}\delta(s-s_m^*) \ .
\end{equation}
Note in particular that $\r_0(s)$ is manifestly real. This is by
no means obvious for the full expectation value of $\r(s,s_m)$
since it must be computed by averaging with respect to a complex
measure. In the planar limit, however, the quantum integral is
essentially localized at a single classical configuration
$\{s_m^*\}$ and this is why $\r_0(s)$ is real. The large $\hN$
limit of the resolvent
$\o_0(x)=\o(x;s_m^*)$ then is still given by (\ref{omegaofrho}). \\
This prescription to compute expectation values of operators in
the large $\hN$ limit is true for all ``microscopic'' operators,
i.e. operators that do not modify the saddle point equations
(\ref{eommatrix}). (Things would be different for
``macroscopic''operators like $e^{\hN\sum_{p=1}^{\hN}V(\l_p)}$.)
In particular, this shows that
expectation values factorise in the large $\hN$ limit.\\

{\bf Riemann surfaces and planar solutions}\\
This factorisation of expectation values shows that in the large
$\hN$ limit the loop equation (\ref{loop}) reduces to the
algebraic equation
\begin{equation}\label{algconstraint}
\o_0(x)^2-{1\over t}W'(x)\o_0(x)-{1\over 4t^2}f_0(x)=0\ ,
\end{equation}
where
\begin{equation}\label{f0matrix}
f_0(x)=-4t\int \d s \ \r_0(s){{W'(x)-W'(\l(s))}\over{x-\l(s)}}
\end{equation}
is a polynomial of degree $n-1$ with leading coefficient $-4t$.
Note that this coincides with the planar limit of equation
(\ref{eomloop}). If we define
\begin{equation}\label{y0omega0}
y_0(x):=W'(x)-{2t}\o_0(x)\ ,
\end{equation}
then $y_0$ is one of the branches of the algebraic curve
\begin{equation}
y^2=W'(x)^2+f_0(x)\ ,\label{algcurve}
\end{equation}
as can be seen from (\ref{algconstraint}). This is in fact an
extremely interesting result. Not only is it quite amazing that a
Riemann surface arises in the large $\hN$ limit of the holomorphic
matrix model, but it is actually the Riemann surface
(\ref{heRiemann}), which we encountered when we were calculating
the integrals of $\O$ over three-cycles in $X_{def}$. This is
important because we can now readdress the problem of calculating
integrals on this Riemann surface. In fact, the matrix model will
provide us with the techniques that are necessary to solve this
problem.

On the curve (\ref{algcurve}) we use the same conventions as in
section \ref{periodint}, i.e. $y_0(x)$ is defined on the upper
sheet and cycles and orientations are chosen as in Fig.
\ref{ABcycles} and Fig. \ref{albecycles}.

Solving a matrix model in the planar limit means to find a
normalised, real, non-negative $\r_0(s)$ and a path $\tilde
\g_{kl}$ which satisfy (\ref{omegaofrho}), (\ref{eomplanar}) and
(\ref{algconstraint}/\ref{f0matrix}) for a given potential $W(z)$
and a given asymptotics $(k,l)$ of $\g$.

Interestingly, for any algebraic curve (\ref{algcurve}) there is a
contour $\tilde \g_{kl}$ supporting a {\it formal} solution of the
matrix model in the planar limit. To construct it start from an
arbitrary polynomial $f_0(x)$ or order $n-1$, with leading
coefficient $-4t$, which is given together with the potential
$W(x)$ of order $n+1$. The corresponding Riemann surface is given
by (\ref{algcurve}), and we denote its branch points by $a_i^\pm$
and choose branch-cuts $\mathcal{C}_i$ between them. We can read
off the two solutions $y_0$ and $y_1=-y_0$ from (\ref{algcurve}),
where we take $y_0$ to be the one with a behaviour
$y_0\stackrel{x\rightarrow\infty}{\rightarrow}+W'(x)$. $\o_0(x)$
is defined as in (\ref{y0omega0}) and we choose a path
$\tilde\g_{kl}$ such that $\mathcal{C}_i\subset \tilde\g_{kl}$ for
all $i$. Then the formal planar spectral density satisfying all
the requirements can be determined from (\ref{solution}) (see
\cite{La03}). However, in general, this will lead to a {\it
complex} distribution $\r_0(s)$. This can be understood from the
fact the we constructed $\rho_0(s)$ from a completely arbitrary
hyperelliptic Riemann surface. However, in the matrix model the
algebraic curve (\ref{algcurve}) is {\it not} general, but the
coefficients $\a_k$ of $f_0(x)$ are constraint. This can be seen
by computing the filling fractions $\langle\s_i(s_m)\rangle$ in
the planar limit where they reduce to $\s_i(s_m^*)$. They are
given by
\begin{equation}\label{nustar}
\n_i^*:=\s_i(s_m^*)={1\over2\pi i}\int_{\partial D_i}\o_0(x)\d
x={1\over4\pi it}\int_{A^i}y_0(x)\d
x=\int_{\g^{-1}(\mathcal{C}_i)}\r_0(s)\d s\ ,
\end{equation}
which must be real and non-negative. Here we used the fact that
the $D_i$ were chosen such that $\g_{p_ip_{i+1}}\subset D_i$ and
therefore $\mathcal{C}_i\subset D_i$, so for $D_i$ on the upper
plane, $\partial D_i$ is homotopic to $-A^i$. Hence, ${\rm
Im}\left(i\int_{A^i}y(x)\d x\right)=0$ which constrains the
$\a_k$. We conclude that to construct distributions $\r_0(s)$ that
are relevant for the matrix model one can proceed along the lines
described above, but one has to impose the additional constraint
that $\r_0(s)$ is {\it real} \cite{BM05}. As for finite $\hN$,
this will impose conditions on the possible paths $\tilde\g_{kl}$
supporting the eigenvalue distributions.

To see this, we assume that the coefficients $\a_k$ in $f_0(x)$
are small, so that the lengths of the cuts are small compared to
the distances between the different critical points:
$|a_i^+-a_i^-|\ll|\m_i-\m_j|$. Then in first approximation the
cuts are straight line segments between $a_i^+$ and $a_i^-$. For
$x$ close to the cut $\mathcal{C}_i$ we have
$y^2\approx(x-a_i^+)(x-a_i^-)\prod_{j\neq
i}(\m_i-\m_j)^2=(x-a_i^+)(x-a_i^-)(W''(\m_i))^2$. If we set
$W''(\m_i)=|W''(\m_i)|e^{i\phi_i}$ and
$a_i^+-a_i^-=r_ie^{i\psi_i}$, then, on the cut $\mathcal{C}_i$,
the path $\g$ is parameterised by
$\l(s)={a_i^+-a_i^-\over|a_i^+-a_i^-|}s=se^{i\psi_i}$, and we find
from (\ref{solution})
\begin{eqnarray}
\r_0(s)&=&{1\over2\pi t }\sqrt{|\l(s)-a_1^+|}\sqrt{|\l(s)-a_1^-|}\ |W''(\m_i)|e^{i(\phi_i+2\psi_i)}\nonumber\\
&=& {1\over2 \pi t }\sqrt{|\l(s)-a_1^+|}\sqrt{|\l(s)-a_1^-|}\
W''(\m_i)(\dot\l(s))^2\ .
\end{eqnarray}
So reality and positivity of $\r_0(s)$ lead to conditions on the
orientation of the cuts in the complex plane, i.e. on the path
$\g$:
\begin{equation}\label{condition}
\psi_i=-{\phi_i/2}\ \ \ \ \ \ ,\ \ \ \ \ W''(\m_i)(\dot\l(s))^2>0\
.
\end{equation}
These are precisely the conditions we already derived for the case
of finite $\hN$. We see that the two approaches are consistent
and, for given $W$ and fixed $\hN_i$ respectively $\n_i^*$, lead
to a unique\footnote{To be more precise the path $\tilde\g_{kl}$
is not entirely fixed. Rather, for every piece
$\tilde\g_{p_ip_{i+1}}$ we have the requirement that
$\mathcal{C}_i\subset\tilde\g_{p_ip_{i+1}}$.} solution
$\{\l(s),\r_0(s)\}$ with real and positive eigenvalue
distribution. Note again that the requirement of reality and
positivity of $\r_0(s)$ constrains the phases of $a_i^+-a_i^-$ and
hence the coefficients $\a_k$ of $f_0(x)$.

\subsection{The saddle point approximation for the partition
function}
Recall that our goal is to calculate the integrals of $\zeta=y\d
x$ on the Riemann surface (\ref{algcurve}) using matrix model
techniques. The tack will be to establish relations for these
integrals which are similar to the special geometry relations
(\ref{SG}). After we have obtained a clear cut understanding of
the function $\mF$ appearing in these relations, we can use the
matrix model to calculate $\mF$ and therefore the integrals. A
natural candidate for this function $\mF$ is the free energy of
the matrix model, or rather, since we are working in the large
$\hN$ limit, its planar component $F_0(t)$. However, $F_0(t)$
depends on $t$ only and therefore it cannot appear in relations
like (\ref{SG}). To remedy this we introduce a set of sources
$J_i$ and obtain a free energy that depends on more variables. In
this subsection we evaluate this source dependent free energy and
its Legendre transform $\mathcal{F}_0(t,S)$ in the planar limit
using a saddle point expansion \cite{BM05}.

We start by coupling sources to the filling
fractions,\footnote{Note that $\exp\left(-{\hN^2\over
t}\sum_{i=1}^{n-1}J_i\s_i(s_m)\right)$ looks like a macroscopic
operator that changes the equations of motion. However, because of
the special properties of $\s_i(s_m)$ we have
${\partial\over\partial s_n}\s_i(s_m)={1\over\hN}\int_{\partial
D_i}{\d x\over2\pi i}{\dot\l(s_n)\over(x-\l(s_n))^2}$. In
particular, for the path $\tilde \g_{kl}$ that will be chosen
momentarily and the corresponding domains $D_i$ the eigenvalues
$\l_m$ cannot lie on $\partial D_i$. Hence,
${\partial\over\partial s_n}\s_i(s_m)=0$ and the equations of
motion remain unchanged.}
\begin{eqnarray}\label{partsources}
Z(g_s,\hN,J)&:=&{1\over{\hN!}}\int_{\g}\d \l_1\ldots\int_{\g}\d
\l_{\hat{N}}\exp\left(-\hN^2S(g_s,\hN;\l_m)-{\hN\over
g_s}\sum_{i=1}^{n-1}J_i\tilde\s_i(\l_m)\right)\nonumber\\
&=&\exp\left(-F(g_s,\hN,J)\right)\ .
\end{eqnarray}
where $J:=\{J_1,\ldots,J_{n-1}\}$. Note that because of the
constraint $\sum_{i=1}^{n}\tilde\s_i(\l_m)=1$,
$\tilde\s_{n}(\l_m)$ is not an independent quantity and we can
have only $n-1$ sources. This differs from the treatment in
\cite{La03} and has important consequences, as we will see below.
We want to evaluate this partition function for
$\hN\rightarrow\infty$, $t=g_s\hN$ fixed, from a saddle point
approximation. We therefore use the path $\tilde\g_{kl}$ from
(\ref{tildegamma}) that was chosen in such a way that the equation
of motion (\ref{eommatrix}) has solutions $s_m^*$ and, for large
$\hN$, $\mathcal{C}_i\subset\g_{p_ip_{i+1}}$. It is only then that
the saddle point expansion converges and makes sense. Obviously
then each integral $\int_{\g}\d\l_m$ splits into a sum
$\sum_{i=1}^{n}\int_{\g_{p_ip_{i+1}}}\d\l_m$. Let
$s^{(i)}\in\mathbb{R}$ be the length coordinate on
$\g_{p_{i}p_{i+1}}$, so that $s^{(i)}$ runs over all of
$\mathbb{R}$. Furthermore, $\tilde\s_i(\l_m)$ only depends on the
number $\hN_i$ of eigenvalues in $\tilde\g_{kl}\cap
D_i=\g_{p_ip_{i+1}}$. Then the partition function
(\ref{partsources}) is a sum of contributions with fixed $\hN_i$
and we rewrite is as
\begin{equation}\label{Z(J)}
Z(g_s,N,J)=\sum_{\stackrel{\sum_i\hN_i=\hN}{\hN_1,\ldots,\hN_{n}}}Z(g_s,\hN,\hN_i)e^{-{1\over
g_s}\sum_{i=1}^{n-1}J_i\hN_i}\ ,
\end{equation}
where now
\begin{eqnarray}
&&Z(g_s,\hN,\hN_i)=\nonumber\\
&&={1\over \hN_1\ldots\hN_n!}\int_{\g_{p_1p_2}}\d\l^{(1)}_1\ldots
\int_{\g_{p_1p_2}}\d\l^{(1)}_{\hN_1}\ldots\int_{\g_{p_np_{n+1}}}
\d\l^{(n)}_1\ldots\int_{\g_{p_np_{n+1}}}\d\l^{(n)}_{\hN_n}\times\nonumber\\
&&\times\exp\left(-{t^2\over
g_s^2}S(g_s,t;\l_k^{(i)})\right)\nonumber\\
&&=:\exp\left(\tilde{\mathcal{F}}(g_s,t,\hN_i)\right)
\end{eqnarray}
is the partition function with the additional constraint that
precisely $\hN_i$ eigenvalues lie on $\g_{p_ip_{i+1}}$. Note that
it depends on $g_s,t=g_s\hN$ and $\hN_1,\ldots\hN_{n-1}$ only, as
$\sum_{i=1}^n\hN_i=\hN$. Now that these numbers have been fixed,
there is precisely one solution to the equations of motion, i.e. a
{\it unique} saddle-point configuration, up to permutations of the
eigenvalues, on {\it each} $\g_{p_ip_{i+1}}$. These permutations
just generate a factor $\prod_i \hN_i!$ which cancels the
corresponding factor in front of the integral. As discussed above,
it is important that we have chosen the $\g_{p_ip_{i+1}}$ to
support this saddle point configuration close to the critical
point $\m_i$ of $W$. Moreover, since $\g_{p_ip_{i+1}}$ runs from
one convergence sector to another and by (\ref{condition}) the
saddle point really is dominant (stable), the ``one-loop" and
other higher order contributions are indeed subleading as
$g_s\rightarrow 0$ with $t=g_s\hN$ fixed. This is why we had to be
so careful about the choice of our path $\g$ as being composed of
$n$ pieces $\g_{p_ip_{i+1}}$. In the planar limit
$\n_i:={\hN_i\over\hN}$ is finite, and
$\tilde{\mathcal{F}}(g_s,t,\n_i)={1\over
g_s^2}\tilde{\mathcal{F}}_{0}(t,\n_i)+\ldots$. The saddle point
approximation gives
\begin{equation}\label{F0Seff}
\tilde{\mathcal{F}}_{0}(t,\n_i)= -t^2
S_{eff}(g_s=0,t;s_a^{(j)*}(\n_i))\ ,
\end{equation}
where (cf. (\ref{effaction})) $S_{eff}(g_s=0,t;s_a^{(j)*}(\nu_i))$
is meant to be the value of $S(0,t;\l(s_a^{(j)*}(\nu_i)))$, with
$\l(s_a^{(j)*}(\nu_i))$ the point on $\g_{p_ip_{i+1}}$
corresponding to the unique saddle point value $s_a^{(j)*}$ with
fixed fraction $\nu_i$ of eigenvalues $\l_m$ in $D_i$. Note that
the ${1\over\hN^2}\sum\dot\l(s_m)$-term in (\ref{effaction})
disappears in the present planar limit. One can go further and
evaluate subleading terms. In particular, the remaining integral
leads to the logarithm of the determinant of the
$\hN\times\hN$-matrix of second derivatives of $S$ at the saddle
point. This is {\it not} order $g_s^0\sim\hN^0$ as naively
expected from the expansion of $\tilde \mF$ in powers of $g_s^2$.
Instead one finds contributions like $-\hN\log\hN,\
{\hN\over2}\log t$. The point is that
$-\hN^2S_{eff}(g_s,t,s^*(\n))$ also has subleading contributions,
which are dropped in the planar limit (\ref{F0Seff}). In
particular, one can check very explicitly for the Gaussian model
that these subleading contributions in $S_{eff}(s^*)$ cancel the
$\hN\log\hN$ and the ${\hN\over2}\log t$ pieces from the
determinant. A remaining $\hN$-dependent (but $\n$- and
$t$-independent) constant could have been absorbed in the overall
normalisation of $Z$. Hence the subleading terms are indeed
$o(g_s^0)\sim o(\hN^0)$.

It remains to sum over the $\hN_i$ in (\ref{Z(J)}). In the planar
limit these sums are replaced by integrals:
\begin{eqnarray}
Z(g_s,t,J) &=&\int_0^1\d\n_1\ldots\int_0^1\d
\n_{n}\ \delta\left(\sum_{i=1}^{n}\n_i-1\right)\nonumber\\
&&\exp\left[-{1\over
g_s^2}\left(t\sum_{i=1}^{n-1}J_i\n_i-\tilde{\mathcal{F}}_0(t,\n_i)\right)+
c(\hN)+o(g_s^{0})\right]\ .
\end{eqnarray}
Once again, in the planar limit, this integral can be evaluated
using the saddle point technique and for the source-dependent free
energy $F(g_s,t,J)={1\over g_s^2} F_0(t,J)+\ldots$ we find
\begin{equation}\label{F0}
F_0(t,J)=\sum_{i=1}^{n-1}J_i\
t\n_i^*-\tilde{\mathcal{F}}_0(t,\n_i^*)\ ,
\end{equation}
where $\n_i^*$ solves the new saddle point equation,
\begin{equation}
tJ_i={\partial\tilde{\mathcal{F}}_0\over\partial\n_i}(t,\n_j)\ .
\end{equation}
This shows that $F_0(t,J)$ is nothing but the Legendre transform
of $\tilde{\mathcal{F}}_0(t,\n_i^*)$ in the $n-1$ latter
variables. If we define
\begin{equation}\label{Snustar}
S_i:=t\n_i^*\ ,\ \mbox{for}\ i=1,\ldots,n-1,
\end{equation}
we have the inverse relation
\begin{equation}\label{SJ}
S_i={\partial F_0\over\partial J_i}(t,J)\ ,
\end{equation}
and with $\mathcal{F}_0(t,S):=\tilde{\mathcal{F}}_0(t,{S_i\over
t})$, where $S:=\{S_1,\ldots,S_{n-1}\}$, one has from (\ref{F0})
\begin{equation}
\mathcal{F}_0(t,S)=\sum_{i=1}^{n-1}J_iS_i-F_0(t,J)\ ,
\end{equation}
where $J_i$ solves (\ref{SJ}). From (\ref{F0Seff}) and the
explicit form of $S_{eff}$, Eq.(\ref{effaction}) with
$\hN\rightarrow\infty$, we deduce that
\begin{equation}\label{mathcalF0}
\mathcal{F}_0(t,S)=t^2\mathcal{P}\int\d s\int\d s'\
\ln(\l(s)-\l(s'))\r_0(s;t,S_i)\r_0(s';t,S_i)-t\int\d s\
W(\l(s))\r_0(s;t,S_i)\ ,
\end{equation}
where $\r_0(s;t,S_i)$ is the eigenvalue density corresponding to
the saddle point configuration $s_a^{(i)*}$ with
${\hN_i\over\hN}=\n_i$ fixed to be $\n_i^*={S_i\over t}$. Hence it
satisfies
\begin{equation}\label{intrhoC}
t\int_{\g^{-1}(\mathcal{C}_i)}\r_0(s;t,S_j)\d s=S_i\ \ \mbox{for}\
i=1,2,\ldots n-1\ ,
\end{equation}
and obviously
\begin{equation}\label{intrhoCn}
t\int_{\g^{-1}(\mathcal{C}_n)}\r_0(s;t,S_j)\d
s=t-\sum_{i=1}^{n-1}S_i\ .
\end{equation}
Note that the integrals in (\ref{mathcalF0}) are convergent and
$\mathcal{F}_0(t,S)$ is a well-defined function.

\section{Special geometry relations}
After this rather detailed study of the planar limit of
holomorphic matrix models we now turn to the derivation of the
special geometry relations for the Riemann surface
(\ref{heRiemann}) and hence the local Calabi-Yau (\ref{Xdef}).
Recall that in the matrix model the $S_i=t\n_i^*$ are real and
therefore $\mathcal{F}_0(t,S)$ of Eq. (\ref{mathcalF0}) is a
function of {\it real} variables. This is reflected by the fact
that one can generate only a subset of all possible Riemann
surfaces (\ref{heRiemann}) from the planar limit of the
holomorphic matrix model, namely those for which the
$\n_i^*={1\over4\pi it}\int_{A^i}\zeta$ are real (recall
$\zeta=y\d x$). We are, however, interested in the special
geometry of the most general surface of the form
(\ref{heRiemann}), which can no longer be understood as a surface
appearing in the planar limit of a matrix model. Nevertheless, for
any such surface we can apply the formal construction of
$\r_0(s)$, which will in general be complex. Then one can use this
complex ``spectral density" to calculate the function
$\mathcal{F}_0(t,S)$ from (\ref{mathcalF0}), that now depends on
{\it complex} variables. Although this is {\it not} the planar
limit of the free energy of the matrix model, it will turn out to
be the prepotential for the general hyperelliptic Riemann surface
(\ref{algcurve}) and hence of the local Calabi-Yau manifold
(\ref{Xdef}).

\subsection{Rigid special geometry}
Let us then start from the general hyperelliptic Riemann surface
(\ref{heRiemann}) which we view as a two-sheeted cover of the
complex plane (cf. Figs. \ref{ABcycles}, \ref{albecycles}), with
its
 cuts $\mathcal{C}_i$ between $a_i^-$ and $a_i^+$. We choose
a path $\g$ on the upper sheet with parameterisation $\l(s)$ in
such a way that $\mathcal{C}_i\subset\g$. The complex function
$\r_0(s)$ is determined from (\ref{solution}) and
(\ref{y0omega0}), as described above. We define the complex
quantities
\begin{equation}
S_i:={1\over4\pi
i}\int_{A^i}\zeta=t\int_{\g^{-1}(\mathcal{C}_i)}\r_0(s) \ \ \
\mbox{for}\ i=1,\ldots,n-1\ ,
\end{equation}
and the prepotential $\mathcal{F}_0(t,S)$ as in (\ref{mathcalF0})
(of course, $t$ is $-{1\over4}$ times the leading coefficient of
$f_0$ and it can now be complex as well).

Following \cite{La03} one defines the ``principal value of $y_0$"
along the path $\g$ (c.f. (\ref{Pvalue}))
\begin{equation}
y_0^p(s):={1\over2}\lim_{\e\rightarrow0}[y_0(\l(s)+i\e\dot{\l}(s))+y_0(\l(s)-i\e\dot{\l}(s))]\
.
\end{equation}
For points $\l(s)\in\g$ outside $\mathcal{C}:=\cup_i\mathcal{C}_i$
we have $y_0^p(s)=y_0(\l(s))$, while $y_0^p(s)=0$ on
$\mathcal{C}$. With
\begin{equation}
\phi(s):=W(\l(s))-2t\mathcal{P}\int\d
s'\ln(\l(s)-\l(s'))\r_0(s';t,S_i)
\end{equation}
one finds, using (\ref{omegaofrho}), (\ref{y0omega0}) and
(\ref{Pvalue}),
\begin{equation}\label{phiy0}
{d\over ds}\phi(s)=\dot{\l}(s)y_0^p(s)\ .
\end{equation}
The fact that $y_0^p(s)$ vanishes on $\mathcal{C}$ implies
\begin{equation}
\phi(s)=\xi_i:=\mbox{constant on}\ \mathcal{C}_i\ .
\end{equation}
Integrating (\ref{phiy0}) between $\mathcal{C}_i$ and
$\mathcal{C}_{i+1}$ gives
\begin{equation}
\xi_{i+1}-\xi_i=\int_{a^+_i}^{a^-_{i+1}}\d\l \ y_0(\l)={1\over
2}\int_{\b_i}\d x \ y(x)={1\over 2}\int_{\b_i}\zeta\ .
\end{equation}
From (\ref{mathcalF0}) we find for $i<n$
\begin{equation}\label{xinxii}
{\partial\over\partial S_{i}}\mathcal{F}_0(t,S)=-t\int\d
s{\partial\r_0(s;t,S_j)\over\partial S_i}\phi(s)=\xi_n-\xi_i\ .
\end{equation}
To arrive at the last equality we used that $\r_0(s)\equiv0$ on
the complement of the cuts, while on the cuts $\phi(s)$ is
constant and we can use (\ref{intrhoC}) and (\ref{intrhoCn}).
Then, for $i<n-1$,
\begin{equation}\label{F0S}
{\partial\over\partial
S_{i}}\mathcal{F}_0(t,S)-{\partial\over\partial
S_{i+1}}\mathcal{F}_0(t,S)={1\over 2}\int_{\b_i}\zeta\ .
\end{equation}
For $i=n-1$, on the other hand, we find
\begin{equation}
{\partial\over\partial
S_{n-1}}\mathcal{F}_0(t,S)=\xi_{n}-\xi_{n-1}={1\over
2}\int_{\b_{n-1}}\zeta\ .
\end{equation}
We change coordinates to
\begin{equation}
\tilde S_i:=\sum_{j=1}^iS_j\ ,\label{tildeS}
\end{equation}
and find the rigid special geometry\footnote{For a review of rigid
special geometry see appendix \ref{rigSG}.} relations
\begin{eqnarray}
\tilde S_i&=&{1\over 4\pi i }\int_{\a_{i}}\zeta\ \label{sgrela},\\
{\partial\over\partial \tilde S_{i}}\mathcal{F}_0(t,\tilde
S)&=&{1\over2}\int_{\b_{i}}\zeta\ .\label{sgrelb}
\end{eqnarray}
for $i=1,\ldots,n-1$. Note that the basis of one-cycles that
appears in these equations is the one shown in Fig.
\ref{albecycles} and differs from the one used in \cite{La03}. The
origin of this difference is the fact that we introduced only
$n-1$ currents $J_i$ in the partition function
(\ref{partsources}).\\
Next we use the same methods to derive the relation between the
integrals of $\zeta$ over the cycles $\hat \a$ and $\hat \b$ and
the planar free energy \cite{BM05}.

\subsection{Integrals over relative cycles}
The first of these integrals encircles all the cuts, and by
deforming the contour one sees that it is given by the residue of
the pole of $\zeta$ at infinity, which is determined by the
leading coefficient of $f_0(x)$:
\begin{equation}
{1\over4\pi i}\int_{\hat\a}\zeta=t\ .\label{SGhat1}
\end{equation}
The cycle $\hat\beta$ starts at infinity of the lower sheet, runs
to the $n$-th cut and from there to infinity on the upper sheet.
The integral of $\zeta$ along $\hat\b$ is divergent, so we
introduce a (real) cut-off $\L_0$ and instead take $\hat\b$ to run
from $\L_0'$ on the lower sheet through the $n$-th cut to $\L_0$
on the upper sheet. We find
\begin{eqnarray}
{1\over2}\int_{\hat\b}\zeta=\int_{a^+_{n}}^{\L_0}
y_0(\l)\d\l&=&\phi(\l^{-1}(\L_0))-\phi(\l^{-1}(a^+_{n}))=\phi(\l^{-1}(\L_0))-\xi_{n}\nonumber\\
&=&W(\L_0)-2t\mathcal{P}\int\d s'\ln(\L_0-\l(s'))\r_0(s';t,\tilde
S_i)-\xi_{n}\ .\nonumber\\
\end{eqnarray}
On the other hand we can calculate
\begin{equation}\label{xin}
{\partial\over\partial t}\mathcal{F}_0(t,\tilde S)=-\int\d s\
\phi(\l(s)){\partial\over\partial t}[t\r_0(s;t,\tilde
S_i)]=-\sum_{i=1}^n\xi_i\int_{\g^{-1}(\mathcal{C}_i)}\d s\
{\partial\over\partial t}[t\r_0(s;t,\tilde S_i)]=-\xi_{n}\ ,
\end{equation}
(where we used (\ref{intrhoC}) and (\ref{intrhoCn})) which leads
to
\begin{eqnarray}
{1\over2}\int_{\hat \b}\zeta&=&{\partial\over\partial
t}\mathcal{F}_0(t,\tilde
S)+W(\L_0)-2t\mathcal{P}\int\d s'\ln(\L_0-\l(s'))\r_0(s';t,\tilde S_i)\nonumber\\
&=&{\partial\over\partial t}\mathcal{F}_0(t,\tilde
S)+W(\L_0)-t\log\L_0^2+o\left({1\over\L_0}\right)\ .\label{SGhat2}
\end{eqnarray}
Together with (\ref{SGhat1}) this looks very similar to the usual
special geometry relation. In fact, the cut-off independent term
is the one one would expect from special geometry. However, the
equation is corrected by cut-off dependent terms. The last terms
vanishes if we take $\L_0$ to infinity but there remain two
divergent terms which we are going to study in detail
below.\footnote{Of course, one could define a cut-off dependent
function $\mathcal{F}^{\L_0}(t,\tilde S):=\mathcal{F}_0(t,\tilde
S)+tW(\L_0)-{t^2\over2}\log\L_0^2$ for which one has
${1\over2}\int_{\hat\b}\zeta={\partial\mathcal{F}^{\L_0}\over\partial
t}+o\left({1\over\L_0}\right)$ similar to \cite{DOV04}. Note,
however, that this is not a standard special geometry relation due
to the presence of the $o\left({1\over\L_0}\right)$-terms.
Furthermore, $\mathcal{F}^{\L_0}$ has no interpretation in the
matrix model and is divergent as $\L_0\rightarrow\infty$.} For a
derivation of (\ref{SGhat2}) in a slightly different context see
\cite{CSW03b}.

\subsection{Homogeneity of the prepotential}
The prepotential on the moduli space of complex structures of a
{\it compact} Calabi-Yau manifold is a holomorphic function that
is homogeneous of degree two. On the other hand, the structure of
the local Calabi-Yau manifold (\ref{Xdef}) is captured by a
Riemann surface and it is well-known that these are related to
rigid special geometry. The prepotential of rigid special
manifolds does not have to be homogeneous (see for example
\cite{CRTP97}), and it is therefore important to explore the
homogeneity structure of $\mathcal{F}_0(t,\tilde S)$. The result
is quite interesting and it can be written in the form
\begin{equation}\label{homogeneity}
\sum_{i=1}^{n-1}\tilde S_i{\partial\mathcal{F}_0\over\partial
\tilde{S_i}}(t,\tilde S_i)+t{\partial\mathcal{F}_0\over\partial
t}(t,\tilde S_i)=2\mathcal{F}_0(t,\tilde S_i)+t\int\d s\
\r_0(s;t,\tilde S_i)W(\l(s))\ .
\end{equation}
To derive this relation we rewrite Eq. (\ref{mathcalF0}) as
\begin{eqnarray}
2\mathcal{F}_0(t,\tilde S_i)&=&-t\int\d s\ \r_0(s;t,\tilde
S_i)\left[\phi(s)+W(\l(s))\right]\nonumber\\
&=&-t\int\d s\ \r_0(s;t,\tilde
S_i)W(\l(s))+\sum_{i=1}^{n-1}(\xi_n-\xi_i)S_i-t\xi_n\ .
\end{eqnarray}
Furthermore, we have $\sum_{i=1}^{n-1}\tilde
S_i{\partial\mathcal{F}_0\over\partial\tilde S_i}(t,\tilde
S_i)=\sum_{i=1}^{n-1} S_i{\partial\mathcal{F}_0\over\partial
S_i}(t, S_i)=\sum_{i=1}^{n-1}S_i(\xi_n-\xi_i)$, where we used
(\ref{xinxii}). The result then follows from (\ref{xin}).\\
Of course, the prepotential was not expected to be homogeneous,
since already for the simplest example, the conifold,
$\mathcal{F}_0$ is known to be non-homogeneous (see section
\ref{exsum}). However, Eq. (\ref{homogeneity}) shows the precise
way in which the homogeneity relation is modified on the local
Calabi-Yau manifold (\ref{Xdef}).

\subsection{Duality transformations}
The choice of the basis $\{\a^i,\b_j,\hat\a,\hat\b\}$ for the
(relative) one-cycles on the Riemann surface was particularly
useful in the sense that the integrals over the compact cycles
$\a^i$ and $\b_j$ reproduce the familiar rigid special geometry
relations, whereas new features appear only in the integrals over
$\hat\a$ and $\hat\b$. In particular, we may perform any
symplectic transformation of the compact cycles $\a^i$ and $\b_j$,
$i,j=1,\ldots n-1$, among themselves to obtain a new set of
compact cycles which we call $a^i$ and $b_j$. Such symplectic
transformations can be generated from (i) transformations that do
not mix $a$-type and $b$-type cycles, (ii) transformations
$a^i=\a^i,\ b_i=\b_i+\a^i$ for some $i$ and (iii) transformations
$a^i=\b_i$, $b_i=-\a^i$ for some $i$. (These are analogue to the
trivial, the $T$ and the $S$ modular transformations of a torus.)
For transformations of the first type the prepotential
$\mathcal{F}$ remains unchanged, except that it has to be
expressed in terms of the new variables $s_i$, which are the
integrals of $\zeta$ over the new $a^i$ cycles. Since the
transformation is symplectic, the integrals over the new $b_j$
cycles then automatically are the derivatives
${\partial\mathcal{F}_0(t,s)\over\partial s_i}$. For
transformations of the second type the new prepotential is given
by $\mathcal{F}_0(t,\tilde S_i)+i\pi\tilde S_i^2$ and for
transformations of the third type the prepotential is a Legendre
transform with respect to $\int_{a^i}\zeta$. In the corresponding
gauge theory the latter transformations realise electric-magnetic
duality. Consider e.g. a symplectic transformation that exchanges
all compact $\a^i$-cycles with all compact $\b_i$ cycles:
\begin{equation}
\left(\begin{array}{c}\a^i\\\b_i\end{array}\right)\rightarrow\left(\begin{array}{c}a^i\\
b_i\end{array}\right)=\left(\begin{array}{c}\b_i\\-\a^i\end{array}\right),
\ \ \ i=1,\ldots n-1 \ .
\end{equation}
Then the new variables are the integrals over the $a_i$-cycles
which are
\begin{equation}
\tilde\pi_i:={1\over2}\int_{\b_i}\zeta={\partial\mathcal{F}_0(t,\tilde
S)\over\partial \tilde S_i}
\end{equation}
and the dual prepotential is given by the Legendre transformation
\begin{equation}
\mathcal{F}_{D}(t,\tilde\pi):=\sum_{i=1}^{n-1}\tilde
S_i\tilde\pi_i-\mathcal{F}_0(t,\tilde S)\ ,
\end{equation}
such that the new special geometry relation is
\begin{equation}
{\partial\mathcal{F}_D(t,\tilde\pi)\over\partial\tilde\pi_i}=\tilde
S_i={1\over4\pi i}\int_{\a^i}\zeta\ .
\end{equation}
Comparing with (\ref{F0}) one finds that
$\mathcal{F}_D(t,\tilde\pi)$ actually coincides with $F_0(t,J)$
where $J_i-J_{i+1}=\tilde\pi_i$ for $i=1,\ldots n-2$ and
$J_{n-1}=\tilde\pi_{n-1}$.

Next, let us see what happens if we also include symplectic
transformations involving the relative cycles $\hat\a$ and
$\hat\b$. An example of a transformation of type (i) that does not
mix $\{\a,\hat\a\}$ with $\{\b,\hat\b\}$ cycles is the one from
$\{\a^i,\b_j,\hat\a,\hat\b\}$ to $\{A^i,B_j\}$, c.f. Figs.
\ref{albecycles}, \ref{ABcycles}. This corresponds to
\begin{eqnarray}
\bar S_1&:=&\tilde S_1\ ,\nonumber\\
\bar S_i&:=&\tilde S_i-\tilde S_{i-1}\ \ \ \mbox{for}\ \ i=2,\ldots n-1\ ,\label{barS}\\
\bar S_n&:=&t-\tilde S_{n-1}\ ,\nonumber
\end{eqnarray}
so that
\begin{equation}
\bar S_i={1\over 4\pi i}\int_{A^i}\zeta\ .
\end{equation}
The prepotential does not change, except that it has to be
expressed in terms of the $\bar S_i$. One then finds for
$B_i=\sum_{j=i}^{n-1}\b_j+\hat\b$
\begin{equation}
{1\over2}\int_{B_i}\zeta={\partial\mathcal{F}_0(\bar
S)\over\partial \bar S_i}+W(\L_0)-\left(\sum_{j=1}^n\bar
S_i\right)\log\L_0^2+o\left({1\over\L_0}\right)\ .
\end{equation}
We see that as soon as one ``mixes" the cycle $\hat\b$ into the
set $\{\b_i\}$ one obtains a number of relative cycles $B_i$ for
which the special geometry relations are corrected by cut-off
dependent terms. An example of transformation of type (iii) is
$\hat\a\rightarrow\hat\b$, $\hat\b\rightarrow-\hat\a$. Instead of
$t$ one then uses
\begin{equation}
\hat\pi:={\partial\mathcal{F}_0(t,\tilde S)\over\partial
t}=\lim_{\L_0\rightarrow\infty}\left[{1\over2}\int_{\hat\b}\zeta-W(\L_0)+t\log\L_0^2\right]
\end{equation}
as independent variable and the Legendre transformed prepotential
is
\begin{equation}
\hat{\mathcal{F}}(\hat\pi,\tilde
S):=t\hat\pi-\mathcal{F}_0(t,\tilde S)\ ,
\end{equation}
so that now
\begin{equation}
{\partial\hat{\mathcal{F}}(\hat\pi,\tilde
S)\over\partial\hat\pi}=t={1\over4\pi i}\int_{\hat\a}\zeta\ .
\end{equation}
Note that the prepotential is well-defined and independent of the
cut-off in all cases (in contrast to the treatment in
\cite{DOV04}). The finiteness of $\hat{\mathcal{F}}$ is due to
$\hat\pi$ being the {\it corrected, finite} integral over the
relative $\hat\b$-cycle.\\
Note also that if one exchanges all coordinates simultaneously,
i.e. $\a^i\rightarrow\b_i,\
\hat\a\rightarrow\hat\b,\b_i\rightarrow-\a^i,\
\hat\b\rightarrow-\hat\a$, one has
\begin{equation}
\hat{\mathcal{F}}_D(\hat\pi,\tilde\pi):=t\hat\pi+\sum_{i=1}^{n-1}\tilde
S_i\tilde\pi_i-\mathcal{F}_0(t,\tilde S_i)\ .
\end{equation}
Using the generalised homogeneity relation (\ref{homogeneity})
this can be rewritten as
\begin{equation}
\hat{\mathcal{F}}_D(\hat\pi,\tilde\pi)=\mathcal{F}_0(t,\tilde
S_i)+t\int\d s\ \r_0(s;t,\tilde S_i)W(\l(s))\ .
\end{equation}

It would be quite interesting to understand the results of this
chapter concerning the parameter spaces of local Calabi-Yau
manifolds in a more geometrical way in the context of (rigid)
special K\"ahler manifolds along the lines of \ref{rigSG}.

\subsection{Example and summary}\label{exsum}
Let us pause for a moment and collect the results that we have
deduced so far. In order to compute the effective superpotential
of our gauge theory we have to study the integrals of $\O$ over
all compact and non-compact three-cycles on the space $X_{def}$,
given by
\begin{equation}
W'(x)^2+f_0(x)+v^2+w^2+z^2=0\ .
\end{equation}
These integrals map to integrals of $y\d x$ on the Riemann surface
$\S$, given by $y^2=W'(x)^2+f_0(x)$ over all the elements of
$H_1(\S,\{Q,Q'\})$. We have shown that it is useful to split the
elements of this set into a set of compact cycles $\a^i$ and
$\b_i$ and a set containing the compact cycle $\hat\a$ and the
non-compact cycle $\hat\b$, which together form a symplectic
basis. The corresponding three-cycles on the Calabi-Yau manifold
are $\G_{\a^i},\G_{\b_j},\G_{\hat\a},\G_{\hat\b}$. This choice of
cycles is appropriate, since the properties that arise from the
non-compactness of the manifold are then captured entirely by the
integral of the holomorphic three-form $\O$ over the non-compact
three-cycle $\G_{\hat\b}$ which corresponds to $\hat\b$. Indeed,
combining (\ref{integrals}), (\ref{sgrela}), (\ref{sgrelb}),
(\ref{SGhat1}) and (\ref{SGhat2}) one finds the following
relations
\begin{eqnarray}
-{1\over2\pi i}\int_{\G_{\a^i}}\O&=&2\pi i\tilde S_i\ ,\label{intalpha}\\
-{1\over2\pi i}\int_{\G_{\b_i}}\O&=&{\partial\mathcal{F}_0(t,\tilde S)\over\partial\tilde S_i}\ ,\label{intbeta}\\
-{1\over2\pi i}\int_{\G_{\hat\a}}\O&=&2\pi it\ ,\\
-{1\over2\pi i}\int_{\G_{\hat\b}}\O&=&
{\partial\mathcal{F}_0(t,\tilde S)\over\partial t}+W(\L_0)-t\log
\L_0^2+o\left({1\over\L_0}\right)\ .\label{intbetahat}
\end{eqnarray}
These relations are useful to calculate the integrals, since
$\mF_0(t,\tilde S)$ is the (Legendre transform of) the free energy
coupled to sources, and it can be evaluated from
(\ref{mathcalF0}). In the last relation the integral is understood
to be over the regulated cycle $\G_{\hat\b}$ which is an
$S^2$-fibration over a line segment running from the $n$-th cut to
the cut-off $\L_0$. Clearly, once the cut-off is removed, the last
integral diverges. This divergence will be studied in more detail
below.

\bigskip
{\bf The conifold}\\
Let us illustrate these ideas by looking at the simplest example,
the deformed conifold. In this case we have $n=1$,
$W(x)={x^2\over2}$ and $f_0(x)=-\m=-4t$, $\m\in\mathbb{R}^+$, and
$X_{def}=C_{def}$ is given by
\begin{equation}
x^2+v^2+w^2+z^2-\m=0\ .
\end{equation}
As $n=1$ the corresponding Riemann surface has genus zero. Then
\begin{equation}
\zeta=y\d x=\left\{\begin{array}{c}\ \ \sqrt{x^2-4t}\ \d x\ \ \
\mbox{on the upper sheet}\\-\sqrt{x^2-4t}\ \d x\ \ \ \mbox{on the
lower sheet}\end{array}\right.\ .
\end{equation}
We have a cut $\mathcal{C}=[-2\sqrt{t},2\sqrt{t}]$ and take
$\l(s)=s$ to run along the real axis. The corresponding $\r_0(s)$
is immediately obtained from (\ref{solution}) and (\ref{y0omega0})
and yields the well-known $\r_0(s)={1\over2\pi t}\sqrt{4t-s^2}$,
for $s\in[-2\sqrt{t},2\sqrt{t}]$ and zero otherwise, and from
(\ref{mathcalF0}) we find the planar free energy
\begin{equation}
\mathcal{F}_0(t)={t^2\over2}\log t-{3\over4}t^2\
.\label{F0conifold}
\end{equation}
Note that $t\int\d s\,\r_0(s)W(\l(s))={t^2\over2}$ and
$\mathcal{F}_0$ satisfies the generalised homogeneity relation
(\ref{homogeneity})
\begin{equation}
t{\partial\mathcal{F}_0\over\partial
t}(t)=2\mathcal{F}_0(t)+{t^2\over2}\ .
\end{equation}

For the deformed conifold the integrals over $\O$ can be
calculated without much difficulty and one obtains
\begin{eqnarray}
-{1\over2\pi i}\int_{\G_{\hat\a}}\O={1\over2}\int_{\hat\a}\zeta&=&2\pi i t=2\pi i\bar S\label{intcon1}\\
-{1\over2\pi
i}\int_{\G_{\hat\b}}\O={1\over2}\int_{\hat\b}\zeta&=&{\L_0\over2}\sqrt{\L_0^2-4t}-2t\log\left({\L_0\over2\sqrt
t}+\sqrt{{\L_0^2\over4t}-1}\right)\ .\label{intcon2}
\end{eqnarray}
On the other hand, using the explicit form of $\mF_0(t)$ we find
\begin{equation}
{\partial\mathcal{F}_0(t)\over\partial t}+W(\L_0)-t\log
\L_0^2={\L_0^2\over2}+t\log\left({t\over\L_0^2}\right)-t
\end{equation}
which agrees with (\ref{intcon2}) up to terms of order
$o\left(1\over\L_0^2\right)$.

\chapter{Superstrings, the Geometric Transition and Matrix
Models}\label{SSGTMM}
We are now in a position to combine all the results obtained so
far, and explain the conjecture of Cachazo, Intriligator and Vafa
\cite{CIV01} in more detail. Recall that our goal is to determine
the effective superpotential, and hence the vacuum structure, of
$\mN=1$ super Yang-Mills theory with gauge group $U(N)$ coupled to
a chiral superfield $\Phi$ in the adjoint representation with
tree-level superpotential
\begin{equation}
W(\Phi)=\sum_{k=1}^\infty{g_k\over k}\tr \Phi^k+g_0\ .
\end{equation}
In a given vacuum of this theory the eigenvalues of $\Phi$ sit at
the critical points of $W(x)$, and the gauge group $U(N)$ breaks
to $\prod_{i=1}^nU(N_i)$, if $N_i$ is the number of eigenvalues at
the $i$-th critical point of $W(x)$. Note that such a $\Phi$ also
satisfies the D-flatness condition (\ref{Dflatness}), since
$\tr([\Phi^\dagger,\Phi]^2)=0$.

At low energies the $SU(N_i)$-part of the group $U(N_i)$ confines
and the remaining gauge group is $U(1)^n$. The good degrees of
freedom in this low energy limit are the massless photons of the
$U(1)\subset U(N_i)$ and the massive chiral
superfields\footnote{For clarity we introduce a superscript $gt$
for gauge theory quantities and $st$ for string theory objects.
These will be identified momentarily, and then the superscript
will be dropped.} $S_i^{gt}$ with the gaugino bilinear
$\l_a^{(i)}\l^{a(i)}$ as their lowest component. We want to
determine the quantities $\la S_i^{gt}\ra$, and these can be
obtained from minimising the effective superpotential,
\begin{equation}
\left.{\partial W_{eff}^{gt}(S_i^{gt})\over\partial
S_i^{gt}}\right|_{\la S_i^{gt}\ra}=0\ .
\end{equation}
Therefore, we are interested in determining the function
$W_{eff}^{gt}(S_i^{gt})$. Since we are in the low energy regime of
the gauge theory, where the coupling constant is large, this is a
very hard problem in field theory. However, it turns out that the
effective superpotential can be calculated in a very
elegant way in the context of string theory.\\

As mentioned in the introduction, the specific vacuum of the
original $U(N)$ gauge theory, in which the gauge group $U(N)$ is
broken to $\prod_{i=1}^nU(N_i)$, can be generated from type IIB
theory by wrapping $N_i$ D5-branes around the $i$-th
$\mathbb{CP}^1$ in $X_{res}$ (as defined in (\ref{Xres})). We will
not give a rigorous proof of this statement, but refer the reader
to \cite{KKLM99} where many of the details have been worked out.
Here we only try to motivate the result, using some more or less
heuristic arguments. Clearly, the compactification of type IIB on
$X_{res}$ leads to an $\mN=2$ theory in four dimensions. The
geometric engineering of $\mN=2$ theories from local Calabi-Yau
manifolds is reviewed in \cite{Ma98}. Introducing the
$N=\sum_iN_i$ D5-branes now has two effects. First, the branes
reduce the amount of supersymmetry. If put at arbitrary positions,
the branes will break supersymmetry completely. However, as shown
in \cite{BBS95}, if two dimensions of the branes wrap holomorphic
cycles in $X_{res}$, which are nothing but the resolved
$\mathbb{CP}^1$, and the other dimensions fill Minkowski space,
they only break half of the supersymmetry, leading to an $\mN=1$
theory. Furthermore, $U(N_i)$ vector multiplets arise from open
strings polarised along Minkowski space, whereas those strings
that start and end on the wrapped branes lead to a
four-dimensional chiral superfield $\Phi$ in the adjoint
representation of the gauge group. A brane wrapped around a
holomorphic cycle $\mathcal{C}$ in a Calabi-Yau manifold $X$ can
be deformed without breaking the supersymmetry, provided there
exist holomorphic sections of the normal bundle $N\mathcal{C}$.
These deformations are the scalar fields in the chiral multiplet,
which therefore describe the position of the wrapped brane. The
number of these deformations, and hence of the chiral fields, is
therefore given by the number of holomorphic sections of
$N\mathcal{C}$. However, on some geometries these deformations are
obstructed, i.e. one cannot construct a finite deformation from an
infinitesimal one (see \cite{KKLM99} for the mathematical
details). These obstructions are reflected in the fact that there
exists a superpotential for the chiral superfield at the level of
the four-dimensional action. In \cite{KKLM99} it is shown that the
geometry of $X_{res}$ is such that the superpotential of the
four-dimensional field theory is nothing but $W(\Phi)$.

We have seen that the local Calabi-Yau manifold $X_{res}$ can go
through a geometric transition, leading to the deformed space
$X_{def}$ (as defined in (\ref{Xdef})). This tells us that there
is another interesting setup, which is intimately linked to the
above, namely type IIB on $X_{def}$ with additional three-form
background flux $G_3$. The background flux is necessary in order
to ensure that the four-dimensional theory is $\mN=1$
supersymmetric. A heuristic argument for its presence has been
given in the introduction. The $\mN=1$ four-dimensional theory
generated by this string theory then contains $n$ $U(1)$ vector
superfields and $n$ chiral superfields\footnote{The bar in $\bar
S_i^{st}$ indicates that in the defining equation (\ref{barSi})
one uses the set of cycles $\G_{A^i},\G_{B_j}$, c.f. Eq.
(\ref{barS}).} $\bar S_i^{st}$, the lowest component of which is
proportional to the size\footnote{$\O$ is a calibration, i.e. it
reduces to the volume form on $\G_{A^i}$, see e.g. \cite{Jo01}.}
of the three-cycles in $X_{def}$,
\begin{equation}\label{barSi}
\bar S_i^{st}:={1\over4\pi^2}\int_{\G_{A^i}}\O\ .
\end{equation}
The superpotential $W_{eff}^{st}$ for these fields is given by the
formula\footnote{The original formula has the elegant form
$W_{eff}^{st}={1\over2\pi i}\int_{X} G_3\w\O$, which can then be
written as (\ref{GVWAB}) using the Riemann bilinear relations for
Calabi-Yau manifolds. However, $X_{def}$ is a {\it local}
Calabi-Yau manifold and we are not aware of a proof that the
bilinear relation does hold for these spaces as well.}
\cite{GVW99}
\begin{equation}\label{GVWAB}
W_{eff}^{st}(\bar S_i^{st})={1\over2\pi
i}\sum_{i=1}^n\left(\int_{\G_{A_i}}G_3\int_{\G_{B_i}}\O-\int_{\G_{B_i}}G_3\int_{\G_{A^i}}\O\right)\
,
\end{equation}
see also \cite{TV99}, \cite{M00}. Of course, $W_{eff}^{st}$ does
not only depend on the $\bar S_i^{st}$, but also on the $g_k$,
since $\O$ is defined in terms of the tree-level superpotential,
c.f. Eq. (\ref{Omega}). Furthermore, as we will see, it also
depends on parameters $\L_i$, which will be identified with the
dynamical scales of the $SU(N_i)$ theories below. It is quite
interesting to compare the derivation of this formula in
\cite{GVW99} with the one of the Veneziano-Yankielowicz formula in
\cite{VY82}. The logic is very similar, and indeed, as we will
see, Eq. (\ref{GVWAB}) gives the effective superpotential in the
Veneziano-Yankielowicz sense. It is useful to define
\begin{equation}
\tilde S^{st}_i:={1\over4\pi^2}\int_{\G_{\a^i}}\O\ \ \ ,\ \ \
t^{st}:={1\over4\pi^2}\int_{\G_{\hat\a}}\O\ ,
\end{equation}
for $i=1,\ldots n-1$, and use the set of cycles
$\{\G_{\a^i},\G_{\b_j},\G_{\hat\a},\G_{\hat\b}\}$ instead (see the
discussion in section \ref{periodint} and Figs. \ref{ABcycles},
\ref{albecycles} for the definition of these cycles). Then we find
\begin{eqnarray}\label{Weff}
&&W_{eff}^{st}(t^{st},\tilde
S^{st}_i)=\nonumber\\
&&={1\over2\pi
i}\sum_{i=1}^{n-1}\left(\int_{\G_{\a^i}}G_3\int_{\G_{\b_i}}
\O-\int_{\G_{\b_i}}G_3\int_{\G_{\a^i}}\O\right)+{1\over2\pi
i}\left(\int_{\G_{\hat\a}}G_3\int_{\G_{\hat\b}}
\O-\int_{\G_{\hat\b}}G_3\int_{\G_{\hat\a}}\O\right)\nonumber\\
&&=-{1\over2}\sum_{i=1}^{n-1}\left(\int_{\G_{\a^i}}G_3\int_{\b_i}\zeta-\int_{\G_{\b_i}}
G_3\int_{\a^i}\zeta\right)-{1\over2}\left(\int_{\G_{\hat\a}}G_3\int_{\hat\b}\zeta-\int_{\G_{\hat\b}}
G_3\int_{\hat\a}\zeta\right)\ .\nonumber\\
\end{eqnarray}
Here we used the fact that, as explained in section
\ref{periodint}, the integrals of $\O$ over three-cycles reduce to
integrals of $\zeta:=y\d x$ over the one-cycles in
$H_1(\S,\{Q,Q'\})$ on the Riemann surface $\S$,
\begin{equation}\label{Sigma}
y^2=W'(x)^2+f_0(x)\ ,
\end{equation}
c.f. Eq. (\ref{integrals}). However, from our analysis of the
matrix model we know that the integral of $\zeta$ over the cycle
$\hat \b$ is divergent, c.f. Eq. (\ref{SGhat2}). Since the
effective superpotential has to be finite we see that (\ref{Weff})
cannot yet be the correct formula. Indeed from inspection of
(\ref{SGhat2}) we find that it contains two divergent terms, one
logarithmic, and one polynomial divergence. The way how these
divergences are dealt with, and how a finite effective
superpotential is generated, will be explained in the first section of this chapter.\\

The claim of Cachazo, Intriligator and Vafa is that the
four-dimensional superpotential (\ref{GVWAB}), generated from IIB
on $\mathbb{R}^4\times X_{def}$ with background flux, is nothing
but the effective superpotential (in the Veneziano-Yankiwlowicz
sense) of the original gauge theory,
\begin{equation}\label{CIVconjecture}
W_{eff}^{gt}(S_i^{gt})\equiv W_{eff}^{st}(\bar S_i^{st})\ .
\end{equation}
Put differently we have
\begin{equation}
\la S^{gt}_i\ra=\la \bar S_i^{st}\ra\ ,
\end{equation}
where the left-hand side is the vacuum expectation value of the
gauge theory operator $S_i^{gt}$, whereas the right-hand side is a
K\"ahler parameter of $X_{def}$ (proportional to the size of
$\G_{A^i}$), that solves
\begin{equation}
\left.\partial W_{eff}^{st}(\bar S_i^{st})\over\partial \bar
S_i^{st}\right|_{\la \bar S_i^{st}\ra}=0\ .
\end{equation}
Therefore, the vacuum structure of a gauge theory can be studied
by evaluating $W_{eff}^{st}$, i.e. by performing integrals in the
geometry $X_{def}$. From now on the superscripts $gt$ and $st$
will be suppressed.

We are going to check the conjecture of Cachazo-Intriligator and
Vafa by looking at simple examples below.

\section{Superpotentials from string theory with fluxes}
In this section we will first analyse the divergences in
(\ref{Weff}) and show that the effective potential is actually
finite if we modify the integration over $\O$ in a suitable way.
We closely follow the analysis of \cite{BM05}, where the necessary
correction terms were calculated. Then we use our matrix model
results to relate the effective superpotential to the planar limit
of the matrix model free energy.

\subsection{Pairings on Riemann surfaces with marked points}
In order to understand the divergences somewhat better, we will
study the meromorphic one-form $\zeta:=y\d x$ on the Riemann
surface $\S$ given by Eq. (\ref{Sigma}) in more detail. Recall
that $Q,Q'$ are those points on the Riemann surface that
correspond to $\infty,\infty'$ on the two-sheets of the
representation (\ref{Sigma}) and that these are the points where
$\zeta$ has a pole. The surface with the points $Q,Q'$ removed is
denoted by $\hat\S$. First of all we observe that the integrals
$\int_{\a^i}\zeta$ and $\int_{\b_j}\zeta$ only depend on the
cohomology class $[\zeta]\in H^1(\hat\S)$, whereas
$\int_{\hat\b}\zeta$  (where $\hat\b$ extends between the poles of
$\zeta$, i.e. from $\infty'$ on the lower sheet, corresponding to
$Q'$, to $\infty$ on the upper sheet, corresponding to $Q$,) is
not only divergent, it also depends on the representative of the
cohomology class, since for $\tilde\zeta= \zeta+\d\rho$ one has
$\int_{\hat\b}\tilde\zeta=\int_{\hat\b}\zeta+
\int_{\partial\hat\b}\r\left(\neq\int_{\hat\b}\zeta\right)$. Note
that the integral would be independent of the choice of the
representative if we constrained $\r$ to be zero at
$\partial\hat\b$. But as we marked $Q,Q'$ on the Riemann surface,
$\r$ is allowed to take finite or even infinite values at these
points and therefore the integrals differ in general.

The origin of this complication is, of course, that our cycles are
elements of the {\it relative} homology group $H_1(\S,\{Q,Q'\})$.
Then, their is a natural map
$\langle.,.\rangle:H_1(\S,\{Q,Q'\})\\\times
H^1(\S,\{Q,Q'\})\rightarrow \mathbb{C}$. $H^1(\S,\{Q,Q'\})$ is the
relative cohomology group dual to $H_1(\S,\\\{Q,Q'\})$. In
general, on a manifold $M$ with submanifold $N$, elements of
relative cohomology can be defined as follows (see for example
\cite{KL87}). Let $\O^k(M,N)$ be the set of $k$-forms on $M$ that
vanish on $N$. Then $H^k(M,N):=Z^k(M,N)/B^k(M,N)$, where
$Z^k(M,N):=\{\o\in\O^k(M,N):\d\o=0\}$ and
$B^k(M,N):=\d\O^{k-1}(M,N)$. For $[\hat\G]\in H_k(M,N)$ and
$[\eta]\in H^k(M,N)$ the pairing is defined as
\begin{equation}
\langle\hat\G,\eta\rangle_0:=\int_{\hat \G}\eta\ .
\end{equation}
This does not depend on the representative of the classes, since
the forms are constraint to vanish on $N$.

Now consider $\xi\in\O^k(M)$ such that $i^*\xi=\d\phi$, where
$i:N\rightarrow M$ is the inclusion mapping. Note that $\xi$ is
not a representative of an element of relative cohomology, as it
does not vanish on $N$. However, there is another representative
in its cohomology class $[\xi]\in H^k(M)$, namely
$\xi_{\phi}=\xi-\d\phi$ which now is also a representative of
$H^k(M,N)$. For elements $\xi$ with this property we can extend
the definition of the pairing to
\begin{equation}
\langle \hat\G,\xi\rangle_0:=\int_{\hat
\G}\left(\xi-\d\phi\right)\ .
\end{equation}
More details on the various possible definitions of relative
(co-)homology can be found in appendix \ref{relhom}.

Clearly, the one-form $\zeta=y\d x$ on $\hat\S$ is not a
representative of an element of $H^1(\S,\{Q,Q'\})$. According to
the previous discussion, one might try to find
$\zeta_{\varphi}=\zeta-\d\varphi$ where $\varphi$ is chosen in
such a way that $\zeta_{\varphi}$ vanishes at $Q,Q'$, so that in
particular $\int_{\hat\b}\zeta_{\varphi}=\mbox{finite}$. In other
words, we would like to find a representative of $[\zeta]\in
H^1(\hat\S)$ which is also a representative of $H^1(\S,\{Q,Q'\})$.
Unfortunately, this is not possible, because of the logarithmic
divergence, i.e. the simple poles at $Q,Q'$, which cannot be
removed by an exact form. The next best thing we can do instead is
to determine $\varphi$ by the requirement that
$\zeta_{\varphi}=\zeta-\d\varphi$ only has simple poles at $Q,Q'$.
Then we define the pairing \cite{BM05}
\begin{equation}\label{pair}
\left\langle\hat\b,\zeta\right\rangle:=\int_{\hat\b}\left(\zeta-\d\varphi\right)=
\int_{\hat\b}\zeta_{\varphi}\ ,
\end{equation}
which diverges only logarithmically. To regulate this divergence
we introduce a cut-off as before, i.e. we take $\hat\b$ to run
from $\L_0'$ to $\L_0$. We will have more to say about this
logarithmic divergence below. So although $\zeta_{\varphi}$ is not
a representative of a class in $H^1(\S,\{Q,Q'\})$ it is as close
as we can get.

We now want to determine $\varphi$ explicitly. To keep track of
the poles and zeros of the various terms it is useful to apply the
theory of divisors, as explained in appendix \ref{div} and  e.g.
in \cite{FK}. The divisors of various functions and forms on $\S$
have already been explained in detail in section \ref{propRS}.
Consider now $\varphi_k:={x^k\over y}$ with $\d\varphi_k={{k
x^{k-1}\d x}\over y}-{x^k {y^2}'\d x\over2 y^3}$. For $x$ close to
$Q$ or $Q'$ the leading term of this expression is
$\pm(k-\gh-1)x^{k-\gh-2}\d x.$ This has no pole at $Q,Q'$ for
$k\leq \gh$, and for $k=\gh+1$ the coefficient vanishes, so that
we do not get simple poles at $Q,Q'$. This is as expected as
$\d\varphi_k$ is exact and cannot contain poles of first order.
For $k\geq\gh+2=n+1$ the leading term has a pole of order $k-\gh$
and so $\d\varphi_k$ contains poles of order
$k-\gh,k-\gh-1,\ldots2$ at $Q,Q'$. Note also that at $P_1,\ldots
P_{2\gh+2}$ one has double poles for all $k$ (unless a zero of $y$
occurs at $x=0$). Next, we set
\begin{equation}
\varphi={\mathcal{P}\over y} ,
\end{equation}
with $\mathcal{P} $ a polynomial of order $2\gh+3$. Then $\d
\varphi$ has poles of order $\gh+3,\gh+2,\ldots 2$ at $Q,Q'$, and
double poles at the zeros of $y$ (unless a zero of $\mathcal{P}_k$
coincides with one of the zeros of $y$). From the previous
discussion it is clear that we can choose the coefficients in
$\mathcal{P}$ such that $\zeta_{\varphi}=\zeta-\d\varphi$ only has
a simple pole at $Q,Q'$ and double poles at $P_1,P_2,\ldots
P_{2\gh+2}$. Actually, the coefficients of the monomials $x^k$ in
$\mathcal{P}$ with $k\leq\gh$ are not fixed by this requirement.
Only the $\gh+2$ highest coefficients will be determined, in
agreement with the fact that we cancel the $\gh+2$ poles of order
$\gh+3,\ldots2$.

It remains to determine the polynomial $\mathcal{P}$ explicitly.
The part of $\zeta$ contributing to the poles of order $\geq2$ at
$Q,Q'$ is easily seen to be $\pm W'(x)\d x$ and we obtain the
condition
\begin{equation}
W'(x)-\left(\mathcal{P}(x)\over\sqrt{W'(x)^2+f(x)}\right)'=
o\left(1\over x^2\right)\ .
\end{equation}
Integrating this equation, multiplying by the square root and
developing the square root leads to
\begin{equation}
W(x)W'(x)-{2t\over {n+1}}x^{n}-\mathcal{P}(x)=cx^{n}+
o\left(x^{n-1}\right)\ ,
\end{equation}
where $c$ is an integration constant. We read off \cite{BM05}
\begin{equation}\label{varphi}
\varphi(x)={W(x)W'(x)-\left({2t\over {n+1}}+c\right)x^{n}+
o\left(x^{n-1}\right)\over y}\ ,
\end{equation}
and in particular, for $x$ close to infinity on the upper or lower
sheet,
\begin{equation}
\varphi(x)\sim\pm\left[W(x)-c+ o\left(1\over x\right)\right]\
.\label{philarge}
\end{equation}
The arbitrariness in the choice of $c$ has to do with the fact
that the constant $W(0)$ does not appear in the description of the
Riemann surface. In the sequel we will choose $c=0$, such that the
full $W(x)$ appears in (\ref{philarge}). As is clear from our
construction, and is easily verified explicitly, close to $Q,Q'$
one has $\zeta_{\varphi}\sim\left(\mp{2t\over x}+o\left({1\over
x^2}\right)\right)\d x$.

With this $\varphi$ we find
\begin{equation}\label{intzetaphi}
\int_{\hat\b}\zeta_{\varphi}=\int_{\hat\b}\zeta-\int_{\hat\b}\d
\varphi=
\int_{\hat\b}\zeta-\varphi(\L_0)+\varphi(\L_0')=\int_{\hat\b}\zeta-2\left(W(\L_0)+o\left(1\over\L_0\right)\right)\
.
\end{equation}
Note that, contrary to $\zeta$, $\zeta_{\varphi}$ has poles at the
zeros of $y$, but these are double poles and it does not matter
how the cycle is chosen with respect to the location of these
poles (as long as it does not go right through the poles). Note
also that we do not need to evaluate the integral of
$\zeta_{\varphi}$ explicitly. Rather one can use the known result
(\ref{SGhat2}) for the integral of $\zeta$ to find from
(\ref{intzetaphi})
\begin{equation}\label{resultpairing}
{1\over2}\left\langle\hat\b,\zeta\right\rangle={1\over2}\int_{\hat\b}\zeta_{\varphi}={\partial\over\partial
t}\mathcal{F}_0(t,\tilde
S)-t\log\L_0^2+o\left({1\over\L_0}\right)\ .
\end{equation}

Let us comment on the independence of the representative of the
class $[\zeta]\in H^1(\hat\S)$. Suppose we had started from
$\tilde\zeta:=\zeta +\d\rho$ instead of $\zeta$. Then determining
$\tilde\varphi$ by the same requirement that
$\tilde\zeta-\d\tilde\varphi$ only has first order poles at $Q$
and $Q'$ would have led to $\tilde\varphi=\varphi+\rho$ (a
possible ambiguity related to the integration constant $c$ again
has to be fixed). Then obviously
\begin{equation}
\left\langle\hat\b,\tilde\zeta\right
\rangle=\int_{\hat\b}\tilde\zeta-\int_{\partial\hat\b}\tilde\varphi=\int_{\hat\b}\zeta-\int_{\partial\hat\b}\varphi=\left\langle\hat\b,\zeta\right\rangle\
,
\end{equation}
and hence our pairing only depends on the cohomology class
$[\zeta]$.

Finally, we want to lift the discussion to the local Calabi-Yau
manifold. There we define the pairing
\begin{equation}\label{pairCY}
\left\langle\G_{\hat\b},\O\right\rangle:=\int_{\G_{\hat\b}}\left(\O-\d\Phi\right)=
(-i\pi)\int_{\hat\b}\left(\zeta-\d\varphi\right)\ ,
\end{equation}
where (recall that $c=0$)
\begin{equation}
\Phi:={W(x)W'(x)-{2t\over{n+1}}x^n\over W'(x)^2+f_0(x)}\cdot{\d
v\w\d w\over 2z}
\end{equation}
is such that
$\int_{\G_{\hat\b}}\d\Phi=-i\pi\int_{\hat\b}\d\varphi$. Clearly,
we have
\begin{equation}
-{1\over2\pi
i}\left\langle\G_{\hat\b},\O\right\rangle={\partial\mathcal{F}_0(t,\tilde
S)\over\partial t}-t\log \L_0^2+o\left({1\over\L_0}\right)\ .
\end{equation}

\subsection{The superpotential and matrix models}
Let us now return to the effective superpotential $W_{eff}$ in
(\ref{Weff}). Following \cite{CIV01} and \cite{DV02c} we have for
the integrals of $G_3$ over the cycles $\G_A$ and $\G_B$:
\begin{equation}
N_i=\int_{\G_{A^i}}G_3\ \ ,\ \ \tau_i:=
\left\langle\G_{B_i},G_3\right\rangle\ \ \ \ \mbox{for}\ \
i=1,\ldots n\ .
\end{equation}
Here we used the fact that the integral over the flux on the
deformed space should be the same as the number of branes in the
resolved space. It follows for the integrals over the cycles
$\G_{\a}$ and $\G_{\b}$
\begin{eqnarray}
\tilde N_i:=\int_{\G_{\a^i}}G_3=\sum_{j=1}^iN_j\ \ &,&\ \ \tilde
\tau_i:=\int_{\G_{\b_i}}G_3=\tau_i-\tau_{i+1}\ \ \ \ \mbox{for}\ \
i=1,\ldots n-1\
,\nonumber\\
N=\sum_{i=1}^nN_i= \int_{\G_{\hat\a}}G_3\ \ &,&\ \
\tilde\tau_0:=\left\langle\G_{\hat\b},G_3\right\rangle=\tau_n\ .
\end{eqnarray}
For the non-compact cycles, instead of the usual integrals, we use
the pairings of the previous section. On the Calabi-Yau, the
pairings are to be understood e.g. as $\tau_i=-i\pi\left\langle
B_i,h\right\rangle$, where $\int_{S^2}G_3=-2\pi i h$ and $S^2$ is
the sphere in the fibre of $\G_{B_i}\rightarrow B_i$. Note that
this implies that the $\tau_i$ as well as $\tilde\tau_0$ have (at
most) a logarithmic divergence, whereas the $\tilde\tau_i$ are
finite. We propose \cite{BM05} that the superpotential should be
defined as
\begin{eqnarray}\label{newsuperpotential}
&&\hspace{-0,7cm}W_{eff}(t,\tilde S_i)=\nonumber\\
&&\hspace{-0,7cm}={1\over2\pi
i}\sum_{i=1}^{n-1}\left(\int_{\G_{\a^i}}G_3\int_{\G_{\b_i}}
\O-\int_{\G_{\b_i}}G_3\int_{\G_{\a^i}}\O\right)+{1\over2\pi
i}\left(\int_{\G_{\hat\a}}G_3\cdot \left\langle\G_{\hat\b},
\O\right\rangle-\left\langle\G_{\hat\b},G_3\right\rangle\int_{\G_{\hat\a}}\O\right)\nonumber\\
&&\hspace{-0,7cm}=-{1\over2}\sum_{i=1}^{n-1}\left(\tilde
N_i\int_{\b_i}\zeta-
\tilde\tau_i\int_{\a^i}\zeta\right)-{1\over2}\left(N
\left\langle\hat\b,\zeta\right\rangle-\tilde\tau_0\int_{\hat\a}\zeta\right)\
.
\end{eqnarray}
This formula is very similar to the one advocated for example in
\cite{LMW02}, but now the pairing (\ref{pair}) is to be used. Note
that Eq. (\ref{newsuperpotential}) is invariant under symplectic
transformations on the basis of (relative) three-cycles on the
local Calabi-Yau manifold, resp. (relative) one-cycles on the
Riemann surface, provided one uses the pairing (\ref{pair}) for
every relative cycle. These include $\a^i\rightarrow\b_i,\
\hat\a\rightarrow\hat\b,\ \b_i\rightarrow-\a^i,\
\hat\b\rightarrow-\hat\a$, which acts as electric-magnetic
duality.

It is quite interesting to note what happens to our formula for
the superpotential in the classical limit, in which we have
$\tilde S_i=t=0$, i.e. $f_0(x)\equiv0$. In this case we have
$\zeta=\d W$ and we find
\begin{eqnarray}
W_{eff}&=&\sum_{i=1}^{n-1}\tilde N_i\int_{\m_i}^{\m_{i+1}}\d W+
{N\over2}\left(2\int_{\m_n}^{\L_0}\d
W-\int_{\L_0'}^{\L_0}\d\left({W(x)W'(x)+o\left(x^{n-1}\right)\over
y}\right)\right)\nonumber\\
&=&\sum_{i=1}^nN_iW(\m_i)\ ,
\end{eqnarray}
where $\m_i$ are the critical points of $W$. But this is nothing
but the value of the tree-level superpotential in the vacuum
in which $U(N)$ is broken to $U(N_i)$.\\

By now it should be clear that the matrix model analysis was
indeed very useful to determine physical quantities. Not only do
we have a precise understanding of the divergences, but we can now
also rewrite the effective superpotential in terms of the matrix
model free energy. Using the special geometry relations
(\ref{sgrela}), (\ref{sgrelb}) for the standard cycles and
(\ref{SGhat1}), (\ref{resultpairing}) for the relative cycles, we
obtain
\begin{eqnarray}
W_{eff}&=&-\sum_{i=1}^{n-1}\tilde N_i{\partial\over\partial \tilde
S_i}\mathcal{F}_0(t,\tilde S_1,\ldots,\tilde S_{n-1})+2\pi
i\sum_{i=1}^{n-1}\tilde\tau_i\tilde
S_i\nonumber\\&&-N{\partial\over\partial t}\mathcal{F}_0(t,\tilde
S_1,\ldots,\tilde S_{n-1})+\left(N\log \L^2_0+2\pi
i\tilde\tau_0\right)t+o\left({1\over\L_0}\right) \ .\label{WL0}
\end{eqnarray}
The limit $\L_0\rightarrow \infty$ can now be taken provided
$N\log\L_0^2+2\pi i\tilde\tau_0$ is finite. In other words, $2\pi
i\tilde\tau_0$ itself has to contain a term $-N\log\L_0^2$ which
cancels against the first one. This is of course quite reasonable,
since $\tilde\tau_0$ is defined via the pairing
$\left\langle\hat\b,h\right\rangle$, which is expected to contain
a logarithmic divergence. Note that $\tilde\tau_0$ is the only
flux number in (\ref{WL0}) that depends on $\L_0$. This
logarithmic dependence on some scale parameter is of course
familiar from quantum field theory, where we know that the
coupling constants depend logarithmically on some energy scale. It
is then very natural to identify \cite{CIV01} the flux number
$\tilde\tau_0$ with the gauge theory bare coupling as
\begin{equation}\label{tau0}
\tilde\tau_0={4\pi i\over g_0^2}+{\Theta\over2\pi}\ .
\end{equation}
In order to see this in more detail note that our gauge theory
with a chiral superfield in the adjoint has a $\b$-function
$\b(g)=-{2N\over16\pi^2}g^3$. This leads to ${1\over
g^2(\m)}={2N\over8\pi^2}\log\left({\m\over|\L|}\right)$. If
$\tilde\L_0$ is the cut-off of the gauge theory we have ${1\over
g_0^2}\equiv{1\over g^2(\tilde\L_0)}={1\over
g^2(\m)}+{2N\over8\pi^2}\log\left(\tilde\L_0\over\m\right)$. We
now have to identify the gauge theory cut-off $\tilde \L_0$ with
the cut-off $\L_0$ on the Riemann surface as
\begin{equation}
\L_0={\tilde\L_0}\label{id}
\end{equation}
to obtain a finite effective superpotential. Indeed, then one gets
\begin{equation}\label{renorm}
N\log\L_0^2+2\pi i\tilde\tau_0=2\pi
i\tilde\tau(\m)+2N\log\m=2N\log\L\ ,
\end{equation}
with finite $\L=|\L|e^{i\Theta/2N}$ and $\tilde\tau(\m)$ as in
(\ref{tau0}), but now with $g(\m)$ instead of $g_0$. Note that
$\L$ is the dynamical scale of the gauge theory with dimension 1,
which has nothing to do with the cut-off $\L_0$.

Eq. (\ref{WL0}) can be brought into the form of \cite{DV02c} if we
use the coordinates $\bar S_i$, as defined in (\ref{barS}) and
such that $\bar S_i={1\over4\pi i}\int_{A^i}\zeta$ for all
$i=1,\ldots n$. We get
\begin{equation}\label{DVsuperpotential}
W_{eff}=-\sum_{i=1}^{n} N_i{\partial\over\partial \bar
S_i}\mathcal{F}_0(\bar S)+\sum_{i=1}^{n-1}\bar S_i\left(2\pi
i\sum_{j=i}^{n-1}\tilde\tau_j+\log\L^{2N}\right)+\bar
S_{n}\log\L^{2N}\ .
\end{equation}
In order to compare to \cite{DV02c} we have to identify
${\partial\mathcal{F}_0(S)\over\partial \bar S_i}$ with
${\partial\mathcal{F}_0^{p}(S)\over\partial \bar S_i}+\bar
S_i\log\bar S_i+\ldots$, where $\mathcal{F}_0^{p}$ is the
perturbative part of the free energy of the matrix model. Indeed,
it was argued in \cite{DV02c} that the $\bar S_i\log \bar S_i$
terms come from the measure and are contained in the
non-perturbative part
${\partial\mathcal{F}_0^{np}\over\partial\bar S_i}$. In fact, the
presence of these terms in
${\partial\mathcal{F}_0^{np}\over\partial\bar S_i}$ can easily be
proven by monodromy arguments \cite{CIV01}. Alternatively, the
presence of the $S_i\log S_i$-terms in $\mathcal{F}_0$ can be
proven by computing $\mF_0$ exactly in the planar limit. We will
discuss some explicit examples below.

We could have chosen $\hat\b$ to run from a point
$\L_0'=|\L_0|e^{i\t/2}$ on the lower sheet to a point
$\L_0=|\L_0|e^{i\t/2}$ on the upper sheet. Then one would have
obtained an additional term $-it\t$, on the right-hand side of
(\ref{resultpairing}), which would have led to
\begin{equation}
\Theta\rightarrow\Theta+N\t
\end{equation}
in (\ref{DVsuperpotential}).\\

Note that (\ref{DVsuperpotential}) has dramatic consequences. In
particular, after inserting $\mF_0=\mF_0^{p}+\mF_0^{np}$ we see
that the effective superpotential, which upon extremisation gives
a non-perturbative quantity in the gauge theory can be calculated
from a perturbative expansion in a corresponding matrix model. To
be more precise, $\mF_0^p$ can be calculated by expanding the
matrix model around a vacuum in which the filling fractions
$\n_i^*$ are fixed in such a way that the pattern of the gauge
group $\prod_{i=1}^n U(N_i)$ is reproduced. This means that
whenever $N_i=0$ we choose $\n_i^*=0$ and whenever $N_i\neq0$ we
have $\n_i^*\neq0$. Otherwise the $\n_i^*$ can be chosen
arbitrarily (and are in particular independent of the $N_i$.)
$\mF_0^p$ is then given by the sum of all planar vacuum
amplitudes.

As was shown in \cite{DV02c} this can be generalised to more
complicated $\mN=1$ theories with different gauge groups and field
contents.

\section{Example: Superstrings on the conifold}
Next we want to illustrate our general discussion by looking at
the simplest example, i.e. $n=1$ and $W={x^2\over2}$. This means
we study type IIB string theory on the resolved conifold. Wrapping
$N$ D5-branes around the single $\mathbb{CP}^1$ generates $\mN=1$
Super-Yang-Mills theory with gauge group $U(N)$ in four
dimensions. The corresponding low energy effective superpotential
is well-known to be the Veneziano-Yankielowicz superpotential. As
a first test of the claim of Cachazo-Intriligator Vafa we want to
reproduce this superpotential.

According to our recipe we have to take the space through a
geometric transition and evaluate Eq. (\ref{newsuperpotential}) on
the deformed space. From our discussion in section \ref{propCY} we
know that this is given by the deformed conifold,
\begin{equation}
x^2+v^2+w^2+z^2-\m=0\ ,
\end{equation}
where we took $f(x)=-\m=-4t$, $\mu\in\mathbb{R}^+$. The integrals
of $\zeta$ on the corresponding Riemann surface have already been
calculated in (\ref{intcon1}) and (\ref{intcon2}). Obviously one
has $\zeta=-2t{\d x\over y}+\d\left(xy\over2\right)$, which would
correspond to $\varphi={xy\over2}$. Comparing with (\ref{varphi})
this would yield $c=t$. The choice $c=0$ instead leads to
$\varphi={xy\over2}+t{x\over y}$ and $\zeta=-2t{\d x\over
y}+4t^2{\d x\over y^3} +\d\varphi$. The first term has a pole at
infinity and leads to the logarithmic divergence, while the second
term has no pole at infinity but second order poles at
$\pm2\sqrt{t}$. One has
\begin{eqnarray}
2\varphi(\L_0)&=&\L_0\sqrt{\L_0^2-4t}+2t{\L_0\over\sqrt{\L_0^2-4t}}\
.
\end{eqnarray}
and
\begin{eqnarray}
{1\over2}\left\langle\hat\b,\zeta\right\rangle
&=&t\log\left({4t\over\L_0^2}\right)-2t\log\left(1+\sqrt{{1-{4t\over\L_0^2}}}\right)-t{1\over\sqrt{1-{4t\over\L_0^2}}}\nonumber\\
&=&{\partial\mathcal{F}_0(t)\over \partial
t}-t\log\L_0^2+o\left({1\over\L_0^2}\right)\ ,
\end{eqnarray}
where we used (\ref{intcon2}) and the explicit form of
$\mathcal{F}_0(t)$, (\ref{F0conifold}). Finally, in the present
case, Eq. (\ref{newsuperpotential}) for the superpotential only
contains the relative cycles,
\begin{equation}
W_{eff}=-{
N\over2}\left\langle{\hat\b},\zeta\right\rangle+{\tilde\tau_0\over2}\int_{\hat\a}\zeta
\end{equation}
or
\begin{equation}
W_{eff}=-N\left(t\log t-t-t\log\L_0^2\right)+2\pi
i\tilde\tau_0t+o\left({1\over\L_0^2}\right)\ .
\end{equation}
For $U(N)$ super Yang-Mills theory the $\b$-function reads
$\b(g)=-{3N\over16\pi^2}g^3$ and one has to use the identification
$\L_0^2=\tilde\L_0^3$ between the geometric cut-off $\L_0$ and the
gauge theory cut-off $\tilde\L_0$. Then $N\log\L_0^2+2\pi
i\tau_0=3N\log|\L|+i\Theta=3N\log\L$. Therefore, sending the
cut-off $\L_0$ to infinity, and  using $\bar S=t$, we indeed find
the Veneziano-Yankielowicz superpotential,
\begin{equation}
W_{eff}=\bar S\log\left({\L^{3N}\over \bar S^{N}}\right)+\bar SN\
.
\end{equation}

\section{Example: Superstrings on local Calabi-Yau manifolds}
After having studied superstrings on the conifold we now want to
extend these considerations to the more complicated local
Calabi-Yau spaces $X_{res}$. In other words, we want to study the
low energy effective superpotential of an $\mN=1$ $U(N)$ gauge
theory coupled to a chiral superfield $\Phi$ in the adjoint, with
tree-level superpotential $W(\Phi)$. The general structure of this
gauge theory has been analysed using field theory methods in
\cite{CIV01}, see also \cite{CSW03a}, \cite{CSW03b}. There the
authors made use of the fact that the gauge theory can be
understood as an $\mN=2$ theory which has been broken to $\mN=1$
by switching on the tree-level superpotential. Therefore, one can
apply the exact results of Seiberg and Witten \cite{SW94} on
$\mN=2$ theories to extract some information about the $\mN=1$
theory. In particular, the form of the low energy effective
superpotential $W_{eff}$ can be deduced \cite{CIV01}. Furthermore,
by studying the monodromy properties of the geometric integrals,
the general structure of the Gukov-Vafa-Witten superpotential was
determined, and the structures of the two superpotentials were
found to agree, which provides strong evidence for the conjecture
(\ref{CIVconjecture}). For the case of the cubic superpotential
the geometric integrals were evaluated approximately, and the
result agreed with the field theory calculations.

\begin{figure}[h]
\centering
\includegraphics[width=0.8\textwidth]{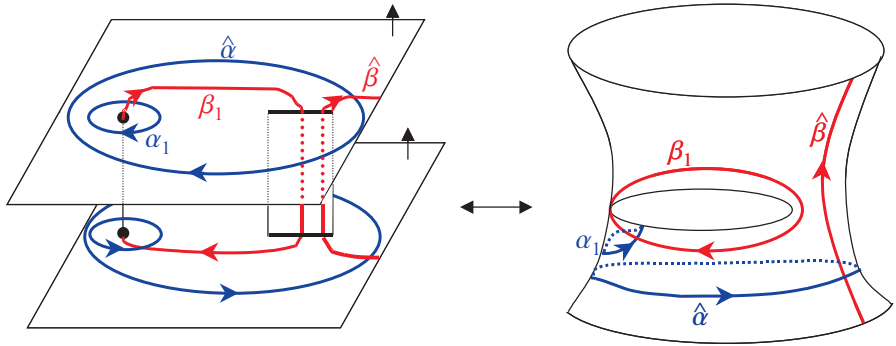}\\
\caption[]{The Riemann surface for a cubic tree level potential,
and $f_0$ such that one of the two cuts collapses to a
point.}\label{cubic}
\end{figure}

\begin{figure}[h]
\centering
\includegraphics[width=0.6\textwidth]{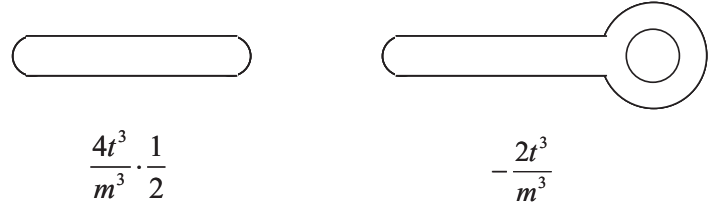}\\
\caption[]{The linear term in $W$ leads to tadpoles in the Feynman
diagrams, each tadpole coming with ${2tg\over mg_s}$, which
contains a factor of $g$. At order $g^2$ only two diagrams
contribute.}\label{tadpoles}
\end{figure}

Unfortunately, even for the simplest case of a cubic
superpotential the calculations are in general quite involved.
Therefore, we are going to consider a rather special case in this
section. We choose $n=2$, i.e. the tree-level superpotential $W$
is cubic, and we have 2 $\mathbb{CP}^1$s in $X_{res}$,
corresponding to the small resolution of the two singularities in
$W'(x)^2+v^2+w^2+z^2=0$. In order for many of the calculations to
be feasible, we choose the specific vacuum in which the gauge
group remains unbroken. In other words, we wrap all the $N$
physical D-branes around one of the two $\mathbb{CP}^1$s, e.g.
$N_1=0$ and $N_2=N$. Therefore, strings can end on only one
$\mathbb{CP}^1$. We saw already that the pattern of the breaking
of the gauge group is mirrored in the filling fractions $\n_i^*$
of the matrix model. In our case we must have $\n_1^*=0$ and
$\n_2^*=1$.\footnote{This can also be understood by looking at the
topological string. As we will see in the next chapter, the B-type
topological string with $\hN$ topological branes wrapping the two
$\mathbb{CP}^1$s calculates terms in the four-dimensional
effective action of the superstring theory. If there are no
D5-branes wrapping $\mathbb{CP}^1$ the related topological strings
will also not be allowed to end on this $\mathbb{CP}^1$, i.e.
there are no topological branes wrapping it. This amounts to
$\hN_1=0$ and $\hN_2=\hN$. In the next chapter we will see that
the number of topological branes and the filling fractions are
related as $\n_i^*=\hN_i/\hN$.} But from
${1\over4\pi^2}\int_{\G_{A^i}}\O=\bar S_i=t\n_i^*$ we learn that
the corresponding deformed geometry $X_{def}$ contains a cycle
$\G_{A^1}$ of vanishing size, together with the finite $\G_{A^2}$.
This situation is captured by a hyperelliptic Riemann surface,
where the two complex planes are connected by one cut and one
point, see Fig. \ref{cubic}
The situation in which one cut collapsed to zero size is described
mathematically by a double zero of the polynomial defining the
Riemann surface. Therefore, the vacuum with unbroken gauge group
leads to the surface
\begin{equation}\label{cubiccurve}
y^2=g^2(x-a)^2(x-b)(x+b)=\left(gx^2-agx-{b^2g\over2}\right)^2+ab^2g^2x-a^2b^2g^2-{b^4g^2\over4}\
.
\end{equation}
Without loss of generality we took the cut to be symmetric around
zero. We want both the cut and the double zero to lie on the real
axis, i.e. we take $g$ real and positive and $a,b\in\mathbb{R}$.
From (\ref{cubiccurve}) we can read off the superpotential,
\begin{equation}
W(x)={g\over3}x^3-{ag\over2}x^2-{b^2g\over2}x\ ,
\end{equation}
where we chose $W(0)=0$. Note that, in contrast to most of the
discussion in chapter \ref{holMM}, the leading coefficient of $W$
is ${g\over3}$ and not ${1\over3}$, because we want to keep track
of the coupling $g$. This implies (c.f. Eq. (\ref{f0matrix})) that
the leading coefficient of $f_0$ is $-4tg$, and therefore
\begin{equation}
t=-{ab^2g\over4}\ .
\end{equation}
Since $t$ is taken to be real and positive, $a$ must be negative.
Furthermore, we define the positive $m:=-ag$. Then
$W(x)={g\over3}x^3+{m\over2}x^2-{2tg\over m}x$. Note that we have
in this special setup $\bar S_1=\tilde
S_1={1\over4\pi^2}\int_{\G_{\a^1}}\O=0$ and $\bar
S_2={1\over4\pi^2}\int_{\G_{\hat\a}}\O={-i\over4\pi}\int_{\hat\a}\zeta={-i\over4\pi}\int_{\hat\a}y(x)\d
x=t$. Using (\ref{solution}) and (\ref{y0omega0}) our curve
(\ref{cubiccurve}) leads to the spectral density
\begin{equation}
\r_0(x)={g\over2\pi t}(x-a)\sqrt{b^2-x^2}\ \ \ \mbox{for}\
x\in[-b,b]
\end{equation}
and zero otherwise. The simplest way to calculate the planar free
energy is by making use of the homogeneity relation
(\ref{homogeneity}). We find
\begin{eqnarray}
\mF_0(t)&=&{t\over2}{\partial\mF_0(t)\over\partial
t}-{t\over2}\int\d s\ \r_0(s)W(s)\nonumber\\
&=&{t\over2}\lim_{\L_0\rightarrow\infty}\left({1\over2}
\int_{\hat\b}\zeta-W(\L_0)+t\log\L_0^2\right)-{t\over2}\int_{-b}^b\d
s\ \r_0(s)W(s)\ .
\end{eqnarray}
Here we used that the cut between $-b$ and $b$ lies on the real
axis, where we can take $\l(s)=s$. In the second line we made use
of (\ref{SGhat2}). The integrals can be evaluated without much
effort, and the result is
\begin{equation}\label{F0cubic}
\mF_0(t)={t^2\over2}\log{t\over m}-{3\over4}t^2+{2\over3}{g^2\over
m^3}t^3\ .
\end{equation}
This result is very interesting, since it contains the planar free
energy of the conifold (\ref{F0conifold}), which captures the
non-perturbative contributions, together with a perturbative term
${2\over3}{t^3\over m^3}g^2$. Note that this is precisely the term
calculated from matrix model perturbation theory in Eq.
(\ref{F0expansion}). (Recall that in our special case
$\mF_0(t)=-F_0(t)$, c.f. Eq. (\ref{F0}), which accounts for the
minus sign.) On the other hand, it seems rather puzzling that
there are no $o(g^4)$-terms in the free energy. In particular,
(\ref{F0expansion}) contains a term $-{8\over3}{t^4\over m^6}g^4$,
coming from fatgraph diagrams containing four vertices. However,
there we had a slightly different potential, namely
$W(x)={g\over3}x^3+{m\over2}x^2$, i.e. there was no linear term.
In our case this linear term is present, leading to tadpoles in
the Feynman diagrams. Each tadpole comes with a factor ${2tg\over
mg_s}$. Then, next to the planar diagrams of Fig.
\ref{fatgraphs2}, there are two more diagrams that contribute to
$F_0$ at order $g^2$, see Fig. \ref{tadpoles}.
However, their contribution cancels, because of different symmetry
factors. This is why, in the presence of tadpoles, we still
reproduce the result of (\ref{Fexpansion}) at order $g^2$.
Furthermore, the explicit expression for $\mF_0(t)$ tells us that
starting from order $g^4$ (as in the case without tadpoles there
are no diagrams that lead to an odd power of $g$) the
contributions coming from tadpole diagrams cancel the
contributions of the diagrams without tadpoles.

It is quite interesting to look at this from a slightly different
perspective. We again start from the potential
$W(x)={g\over3}x^3+{m\over2}x^2-{2tg\over m}x$ and perform the
shift $x=y+\a:=y+{m\over 2g}\left(\sqrt{1+{8tg^2\over
m^3}}-1\right)$. Then we have $W(x)=\tilde W(y)$ with
\begin{equation}
\tilde W(y)={g\over3}y^3+{m\over2}\Delta y^2 +\tilde W_0\ ,
\end{equation}
where $\Delta:=\sqrt{1+{8tg^2\over m^3}}$ and $W_0={m^3\over
12g^2}\left(-{1\over2}+{3\over2}\Delta^2-\Delta^3\right)$. The
shift of $x$ was chosen in such a way that $\tilde W(y)$ does not
contain a term linear in y.

Next consider what happens at the level of the partition function.
We have
\begin{eqnarray}
Z&=&\int DMe^{-{1\over g_s}\tr W(M)}=\int D\tilde Me^{-{1\over
g_s}\tr \tilde W(\tilde M)}=e^{-{N\tilde W_0\over g_s}}\int
D\tilde M e^{-{1\over g_s}\tr\left({g\over3}\tilde
M^3+{m\D\over2}\tilde M^2\right)}\nonumber\\
&=&\exp\left(-{t\tilde W_0\over
g_s^2}-{t^2\over2g_s^2}\log\D\right)\int D\hat M\exp\left(-{1\over
g_s}\tr\left({g\D^{-3/2}\over3}\hat M^3+{m\over2}\hat
M^2\right)\right)
\end{eqnarray}
This shows that partition functions with different coefficients in
the defining polynomial potential are related as
\begin{equation}
Z\left(g,m,-{2tg\over m}\right)=e^{-{1\over g_s^2}\left(t\tilde
W_0+{t^2\over2}\log\Delta\right)}Z\left(g\D^{-3/2},m,0\right)\ .
\end{equation}
In terms of the planar free energy this is
\begin{eqnarray}
\mF_0(g\D^{-3/2},m,0)&=&\mF_0\left(g,m,-{2tg\over
m}\right)+t\tilde
W_0(g,m,t)+{t^2\over2}\log\D(g,m,t)\nonumber\\
&=&\mF_0\left(g,m,-{2tg\over
m}\right)+t^2\left({-1+3\D^2-2\D^3\over3(\D^2-1)}+{1\over2}\log\D\right)
\end{eqnarray}
The first term on the right-hand side of this equation is the one
we calculated in (\ref{F0cubic}). Therefore,
\begin{eqnarray}
\mF_0(g\D^{-3/2},m,0)&=&{t^2\over2}\log {t\over
m}-{3\over4}t^2+{t^2\over12}(\D^2-1)+{t^3\over3}\left({1-3\D^2+2\D^3\over1-\D^2}+{3\over2}\log\D\right)\nonumber\\
&=&{t^2\over2}\log {t\over m}-{3\over4}t^2+\mF_0^p\ .
\end{eqnarray}
Note that this expression is exact, and it contains both the
perturbative and the non-perturbative part of $\mF_0$ for a
potential that only contains a cubic and a quadratic term. In
other words, expanding this expression in terms of $\tilde
g^2:=g^2\D^{-3}$ we reproduce the contributions of all the planar
fatgraph Feynman diagrams. This can be done by using
$\D^2=1+{8t\over m^3}\tilde g^2\D^3$ to express $\D$ in terms of
$\tilde g^2$. The result is
\begin{eqnarray}
\mF_0(\tilde g,m,0)&=&t^2\left\{{1\over2}\log{t\over
m}-{3\over4}+{2\over3}{t\tilde g^2\over
m^3}+{8\over3}\left({t\tilde g^2\over
m^3}\right)^2+{56\over3}\left({t\tilde g^2\over
m^3}\right)^3+{512\over3}\left({t\tilde g^2\over
m^3}\right)^4\right.\nonumber\\
&&\left.+{9152\over5}\left({t\tilde g^2\over
m^3}\right)^5+\ldots\right\}\nonumber\\
&=&t^2\left\{{1\over2}\log{t\over
m}-{3\over4}+{1\over2}\sum_{k=1}^\infty{(8{t\tilde g^2\over
m^3})^k\over(k+2)!}{\G(3k/2)\over\G(k/2+1)}\right\}\ .
\end{eqnarray}
This is precisely the expression found in \cite{BIPZ78}.\\

Coming now back to our original form of the potential
$W(x)={g\over3}x^3+{m\over2}x^2-{2tg\over m}x$, it remains to
write down the effective superpotential. For that purpose we write
$S:=t$. From (\ref{F0cubic}) we find
\begin{equation}
{\partial\mF_0(S)\over\partial S}=S\log S-S\log m-S+2{g^2\over
m^3}S^2\ .
\end{equation}
As promised below (\ref{DVsuperpotential}), this contains a term
$S\log S$. Plugging this into (\ref{DVsuperpotential}) gives
\begin{equation}
W_{eff}(S)=NS+S\log\left({\L^{2N}m^N\over S^N}\right)-2N{g^2\over
m^3}S^2\ .
\end{equation}
Note that we also have a term $S\log m$ in the derivative of
$\mF_0$. This is interesting for various reasons. Firstly, so far
we did not keep the dimensions of the various fields in our
theory. The argument of the logarithm in our final result should,
however, be dimensionless and since $m$ and $\L$ have dimension
one, and $S$ has dimension three this is indeed the case.
Furthermore, one can use the threshold matching condition
\cite{CIV01} $\L_L^{3N}=\L^{2N}m^N$ in order to relate our result
to the dynamical scale $\L_L$ of the low-energy pure $\mN=1$ super
Yang-Mills theory with gauge group $U(N)$. Similar threshold
matching conditions also hold for the more general case, see
\cite{CIV01} for a discussion.

\chapter{B-Type Topological Strings and Matrix Models}\label{TSMM}
In the last chapters we learned how the holomorphic matrix model
can be used to calculate the integrals of the holomorphic
$(3,0)$-form $\O$ over three-cycles in $X_{def}$. These integrals
in turn are the central building blocks of the effective
superpotential of our $U(N)$ gauge theory. The starting point of
the argument was the fact that the Riemann surface
(\ref{heRiemann}), that appears when calculating the integrals,
and the one of (\ref{algcurve}), that arises in the 't Hooft limit
of the matrix model, are actually the same. However, so far we
have not explained the reason {\it why} these two surfaces agree.
What motivated us to study the holomorphic matrix model in the
first place? In this final chapter we want to fill this gap and
show that the holomorphic matrix model is actually nothing but the
string field theory of the open B-type topological string on
$X_{res}$. We will not be able to present a self-contained
discussion of this fact, since many of the theories involved are
quite complicated and it would take us too far to explain them in
detail. The goal of this chapter is rather to familiarise the
reader with the central ideas and the main line of argument,
without spelling out the mathematics.

Some elementary background material on topological strings is
given in appendix \ref{TST}, see \cite{Vo05} and \cite{Horietal}
for more details. A recent review of string field theory appeared
in \cite{Ra05}. Central for the discussion are the results of the
classic article \cite{Wi95a}, and the holomorphic matrix model
first occurred in \cite{DV02a}. A recent review of the entire
setup can be found in
\cite{M04b}.\\

{\bf The open B-type topological string and its string field
theory}\\
It has been known for a long time that in the case of Calabi-Yau
compactifications of type II string theory certain terms of the
four-dimensional effective action can be calculated by studying
topological string theory on the Calabi-Yau manifold
\cite{BCOV93b}. To be specific, consider the case of type IIB
string theory on a compact Calabi-Yau manifold $X$. Then there are
terms in the four-dimensional effective action, which (when
formulated in $\mN=2$ superspace language) have the form
$\int\d^4x\d^4\t\ F_\gh(X^I)(\mathcal{W}^2)^\gh$, where $X^I$ are
the $\mN=2$ vector multiplets and
$\mathcal{W}^2:=\mathcal{W}_{\a\b}\mathcal{W}^{\a\b}$ is built
from the chiral $\mN=2$ Weyl multiplet $\mathcal{W}_{\a\b}$, which
contains the graviphoton. The function $F_\gh$ now turns out
\cite{BCOV93b} to be nothing but the free energy of the B-type
topological string with target space $X$ at genus $\gh$. For
example, it is well known that the prepotential governing the
structure of vector multiplets in an $\mN=2$ supersymmetric theory
is nothing but the genus zero free energy of the topological
string on the Calabi-Yau manifold.

Our setup is slightly more complicated, since we are interested in
type IIB theory on $M_4\times X_{res}$, with additional D-branes
wrapping the resolved two-cycles. The D-branes lead to an open
string sector, and therefore we expect that we have to study the
open B-type topological string on $X_{res}$. In other words, we
allow for Riemann surfaces with boundaries as the world-sheet of
the topological string. It can be shown that in the B-type
topological string, when mapping the world-sheet into the target
space, the boundaries have to be mapped to holomorphic cycles in
the target space. The appropriate boundary conditions are then
Dirichlet along these holomorphic cycles and Neumann in the
remaining directions. These boundary conditions amount to
introducing ``topological branes", which wrap the cycles. See
\cite{Wi95a}, \cite{M04a}, \cite{M04b} for a discussion of open
topological strings and more references. Since the various
$\mathbb{CP}^1$s in $X_{res}$ around which the physical D-branes
are wrapped are all holomorphic (a fact that we have not proven)
we expect that the relevant topological theory is the open B-type
topological string with topological branes around the various
$\mathbb{CP}^1$s. An obvious question to ask is then whether this
topological theory calculates terms in the four-dimensional
theory.

As to answer this question we have to take a little detour, and
note that the open B-type topological string can actually be
described $\cite{Wi95a}$ in terms of a cubic string field theory,
first introduced in \cite{Wi86}. Usually in string theory the
S-matrix is given in terms of a sum over two-dimensional
world-sheets embedded in space-time. The corresponding string
field theory, on the other hand, is a theory which reproduces this
S-matrix from the Feynman rules of a space-time action $S[\Psi]$.
$\Psi$, called the string field, is the fundamental dynamical
variable, and it contains infinitely many space-time fields,
namely one for each basis state of the standard string Fock space.
Writing down the string field theory of a given string theory is a
difficult task, and not very much is known about the string field
theories of superstrings. However, one does know the string field
theory for the open bosonic string \cite{Wi86}. Its action reads
\begin{equation}\label{SFT}
S={1\over g_s}\int\tr\left({1\over2}\Psi\star
Q_{BRST}\Psi+{1\over3}\Psi\star\Psi\star\Psi\right)\ ,
\end{equation}
where $g_s$ is the string coupling. The trace comes from the fact
that, once we add Chan-Paton factors, the string field is promoted
to a $U(\hN)$ matrix of string fields. We will not need the
detailed structure of this action, so we just mention that $\star$
is some associative product on the space of string fields and
$\int$ is a linear map from the space of string fields to
$\mathbb{C}$.\footnote{To be more precise, in open string field
theory one considers the world-sheet of the string to be an
infinite strip parameterised by a spacial coordinate
$0\leq\s\leq\pi$ and a time coordinate $-\infty<\tau<\infty$ with
flat metric $\d s^2=\d \s^2+\d \tau^2$. One then considers maps
$x:[0,\pi]\rightarrow X$ into the target space $X$. The string
field is a functional $\Psi[x(\s),\ldots]$, where $\ldots$ stands
for the ghost fields $c,\tilde c$ in the case of the bosonic
string and for $\eta,\theta$ (c.f. appendix \ref{TST}) in the
B-topological string. In \cite{Wi86} Witten defined two operations
on the space of functionals, namely integration, as well as an
associative, non-commutative star product
\begin{eqnarray}
\int\Psi&:=&\int
Dx(\s)\prod_{0\leq\s\leq\pi/2}\delta[x(\s)-x(\pi-\s)]\Psi[x(\s)]\
,\\
\int\Psi_1\star\ldots\star\Psi_p&:=&\int\prod_{i=1}^pDx_i(\s)
\prod_{i=1}^p\prod_{0\leq\s\leq\pi/2}\delta[x_i(\s)-x_{i+1}(\pi-\s)]\Psi_i[x_i(\s)]\
,
\end{eqnarray}
where $x_{p+1}\equiv x_1$. The integration can be understood as
folding the string around its midpoint and gluing the two halves,
whereas the star product glues two strings by folding them around
their midpoints and gluing the second half of one with the first
half of the following. See \cite{Pol}, \cite{Wi95a}, \cite{M04b},
\cite{Ra05} for more details and references.}

The structure of the topological string (see appendix \ref{TST}
for some of its properties) is very similar to the bosonic string.
In particular there exists a generator $Q_B$ with $Q_B^2=0$, the
role of the ghost fields is played by some of the particles
present in the super multiplets, and the ghost number is replaced
by the R-charge. So it seems plausible that an action similar to
(\ref{SFT}) should be the relevant string field theory for the
B-type topological string. However, in the case of the open
bosonic string the endpoints of the strings are free to move in
the entire target space. This situation can be generated in the
topological string by introducing topological branes which fill
the target space completely. This is reasonable since any
Calabi-Yau $X$ is a holomorphic submanifold of itself, so the
space filling topological branes do wrap holomorphic cycles. This
completes the analogy with the bosonic string, and it was indeed
shown in \cite{Wi95a} that the string field theory action for the
open B-type topological string with space filling topological
branes is given by (\ref{SFT}) with $Q_B$ instead of $Q_{BRST}$.

In the same article Witten showed that this action does actually
simplify enormously. In fact, the string functional is a function
of the zero mode of the string, corresponding to the position of
the string midpoint, and of oscillator modes. If we decouple all
the oscillators the string functional becomes an ordinary function
of (target) space-time, the $\star$-product becomes the usual
product of functions and the integral becomes the usual integral
on target space. In \cite{Wi95a} it was shown that this decoupling
does indeed take place in the open B-type topological string. This
comes from the fact that in the B-model the classical limit is
exact (because the Lagrangian is independent of the coupling
constant $t$ and therefore one can take $t\rightarrow\infty$,
which is the classical limit, c.f. the discussion in appendix
\ref{TST}). We will not discuss the details of this decoupling but
only state the result: the string field theory of the open
topological B-model on a Calabi-Yau manifold $X$ with $\hN$
space-time filling topological branes is given by holomorphic
Chern-Simons theory on $X$, with the action
\begin{equation}\label{holCS}
S={1\over2g_s}\int_{X}\O\w\tr\left(\bar A\w\bar\partial\bar
A+{2\over3}\bar A\w \bar A\w \bar A\right)\ ,
\end{equation}
where $\bar A$ is the $(0,1)$-part of a $U(\hN)$ gauge connection
on the target manifold $X$.\\

\bigskip
{\bf Holomorphic matrix models from holomorphic Chern-Simons
theory}\\
The holomorphic Chern-Simons action (\ref{holCS}) is the string
field theory in the case where we have $\hN$ space-time filling
topological branes. However, we are interested in the situation,
where the branes only wrap holomorphic two-cycles in $X_{res}$.
Here we will see that in this specific situation the action
(\ref{holCS}) simplifies further.

Before we attack the problem of the open B-type topological string
on $X_{res}$ let us first study the slightly simpler target space
$\mO(-2)\oplus\mO(0)\rightarrow\mathbb{CP}^1$ (c.f. definition
\ref{OmOn}) with $\hN$ topological branes wrapped around the
$\mathbb{CP}^1$. The corresponding string field theory action can
be obtained from a dimensional reduction \cite{KKLM99},
\cite{DV02a}, \cite{M04b}, of the action (\ref{holCS}) on
$\mO(-2)\oplus\mO(0)\rightarrow\mathbb{CP}^1$ down to
$\mathbb{CP}^1$. Clearly, the original gauge connection $\bar A$
leads to a $(0,1)$ gauge field $\bar a$ on $\mathbb{CP}^1$,
together with two fields in the adjoint, $\Phi_0$ and
$\Phi_1^{\bar y}$, which are sections of $\mO(0)$ and $\mO(-2)$
respectively. In other words we take $z,\bar z$ to be coordinates
on the $\mathbb{CP}^1$, $y,\bar y$ are coordinates on the fibre
$\mO(-2)$ and $x,\bar x$ are coordinates on $\mO(0)$, and write
$\bar A(x,\bar x,y,\bar y,z,\bar z)=\bar a(z,\bar z)+\Phi_1^{\bar
y}(z,\bar z)\d\bar y+\Phi_0(z,\bar z)\d\bar x$, where $\bar
a:=\bar a_{\bar z}\d\bar z$. Of course $\Phi_0$ and $\Phi_1^{\bar
y}$ are in the adjoint representation. The index $\bar y$ in
$\Phi_1^{\bar y}$ reminds us of the fact that $\Phi_1^{\bar y}$
transform if we go from one patch of the $\mathbb{CP}^1$ to
another, whereas $\Phi_0$ does not. In fact we can build a
$(1,0)$-form $\Phi_1:=\Phi_1^{\bar y}\d z$. Plugging this form of
$\bar A$ into (\ref{holCS}) gives
\begin{equation}
S={1\over g_s}\int_X\O_{xyz}(x,y,z)\d x\w\d y\w\d\bar x\w\d\bar
y\w\tr\left(\Phi_1\bar D_{\bar a}\Phi_0\right)\ ,
\end{equation}
where $\bar D_{\bar a}:=\bar\partial+[\bar a,\cdot]$ and
$\bar\partial:=\bar\partial_{\bar z}\d\bar z$. We integrate this
over the two line bundles to obtain
\begin{equation}
S={1\over g_s}\int_{\mathbb{CP}^1}f(z)\tr\left(\Phi_1\bar D_{\bar
a}\Phi_0\right)\ .
\end{equation}
But $\tr\left(\Phi_1\bar D_{\bar a}\Phi_0\right)$ is a (1,1)-form
on $\mathbb{CP}^1$ that does not transform if we change the
coordinate system. From the invariance of $S$ we deduce that
$f(z)$ must be a (holomorphic) function. Since holomorphic
functions on $\mathbb{CP}^1$ are constants we have \cite{DV02a}
\begin{equation}
S={1\over g_s}\int_{\mathbb{P}^1}\tr\left(\Phi_1\bar D_{\bar
a}\Phi_0\right)\ .
\end{equation}
Here we suppressed a constant multiplying the right-hand side.

We are interested in the more general situation in which the
target space is $X_{res}$, which can be understood as a
deformation of $\mO(-2)\oplus\mO(0)\rightarrow\mathbb{CP}^1$ by
$W$, see the discussion in section \ref{locCY}. In particular, we
want to study the case in which $\hN_i$ topological branes wrap
the $i$-th $\mathbb{CP}^1$. The string field theory action
describing the dynamics of the branes in this situation reads
\cite{KKLM99}, \cite{DV02a}, \cite{M04b}
\begin{equation}\label{redholCS}
S={1\over g_s}\int_{\mathbb{P}^1}\tr\left(\Phi_1\bar D_{\bar
a}\Phi_0+W(\Phi_0)\o\right)\ ,
\end{equation}
where $\o$ is the K\"ahler class on $\mathbb{CP}^1$ with
$\int_{\mathbb{P}^1}\o=1$. We will not prove that this is indeed
the correct action, but we can at least check whether the
equations of motion lead to the geometric picture of branes
wrapping the $n$ $\mathbb{CP}^1$s. As to do so note that the field
$\bar a$ is just a Lagrange multiplier and it enforces
\begin{equation}
[\Phi_0,\Phi_1]=0\ ,
\end{equation}
i.e. we can diagonalise $\Phi_0$ and $\Phi_1$ simultaneously.
Varying with respect to $\Phi_1$ gives
\begin{equation}
\bar\partial\Phi_0=0\ ,
\end{equation}
which implies that $\Phi_0$ is constant, as $\mathbb{P}^1$ is
compact. Finally, the equation for $\Phi_0$ reads
\begin{equation}\label{em}
\bar\partial \Phi_1=W'(\Phi_0)\o\ .
\end{equation}
Integrating both sides over $\mathbb{CP}^1$ gives
\begin{equation}
W'(\Phi_0)=0\ ,
\end{equation}
for non-singular $\Phi_1$. Plugging this back into (\ref{em})
gives $\bar\partial \Phi_1=0$. But there are no holomorphic one
forms on $\mathbb{CP}^1$, implying $\Phi_1\equiv0$. All this tells
us that classical vacua are described by $\Phi_1=0$ and a diagonal
$\Phi_0$, where the entries on the diagonal are constants, located
at the critical points of $W$. But of course, from our discussion
of $X_{res}$ we know that these critical points describe the
positions of the various $\mathbb{CP}^1$s in $X_{res}$. Since the
eigenvalues of $\Phi_0$ describe the position of the topological
branes we are indeed led to our picture of $\hN_i$ topological
branes wrapping the $i$-th $\mathbb{CP}^1$.

After having seen that the classical configurations of our string
field theory action do indeed reproduce our geometric setup we now
turn back to the action itself. We note that both $\bar a$ and
$\Phi_1$ appear linearly in (\ref{redholCS}), and hence they can
be integrated out. As we have seen, this results in the constraint
$\bar\partial\Phi_0=0$, which means that $\Phi_0$ is a {\it
constant $\hN\times \hN$ matrix},
\begin{equation}
\Phi_0(z)\equiv\Phi={\rm const}\ .
\end{equation}
Now we can plug this solution of the equations of motion back into
the action, which then reduces to
\begin{equation}
S={1\over g_s}\tr W(\Phi).
\end{equation}
But since $\Phi$ is a constant $\hN\times\hN$ matrix we find that
the string field theory partition function is nothing but a
holomorphic matrix model with potential $W(x)$. An alternative
derivation of this fact has been given in \cite{M04b}. It is quite
important to note that the number of topological branes $\hN$ is
unrelated to the number $N$ of physical D-branes. Indeed, $\hN$ is
the size of the matrices in the matrix model and we have seen that
interesting information about the physical $U(N)$ gauge theory can
be obtained by taking $\hN$ to infinity. This now completes the
logic of our reasoning and finally tells us why the holomorphic
matrix model can be used in order to extract information about our
model.\\

{\bf Open-closed duality}\\
Let us now analyse what the above discussion implies for the
relations between the various free energies involved in our setup.
Like any gauge theory, the free energy of the matrix model can be
expanded in terms of fatgraphs, as discussed in the introduction
and in section \ref{fatgraphs}. Such an expansion leads to
quantities $\mF^{mm,p}_{\gh,h}$ from fatgraphs with $h$ index
loops on a Riemann surface of genus $\gh$, where the superscript
$p$ denotes the perturbative part. The statement that the
holomorphic matrix model is the string field theory of the open
B-type topological string on $X_{res}$ means that the matrix model
free energy coincides with the free energy of the open B-type
topological string. To be more precise, we consider the open
B-type topological string on $X_{res}$ with $\hN_i$ topological
branes wrapped around the $i$-th $\mathbb{CP}^1$ with coupling
constant $g_s$. When mapping a Riemann surface with boundaries
into the target space we know that the boundaries have to be
mapped onto the holomorphic cycles. We denote the free energy for
the case in which $h_i$ boundaries are mapped to the $i$-th cycle
by $\mF^{oB,p}_{\gh,h_1,\ldots h_n}$. In the corresponding matrix
model with coupling constant $g_s$, on the other hand, one also
has to choose a vacuum around which one expands to calculate the
free energy. But from our analysis it is obvious that the
corresponding vacuum of the matrix model is the one in which the
filling fraction $\n_i^*$ is given by the number of topological
branes as
\begin{equation}
t\n_i^*=\hN_ig_s=\bar S_i\ .
\end{equation}
On can now expand the matrix model around this particular vacuum
(see e.g. \cite{KMT02}, \cite{M04b} for explicit examples) and
from this expansion one can read off the quantities
$\mF_{\gh,h_1,\ldots h_n}^{mm,p}$. The statement that the matrix
model is the string field theory of the open B-topological string
implies then that
\begin{equation}
\mF^{mm,p}_{\gh,h_1,\ldots h_n}=\mF_{\gh,h_1,\ldots h_n}^{oB,p}\ .
\end{equation}
Let us define the quantities
\begin{eqnarray}
\mathcal{F}^{mm,p}_{\gh}(\bar S_i)&:=&\sum_{h_1=1}^\infty\ldots
\sum_{h_n=1}^\infty\mathcal{F}^{mm,p}_{\gh,h_1,\ldots h_n}\bar
S_1^{h_1}\ldots\bar S_n^{h_n}\nonumber\\
&=&\sum_{h_1=1}^\infty\ldots
\sum_{h_n=1}^\infty\mathcal{F}^{oB,p}_{\gh,h_1,\ldots h_n}\bar
S_1^{h_1}\ldots\bar S_n^{h_n}=:\mF_{\gh}^{oB,p}(\bar S_i)\ .
\end{eqnarray}
Next we look back at Eqs. (\ref{intalpha}) and (\ref{intbeta}).
These are the standard special geometry relations on $X_{def}$,
with $\mF_0^{mm}$ as prepotential. On the other hand, the
prepotential is known \cite{Horietal} to be the free energy of the
{\it closed} B-type topological string at genus zero:
$\mF_0^{mm}(\bar S_i)=\mF_0^B(\bar S_i)$. This led Dijkgraaf and
Vafa to the conjecture \cite{DV02a} that this equality remains
true for all $\gh$, so that
\begin{equation}
\mF^{mm}(\bar S_i)=\mF^{B}(\bar S_i)\ ,
\end{equation}
where the left-hand side is the matrix model free energy with
coupling constant $g_s$, and the right-hand side denotes the free
energy of the {\it closed} B-type topological string on $X_{def}$
with coupling constant $g_s$. On the left-hand side $\bar S_i=g_s
\hN_i$, whereas on the right-hand side $\bar
S_i={1\over4\pi^2}\int_{\G_{A^i}}\O$.

Note that this implies that we have an open-closed duality
\begin{equation}
\mF^B_\gh(\bar S_i)=\mF^{mm}_\gh(\bar S_i)=\mF^{oB}_\gh(\bar S_i)\
.\label{openclosed}
\end{equation}
In other words, we have found a closed string theory which
calculates the free energy of the gauge theory, and thus we have
found an example in which the old idea of 't Hooft \cite{tH74} has
become true.\\

{\bf The effective superpotential revisited}\\
Let us finally conclude this chapter with some remarks on the
effective superpotential.

So far our general philosophy has been as follows: we
geometrically engineered a certain gauge theory from an open
string theory on some manifold, took the manifold through a
geometric transition, studied closed string theory with flux on
the new manifold and found that the four-dimensional effective
action generated from this string theory is nothing but the low
energy effective action of the geometrically engineered gauge
theory.

One might now ask whether it is also possible to find the low
energy effective superpotential $W_{eff}$ {\it before} going
through the geometric transition from the open string setup. It
was shown in \cite{BCOV93b}, \cite{V01} that this is indeed
possible, and that the low energy effective superpotential is
schematically given by
\begin{equation}
W_{eff}\sim\sum_{h=1}^\infty \mF_{0,h}^{oB}NhS^{h-1}+\a S\ ,
\end{equation}
where we introduced only one chiral superfield $S$ and $N$
D-branes (i.e. the gauge group is unbroken). To derive this
formula one uses arguments that are similar to those leading to
the $\mF_\gh(\mathcal{W}^2)^\gh$-term in the case of the closed
topological string \cite{BCOV93b}. Introducing the formal sum
\begin{equation}
\mF_{\gh}^{oB}(S):=\sum_{h=1}^\infty \mF_{\gh,h}S^h
\end{equation}
this can be written as
\begin{equation}
W_{eff}\sim N{\partial \mF_0^{oB}(S)\over\partial S}+\a S\ .
\end{equation}
If we now use the open-closed duality (\ref{openclosed}) this has
precisely the form of (\ref{DVsuperpotential}). So, in principle,
one can calculate the effective superpotential from the open
topological string. However, there one has to calculate all the
terms $\mF_{0,h}^{oB}$ and sum them over $h$. In practice this
task is not feasible explicitly. The geometric transition is so
useful, because it does this summation for us by mapping the sum
to the quantity $\mF^{B}_\gh$ in closed string theory, which is
much more accessible.

\chapter{Conclusions}
Although we have covered only a small part of a vast net of
interdependent theories, the picture we have drawn is amazingly
rich and beautiful. We saw that string theory can be used to
calculate the low energy effective superpotential, and hence the
vacuum structure, of four-dimensional $\mN=1$ supersymmetric gauge
theories. This effective superpotential can be obtained from
geometric integrals on a suitably chosen Calabi-Yau manifold,
which reduce to integrals on a hyperelliptic Riemann surface. This
Riemann surface also appears in the planar limit of a holomorphic
matrix model, and the integrals can therefore be related to the
matrix model free energy. The free energy consists of a
perturbative and a non-perturbative part, and the perturbative
contributions can be easily evaluated using matrix model Feynman
diagrams. Therefore, after adding the non-perturbative $S\log S$
term, the effective superpotential can be obtained using matrix
model perturbation theory. This is quite surprising, since vacuum
expectation values like $\la S\ra^N=\L^3$ in super Yang-Mills
theory are non-perturbative in the gauge coupling. In other words,
non-perturbative gauge theory quantities can be calculated from a
perturbative expansion in a matrix model.

Our analysis has been rather ``down to earth", in the sense that
we had an explicit manifold, $X_{def}$, on which we had to
calculate very specific integrals. After having obtained a
detailed understanding of the matrix model it was not too
difficult to relate our integrals to the matrix model free energy.
Plugging the resulting expressions into the Gukov-Vafa-Witten
formula expresses the superpotential in terms of matrix model
quantities. However, this technical approach does not lead to a
physical understanding of why all these theories are related, and
why the Gukov-Vafa-Witten formula gives the correct
superpotential. As we tried to explain in chapter \ref{TSMM}, the
deeper reason for these relations can be understood from
properties of the topological string. It is well known that both
the open and the closed topological string calculate terms in the
four-dimensional effective action of Calabi-Yau compactifications.
In particular, the low energy effective superpotential $W_{eff}$
can be calculated by summing infinitely many quantities
$\mF_{0,h}^{oB}$ of the open topological string on $X_{res}$.
Quite interestingly, there exists a dual closed topological string
theory on $X_{def}$, in which this sum is captured by $\mF_0^{B}$,
which can be calculated from geometric integrals. The relation
between the target spaces of the open and the dual closed
topological string is amazingly simple and given by a geometric
transition. In a sense, this duality explains why the
Gukov-Vafa-Witten formula is valid. The appearance of the
holomorphic matrix model can also be understood from an analysis
of the topological string, since it is nothing but the string
field theory of the open topological string on $X_{res}$.
Interestingly, the matrix model also encodes the deformed
geometry, which appears once we take the large $\hN$ limit.\\

These relations have been worked out in the articles \cite{V01},
\cite{CIV01}, \cite{DV02a}, \cite{DV02b} and \cite{DV02c}.
However, in these papers some fine points have not been discussed.
In particular, the interpretation  and cut-off dependence of the
righthand side of
\begin{equation}
\int_{\G_{B_i}}\O\sim{\partial \mF_0\over\partial S_i}
\end{equation}
was not entirely clear. Over the last chapters and in \cite{BM05}
we improved this situation by choosing a symplectic basis
$\{\G_{\a^i},\G_{\b_i},\G_{\hat\a},\G_{\hat\b}\}$ of the set of
compact and non-compact three-cycles in $X_{def}$, given by
$W'(x)^2+f_0(x)+v^2+w^2+z^2=0$. These map to a basis
$\{\a^i,\b_i,\hat\a,\hat\b\}$ of the set of relative one-cycles
$H_1(\S,\{Q,Q'\})$ on the Riemann surface $\S$, given by
$y^2=W'(x)^2+f_0(x)$, with two marked points $Q,Q'$. Then we
showed that the precise form of the special geometry relations on
$X_{def}$ reads
\begin{eqnarray}
-{1\over2\pi i}\int_{\G_{\a^i}}\O&=&2\pi i\tilde S_i\ ,\\
-{1\over2\pi i}\int_{\G_{\b_i}}\O&=&{\partial\mathcal{F}_0(t,\tilde S)\over\partial\tilde S_i}\ ,\\
-{1\over2\pi i}\int_{\G_{\hat\a}}\O&=&2\pi it\ ,\\
-{1\over2\pi i}\int_{\G_{\hat\b}}\O&=&
{\partial\mathcal{F}_0(t,\tilde S)\over\partial t}+W(\L_0)-t\log
\L_0^2+o\left({1\over\L_0}\right)\ .\label{last}
\end{eqnarray}
where $\mF_0(t,\tilde S_i)$ is the Legendre transform of the free
energy of the matrix model with potential $W$, coupled to sources.
In the last relation the integral is understood to be over the
regulated cycle $\G_{\hat\b}$ which is an $S^2$-fibration over a
line segment running from the $n$-th cut to the cut-off $\L_0$.
Clearly, once the cut-off is removed, the last integral diverges.
These relations show that the choice of basis
$\{\G_{\a^i},\G_{\b_i},\G_{\hat\a},\G_{\hat\b}\}$, although
equivalent to any other choice, is particularly useful. The
integrals over the compact cycles lead to the familiar rigid
special geometry relations, whereas the new features, related to
the non-compactness of the manifold, only show up in the remaining
two integrals. We further improved these formulae by noting that
one can get rid of the polynomial divergence by introducing
\cite{BM05} a paring on $X_{def}$ defined as
\begin{equation}\label{pairconc}
\left\langle\G_{\hat\b},\O\right\rangle:=\int_{\G_{\hat\b}}\left(\O-\d\Phi\right)=
(-i\pi)\int_{\hat\b}\left(\zeta-\d\varphi\right)\ ,
\end{equation}
where
\begin{equation}
\Phi:={W(x)W'(x)-{2t\over{n+1}}x^n\over W'(x)^2+f_0(x)}\cdot{\d
v\w\d w\over 2z}
\end{equation}
is such that
$\int_{\G_{\hat\b}}\d\Phi=-i\pi\int_{\hat\b}\d\varphi$, with
$\varphi$ as in (\ref{varphi}). This pairing is very similar in
structure to the one appearing in the context of relative
(co-)homology and we proposed that one should use this pairing so
that Eq. (\ref{last}) is replaced by
\begin{equation}
-{1\over2\pi
i}\left\langle\G_{\hat\b},\O\right\rangle={\partial\mathcal{F}_0(t,\tilde
S)\over\partial t}-t\log \L_0^2+o\left({1\over\L_0}\right)\ .
\end{equation}
At any rate, whether one uses this pairing or not, the integral
over the non-compact cycle $\G_{\hat\b}$ is {\it not} just given
by the derivative of the prepotential with respect to $t$, as is
often claimed in the literature.

Using this pairing the modified Gukov-Vafa-Witten formula for the
effective superpotential is proposed to read
\begin{eqnarray}
W_{eff}&=&{1\over2\pi
i}\sum_{i=1}^{n-1}\left(\int_{\G_{\a^i}}G_3\int_{\G_{\b_i}}\O-\int_{\G_{\b_i}}
G_3\int_{\G_{\a^i}}\O\right)\nonumber\\
&&+{1\over2\pi
i}\left(\int_{\G_{\hat\a}}G_3\left\langle\G_{\hat\b},
\O\right\rangle-\left\langle\G_{\hat\b},G_3\right\rangle\int_{\G_{\hat\a}}\O\right)\
.\label{Wconc}
\end{eqnarray}
We emphasize that, although the commonly used formula
$W_{eff}\sim\int G_3\w \O$ is very elegant, it should rather be
considered as a mnemonic for (\ref{Wconc}) because the Riemann
bilinear relations do not necessarily hold on non-compact
Calabi-Yau manifolds.  Note that although the introduction of the
pairing did not render the integrals of $\O$ and $G_3$ over the
$\G_{\hat\b}$-cycle finite, since they are still logarithmically
divergent, these divergences cancel in (\ref{Wconc}) and the
effective superpotential is well-defined.

The detailed analysis of the holomorphic matrix model also led to
some new results. In particular, we saw that, in order to
calculate the matrix model free energy from a saddle point
expansion, the contour $\g$ has to be chosen in such a way that it
passes through (or at least close to) all critical points of the
matrix model potential $W$, and the tangent vectors of $\g$ at the
critical points are such that the critical points are local minima
along $\g$. This specific form of $\g$ is dictated by the
requirement that the planar limit spectral density $\r_0(s)$ has
to be real. $\r_0$ is given by the discontinuity of
$y_0=W'(x)-2t\o_0$, which is one of the branches of the Riemann
surface $y^2=W'(x)^2+f_0(x)$. The reality of $\r_0$ therefore puts
constraints on the coefficients in $f_0$ and hence on the form of
the cuts in the Riemann surface. Since the curve $\g$ has to go
through all the cuts of the surface, the reality of $\r_0$
constrains the form of the contour. This guarantees that one
expands around a configuration for which the first derivatives of
the effective action indeed vanish. To ensure that saddle points
are really stable we were led to choose $\g$ to consist of $n$
pieces where each piece contains one cut and runs from infinity in
one convergence domain to infinity of another domain. Then the
``one-loop" term is a convergent, subleading Gaussian integral.

\part{M-theory Compactifications, $G_2$-Manifolds and Anomalies}

\chapter{Introduction}
In the middle of the nineteen nineties it became clear that the
five consistent ten-dimensional string theories, Type IIA, Type
IIB, Type I, $SO(32)$-heterotic and $E_8\times E_8$-heterotic, are
not independent, but are related by duality transformations.
Furthermore, a relation of these string theories to
eleven-dimensional supergravity was found, and this web of
interrelated theories was dubbed {\it M-theory} \cite{To95},
\cite{Wi95b}. One of the intriguing new features of M-theory is
the appearance of an additional, eleventh dimension, which implies
that the old constructions of string compactifications
\cite{CHSW85} had to be generalised. In fact, one is immediately
led to the question on which manifolds one has to compactify
eleven-dimensional supergravity, in order to obtain a physically
interesting four-dimensional $\mN=1$ effective field theory. It
turns out that the mechanism of these compactifications is quite
similar to the one of Calabi-Yau compactifications, and the
compact seven-dimensional manifold has to be a so-called {\it
$G_2$-manifold}. The Kaluza-Klein reduction of eleven-dimensional
supergravity on these manifolds was first derived in \cite{PT95}.
However, one finds that four-dimensional standard model like
theories, containing non-Abelian gauge groups and charged chiral
fermions, can only be obtained from $G_2$-compactifications if we
allow the seven-manifold to be singular (see for example
\cite{AG04} for a review). To be more precise, if the
$G_2$-manifold carries conical singularities, four-dimensional
charged chiral fermions occur which are localised at these
singularities. Non-Abelian gauge groups arise from
ADE-singularities on the $G_2$-manifold. Clearly, once a theory
with charged chiral fermions is constructed, one has to check
whether it is also free of anomalies. Two different notions of
anomaly cancellation occur in this context. {\it Global anomaly
cancellation} basically is the requirement that the
four-dimensional theory is anomaly free after summation of the
anomaly contributions from all the singularities of the internal
manifold. {\it Local anomaly cancellation} on the other hand
imposes the stronger condition that the contributions to the
anomalies associated with each singularity have to cancel
separately. We will study these issues in more detail below. What
we find is that, in the case of singular $G_2$-compactifications,
the anomalies present at a given singularity are cancelled locally
by a contribution which ``flows" into the singularity from the
bulk, provided one modifies the fields close to the singularity
\cite{Wi01}, \cite{BM03b}.

Compactifications on $G_2$-manifolds lead to four-dimensional
Minkowski space, since the metric on a $G_2$-manifold is
Ricci-flat. There are also other solutions of the equations of
motion of eleven-dimensional supergravity. One of them is the
direct product of $AdS_4$ with a seven-dimensional compact
Einstein space with positive curvature. Since the metric on the
seven-manifold is Einstein rather than Ricci-flat, these manifolds
cannot be $G_2$-manifolds. However, a generalisation of the
concept of $G_2$-manifolds to the case of Einstein manifolds
exists. These manifolds are known to be {\it weak
$G_2$-manifolds}. Quite interestingly, we were able to write down
for the first time a family of explicit metrics for such weak
$G_2$-manifolds that are compact and have two conical
singularities \cite{BM03a}. Although these manifolds have weak
$G_2$ rather than $G_2$ metrics, they are quite similar to
$G_2$-manifolds, and hence provide a framework in which many of
the features of compact, singular $G_2$-manifolds can be studied
explicitly.

Another context in which the cancellation of anomalies plays a
crucial role is the M5-brane. It carries chiral fields on its
six-dimensional world-volume and this field theory on its own
would be anomalous. However, once embedded into eleven-dimensional
supergravity one finds that a contribution to the anomaly flows
from the bulk into the brane, exactly cancelling the anomaly. In
fact, from these considerations a first correction term to
eleven-dimensional supergravity has been deduced already ten years
ago \cite{DLM95}, \cite{Wi96}, \cite{FHMM98}. The mechanism of
anomaly cancellation for the M5-brane has been reviewed in detail
in \cite{BM03c} and \cite{Me03} and we will not cover it here.

Finally, our methods of local anomaly cancellation and inflow from
the bulk can be applied to eleven-dimensional supergravity on the
interval \cite{HW95}, \cite{BM03c}. In this context an intriguing
interplay of new degrees of freedom living on the boundaries, a
modified Bianchi identity and anomaly inflow leads to a complete
cancellation of anomalies. The precise mechanism has been the
subject of quite some controversy in the literature (see for
example \cite{BDSa99} for references). Our treatment in
\cite{BM03c} finally provides a clear proof of {\it local} anomaly
cancellation.

In the following I am going to make this discussion more precise
by summarising the results of the publications \cite{BM03a},
\cite{BM03b} and \cite{BM03c}. The discussion will be rather
brief, since many of the details can be found in my work
\cite{Me03}. In the remainder of the introduction I quickly review
the concept of $G_2$-manifolds, explain the action of
eleven-dimensional supergravity and introduce the important
concept of anomaly inflow. The notation is explained in appendix
\ref{notation}.

\bigskip
\begin{center}
{\bf $G_2$-manifolds}
\end{center}
\begin{definition}
{\em Let $(x_1,\ldots ,x_7)$ be coordinates on $\mathbb{R}^7$.
Write $\d{\bf x}_{ij\ldots l}$ for the exterior form $\d x_i\wedge
\d x_j\wedge\ldots \wedge \d x_l$ on $\mathbb{R}^7$. Define a
three-form $\Phi_0$ on $\mathbb{R}^7$ by
\begin{equation}
\Phi_0:=\d\textbf{x}_{123}+\d\textbf{x}_{516}+\d\textbf{x}_{246}+
\d\textbf{x}_{435}+\d\textbf{x}_{147}+\d\textbf{x}_{367}+\d\textbf{x}_{257}\
.\label{G2form}
\end{equation}
The subgroup of $GL(7,\mathbb{R})$ preserving $\Phi_0$ is the
exceptional Lie group $G_2$. It is compact, connected, simply
connected, semisimple and 14-dimensional, and it also fixes the
four-form
\begin{equation}
*\Phi_0=\d{\bf x}_{4567}+\d{\bf x}_{2374}+\d{\bf x}_{1357}+\d{\bf
x}_{1276}+\d{\bf x}_{2356}+\d{\bf x}_{1245}+\d{\bf x}_{1346}\ ,
\end{equation}
the Euclidean metric $g_0=dx_1^2+\ldots dx_7^2$, and the
orientation on $\mathbb{R}^7$.}
\end{definition}

\begin{definition}
{\em A} $G_2$-structure {\em on a seven-manifold $M$ is a
principal subbundle of the frame bundle of $M$ with structure
group $G_2$. Each $G_2$-structure gives rise to a 3-form $\Phi$
and a metric $g$ on $M$, such that every tangent space of $M$
admits an isomorphism with $\mathbb{R}^7$ identifying $\Phi$ and
$g$ with $\Phi_0$ and $g_0$, respectively. We will refer to
$(\Phi,g)$ as a $G_2$-structure. Let $\nabla$ be the Levi-Civita
connection, then $\nabla\Phi$ is called the {\em torsion} of
$(\Phi,g)$. If $\nabla\Phi=0$ then $(\Phi,g)$ is called} torsion
free. {\em A} $G_2$-manifold {\em is defined as the triple
$(M,\Phi,g)$, where $M$ is a seven-manifold, and $(\Phi,g)$ a
torsion-free $G_2$-structure on $M$.}
\end{definition}

\begin{proposition}\label{G2prop}
Let $M$ be a seven-manifold and $(\Phi,g)$ a $G_2$-structure on
$M$.
Then the following are equivalent:\\
\\
\mbox{
\begin{tabular}{ll}
(i) & $\nabla\Phi=0$ ,\\
(ii)& $\d\Phi=\d*\Phi=0$ ,\\
(iii) & ${\rm Hol}(g)\subseteq G_2$.\\
\end{tabular}}
\end{proposition}
Note that the holonomy group of a $G_2$-manifold may be a proper
subset of $G_2$. However, we will mean a manifold with holonomy
group $G_2$ whenever we speak of a $G_2$-manifold in the
following. Let us list some properties of compact Riemann
manifolds $(M,g)$ with ${\rm Hol}(g)=G_2$.
\begin{itemize}
\item  $M$ is a spin manifold and there exists exactly one
covariantly constant spinor,\footnote{$\nabla^S$ contains the spin
connection, see (\ref{nablaS}) or appendix \ref{notation}.}
$\nabla^S\theta=0$\ .

\item $g$ is Ricci-flat.

\item The Betti numbers are $b^0=b^7=1$, $b^1=b^6=0$ and $b^2=b^5$
and $b^3=b^4$ arbitrary.
\end{itemize}
Many more details on $G_2$-manifolds can be found in \cite{Me03}.
A thorough mathematical treatment of $G_2$-manifolds, which also
contains the proof of proposition \ref{G2prop} can be found in
\cite{Jo00}.

\begin{center}
{\bf Eleven-dimensional supergravity}
\end{center}
It is current wisdom in string theory \cite{Wi95b} that the low
energy limit of M-theory is eleven-dimensional supergravity
\cite{CJS78}. Therefore, some properties of M-theory can be
deduced from studying this well understood supergravity theory.
Here we review the basic field content, the Lagrangian and its
equations of motion. More details can be found in \cite{WB92},
\cite{Du99}, \cite{DNP86} and \cite{West98}. For a recent review
see \cite{MS05}. The field content of eleven-dimensional
supergravity is remarkably simple. It consists of the metric
$g_{MN}$, a Majorana spin-${3\over 2}$ fermion $\psi_M$ and a
three-form $C={1\over {3!}}C_{MNP}\d z^M\wedge \d z^N\wedge \d
z^P$, where $z^M$ is a set of coordinates on the space-time
manifold $M_{11}$. These fields can be combined to give the unique
$\mathcal{N}=1$ supergravity theory in eleven dimensions. The full
action is\footnote{We define
$\bar\psi_M:=i\psi_M^{\dagger}\Gamma^0$, see appendix \ref{notation}.}\\
\parbox{14cm}{
\begin{eqnarray}
S&=&{1\over {2\kappa_{11}^2}}\int
[\mathcal{R}*1-{1\over2}G\wedge*G-{1\over 6}
C\wedge G\wedge G ]\nonumber\\
&& +{1\over 2\kappa_{11}^2}\int d^{11}z
\sqrt{g}\bar{\psi}_M\Gamma^{MNP}\nabla^S_N\left({\omega+\hat{\omega}\over
2}\right)\psi_P\nonumber\\
&&-{1\over 2\kappa_{11}^2}{1\over 192}\int d^{11}z
\sqrt{g}\left(\bar{\psi}_M\Gamma^{MNPQRS}\psi_N+12\bar{\psi}^P\Gamma^{RS}\psi^Q\right)(G_{PQRS}+\hat{
G}_{PQRS})\ .\nonumber
\end{eqnarray}}\hfill\parbox{8mm}{\begin{eqnarray}\label{SUGRAaction}\end{eqnarray}}\\
To explain the contents of the action, we start with the
commutator of the vielbeins, which defines the {\em anholonomy
coefficients} $\Omega_{AB}^{\ \ \ \ C}$
\begin{equation}
[e_A,e_B]=[e^{\ \ M}_A\partial_M,e^{\ \
N}_B\partial_N]=\Omega_{AB}^{\ \ \ \ C}e_C\ .
\end{equation}
Relevant formulae for the spin connection are
\begin{eqnarray}
\omega_{MAB}(e)&=&{1\over2}(-\Omega_{MAB}+\Omega_{ABM}-\Omega_{BMA})\ ,\nonumber\\
\omega_{MAB}&=&\omega_{MAB}(e)+{1\over 8}
[-\bar{\psi}_P\Gamma_{MAB}^{\ \ \ \ \ \
PQ}\psi_Q+2(\bar{\psi}_M\Gamma_B\psi_A-\bar{\psi}_M\Gamma_A\psi_B+\bar{\psi}_B\Gamma_M\psi_A
)]\ ,\nonumber\\
\hat{\omega}_{MAB}&:=&\omega_{MAB}+{1\over
8}\bar{\psi}_P\Gamma_{MAB}^{\ \ \ \ \ \ PQ}\psi_Q\ .
\end{eqnarray}
$\psi_M$ is a Majorana vector-spinor. The Lorentz covariant
derivative reads
\begin{equation}
\nabla^S_M(\omega)\psi_N:=\partial_M\psi_N+{1\over
4}\omega_{MAB}\Gamma^{AB}\psi_N\ .\label{nablaS}
\end{equation}
For further convenience we set
\begin{equation}
\widetilde{\nabla}^S_M(\omega)\psi_N:=\nabla^S_M(\omega)\psi_N-{1\over288}\left(\Gamma_M^{\
\  PQRS}-8\delta_M^P\Gamma^{QRS}\right)\hat{G}_{PQRS}\psi_N\ .
\end{equation}
\begin{equation}
G:=dC \ \ \mbox{i.e.}\ \ G_{MNPQ}=4\partial_{[M}C_{NPQ]}\ .
\end{equation}
$\hat{G}_{MNPQ}$ is defined as
\begin{equation}
\hat{G}_{MNPQ}:=G_{MNPQ}+3\bar{\psi}_{[M}\Gamma_{NP}\psi_{Q]}\ .
\end{equation}

The action is invariant under the supersymmetry transformations
\begin{eqnarray}
\delta e_{\ \ M}^A&=&-{1\over2}\bar{\eta}\Gamma^A\psi_M\ ,\nonumber\\
\delta C_{MNP}&=&-{3\over2}\bar{\eta}\Gamma_{[MN}\psi_{P]}\ ,\label{susy}\\
\delta\psi_M&=&\widetilde{\nabla}^S_M(\hat{\omega})\eta\
.\nonumber
\end{eqnarray}

Next we turn to the equations of motion. We will only need
solutions of the equations of motion with the property that
$\psi_M\equiv0$. Since $\psi_M$ appears at least bilinearly in the
action, we can set $\psi_M$ to zero before varying the action.
This leads to an enormous simplification of the calculations. The
equations of motion with vanishing fermion field
read\\
\parbox{14cm}{
\begin{eqnarray}
\mathcal{R}_{MN}(\omega)-{1\over2}g_{MN}\mathcal{R}(\omega)&=&{1\over
12}\left(G_{MPQR}G_N^{\ \ PQR}-{1\over8}g_{MN}G_{PQRS}G^{PQRS}\right)\ ,\nonumber\\
d*G+{1\over2}G\wedge G&=&0\ .\nonumber
\end{eqnarray}}\hfill\parbox{8mm}{\begin{eqnarray}\end{eqnarray}}\\
In addition to those field equations we also know that G is
closed, as it is exact,
\begin{equation}
dG=0\ .\label{Gclosed}
\end{equation}
A solution $(M, \langle g\rangle, \langle C\rangle,
\langle\psi\rangle)$ of the equations of motion is said to be {\em
supersymmetric} if the variations (\ref{susy}) vanish at the point
$e^A_{\ \ M}=\langle e^A_{\ \ M}\rangle,\ C_{MNP}=\langle
C_{MNP}\rangle,\ \psi_M=\langle\psi_M\rangle$. All the vacua we
are going to study have vanishing fermionic background,
$\langle\psi_M\rangle=0$, so the first two equations are trivially
satisfied and the last one reduces to
\begin{equation}
\widetilde{\nabla}^S_M(\omega)\eta=0\ ,
\end{equation}
evaluated at $C_{MNP}=\langle C_{MNP}\rangle$, $e^A_{\ \
M}=\langle e^A_{\ \ M} \rangle$ and $\psi_M=0$. We see that
$e^A_{\ \ M}$ and $C_{MNP}$ are automatically invariant and we
find that the vacuum is supersymmetric if and only if there exists
a spinor $\eta$ s.t. $\forall M$
\begin{equation}
\nabla^S_M\eta-{1\over 288}\left(\Gamma_M^{\ \
PQRS}-8\delta_M^P\Gamma^{QRS}\right)G_{PQRS}\eta=0\
.\label{SUSYvac}
\end{equation}

\bigskip
\begin{center}
{\bf Solutions of the equations of motion}\\
\end{center}
Given the explicit form of the equations of motion, it is easy to
see that $\la\psi_M\ra=0$, $\la C\ra=0$, together with any
Ricci-flat metric on the base manifold $M_{11}$ is a solution. In
particular, this is true for $(M_{11},g)=(\mathbb{R}^4\times
M,\eta\times g)$, where $(\mathbb{R}^4,\eta)$ is Minkowski space
and $(M,g)$ is a $G_2$-manifold. For such a vacuum the condition
(\ref{SUSYvac}), reduces to
\begin{equation}
\nabla^S\eta=0\ .\label{covcon}
\end{equation}
The statement that the effective four-dimensional theory should be
$\mN=1$ supersymmetric translates to the requirement that
(\ref{covcon}) has exactly four linearly independent solutions.
After the compactification the original Poincar\'{e} group
$P(10,1)$ is broken to $P(3,1)\times P(7)$. The ${\bf 32}$ of
$SO(10,1)$ decomposes as ${\bf 32}={\bf 4}\otimes {\bf 8}$, thus,
for a spinor in the compactified theory we have
\begin{equation}
\eta(x,y)=\epsilon(x)\otimes \theta(y)\ ,
\end{equation}
with $\epsilon$ a spinor in four and $\theta$ a spinor in seven
dimensions. The $\Gamma$-matrices can be rewritten as\\
\parbox{14cm}{
\begin{eqnarray}
\Gamma^{a}&=&\gamma^{a}\otimes \opone\ ,\nonumber\\
\Gamma^m&=&\gamma_5\otimes \gamma^m\ ,\nonumber
\end{eqnarray}}\hfill\parbox{8mm}{\begin{eqnarray}\label{gammasplit}\end{eqnarray}}
with $\{\gamma^m\}$ the generators of a Clifford algebra in seven
dimensions. Then it is not hard to see that for
$\nabla^S=\nabla_M^Sdz^M$ one has
\begin{equation}
\nabla^S =\nabla^S_4\otimes\opone+\opone\otimes\nabla_7^S\ .
\end{equation}
Therefore, (\ref{covcon}) reads
\begin{equation}
(\nabla^S_4\otimes\opone+\opone\otimes\nabla_7^S)\epsilon(x)\otimes
\theta(y)=\nabla^S_4\epsilon(x)\otimes\theta(y)+\epsilon(x)\otimes\nabla_7^S\theta(y)=0\
.
\end{equation}
On Minkowski space we can find a basis of four constant spinors
$\epsilon^i$. The condition we are left with is
\begin{equation}
\nabla^S_7\theta(y)=0\ .\label{covcon7}
\end{equation}
Thus, the number of solutions of (\ref{covcon}) is four times the
number of covariantly constant spinors on the compact seven
manifold. Since we already saw that a $G_2$-manifold carries
precisely one covariantly constant spinor, we just proved our
statement that compactifications on $G_2$-manifolds lead to
four-dimensional $\mN=1$ theories.

\bigskip
Next consider what is known as the {\it Freund-Rubin solution} of
eleven-dimensional gravity. Here $(M_{11},g)$ is given by a
Riemannian product, $(M_{11},g)=(M_4\times M_7,g_1\times g_2)$,
and
\footnote{Recall that the topology of $AdS_4$ is $S^1\times \mathbb{R}^3$.}\\
\parbox{14cm}{
\begin{eqnarray}
M_{11}&=&S^1\times\mathbb{R}^{3}\times M_7, \ \ M_7\ \mbox{compact}\ ,\nonumber\\
\langle\psi_M\rangle&=&0\ ,\nonumber\\
\langle g_1\rangle&=& g(AdS_4)\ ,\nonumber\ \\
\langle g_2\rangle&\ \ & \mbox{Einstein, s.t. $\mathcal{R}_{mn}={1\over6}f^2\langle {g_2}_{mn}\rangle$}\ ,\nonumber\\
\langle G_{\mu\nu\rho\sigma}\rangle&=&f\sqrt{\langle g_1\rangle}\
\widetilde{\epsilon}_{\mu\nu\rho\sigma}\ .\nonumber
\end{eqnarray}}\hfill\parbox{8mm}{\begin{eqnarray}\end{eqnarray}}

Next we want to analyze the consequences of (\ref{SUSYvac}) for
the Freund-Rubin solutions. We find\\
\parbox{14cm}{
\begin{eqnarray}
\nabla^S_{\mu}\eta&=&-{if\over6}(\gamma_{\mu}\gamma_5\otimes\opone)\eta\ ,\nonumber\\
\nabla^S_m\eta&=&{if\over12}(\opone\otimes\gamma_m)\eta\
.\nonumber
\end{eqnarray}}\hfill\parbox{8mm}{\begin{eqnarray}\label{Deta}\end{eqnarray}}
Again we have the decomposition ${\bf 32}={\bf 4}\otimes {\bf 8}$
and hence $\eta(x,y)=\epsilon(x)\otimes \theta(y)$. Then
(\ref{Deta}) reduce to
\begin{eqnarray}
\nabla^S_{\mu}\epsilon&=&-{if\over6}\gamma_{\mu}\gamma_5\epsilon,\label{D4e}\\
\nabla^S_m\theta&=&{if\over12}\gamma_m\theta\label{D7t}.
\end{eqnarray}
On $AdS_4$ one can find four spinors satisfying (\ref{D4e}).
Therefore, the number of spinors $\eta$, satisfying (\ref{Deta})
is four times the number of spinors $\theta$ which are solutions
of (\ref{D7t}). In other words, to find Freund-Rubin type
solutions with $\mathcal{N}=k$ supersymmetry we need to find
compact seven-dimensional Einstein spaces with positive curvature
and exactly $k$ Killing spinors. One possible space is the
seven-sphere which admits eight Killing spinors, leading to
maximal supersymmetry in four dimensions. A seven-dimensional
Einstein manifold with exactly one Killing spinor is known as a
{\it weak $G_2$-manifold}.

\bigskip
\begin{center}
{\bf Kaluza-Klein compactification on a smooth $G_2$-manifolds}\\
\end{center}
We already mentioned that one has to introduce singularities into
the compact $G_2$-manifold in order to generate interesting
physics. Indeed, for smooth $G_2$-manifolds we have the following
proposition.
\begin{proposition}\label{KKG2}
The low energy effective theory of M-theory on
$(\mathbb{R}^4\times X, \eta\times g)$ with $(X,g)$ a smooth
$G_2$-manifold is an $\mathcal{N}=1$ supergravity theory coupled
to $b^2(X)$ Abelian vector multiplets and $b^3(X)$ massless
neutral chiral multiplets.
\end{proposition}
This field content was determined in \cite{PT95}, the Kaluza-Klein
compactification procedure is reviewed in \cite{Me03}. Note that
although there are chiral fields in the effective theory these are
not very interesting, since they do not couple to the gauge
fields.\\

\bigskip
\begin{center}
{\bf Anomaly inflow}
\end{center}
Before we embark on explaining the details of the mechanism of
anomaly cancellation on singular $G_2$-manifolds, we want to
comment on a phenomenon known as anomaly inflow. A comprehensive
discussion of anomalies and many references can be found in
\cite{Me03}, the most important results are listed in appendix
\ref{anomalies}, to which we refer the reader for further details.
The concept of anomaly inflow in effective theories was pioneered
in \cite{CH84} and further studied in \cite{Nak87}. See
\cite{Ha05} for a recent review. Here we analyse the extension of
these ideas to the context of M-theory, as studied in \cite{BM03c}.\\

Consider a theory in $d=2n$ dimensions containing a massless
fermion $\psi$ coupled to a non-Abelian external gauge field
$A=A_aT_a$ with gauge invariant (Euclidean) action $S^E[\psi,A]$.
The current
\begin{equation}
J_a^M(x):={\delta S^E[\psi,A]\over\delta A_{aM}(x)}\ ,
\end{equation}
is conserved, because of the gauge invariance of the action,
$D_MJ_a^M(x)=0$. Next we define the functional
\begin{equation}
\exp\left(-X[A]\right):=\int D\psi
D\bar\psi\exp\left(-S^E[\psi,A]\right)\ .\label{functional}
\end{equation}
Under a gauge variation $A(x)\rightarrow A'(x)=A(x)+D\e(x)$ with
$\e(x)=\e_a(x)T_a$ the Euclidean action is invariant, but, in
general, the measure transforms as
\begin{equation}
D\psi
D\bar\psi\rightarrow\exp\left(i\int(\d^dx)_E\e_a(x)G_a[x;A]\right)
D\psi D\bar\psi\ .\label{measure}
\end{equation}
Here $(d^dx)_E$ is the Euclidean measure, and the quantity
$G_a[x;A]$, called the anomaly function, depends on the theory
under consideration (see \cite{Me03} for some explicit examples).
Variation of (\ref{functional}) then gives
\begin{equation*}
\exp(-X[A])\int (d^dx)_ED_{M}\langle
J^{M}_a(x)\rangle\epsilon_a(x)=\int (d^dx)_E \int D\psi
D\bar\psi[iG_a[x;A]\epsilon_a(x)]\exp(-S^E)\ ,
\end{equation*}
where we used the invariance of the Euclidean action $S^E$ under
local gauge transformations. Therefore,
\begin{equation}
D_M\la J^M_a(x)\ra=iG_a[x;A]\ ,
\end{equation}
and we find that the quantum current $\la J^M_a(x)\ra$ is not
conserved. The anomaly $G_a[x;A]$ can be evaluated from studying
the transformation properties of the path integral measure.

Note that (\ref{measure}) implies
\begin{equation}
\delta X=-i\int(\d^dx)_E\e_a(x)G_a[x;A]=:i\int I_{2n}^1\
,\label{masteran}
\end{equation}
where we defined a $2n$-form $I_{2n}^1$. We see that a theory is
free of anomalies if the variation of the functional $X$ vanishes.
This variation is captured by the form $I_{2n}^1$ and it would be
nice if we could find a simple way to derive this form for a given
theory. This is in fact possible, as explained in some detail in
the appendix. It turns out \cite{St84}, \cite{Zu84}, \cite{MSZ85}
that the $2n$-form $I_{2n}^1$ is related to a $2n+2$-form
$I_{2n+2}$ via the so called decent equations,
\begin{equation}
\d  I^1_{2n}=\delta I_{2n+1}\ \ \ ,\ \ \ \d I_{2n+1}= I_{2n+2}\ ,
\end{equation}
where $I_{2n+2}$ is a polynomial in the field strengths.
Furthermore, the anomaly polynomial $I_{2n+2}$ depends only on the
field content of the theory. It can be shown that the only fields
leading to anomalies are spin-$1\over2$ fermions, spin-$3\over2$
fermions and forms with (anti-)self-dual field strength. The
anomaly polynomials corresponding to these fields are given by
(see \cite{AGG84}, \cite{AG85} and references therein)
\begin{eqnarray}
I_{2n+2}^{(1/2)}&=&-2\pi\left[\hat A(M_{2n})\ {\rm
ch}(F)\right]_{2n+2}\label{Ihat1/2}\ ,\\
I_{2n+2}^{(3/2)}&=&-2\pi\left[\hat A(M_{2n})\ \left(\tr
\exp\left({i\over2\pi}R\right)-1\right)\ {\rm
ch}(F)\right]_{2n+2}\label{Ihat3/2}\ ,\\
I_{2n+2}^{A}&=&-2\pi\left[\left(-{1\over2}\right){1\over4}\
L(M_{2n})\right]_{2n+2}.\label{IhatA}
\end{eqnarray}
To be precise these are the anomalies of spin-${1\over2}$ and
spin-${3\over 2}$ particles of positive chirality and a self-dual
form in Euclidean space, under the gauge transformation $\delta
A=D\epsilon$ and the local Lorentz transformations
$\delta\omega=D\epsilon$. All the quantities appearing in these
formulae are explained in appendices \ref{IT} and \ref{anomalies}.
The polynomials of the spin-${1\over2}$ fields,
$I_{2n+2}^{(1/2)}$, can be written as a sum of terms containing
only the gauge fields, terms containing only the curvature tensor
and terms containing both. These terms are often referred to as
the gauge, the gravitational and the mixed anomaly, respectively.
As to determine whether a theory is anomalous or not, is is then
sufficient to add all the anomaly polynomials $I_{2n+2}$. If they
sum up to zero, the variation of the quantum effective action
(\ref{masteran}) vanishes\footnote{This is in fact not entirely
true. It can happen that the sum of the polynomials $I_{2n+2}$ of
a given theory vanishes, but the variation of $X$ is non-zero.
However, in these cases one can always add a local counterterm to
the action, such that the variation of $X$ corresponding to the
modified action vanishes.} as well, and the theory is anomaly
free.

It turns out that this formalism has to be generalized, since we
often encounter problems in M-theory in which the classical action
is not fully gauge invariant. One might argue that in this case
the term ``anomaly" loses its meaning, but this is in fact not
true. The reason is that in many cases we study theories on
manifolds with boundary which are gauge invariant in the bulk, but
the non-vanishing boundary contributes to the variation of the
action. So in a sense, the variation is nonzero because of global
geometric properties of a given theory. If we studied the same
Lagrangian density on a more trivial manifold, the action would be
perfectly gauge invariant. This is why it still makes sense to
speak of an anomaly. Of course, if we vary the functional
(\ref{functional}) in theories which are not gauge invariant we
obtain an additional contribution on the right-hand side. This
contribution is called an {\em anomaly inflow term}, for reasons
which will become clear presently.

Consider for example a theory which contains the topological term
of eleven-dimensional supergravity. In fact, all the examples we
are going to study involve either this term or terms which can be
treated similarly. Clearly, $\int_{M_{11}}C\wedge dC\wedge dC$ is
invariant under $C\rightarrow C+\d\L$ as long as $M_{11}$ has no
boundary. In the presence of a boundary we get the non-vanishing
result $\int_{\partial{M_{11}}}\Lambda\wedge dC\wedge dC$. Let us
study what happens in such a case to the variation of our
functional. To do so we first need to find out how our action can
be translated to Euclidean space. The rules are as follows (see
\cite{BM03c} for a detailed discussion of the transition from
Minkowski to Euclidean space)
\begin{eqnarray}
x^1_E&:=&ix^0_M\ ,\ \ x^2_E:=x^1_M\ ,\ldots\nonumber\\
(d^{11}x)_E&:=&id^{11}x\ ,\nonumber\\
C_{1MN}^E&:=&-iC_{0(M-1)(N-1)}\ \ ,\ \ M,N\ldots\in\{2,\ldots,11\}\nonumber\\
C_{MNP}^E&=&C_{(M-1)(N-1)(P-1)}\ ,\nonumber\\
\tilde\epsilon^E_{123\ldots 11}&=&+1\ .
\end{eqnarray}
We know that $S^M=iS^E$, where $S^M$ is the Minkowski action, but
explicitly we have\footnote{Our conventions are such that
$\e^{M_1\ldots M_d}={\rm
sig}(g){1\over\sqrt{g}}\tilde\e^{M_1\ldots M_d}$, and
$\tilde\e^{M_1\ldots M_d}$ is totally anti-symmetric with
$\tilde\e^{01\ldots d}=+1$. See \cite{BM03c} and appendix
\ref{notation} for more details.}
\begin{eqnarray}
S_{kin}^M&=&-{1\over4\k_{11}^2}\int d^{11}x\sqrt{g}\
{1\over4!}G_{MNPQ}G^{MNPQ}\nonumber\\
&=&\ {i\over4\k_{11}^2}\int
(d^{11}x)_E\sqrt{g}\ {1\over4!}G^E_{MNPQ}(G^E)^{MNPQ}\ ,\nonumber\\
S_{top}^M&=&{1\over12\k_{11}^2}\int d^{11}x\sqrt{g}\
{1\over3!4!4!}\epsilon^{M_0\ldots M_{10}}C_{M_0M_1M_2}G_{M_3M_4M_5M_6}G_{M_7M_8M_9M_{10}}\nonumber\\
&=&-{1\over12\k_{11}^2}\int (d^{11}x)_E\sqrt{g}\
{1\over3!4!4!}(\epsilon^E)^{M_1\ldots
M_{11}}C^E_{M_1M_2M_3}G^E_{M_4M_5M_6M_7}G^E_{M_8M_9M_{10}M_{11}}\
.\nonumber\\
\end{eqnarray}
But then we can read off
\begin{eqnarray}
S^E&=&{1\over4\k_{11}^2}\int(d^{11}x)_E\sqrt{g}\ {1\over4!}G^E_{MNPQ}(G^E)^{MNPQ}\nonumber\\
&&+{i\over12\k_{11}^2}\int (d^{11}x)_E\sqrt{g}\
{1\over3!4!4!}(\epsilon^E)^{M_1\ldots
M_{11}}C^E_{M_1M_2M_3}G^E_{M_4M_5M_6M_7}G^E_{M_8M_9M_{10}M_{11}}\
,\nonumber
\end{eqnarray}
where a crucial factor of $i$ turns up. We write
$S^E=S^E_{kin}+S^E_{top}=:S^E_{kin}+i\widetilde{S}^E_{top}$,
because $S^E_{top}$ is imaginary, so $\widetilde{S}^E_{top}$ is
real. Then, for an eleven-manifold with boundary, we find $\delta
S^E=i\delta\tilde S_{top}^E={i\over12\k^2_{11}}\int_{\partial
M_{11}}\L^E\w G^E\w G^E$. But this means that $\delta S^E$ has
precisely the right structure to cancel an anomaly on the
ten-dimensional space $\partial M_{11}$. This also clarifies why
one speaks of anomaly inflow. A contribution to an anomaly on
$\partial M_{11}$ is obtained by varying an action defined in the
bulk $M_{11}$.

Clearly, whenever one has a theory with $\delta S^M\neq0$ we find
the master formula
\begin{equation}\label{master}
\delta X=\delta S^E+ i\int I_{2n}^1=-i\delta S^M+ i\int I_{2n}^1\
.
\end{equation}
The theory is anomaly free if and only if the right-hand side
vanishes. In other words, to check whether a theory is free of
anomalies we have to rewrite the action in Euclidean space,
calculate its variation and the corresponding $2n+2$-form and add
$i$ times the anomaly polynomials corresponding to the fields
present in the action. If the result vanishes the theory is free
of anomalies. In doing so one has to be careful, however, since
the translation from Minkowski to Euclidean space is subtle. In
particular, one has to keep track of the chirality of the
particles involved. The reason is that with our conventions
(\ref{Gamma}) for the matrix $\G_{d+1}$ we have
$\Gamma_{d+1}^E=-\Gamma_{d+1}^M$ for $d=4k+2$, but
$\Gamma_{d+1}^E=\Gamma_{d+1}^M$ for $4k$. In other words a fermion
of positive chirality in four dimensional Minkowski space
translates to one with positive chirality in four-dimensional
Euclidean space. In six or ten dimensions, however, the chirality
changes. To calculate the anomalies one has to use the polynomials
after having translated everything to Euclidean space.

\chapter{Anomaly Analysis of M-theory on Singular
$G_2$-Manifolds}\label{G2anom}
It was shown in \cite{AtW01}, \cite{AcW01} that compactifications
on $G_2$-manifolds can lead to charged chiral fermions in the low
energy effective action, if the compact manifold has a conical
singularity. Non-Abelian gauge fields arise \cite{Ac98},
\cite{A00} if we allow for ADE singularities on a locus $Q$ of
dimension three in the $G_2$-manifold. We will not review these
results but refer the reader to the literature \cite{AG04}. We are
more interested in the question whether, once the manifold carries
conical singularities, the effective theory is free of anomalies.
This chapter is based on the results of \cite{BM03b}.

\section{Gauge and mixed anomalies}
Let then $X$ be a compact $G_2$-manifold that is smooth except for
conical singularities\footnote{Up to now it is not clear whether
such $G_2$-manifolds exist, however, examples of non-compact
spaces with conical singularities are known \cite{GPP90}, and
compact weak $G_2$-manifolds with conical singularities were
constructed in \cite{BM03a}.} $P_{\alpha}$, with $\alpha$ a label
running from one to the number of singularities in $X$. Then there
are chiral fermions sitting at a given singularity $P_\a$. They
have negative\footnote{Recall that this is true both in Euclidean
and in Minkowski space, since $\g_5^E=\g_5^M$, such that the
chirality does not change if we translate from Minkowski to
Euclidean space.} chirality and are charged under the gauge group
$U(1)^{b^2(X)}$ (c.f. proposition \ref{KKG2}). Their contribution
to the variation of $X$ is given by
\begin{equation}
\delta X|_{anomaly}= i I_4^1\ \ \mbox{with} \ \ I_6=-2\pi[(-1)\hat
A(M_4){\rm ch}(F)]_6\ .
\end{equation}
Here the subscript ``anomaly" indicates that these are the
contributions to $\delta X$ coming from a variation of the
measure. Later on, we will have to add a contribution coming from
the variation of the Euclidean action. The sign of $I_6$ is
differs from the one in (\ref{Ihat1/2}) because the fermions have
negative chirality. The anomaly polynomials corresponding to gauge
and mixed anomalies localized at $P_{\alpha}$ are then given by
(see appendices \ref{notation} and \ref{anomalies} for the
details, in particular $F=iF^iq^i$),
\begin{equation}\label{abanom}
I_{\alpha}^{(gauge)}=-{1\over(2\pi)^23!}\sum_{\sigma\in
T_{\alpha}}\left(\sum_{i=1}^{b^2(X)}q_{\sigma}^iF^i\right)^3\ \ \
,\ \ \ I_{\alpha}^{(mixed)}={1\over24}\sum_{\sigma\in
T_{\alpha}}\left(\sum_{i=1}^{b^2(X)}q_{\sigma}^iF^i\right)p_1'\ .
\end{equation}
$\sigma$ labels the four dimensional chiral multiplets
$\Phi_{\sigma}$ which are present at the singularity $P_\alpha$.
$T_{\alpha}$ is simply a set containing all these labels.
$q_{\sigma}^i$ is the charge of $\Phi_{\sigma}$ with respect to
the $i$-th gauge field $A^i$. As all the gauge fields come from a
Kaluza-Klein expansion of the $C$-field we have $b^2(X)$ of them.
$p_1'=-{1\over8\pi^2}\tr R\wedge R$ is the first Pontrjagin class
of four dimensional space-time $\mathbb{R}^4$. Our task is now to
cancel these anomalies locally, i.e. separately at each
singularity.

So far we have only been using eleven-dimensional supergravity,
the low energy limit of M-theory. However, in the neighbourhood of
a conical singularity the curvature of $X$ blows up. Close to the
singularity $P_\a$ the space $X$ is a cone on some manifold $Y_\a$
(i.e. close to $P_\a$ we have $\d s^2_X\simeq\d r^2_\a+r_\a^2\d
s_{Y_\a}^2$). But as $X$ is Ricci-flat $Y_\a$ has to be Einstein
with $\mathcal{R}^{Y_\a}_{mn}=5\delta_{mn}$. The Riemann tensor on
$X$ and $Y_\a$ are related by $R^{Xmn}_{\ \ \ \ \ pq}={1\over
r^2_\a} (R^{Y_\a mn}_{\ \ \ \ \ \ pq}-
\delta^m_p\delta^n_q+\delta^m_q\delta^n_p)$, for $m\in\{1,2,\ldots
,6\}$. Thus, the supergravity description is no longer valid close
to a singularity and one has to resort to a full M-theory
calculation, a task that is currently not feasible.

To tackle this problem we use an idea that has first been
introduced in \cite{FHMM98} in the context of anomaly cancellation
on the M5-brane. The world-volume $W_6$ of the M5-brane supports
chiral field which lead to an anomaly. Quite interestingly, one
can cancel these anomalies using the inflow mechanism, but only if
the topological term of eleven-dimensional supergravity is
modified in the neighbourhood of the brane. Since we will proceed
similarly below, let us quickly motivate these modifications. The
five-brane acts as a source for the field $G$, i.e. the Bianchi
identity, $\d G=0$, is modified to $\d G \sim \dd^{(5)}(W_6)$. In
the treatment of \cite{FHMM98} a small neighbourhood of the
five-brane world-volume $W_6$ is cut out, creating a boundary.
Then one introduces a smooth function $\r$ which is zero in the
bulk but drops to $-1$ close to the brane, in such a way that
$\d\r$ has support only in the neighbourhood of the boundary. This
function is used to smear out the Bianchi identity by writing it
as $\d G\sim\ldots\w\d\r$. The solution to this identity is given
by $G=\d C+\ldots\w\d\r$, i.e. the usual identity $G=\d C$ is
corrected by terms localised on the boundary. Therefore, it is not
clear a priori how the topological term of eleven-dimensional
supergravity should be formulated (since for example the terms
$C\w\d C\w\d C$ and $C\w G\w G$ are now different). It turns out
that all the anomalies of the M5-brane cancel if the topological
term reads $\St_{\rm CS}=-{1\over 12\k_{11}^2} \int \Ct\w
\Gt\w\Gt$, where $\Ct$ is a field that is equal to $C$ far from
the brane, but is modified in the neighbourhood of the brane.
Furthermore, $\Gt=\d\Ct$. This mechanism of anomaly cancellation
in the context of the M5-brane is reviewed in detail in
\cite{BM03c} and \cite{Me03}.

Now we show that a similar treatment works for conical
singularities. We first concentrate on the neighbourhood of a
given conical singularity $P_\a$ with a metric locally given by
$\d s_X^2\simeq\d r_\a^2+r_\a^2 \d s_{Y_\a}^2$. The local radial
coordinate obviously is $r_\a\ge 0$, the singularity being at
$r_\a=0$. As mentioned above, there are curvature invariants of
$X$ that diverge as $r_\a\to 0$. In particular, supergravity
cannot be valid down to $r_\a=0$. Motivated by the methods used in
the context of the M5-brane, we want to modify our fields close to
the singularity. More precisely, we want to cut of the fluctuating
fields using a smooth function $\r$, which equals one far from the
singularity but is zero close to it. The geometry itself is kept
fixed, and in particular we keep the metric and curvature on $X$.
Said differently, we cut off all fields that represent the quantum
fluctuations, but keep the background fields (in particular the
background geometry) as before. To be specific, we introduce a
small but finite regulator $\e$, and the regularised step function
$\r_\a$ such that
\begin{equation}\label{rho}
\rho_{\alpha}(r_{\alpha})= \left\{\begin{matrix}
                                                0&\mbox{for}&\ \ 0\leq r_{\alpha}\leq R-\epsilon\\
                                                1&\mbox{for}&r_{\alpha}\geq R+\epsilon
                                  \end{matrix}\right.
\end{equation}
where $\epsilon/R$ is small. Using a partition of unity we can
construct a smooth function $\rho$ on $X$ from these
$\rho_{\alpha}$ in such a way that $\rho$ vanishes for points with
a distance to a singularity which is less than $R-\epsilon$ and is
one for distances larger than $R+\epsilon$. We denote the points
of radial coordinate $R$ in the chart around $P_{\alpha}$ by
$Y_{\alpha}$, where the orientation of $Y_\a$ is defined in such a
way that its normal vector points away from the singularity. All
these conventions are chosen in such a way that $\int_X(\ldots
)\wedge d\rho=\sum_{\alpha}\int_{Y_{\alpha}}(\ldots )$. The shape
of the function $\rho$ is irrelevant, in particular, one might use
$\rho^2$ instead of $\rho$, i.e. $\rho^n\simeq\rho$. However, when
evaluating integrals one has to be careful since
$\r^n\d\r={1\over{n+1}}\d\r^{n+1}\simeq{1\over{n+1}}\d\r$, where a
crucial factor of $1\over{n+1}$ appeared. In particular, for any
ten-form $\phi_{(10)}$, not containing $\r$'s or $\d\r$ we have
\begin{equation}
\int_{M_4\times
X}\phi_{(10)}\r^n\d\r=\sum_\a{1\over{n+1}}\int_{M_4\times
Y_\a}\phi_{(10)}\ .
\end{equation}
Using this function $\rho$ we can now ``cut off" the quantum
fluctuations by simply defining
\begin{equation}
\widehat C:=C\rho\ \ \ ,\ \ \ \widehat G=G\rho\ .
\end{equation}
The gauge invariant kinetic term of our theory is constructed from
this field
\begin{equation}
S_{kin}=-{1\over4\kappa_{11}^2}\int \widehat G\wedge \star
\widehat G=-{1\over4\kappa_{11}^2}\int_{r\geq R} \d C\wedge \star
\d C\ .
\end{equation}
However, the new field strength $\widehat G$ no longer is closed.
This can be easily remedied by defining
\begin{equation}
\widetilde C:=C\rho + B\wedge d\rho\ \ \ ,\ \ \
\widetilde{G}:=\d\widetilde C
\end{equation}
Note that $\widetilde G=\widehat G+(C+\d B)\w\d\rho$, so we only
modified $\hat G$ on the $Y_\a$. The auxiliary field $B$ living on
$Y_\a$ has to be introduced in order to maintain gauge invariance
of $\tilde G$. Its transformation law reads
\begin{equation}
\delta B=\L \ ,
\end{equation}
which leads to
\begin{equation}
\delta \widetilde C=d(\Lambda\rho) \ .
\end{equation}
Using these fields we are finally in a position to postulate the
form of a modified topological term \cite{BM03b},
\begin{eqnarray}
\widetilde{S}_{top}&:=&-{1\over12\kappa_{11}^2}\int_{\mathbb{R}^4\times
X}\widetilde{C}\wedge\widetilde{G}\wedge\widetilde{G}\ .
\end{eqnarray}
To see that this form is indeed useful for our purposes, one
simple has to calculate its gauge transformation. After plugging
in a Kaluza-Klein expansion of the fields,
$C=\sum_iA^i\w\o^i+\ldots$, $\L=\sum_i\e^i\o^i+\ldots$, where the
$\o^i$ are harmonic two-forms on $X$ (see \cite{BM03b} and
\cite{Me03} for details), we arrive at
\begin{equation}
\delta \widetilde{S}_{top}
=-\sum_{\alpha}{1\over(2\pi)^23!}\int_{\mathbb{R}^4}\epsilon^i
F^jF^k\int_{Y_{\alpha}}(T_2)^3\omega^i\wedge\omega^j\wedge\omega^k\
.
\end{equation}
Here we used $T_2:=\left({2\pi^2\over\k^2_{11}}\right)^{1/3}$. The
quantity $T_2$ can be interpreted as the M2-brane tension
\cite{BM03c}. Note that the result is a sum of terms which are
localized at $Y_{\alpha}$. The corresponding Euclidean anomaly
polynomial is given by
\begin{equation}
I^{(top)}_E=\sum_\a I^{(top)}_{E,\a}=-i
\sum_{\alpha}I^{(top)}_{M,\alpha}=\sum_{\alpha}{i\over(2\pi)^23!}
F^iF^jF^k\int_{Y_{\alpha}}(T_2)^3\omega^i\wedge\omega^j\wedge\omega^k\
.
\end{equation}
This is very similar to the gauge anomaly $I_{\alpha}^{(gauge)}$
and we do indeed get a local cancellation of the anomaly, provided
we have
\begin{equation}
\int_{Y_{\alpha}}(T_2)^3\omega^i\wedge\omega^j\wedge\omega^k=\sum_{\sigma\in
T_{\alpha}}q_{\sigma}^iq_{\sigma}^jq_{\sigma}^k\
.\label{chargecondition1}
\end{equation}
(Note that the condition of local anomaly cancellation is
$iI_{\alpha}^{(gauge)}+I^{(top)}_{E,\alpha}=0$, from
(\ref{master}).) In \cite{Wi01} it was shown that this equation
holds for all known examples of conical singularities. It is
particularly important that our modified topological term gives a
sum of terms localized at $Y_{\alpha}$ without any integration by
parts on $X$. This is crucial, because local quantities are no
longer well-defined after an integration by
parts\footnote{Consider for example $\int_a^b
df=f(b)-f(a)=(f(b)+c)-(f(a)+c)$. It is impossible to infer the
value of $f$ at the boundaries $a$ and $b$.}.

\bigskip
After having seen how anomaly cancellation works in the case of
gauge anomalies we turn to the mixed anomaly. In fact, it cannot
be cancelled through an inflow mechanism from any of the terms in
the action of eleven-dimensional supergravity. However, it was
found in \cite{VW95}, \cite{DLM95} that there is a first
correction term to the supergravity action, called the {\it
Green-Schwarz term}. On a smooth manifold $\mathbb{R}^4\times X$
it reads
\begin{equation}
S_{GS}=-{T_2\over2\pi}\int_{\mathbb{R}^4\times X}G\wedge
X_7=-{T_2\over2\pi}\int_{\mathbb{R}^4\times X}C\wedge X_8,
\end{equation}
with
\begin{equation}\label{X8}
X_8:={1\over(2\pi)^34!}\left({1\over8}{\rm tr}R^4-{1\over32}({\rm
tr}R^2)^2\right)\ .
\end{equation}
and $X_8=\d X_7$. The precise coefficient of the Green-Schwarz
term was determined in \cite{BM03c}. Then, there is a natural
modification\footnote{The reader might object that in fact one
could also use $\int\tilde C\w\tilde X_8$, $\int\tilde G\w X_8$ or
$\int\tilde G\w\tilde X_8$. However, all these terms actually lead
to the same result \cite{BM03b}.} on our singular manifold
\cite{BM03b},
\begin{equation}
\widetilde{S}_{GS}:=-{T_2\over2\pi}\int_{\mathbb{R}^4\times
X}\widetilde{C}\wedge X_8\ .
\end{equation}
and its variation reads
\begin{equation}
\delta\widetilde{S}_{GS}=-{T_2\over2\pi}\sum_{\alpha}\int_{\mathbb{R}^4\times
Y_{\alpha}}\epsilon^i\omega^i\wedge X_8\ .\label{mixedinflow}
\end{equation}
To obtain this result we again used a Kaluza-Klein expansion, and
we again did not integrate by parts. $X_8$ can be expressed in
terms of the first and second Pontrjagin classes,
$p_1=-{1\over2}\left({1\over2\pi}\right)^2{\rm tr}R^2$ and
$p_2={1\over8}\left(1\over2\pi\right)^4[({\rm tr}R^2)^2-2{\rm
tr}R^4]$, as
\begin{equation}
X_8={\pi\over4!}\left[{p_1^2\over4}-p_2\right]\ .
\end{equation}
The background we are working in is four-dimensional Minkowski
space times a $G_2$-manifold. In this special setup the Pontrjagin
classes can easily be expressed in terms of the Pontrjagin classes
$p_i'$ on $(\mathbb{R}^4,\eta)$ and those on $(X,g)$, which we
will write as $p_i''$. We have $ p_1=p_1'+p_1''$ and
$p_2=p_1'\wedge p_1''$. Using these relations we obtain a
convenient expression for the inflow (\ref{mixedinflow}),
\begin{eqnarray}\label{GSmod}
\delta\widetilde{S}_{GS} ={T_2\over2\pi}{\pi\over
48}\sum_{\alpha}\int_{\mathbb{R}^4}\epsilon^i
p_1'\int_{Y_{\alpha}}\omega^i\wedge p_1''\ .
\end{eqnarray}
The corresponding (Euclidean) anomaly polynomial is given by
\begin{equation}
I^{(GS)}_E=\sum_{\alpha}I^{(GS)}_{E,\alpha}=-i\sum_{\alpha}{1\over24}F^ip_1'\int_{Y_{\alpha}}{T_2\over4}\omega^i\wedge
p_1''\ ,
\end{equation}
and we see that the mixed anomaly cancels locally provided
\begin{equation}
\int_{Y_{\alpha}}{T_2\over4}\omega^i\wedge p_1''=\sum_{\sigma\in
T_{\alpha}}q_{\sigma}^i\ .\label{chargecondition2}
\end{equation}
All known examples satisfy this requirement \cite{Wi01}.

\section{Non-Abelian gauge groups and anomalies}
Finally we also want to comment on anomaly cancellation in the
case of non-Abelian gauge groups. The calculations are relatively
involved and we refer the reader to \cite{Wi01} and \cite{BM03b}
for the details. We only present the basic mechanism. Non-Abelian
gauge fields occur if $X$ carries ADE singularities. The enhanced
gauge symmetry can be understood to come from M2-branes that wrap
the vanishing cycles in the singularity. Since ADE singularities
have codimension four, the set of singular points is a
three-dimensional submanifold $Q$ of $X$. Chiral fermions which
are charged under the non-Abelian gauge group are generated if the
$Q$ itself develops a conical singularity. Close to such a
singularity $P_{\alpha}$ of $Q$ the space $X$ looks like a cone on
some $Y_\alpha$. If $U_{\alpha}$ denotes the intersection of $Q$
with $Y_{\alpha}$ then, close to $P_{\alpha}$, $Q$ is a cone on
$U_{\alpha}$. In this case there are $ADE$ gauge fields on
$\mathbb{R}^4\times Q$ which reduce to non-Abelian gauge fields on
$\mathbb{R}^4$ if we perform a Kaluza-Klein expansion on $Q$. On
the $P_{\alpha}$ we have a number of chiral multiplets
$\Phi_{\sigma}$ which couple to both the non-Abelian gauge fields
and the Abelian ones, coming from the Kaluza-Klein expansion of
the $C$-field. Thus, we expect to get a $U(1)^3$, $U(1)H^2$ and
$H^3$ anomaly\footnote{The $U(1)^2G$ anomaly is not present as
${\rm tr}\  T_a$ vanishes for all generators of $ADE$ gauge
groups, and the $H^3$-anomaly is only present for $H=SU(n)$.},
where $H$ is the relevant $ADE$ gauge group. The relevant anomaly
polynomial for this case is (again taking into account the
negative chirality of the fermions)
\begin{equation}
I_6=-2\pi [(-1)\hat A(M){\rm ch}(F^{(Ab)}){\rm ch}(F)]_6
\end{equation}
where $F^{(Ab)}:=iq^iF^i$ denotes the Abelian and $F$ the
non-Abelian gauge field. Expansion of this formula gives four
terms namely (\ref{abanom}) and
\begin{equation}\label{nonabpolyn}
I^{(H^3)}=-{i\over(2\pi)^23!}{\rm tr}F^3\ \ \ ,\ \ \
I^{(U(1)_iH^2)}={1\over(2\pi)^22}q_iF_i{\rm tr}F^2\ .
\end{equation}
It turns out \cite{Wi01}, \cite{BM03b} that our special setup
gives rise to two terms on $\mathbb{R}^4\times Q$, which read
\begin{eqnarray}
\widetilde{S}_1&=&-{i\over6(2\pi)^2}\int_{\mathbb{R}^4\times Q}K\wedge {\rm tr}(\widetilde{A}\widetilde{F}^2)\ ,\\
\widetilde{S}_2&=&{T_2\over2(2\pi)^2}\int_{\mathbb{R}^4\times
Q}\widetilde{C}\wedge {\rm tr}\widetilde{F}^2\ .
\end{eqnarray}
Here $K$ is the curvature of a certain line bundle described in
\cite{Wi01}, and $\widetilde{A}$ and $\widetilde{F}$ are modified
versions of $A$ and $F$, the gauge potential and field strength of
the non-Abelian $ADE$ gauge field living on $\mathbb{R}^4\times
Q$. The variation of these terms leads to contributions which are
localised at the various conical singularities, and after
continuation to Euclidean space the corresponding polynomials
cancel the anomalies (\ref{nonabpolyn}) locally, i.e. separately
at each conical singularity. For the details of this mechanism the
reader is referred to \cite{BM03b}. The main steps are similar to
what we did in the last chapter, and the only difficulty comes
from the non-Abelian nature of the fields which complicates the
calculation.

\chapter{Compact Weak $G_2$-Manifolds}\label{weakG2}
We have seen already that one possible vacuum of
eleven-dimensional supergravity is given by the direct product of
$AdS_4$ with a compact Einstein seven-manifold of positive
curvature, together with a flux
$G_{\m\n\r\s}\propto\e_{\m\n\r\s}$. Furthermore, if the compact
space carries exactly one Killing spinor we are left with $\mN=1$
in four dimensions. Such manifolds are known as weak
$G_2$-manifolds. As for the case of $G_2$-manifolds one expects
charged chiral fermions to occur, if the compact manifold carries
conical singularities. Unfortunately, no explicit metric for a
compact $G_2$-manifold with conical singularities is known.
However, in \cite{BM03a} explicit metrics for compact weak
$G_2$-manifolds with conical singularities have been constructed.
These spaces are expected to share many properties of singular
compact $G_2$-manifolds, and are therefore useful to understand
the structure of the latter.

The strategy to construct the compact weak $G_2$-holonomy
manifolds is the following: we begin with any non-compact
$G_2$-holonomy manifold $X$ that asymptotically, for ``large $r$''
becomes a cone on some 6-manifold $Y$. Manifolds of this type have
been constructed in \cite{GPP90}. The $G_2$-holonomy of $X$
implies certain properties of the 6-manifold $Y$ which we deduce.
In fact, $Y$ can be any Einstein space of positive curvature with
weak $SU(3)$-holonomy. Then we use this $Y$ to construct a compact
weak $G_2$-holonomy manifold $X_\l$ with two conical singularities
that, close to the singularities, looks like a cone on $Y$.

\section{Properties of weak $G_2$-manifolds}
On a weak $G_2$-manifold there exists a unique Killing
spinor\footnote{ Note that $a,b,c$ are ``flat'' indices with
Euclidean signature, and upper and lower indices are equivalent.},
\begin{equation}\label{Killingspinor}
\left(\partial_j+{1\over 4} \o_j^{ab} \g^{ab}\right) \theta =
i{\l\over 2} \g_j \theta \ ,
\end{equation}
from which one can construct a three-form
\begin{equation}\label{threeform}
\Phi_\l:={i\over6}\theta^\tau\g_{abc}\theta\ e^a\w e^b\w e^c\ ,
\end{equation}
which satisfies
\begin{equation}\label{f4}
\d\Phi_\l = 4 \l\ \st\Phi_\l \ .
\end{equation}
Furthermore, (\ref{Killingspinor}) implies that $X_\l$ has to be
Einstein,
\begin{equation}\label{f5}
\R_{ij}=6\l^2 g_{ij} \ .
\end{equation}
It can be shown that the converse statement is also true, namely
that Eq. (\ref{f4}) implies the existence of a spinor satisfying
(\ref{Killingspinor}). Note that for $\l\to 0$, at least formally,
weak $G_2$ goes over to $G_2$-holonomy.

To proceed we define the quantity $\psi_{abc}$ to be totally
antisymmetric with $\psi_{123} = \psi_{516} = \psi_{624} =
\psi_{435} = \psi_{471} = \psi_{673} = \psi_{572} = 1$, and its
dual
\begin{equation}\label{fa5}
\hat\psi_{abcd}:= {1\over3!}\tilde\epsilon^{abcdefg}\psi_{efg}\ .
\end{equation}
Using these quantities we note that every antisymmetric tensor
$A^{ab}$ transforming as the {\bf 21} of SO(7) can always be
decomposed \cite{BDS01} into a piece $A_+^{ab}$ transforming as
the {\bf 14} of $G_2$ (called self-dual) and a piece $A_-^{ab}$
transforming as the {\bf 7} of $G_2$ (called anti-self-dual):
\begin{eqnarray}\label{fa0}
A^{ab}&=&A^{ab}_+ + A^{ab}_-\\
A^{ab}_+&=&{2\over 3} \left( A^{ab} +{1\over 4} \hat\psi^{abcd}
A^{cd}\right) =:\mathcal{P}_{\bf 14}A^{ab}\  ,\\
A^{ab}_-&=&{1\over 3} \left( A^{ab} -{1\over 2} \hat\psi^{abcd}
A^{cd}\right)=:\mathcal{P}_{\bf 7}A^{ab}\ ,
\end{eqnarray}
with orthogonal projectors $(\mathcal{P}_{\bf
14})_{ab}^{cd}:={2\over3}\left(\delta_{ab}^{cd}+{1\over4}\hat\psi_{ab}^{\
\ cd}\right)$ and $(\mathcal{P}_{\bf
7})_{ab}^{cd}:={1\over3}\left(\delta_{ab}^{cd}-{1\over2}\hat\psi_{ab}^{\
\ cd}\right)$,
$\delta_{ab}^{cd}:={1\over2}(\delta_a^c\delta_b^d-\delta_a^d\delta_b^c)$.
Using the identity\footnote{Many useful identities of this type
are listed in the appendix of \cite{BDS01}.}
\begin{equation}
\hat\psi_{abde}\psi_{dec}=-4\psi_{abc}
\end{equation}
we find that the self-dual\footnote{The reader might wonder how
this name is motivated, since so far we did not encounter a
self-duality condition. As a matter of fact one can show that for
any anti-symmetric tensor $B^{ab}$ the following three statements
are equivalent
\begin{eqnarray}
\psi_{abc}B^{bc}&=&0\ ,\nonumber\\
B_-^{ab}&=&0\ ,\\
B^{ab}&=&{1\over2}\hat\psi_{abcd}B^{cd}\ ,\nonumber
\end{eqnarray}
and the last equation now explains the nomenclature.} part
satisfies
\begin{equation}
\psi^{abc} A_+^{bc}=0 \ .
\end{equation}
In particular, one has
\begin{equation}\label{fa0q}
\o^{ab} \g^{ab}= \o_+^{ab} \g_+^{ab} + \o_-^{ab} \g_-^{ab} \ .
\end{equation}
The importance of (anti-)self-duality will become clear in the
following theorem.

\begin{theorem}
A manifold $(M,g)$ is a (weak) $G_2$-manifold, if and only if
there exists a frame in which the spin connection satisfies
\begin{equation}\label{f9}
\psi_{abc} \o^{bc} =-2\l \, e^a\ ,
\end{equation}
where $e^a$ is the 7-bein on $X$.
\end{theorem}
We will call such a frame a {\it self-dual frame}. The proof can
be found in \cite{BDS01}. In the case of a $G_2$-manifold one
simply has to take $\l=0$. We also need the following:
\begin{proposition}
The three-form $\Phi_\l$ of Eq.(\ref{threeform}) can be written as
\begin{equation}
\Phi_\l={1\over 6} \psi_{abc}\ e^a\w e^b\w e^c\label{sdthreeform}
\end{equation}
if and only if the 7-beins $e^a$ are a self-dual frame. This holds
for $\l=0$ ($G_2$) and $\l\ne 0$ (weak $G_2$).
\end{proposition}
Later, for weak $G_2$, we will consider a frame which is not
self-dual and thus the 3-form $\Phi_\l$ will be slightly more
complicated than (\ref{sdthreeform}). To prove the proposition it
will be useful to have an  explicit representation for the
$\g$-matrices in 7 dimensions. A convenient representation is in
terms of the $\psi_{abc}$ as \cite{BDS01}
\begin{equation}\label{fa3}
(\gamma_a)_{AB} = i(\psi_{aAB} +\delta_{aA}\delta_{8B}
-\delta_{aB}\delta_{8A}) \ .
\end{equation}
Here $a=1, \ldots 7$ while $A, B =1, \ldots 8$ and it is
understood that $\psi_{aAB}=0$ if $A$ or $B$ equals 8. One then
has \cite{BDS01}
\begin{eqnarray}
(\gamma_{ab})_{AB}&=& \hat\psi_{abAB}
+\psi_{abA}\delta_{8B}-\psi_{abB}\delta_{8A}
+\delta_{aA}\delta_{bB}-\delta_{aB}\delta_{bA}\ ,\\
(\gamma_{abc})_{AB} &=& i\psi_{abc}
(\delta_{AB}-2\delta_{8A}\delta_{8B}) -3i\psi_{A[ab}\delta_{c]B}
-3i\psi_{B[ab}\delta_{c]A} \nonumber\\
& &-i\hat\psi_{abcA}\delta_{8B} -i\hat\psi_{abcB}\delta_{8A} \ .
\end{eqnarray}
In order to see under which condition (\ref{threeform}) reduces to
(\ref{sdthreeform}) we use the explicit representation for the
$\g$-matrices (\ref{fa3}) given above. It is then easy to see that
$\theta^T\g_{abc}\theta \sim \psi_{abc}$ if and only if
$\theta_A\sim \delta_{8A}$. This means that our 3-form $\Phi$ is
given by (\ref{sdthreeform}) if and only if the covariantly
constant, resp. Killing spinor $\theta$ only has an eighth
component, which then must be a constant which we can take to be
1. With this normalisation we have
\begin{equation}\label{fa6}
\theta^T\g_{abc}\theta = -i\psi_{abc} \ ,
\end{equation}
so that $\Phi$ is correctly given by (\ref{sdthreeform}). From the
above explicit expression for $\g_{ab}$ one then deduces that
$(\g_{ab})_{AB}\theta_B = \psi_{abA}$ and
$\o^{ab}(\g_{ab})_{AB}\theta_B = \o^{ab} \psi_{abA}$. Also,
$i(\g_c)_{AB}\theta_B = - \delta_{cA}$, so that Eq.
(\ref{Killingspinor}) reduces to $\o^{ab}\psi_{abc} =-2\l e^c$.

\section{Construction of weak $G_2$-holonomy manifolds with
singularities}
Following \cite{BM03a} we start with any (non-compact)
$G_2$-manifold $X$ which asymptotically is a cone on a compact
6-manifold $Y$,
\begin{equation}\label{f10}
\d s_X^2 \sim \d r^2 +r^2 \d s_Y^2 \ .
\end{equation}
Since $X$ is Ricci flat, $Y$ must be an Einstein manifold with
$\R_{\a\b}=5\delta_{\a\b}$. In practice \cite{GPP90},
$Y=\mathbb{CP}^3,\ S^3\times S^3$ or $SU(3)/U(1)^2$, with
explicitly known metrics. On $Y$ we introduce 6-beins
\begin{equation}\label{f11}
\d s_Y^2 = \sum_{\a=1}^6 \et^{\,\a} \otimes \et^{\,\a} \ ,
\end{equation}
and similarly on $X$
\begin{equation}\label{f12}
\d s_X^2 =\sum_{a=1}^7 \hat e^a \otimes \hat e^a \ .
\end{equation}
Our conventions are that $a,b,\ldots$ run from 1 to 7 and
$\a,\b,\ldots$ from 1 to 6. The various manifolds and
corresponding viel-beins are summarized in the table below. Since
$X$ has $G_2$-holonomy we may assume that the 7-beins $\hat e^a$
are chosen such that the $\o^{ab}$ are self-dual, and hence we
know from the above remark that the closed and co-closed 3-form
$\Phi$ is simply given by Eq. (\ref{sdthreeform}), i.e.
$\Phi={1\over 6} \psi_{abc}\, \hat e^a \w \hat e^b \w \hat e^c$.
Although there is such a self-dual choice, in general, we are not
guaranteed that this choice is compatible with the natural choice
of 7-beins on $X$ consistent with a cohomogeneity-one metric as
(\ref{f10}). (For any of the three examples cited above, the
self-dual choice actually is compatible with a cohomogeneity-one
metric.)

\vskip 10.mm
\begin{center}
\begin{tabular}{|c|c|c|c|}\hline
$X$ &  $Y$ &$X_c$ & $X_\l$ \\
\hline \hline
$\hat e^a$ &  $\et^\a$ &$\eb^a$ & $e^a$\\
\hline
$\Phi$ &  &$\phi$ & $\Phi_\l$ \\
\hline
\end{tabular}
\end{center}
\vskip 5.mm \centerline{\it Table 1 : The various manifolds,
corresponding viel-beins \ \ \ \  } \centerline{\it and 3-forms
that enter our construction.} \vskip 5.mm

Now we take the limit $X\to X_c$ in which the $G_2$-manifold
becomes exactly a cone on $Y$ so that $\hat e^a\to \eb^a$ with
\be\label{f14} \eb^\a = r \et^\a \quad , \quad \eb^7 = \d r \ .
\ee In this limit the cohomogeneity-one metric can be shown to be
compatible with the self-dual choice of frame (see \cite{BM03a}
for a proof) so that we may assume that (\ref{f14}) is such a
self-dual frame. More precisely, we may assume that the original
frame $\hat e^a$ on $X$ was chosen in such a way that after taking
the conical limit the $\eb^a$ are a self-dual frame. Then we know
that the 3-form $\Phi$ of $X$ becomes a 3-form $\f$ of $X_c$ given
by the limit of (\ref{sdthreeform}), namely
\begin{equation}\label{f15}
\f=r^2 \d r \w \xi + r^3 \zeta\ ,
\end{equation}
with the 2- and 3-forms on $Y$ defined by
\begin{equation}\label{f16}
\xi={1\over 2} \psi_{7\a\b}\, \et^{\,\a} \w \et^{\,\b}\ \ \ ,\ \ \
\zeta={1\over 6} \psi_{\a\b\g}\,  \et^{\,\a} \w \et^{\,\b} \w
\et^{\,\g} \ .
\end{equation}
The dual 4-form is given by
\begin{equation}\label{f17}
\st\f=r^4 \sty\xi - r^3 \d r \w \sty\zeta\ ,
\end{equation}
where $\sty\xi$ is the dual of $\xi$ in $Y$.\footnote{We need to
relate Hodge duals on the 7-manifolds $X$, $X_c$ or $X_\l$ to the
Hodge duals on the 6-manifold $Y$. To do this, we do not need to
specify the 7-manifold and just call it $X_7$. We assume that the
7-beins of $X_7$, called $e^a$, and the 6-beins of $Y$, called
$\et^\a$ can be related by
\begin{equation}\label{fa20}
e^7=\d r \ , \quad e^\a= h(r) \et^{\,\a} \ .
\end{equation}
We denote the Hodge dual of a form $\pi$ on $X_7$ simply by
$\st\pi$ while the 6-dimensional Hodge dual of a form $\sigma$ on
$Y$ is denoted $\sty\sigma$. The duals of $p$-forms on $X_7$ and
on $Y$ are defined in terms of their respective viel-bein basis,
namely
\begin{equation}\label{fa21}
\st\left( e^{a_1} \w \ldots \w e^{a_p}\right) ={1\over (7-p)!}
\et^{\,a_1\ldots a_p}_{\phantom{a_1\ldots a_p} b_1\ldots b_{7-p}}
e^{b_1} \w \ldots \w e^{b_{7-p}}
\end{equation}
and
\begin{equation}\label{fa22}
\sty\left( \et^{\,\a_1} \w \ldots \w \et^{\,\a_p}\right) ={1\over
(6-p)!} \et^{\,\a_1\ldots \a_p}_{\phantom{\a_1\ldots \a_p}
\b_1\ldots \b_{6-p}} \et^{\,\b_1} \w \ldots \w \et^{\,\b_{6-p}} \
.
\end{equation}
Here the $\et$-tensors are the ``flat'' ones that equal $\pm 1$.
Expressing the $e^a$ in terms of the $\et^\a$ provides the desired
relations. In particular, for a $p$-form $\o_p$ on $Y$ we have
\begin{eqnarray}\label{fa23}
\st\left( \d r \w \o_p\right) &=& h(r)^{6-2p}
\sty\o_p\ , \nonumber\\
\st\o_p &=& (-)^p h(r)^{6-2p} \d r \w \sty\o_p \ ,
\end{eqnarray}
where we denote both the form on $Y$ and its trivial extension
onto $X_\l$ by the same symbol $\o_p$.} As for the original
$\Phi$, after taking the conical limit, we
still have $\d\f=0$ and $\d\st\f=0$. This is equivalent to\\
\parbox{14cm}{
\begin{eqnarray}
\d\xi&=& 3\zeta\ ,\nonumber \\
\d \sty\zeta &=& - 4\sty\xi \ .\nonumber
\end{eqnarray}}\hfill\parbox{8mm}{\begin{eqnarray}\label{f18}\end{eqnarray}}
These are properties of appropriate forms on $Y$, and they can be
checked to be true for any of the three standard $Y$'s. Actually,
these relations show that $Y$ has weak $SU(3)$-holonomy.
Conversely, if $Y$ is a 6-dimensional manifold with weak
$SU(3)$-holonomy, then we know that these forms exist. This is
analogous to the existence of the 3-form $\Phi_\l$ with
$\d\Phi_\l=4 \l \st\Phi_\l$ for weak $G_2$-holonomy. These issues
were discussed e.g. in \cite{Hi01}. Combining the two relations
(\ref{f18}), we see that on $Y$ there exists a 2-form $\xi$
obeying
\begin{equation}\label{f19}
\d\sty\d\xi + 12 \sty\xi=0 \quad , \quad \d\sty\xi=0 \ .
\end{equation}
This implies $\Delta_Y\xi = 12 \xi$, where $\Delta_Y =
-\sty\d\sty\d - \d\sty\d\sty$ is the Laplace operator on forms on
$Y$. Note that with $\zeta={1\over 3}\d\xi$ we actually have
$\f=\d\left( {r^3\over 3}\xi\right)$ and $\f$ is cohomologically
trivial. This was not the case for the original $\Phi$.

We now construct a manifold $X_\l$ with a 3-form $\Phi_\l$ that is
a deformation  of this 3-form $\f$  and that will satisfy the
condition (\ref{f4}) for weak $G_2$-holonomy. Since weak
$G_2$-manifolds are Einstein manifolds we need to introduce some
scale $r_0$ and make the following ansatz for the metric on $X_\l$
\be\label{f20} \d s_{X_\l}^2 = \d r^2 +r_0^2 \sin^2 \rh\ \d s_Y^2
\ , \ee with \be\label{f21} \rh = {r\over r_0} \quad , \quad 0\le
r \le \pi r_0 \ . \ee Clearly, this metric has two conical
singularities, one at $r=0$ and the other at $r=\pi r_0$.

We see from (\ref{f20}) that we can choose 7-beins $e^a$ on $X_\l$
that are expressed in terms of the 6-beins $\et^\a$ of $Y$ as
\begin{equation}\label{f24}
e^\a=r_0 \sin \rh \ \et^{\,\a} \quad , \quad e^7=\d r \ .
\end{equation}
Although this is the natural choice, it should be noted that it is
{\it not} the one that leads to a self-dual spin connection
$\o^{ab}$ that satisfies Eq. (\ref{f9}). We know from \cite{BDS01}
that such a self-dual choice of 7-beins must exist if the metric
(\ref{f20}) has weak $G_2$-holonomy but, as noted earlier, there
is no reason why this choice should be compatible with
cohomogeneity-one, i.e choosing $e^7=\d r$. Actually, it is easy
to see that for weak $G_2$-holonomy, $\l\ne 0$, a
cohomogeneity-one choice of frame and self-duality are
incompatible: a cohomogeneity-one choice of frame means $e^7=\d r$
and $e^\a=h_{(\a)}(r) \et^\a$ so that $\o^{\a\b}={ h_{(\a)}(r)
\over h_{(\b)}(r)} \wt\o^{\a\b}$ and $\o^{\a 7}=h_{(\a)}'(r)
\et^\a$. But then the self-duality condition for $a=7$ reads
$\psi_{7\a\b}{ h_{(\a)}(r) \over h_{(\b)}(r)} \wt\o^{\a\b}=-2\l \d
r$. Since $\wt\o^{\a\b}$ is the spin connection on $Y$, associated
with $\et^\a$, it contains no $\d r$-piece, and the self-duality
condition cannot hold unless $\l=0$.

Having defined the 7-beins on $X_\l$ in terms of the 6-beins on
$Y$, the Hodge duals on $X_\l$ and on $Y$ are related accordingly.
If $\o_p$ is a $p$-form on $Y$, we have\\
\parbox{14cm}{
\begin{eqnarray}
\st\left( \d r \w \o_p\right) &=&\left( r_0 \sin\rh\right)^{6-2p}
\sty\o_p \nonumber\\
\st\o_p &=& (-)^p \left( r_0 \sin\rh\right)^{6-2p} \d r \w
\sty\o_p \ ,\nonumber
\end{eqnarray}}\hfill\parbox{8mm}{\begin{eqnarray}\label{f27}\end{eqnarray}}
where we denote both the form on $Y$ and its trivial
($r$-independent) extension onto $X_\l$ by the same symbol $\o_p$.

Finally, we are ready to determine the 3-form $\Phi_\l$ satisfying
$\d\Phi_\l =\l \st\Phi_\l$. We make the ansatz \cite{BM03a}
\begin{equation}\label{f28}
\Phi_\l= (r_0\sin\rh)^2 \, \d r\w\xi + (r_0\sin\rh)^3 \left(
\cos\rh\ \zeta + \sin\rh\ \rho\right) \ .
\end{equation}
Here, the 2-form $\xi$ and the 3-forms $\zeta$ and $\r$ are forms
on $Y$ which are trivially extended to forms on $X_\l$ (no
$r$-dependence). Note that this $\Phi_\l$ is not of the form
(\ref{sdthreeform}) as the last term is not just $\zeta$ but
$\cos\rh\ \zeta + \sin\rh\ \r$. This was to be expected since the
cohomogeneity-one frame cannot be self-dual. The Hodge dual of
$\Phi_\l$ then is  given by
\begin{equation}\label{f29}
\st\Phi_\l= (r_0\sin\rh)^4 \sty\xi - (r_0\sin\rh)^3 \, \d r \w
\left( \cos\rh \sty\zeta + \sin\rh \sty\r\right)
\end{equation}
while
\begin{eqnarray}\label{f30}
\d\Phi_\l&=& (r_0\sin\rh)^2\, \d r\w (-\d\xi+ 3 \zeta) +
(r_0\sin\rh)^3 \, \left( \cos\rh\ \d\zeta
+ \sin\rh\ \d\r\right) \nonumber\\
&+&{4\over r_0} (r_0\sin\rh)^3\, \d r \w \left( \cos\rh\ \r -
\sin\rh\ \zeta\right) \ .
\end{eqnarray}
In the last term, the derivative $\partial_r$ has exchanged
$\cos\rh$ and $\sin\rh$ and this is the reason why both of them
had to be present in the first place.

Requiring $\d \Phi_\l = 4\l \st\Phi_\l$ leads to the following
conditions
\begin{eqnarray}
\d\xi&=& 3\zeta\ ,\label{f31}\\
\d\r &=& 4\l r_0 \sty\xi\ ,\label{f32}\\
\r &=& -\l r_0 \sty\zeta\ ,\label{f33}\\
\zeta &=& \l r_0 \sty \r \ . \label{f34}
\end{eqnarray}
Equations (\ref{f33}) and (\ref{f34}) require
\begin{equation}\label{f35}
r_0={1\over \l}
\end{equation}
and $\zeta=\sty\r\ \Leftrightarrow\ \r = - \sty\zeta$ (since for a
3-form $\sty(\sty\o_3)=-\o_3$). Then (\ref{f32}) is
$\d\r=4\sty\xi$, and inserting $\r=-\sty\zeta$ and Eq. (\ref{f31})
we get
\begin{equation}\label{f36}
\d\sty\d\xi + 12 \sty\xi =0 \ \ \ \mbox{and}\ \ \ \d\sty\xi=0\ .
\end{equation}
But we know from (\ref{f19}) that there is such a two-form $\xi$
on $Y$. Then pick such a $\xi$ and let $\zeta={1\over 3} \d\xi$
and $\r=-\sty\zeta=-{1\over 3} \sty \d\xi$. We conclude that
\begin{equation}\label{f37}
\Phi_\l =\left({\sin \l r\over \l}\right)^2 \, \d r\w\xi +{1\over
3} \left({\sin \l r\over \l}\right)^3 \left(\cos\l r \, \d\xi -
\sin\l r \, \sty\d\xi\right)
\end{equation}
satisfies $\d\Phi_\l = 4\l \st\Phi_\l$ and that the manifold with
metric (\ref{f20}) has weak $G_2$-holonomy. Thus we have succeeded
to construct, for every non-compact $G_2$-manifold that is
asymptotically (for large $r$) a cone on $Y$, a corresponding
compact weak $G_2$-manifold $X_\l$ with two conical singularities
that look, for small $r$, like cones on the same $Y$. Of course,
one could start directly with any 6-manifold $Y$ of weak
$SU(3)$-holonomy.

The quantity $\l$ sets the scale of the weak $G_2$-manifold $X_\l$
which has a size of order ${1\over \l}$. As $\l\to 0$, $X_\l$
blows up and, within any fixed finite distance from $r=0$, it
looks like the cone on $Y$ we started with.

\newpage
\begin{center}
{\bf Cohomology of the weak $G_2$-manifolds}
\end{center}
As mentioned above, what one wants to do with our weak
$G_2$-manifold at the end of the day is to compactify M-theory on
it, and generate an interesting four-dimensional effective action.
This is done by a Kaluza-Klein compactification (reviewed for
instance in \cite{Me03}), and it is therefore desirable to know
the cohomology groups of the compact manifold. To be more precise,
there are various ways to define harmonic forms which are all
equivalent on a compact manifold without singularities where one
can freely integrate by parts. Since $X_\l$ has singularities we
must be more precise about the definition we adopt and about the
required behaviour of the forms as the singularities are
approached.

Physically, when one does a Kaluza-Klein reduction of an
eleven-dimensional $k$-form $C_k$ one first writes a double
expansion $C_k=\sum_{p=0}^k \sum_i A^i_{k-p}\w  \f_p^i$ where
$A^i_{k-p}$ are $(k-p)$-form fields in four dimensions  and the
$\f_p^i$ constitute, for each $p$, a basis of $p$-form fields on
$X_\l$. It is convenient to expand with respect to a basis of
eigenforms of the Laplace operator on $X_\l$. Indeed, the standard
kinetic term for $C_k$ becomes
\begin{eqnarray}\label{c1}
\int_{{\cal M}_4\times X_\l} \d C_k \w \st \d C_k
&=&\sum_{p,i}\Bigg(
 \int_{{\cal M}_4} \d A^i_{k-p}\w \st  \d A^i_{k-p} \ \
\int_{X_\l} \f_p^i \w \st\f_p^i \nonumber
\\
&+& \int_{{\cal M}_4}  A^i_{k-p}\w \st   A^i_{k-p} \ \ \int_{X_\l}
\d\f_p^i \w \st\d\f_p^i \Bigg) \ .
\end{eqnarray}
Then a massless field $A_{k-p}$ in four dimensions arises for
every closed $p$-form $\f_p^i$ on $X_\l$ for which $\int_{X_\l}
\f_p \w \st\f_p$ is finite. Moreover, the usual gauge condition
$\d\st C_k =0$ leads to the analogous four-dimensional condition
$\d\st A_{k-p}=0$ provided we also have $\d \st\f_p=0$. We are led
to the following definition:
\begin{definition}
{\em An $L^2$-harmonic $p$-form $\f_p$ on $X_\l$ is a $p$-form
such that
\begin{eqnarray}\label{c2}
(i) \quad & &|| \f_p ||^2
\equiv \int_{X_\l}\ \f_p \w \st \f_p < \infty \ , \quad {\rm and}\\
\label{c3} (ii) \quad & &\d\f_p =0 \quad {\rm and} \quad
\d\st\f_p=0 \ .
\end{eqnarray}}
\end{definition}
Then one can prove \cite{BM03a} the following:
\begin{proposition}
Let $X_\l$ be a 7-dimensional manifold with metric given by
(\ref{f20}), (\ref{f21}). Then all
 $L^2$-harmonic $p$-forms $\f_p$ on $X_\l$ for $p\le 3$
are given by the trivial ($r$-independent) extensions to $X_\l$ of
the $L^2$-harmonic $p$-forms $\o_p$ on $Y$. For $p\ge 4$ all
$L^2$-harmonic $p$-forms on $X_\l$ are given by $\st\f_{7-p}$.
\end{proposition}
Since there are no harmonic 1-forms on $Y$ we immediately have the

\noindent {\bf Corollary :} The Betti numbers on $X_\l$ are given
by those of $Y$ as
\begin{eqnarray}\label{c3b}
b^0(X_\l)=b^7(X_\l)=1\quad &,& \quad b^1(X_\l)=b^6(X_\l)=0\ ,\nonumber\\
b^2(X_\l)=b^5(X_\l)=b^2(Y)\quad &,& \quad
b^3(X_\l)=b^4(X_\l)=b^3(Y) \ .
\end{eqnarray}
The proof of the proposition is lengthy and rather technical and
the reader is referred to \cite{BM03a} for details.

\chapter{The Ho\v{r}ava-Witten Construction}\label{HW}
The low-energy effective theory of M-theory is eleven-dimensional
supergravity. Over the last years various duality relations
involving string theories and eleven-dimensional supergravity have
been established, confirming the evidence for a single underlying
theory. One of the conjectured dualities, discovered by Ho\v{r}ava
and Witten \cite{HW95}, relates M-theory on the orbifold
$M_{10}\times S^1/\mathbb{Z}_2$ to $E_8\times E_8$ heterotic
string theory on the manifold $M_{10}$. In \cite{HW95} it was
shown that the gravitino field $\psi_M$, $M,N,\ldots=0,1,\ldots
10$, present in the eleven-dimensional bulk $M_{10}\times
S^1/\mathbb{Z}_2$ leads to an anomaly on the ten-dimensional fixed
``planes" of this orbifold. Part of this anomaly can be cancelled
if we introduce a ten-dimensional $E_8$ vector multiplet on each
of the two fixed planes. This does not yet cancel the anomaly
completely. However, once the vector multiplets are introduced
they have to be coupled to the eleven-dimensional bulk theory. In
\cite{HW95} it was shown that this leads to a modification of the
Bianchi identity to $\d G\neq0$, which in turn leads to yet
another contribution to the anomaly, coming from the
non-invariance of the classical action. Summing up all these terms
leaves us with an anomaly free theory. However, the precise way of
how all these anomalies cancel has been the subject of quite some
discussion in the literature. Using methods similar to the ones we
described in chapter \ref{G2anom}, and building on the results of
\cite{BDSa99}, we were able to prove for the first time
\cite{BM03c} that the anomalies do actually cancel locally, i.e.
separately on each of the two fixed planes. In this chapter we
will explain the detailed mechanism that leads to this local
anomaly cancellation.

\begin{center}
{\bf The orbifold $\mathbb{R}^{10}\times S^1/\mathbb{Z}_2$}
\end{center}
Let the eleven-dimensional manifold $M_{11}$ be the Riemannian
product of ten-manifold $(M_{10},g)$ and a circle $S^1$ with its
standard metric. The coordinates on the circle are taken to be
$\phi\in[-\pi,\pi]$ with the two endpoints identified. In
particular, the radius of the circle will be taken to be one. The
equivalence classes in $S^1/\mathbb{Z}_2$ are the pairs of points
with coordinate $\phi$ and $-\phi$, i.e. $\mathbb{Z}_2$ acts as
$\phi\rightarrow-\phi$. This map has the fixed points 0 and $\pi$,
thus the space $M_{10}\times S^1/\mathbb{Z}_2$ contains two
singular ten-dimensional spaces. In the simplest case we have
$M_{10}=\mathbb{R}^{10}$, which is why we call these spaces fixed
planes.

Before we proceed let us introduce some nomenclature. Working on
the space $M_{10}\times S^1$, with an additional
$\mathbb{Z}_2$-projection imposed, is called to work in the
``upstairs" formalism. Equivalently, one might work on the
manifold $M_{10}\times I\cong M_{10}\times S^1/\mathbb{Z}_2$ with
$I=[0,\pi]$. This is referred to as the ``downstairs" approach. It
is quite intuitive to work downstairs on the interval but for
calculational purposes it is more convenient to work on manifolds
without boundary. Otherwise one would have to impose boundary
conditions for the fields. Starting from supergravity on
$M_{10}\times I$ it is easy to obtain the action in the upstairs
formalism. One simply has to use
$\int_I\ldots={1\over2}\int_{S^1}\ldots$.

Working upstairs one has to impose the $\mathbb{Z}_2$-projection
by hand. By inspection of the topological term of
eleven-dimensional supergravity (c.f. Eq. (\ref{SUGRAaction})) one
finds that $C_{\m\n\r}$, with $\m,\n,\ldots=0,1,\ldots9$, is
$\mathbb{Z}_2$-odd, whereas $C_{\m\n10}$ is $\mathbb{Z}_2$-even.
This implies that $C_{\m\n\r}$ is projected out and $C$ can be
written as $C=\tilde B\w d\phi$.

Following \cite{BDSa99} we define for further convenience\\
\parbox{14cm}{
\begin{eqnarray}
\delta_1:=\delta(\phi)d\phi\ \ \ &,&\ \ \ \delta_2:=\delta(\phi-\pi)d\phi\ ,\nonumber\\
\epsilon_1(\phi):={\rm sig}(\phi)-{\phi\over\pi}\ \ \ &,&\ \ \
\epsilon_2(\phi):=\epsilon_1(\phi-\pi)\ ,\nonumber
\end{eqnarray}}\hfill\parbox{8mm}{\begin{eqnarray}\end{eqnarray}}
which are well-defined on $S^1$ and satisfy
\begin{equation}
d\epsilon_i=2\delta_i-{d\phi\over\pi}\ .\label{depsilon}
\end{equation}
After regularization we get \cite{BDSa99}
\footnote{Of course $\delta_{ij}$ is the usual Kronecker symbol, not to be confused with $\delta_i$.}\\
\begin{equation}\label{reg}
\delta_i\epsilon_j\epsilon_k\rightarrow{1\over3}(\delta_{ji}\delta_{ki})\delta_i\
.
\end{equation}

\begin{center}
{\bf Anomalies of M-theory on $M_{10}\times S^1/\mathbb{Z}_2$}
\end{center}
Next we need to study the field content of eleven-dimensional
supergravity on the given orbifold, and analyse the corresponding
anomalies. Compactifying eleven-dimensional supergravity on the
circle leads to a set of (ten-dimensional) massless fields which
are independent of the coordinate $\phi$ and other
(ten-dimensional) massive modes. Only the former can lead to
anomalies in ten-dimensions. For instance, the eleven-dimensional
Rarita-Schwinger field reduces to a sum of infinitely many massive
modes, and two massless ten-dimensional gravitinos of opposite
chirality. On our orbifold we have to impose the
$\mathbb{Z}_2$-projection on these fields. Only one of the two
ten-dimensional gravitinos is even under $\phi\rightarrow-\phi$,
and the other one is projected out. Therefore, after the
projection we are left with a chiral theory, which, in general, is
anomalous. Note that we have not taken the radius of the
compactification to zero, and there is a ten-plane for every point
in $S^1/\mathbb{Z}_2$. Therefore it is not clear a priori on which
ten-dimensional plane the anomalies should occur. There are,
however, two very special ten-planes, namely those fixed by the
$\mathbb{Z}_2$-projection. It is therefore natural to assume that
the anomalies should localise on these planes. Clearly, the entire
setup is symmetric and the two fixed planes should carry the same
anomaly. In the case in which the radius of $S^1$ reduces to zero
we get the usual ten-dimensional anomaly of a massless gravitino
field. We conclude that in the case of finite radius the anomaly
which is situated on the ten-dimensional planes is given by
exactly one half of the usual gravitational anomaly in ten
dimensions. This result has first been derived in \cite{HW95}.
Since the $\mathbb{Z}_2$-projection gives a positive chirality
spin-${3\over2}$ field and a negative chirality spin-$3\over2$
field in the Minkowskian, (recall $\G^E=-\G^M$), we are led to a
negative chirality spin-$3\over2$ and a positive chirality
spin-$1\over2$ field in the Euclidean. The anomaly polynomial of
these so-called untwisted fields on a single fixed plane reads
\begin{equation}
I_{12(i)}^{(untwisted)}={1\over2}\left\{{1\over2}\left(-I_{grav}^{(3/2)}(R_i)+I_{grav}^{(1/2)}(R_i)\right)\right\}\
.\label{I12untwisted}
\end{equation}
The second factor of ${1\over2}$ arises because the fermions are
Majorana-Weyl. $i=1,2$ denotes the two planes, and $R_i$ is the
curvature two-form on the $i$-th plane. $I_{grav}^{(3/2)}$ is
obtained from $I^{(3/2)}_{12}$ of Eq. (\ref{Ihat3/2}) by setting
$F=0$, and similarly for $I_{grav}^{(1/2)}$.

So eleven-dimensional supergravity on $M_{10}\times
S^1/\mathbb{Z}_2$ is anomalous and has to be modified in order to
be a consistent theory. An idea that has been very fruitful in
string theory over the last years is to introduce new fields which
live on the singularities of the space under consideration.
Following this general tack we introduce massless modes living
only on the fixed planes of our orbifold. These so-called twisted
fields have to be ten-dimensional vector multiplets because the
vector multiplet is the only ten-dimensional supermultiplet with
all spins $\leq 1$. In particular, the multiplets can be chosen in
such a way that the gaugino fields have positive chirality (in the
Minkowskian). Then they contribute to the pure and mixed gauge
anomalies, as well as to the gravitational ones. The corresponding
anomaly polynomial reads
\begin{equation}\label{I12twisted}
I_{12(i)}^{(twisted)}=-{1\over2}\left(n_iI_{grav}^{(1/2)}(R_i)+I_{mixed}^{(1/2)}(R_i,F_i)+I_{gauge}^{(1/2)}(F_i)\right)\
,
\end{equation}
where the minus sign comes from the fact that the gaugino fields
have negative chirality in the Euclidean. $n_i$ is the dimension
of the adjoint representation of the gauge group $G_i$. Adding all
the pieces gives
\begin{eqnarray}
I_{12(i)}^{(fields)}&=&I_{12(i)}^{(untwisted)}+I_{12(i)}^{(twisted)}\nonumber\\
&=&-{1\over{2(2\pi)^5 6!}}\left[{{496-2n_i}\over1008}{\rm
tr}R_i^6+{{-224-2n_i}\over768}{\rm tr}R_i^4
{\rm tr}R_i^2 +{{320-10n_i}\over{9216}}({\rm tr}R_i^2)^3\right. \nonumber\\
&&+\left. {1\over{16}} {\rm tr}R_i^4 {\rm Tr}F_i^2
+{5\over64}({\rm tr}R_i^2)^2 {\rm Tr}F_i^2-{5\over8}{\rm
tr}R_i^2{\rm Tr}F_i^4+{\rm Tr}F_i^6\right]\ ,
\end{eqnarray}
where now $\Tr$ denotes the adjoint trace. To derive this formula
we made use of the general form of the anomaly polynomial as given
in appendix \ref{anomalies}. The anomaly cancels only if several
conditions are met. First of all it is not possible to cancel the
${\rm tr}R^6$ term by a Green-Schwarz type mechanism. Therefore,
we get a restriction on the gauge group $G_i$, namely
\begin{equation}
n_i=248.
\end{equation}
Then we are left with
\begin{eqnarray}
I_{12(i)}^{(fields)} &=&-{1\over{2(2\pi)^5
6!}}\left[-{15\over16}{\rm tr}R_i^4 {\rm tr}R_i^2
-{15\over{64}}({\rm tr}R_i^2)^3+{1\over{16}} {\rm tr}R_i^4
{\rm Tr}F_i^2\right. \nonumber\\
&&+\left. {5\over64}({\rm tr}R_i^2)^2 {\rm Tr}F_i^2-{5\over8}{\rm
tr}R_i^2{\rm Tr}F_i^4+{\rm Tr}F_i^6\right]\ .
\end{eqnarray}
In order to cancel this remaining part of the anomaly we will
apply a sort of Green-Schwarz mechanism. This is possible if and
only if the anomaly polynomial factorizes into the product of a
four-form and an eight-form. For this factorization to occur we
need
\begin{equation}
{\rm Tr}F_i^6={1\over24}{\rm Tr}F_i^4{\rm
Tr}F_i^2-{1\over3600}({\rm Tr}F^2_i)^3\ .
\end{equation}
There is exactly one non-Abelian Lie group with this property,
which is the exceptional group $E_8$. Defining ${\rm
tr}:={1\over30}{\rm Tr}$ for $E_8$ and making use of the
identities
\begin{eqnarray}
{\rm Tr}F^2&=:&30\ {\rm tr}F^2\ ,\\
{\rm Tr}F^4&=&{1\over100}({\rm Tr}F^2)^2\ ,\\
{\rm Tr}F^6&=&{1\over7200}({\rm Tr}F^2)^3\ ,
\end{eqnarray}
which can be shown to hold for $E_8$, we can see that the anomaly
factorizes,
\begin{equation}
I_{12(i)}^{(fields)}=-{\pi\over3}(I_{4(i)})^3-I_{4(i)}\wedge
X_8\label{I12i}\ ,
\end{equation}
with
\begin{eqnarray}
I_{4(i)}:={1\over 16\pi^2}\left({\rm tr}F_i^2-{1\over2}{\rm
tr}R_i^2\right)\ ,
\end{eqnarray}
and $X_8$ as in (\ref{X8}). $X_8$ is related to forms $X_7$ and
$X_6$ via the usual descent mechanism $X_8=dX_7,\ \delta
X_7=dX_6$.

\begin{center}
{\bf The modified Bianchi identity}
\end{center}
So far we saw that M-theory on $S^1/\mathbb{Z}_2$ is anomalous and
we added new fields onto the fixed planes to cancel part of that
anomaly. But now the theory has changed. It no longer is pure
eleven-dimensional supergravity on a manifold with boundary, but
we have to couple this theory to ten-dimensional super-Yang-Mills
theory, with action
\begin{equation}
S_{SYM}=-{1\over4\lambda^2}\int d^{10}x \sqrt{g_{10}}\ \tr
F_{\mu\nu}F^{\mu\nu},
\end{equation}
where $\l$ is an unknown coupling constant. The explicit coupling
of these two theories was determined in \cite{HW95}. The crucial
result of this calculation is that the Bianchi identity, $\d G=0$,
needs to be modified. It reads\footnote{This differs by a factor 2
from \cite{HW95} which comes from the fact that our $\k_{11}$ is
the ``downstairs" $\k$. \cite{HW95} use its ``upstairs" version
and the relation between the two is
$2\k_{downstairs}\equiv2\k_{11}=\k_{upstairs}$. See \cite{BDSa99}
and \cite{BM03c} for a careful discussion.}
\begin{equation}
\d G=-{2\kappa_{11}^2\over\lambda^2}\sum_i\delta_i\wedge\left({\rm
tr}F^2_i-{1\over2}{\rm tr}R^2\right)
=-(4\pi)^2{2\kappa_{11}^2\over\lambda^2}\sum_i\delta_i\wedge
I_{4(i)}\ .
\end{equation}
Since $\delta_i$ has support only on the fixed planes and is a
one-form $\sim \d\phi$, only the values of the smooth four-form
${I_4}_{(i)}$ on this fixed plane are relevant and only the
components not including $\d\phi$ do not vanish. The gauge part
${\rm tr}F_i^2$ always satisfies these conditions but for the
${\rm tr}R^2$ term this is non-trivial. In the following a bar on
a form will indicate that all components containing $\d\phi$ are
dropped and the argument is set to $\phi=\phi_i$. Then the
modified Bianchi identity reads
\begin{equation}
\d G=\gamma\sum_i\delta_i\wedge \bar{I}_{4(i)}\ ,\label{BI}
\end{equation}
where we introduced
\begin{equation}
\gamma:=-(4\pi)^2{2\kappa_{11}^2\over\lambda^2}\ .\label{gamma}
\end{equation}
Define the Chern-Simons form
\begin{equation}
\bar{\omega}_i:= {1\over(4\pi)^2}\left({\rm tr}(A_i \d
A_i+{2\over3}A_i^3)-{1\over2}{\rm tr}(\bar{\Omega}_i
\d\bar{\Omega}_i +{2\over3}\bar{\Omega}_i^3 )\right)\ ,
\end{equation}
so that
\begin{equation}
\d\bar{\omega}_i=\bar{I}_{4(i)}\ .
\end{equation}
Under a gauge and local Lorentz transformation with parameters
$\Lambda^g$ and $\Lambda^L$ independent of $\phi$ one has
\begin{equation}
\delta\bar{\omega}_i=\d\bar{\omega}_i^1\ ,
\end{equation}
where
\begin{equation}
\bar{\omega}_i^1:={1\over(4\pi)^2}\left({\rm tr}\Lambda^g \d
A_i-{1\over2}{\rm tr}\Lambda^L\d\bar{\Omega_i}\right).
\end{equation}
Making use of (\ref{depsilon}) we find that the Bianchi identity
(\ref{BI}) is solved by
\begin{equation}
G=\d C-(1-b)\gamma\sum_i\delta_i\wedge\bar{\omega_i}
+b\gamma\sum_i{\epsilon_i\over2}\bar{I}_{4(i)}
-b\gamma\sum_i{\d\phi\over2\pi}\wedge\bar{\omega_i}\ ,\label{G}\
\end{equation}
where $b$ is an undetermined (real) parameter. As $G$ is a
physical field it is taken to be gauge invariant, $\delta G=0$.
Hence we get the transformation law of the $C$-field,
\begin{equation}
\delta C=\d
B_2^1-\gamma\sum_i\delta_i\wedge\bar{\omega}_i^1-b\gamma\sum_i
{\epsilon_i\over2}\d\bar{\omega}_i^1\ ,
\end{equation}
with some two-form $B_2^1$. Recalling that $C_{\mu\nu\rho}$ is
projected out, this equation can be solved, because
$C_{\mu\nu\rho}=0$ is only reasonable if we also have $\delta
C_{\mu\nu\rho}=0$. This gives
\begin{equation}
(\d B_2^1)_{\mu\nu\rho}={b\gamma\over2}\sum_i(\epsilon_i
\d\bar{\omega}_i^1)_{\mu\nu\rho}\ ,
\end{equation}
which is solved by
\begin{equation}
(B_2^1)_{\mu\nu}=\gamma{b\over2}\sum_i\epsilon_i(\bar{\omega}_i^1)_{\mu\nu}\
.
\end{equation}
So we choose
\begin{equation}
B_2^1=\gamma{b\over2}\sum_i\epsilon_i\bar{\omega}_i^1\ ,
\end{equation}
and get
\begin{equation}
\delta
C=\gamma\sum_i\left[(b-1)\delta_i\wedge\bar{\omega}_i^1-{b\over2\pi}\d\phi\wedge\bar{\omega}_i^1\right]\
.\label{deltaC}
\end{equation}

\begin{center}
{\bf Inflow terms and anomaly cancellation}
\end{center}
In the last sections we saw that introducing a vector
supermultiplet cancels part of the gravitational anomaly that is
present on the ten-dimensional fixed planes. Furthermore, the
modified Bianchi identity led to a very special transformation law
for the $C$-field. In this section we show that this modified
transformation law allows us to cancel the remaining anomaly,
leading to an anomaly free theory. We start from supergravity on
$M_{10}\times I\cong M_{10}\times S^1/\mathbb{Z}_2$ and rewrite it
in the upstairs formalism,
\begin{equation}
S_{top}=-{1\over12\k^2_{11}}\int_{M_{10}\times I} C\w \d C\w \d
C=-{1\over24\k^2_{11}}\int_{M_{10}\times S^1} C\w \d C\w \d C\ .
\end{equation}
However, we no longer have $G=\d C$ and thus it is no longer clear
whether the correct topological term is $C\d C\d C$ or rather
$CGG$. It turns out that the correct term is the one which
maintains the structure $\Ct \d\Ct \d\Ct$ everywhere except on the
fixed planes. However, the field $C$ has to be modified to a field
$\Ct$, similarly to what we did in chapter \ref{G2anom}. To be
concrete let us study the structure of $G$ in more detail. It is
given by
\begin{equation}\label{GdtC}
G=\d\left(C+{b\over2}\g\sum_i\e_i\bar\o_i\right)-\g\sum_i\delta_i\w\bar\o_i=:\d\Ct-\g\sum_i\delta_i\w\bar\o_i\
.
\end{equation}
That is we have $G=\d\Ct$ except on the fixed planes where we get
an additional contribution. Thus, in order to maintain the
structure of the topological term almost everywhere we postulate
\cite{BM03c} it to read
\begin{eqnarray}
\St_{top}&=&-{1\over24\k^2_{11}}\int_{M_{10}\times S^1} \Ct\w G\w
G\nonumber\\
&=&-{1\over24\k^2_{11}}\int_{M_{10}\times S^1}\left( \Ct\w \d\Ct\w
\d\Ct-2\Ct\w \d\Ct\w\g\sum_i\delta_i\w\bar\o_i\right)\ .
\end{eqnarray}
To see that this is reasonable let us calculate its variation
under gauge transformations. From (\ref{deltaC}) we have
\begin{equation}
\delta \Ct=\d\left({\g
b\over2}\sum_i\e_i\bar\o_i^1\right)-\gamma\sum_i\delta_i\w\bar\omega_i^1\
,
\end{equation}
and we find
\begin{equation}
\delta \St_{top} ={\g^3b^2\over96\k_{11}^2}\int_{M_{10}\times
S^1}\sum_{i,j,k}(\delta_i\e_j\e_k+2\e_i\e_j\delta_k)\bar\o_i^1\w\bar
I_{4(j)}\w\bar
I_{4(k)}={\g^3b^2\over96\k^2_{11}}\sum_i\int_{}\bar\o_i^1\w \bar
I_{4(i)}\w \bar I_{4(i)}\ ,
\end{equation}
where we used (\ref{reg}). This $\delta \tilde S_{top}$ is a sum
of two terms, and each of them is localised on one of the fixed
planes. The corresponding (Minkowskian) anomaly polynomial reads
\begin{equation}
I_{12}^{(top)}=\sum_i{\g^3b^2\over96\k^2_{11}}(\bar
I_{4(i)})^3=:\sum_iI_{12(i)}^{(top)}\ .\label{I12top}
\end{equation}
If we choose $\g$ to be
\begin{equation}\label{g1}
\g=-\left(32\pi\k^2_{11}\over b^2\right)^{1/3}
\end{equation}
and use (\ref{master}) we see that this cancels the first part of
the anomaly (\ref{I12i}) through inflow. Note that this amounts to
specifying a certain choice for
the coupling constant $\l$.\\

This does not yet cancel the anomaly entirely. However, we have
seen, that there is yet another term, which can be considered as a
first M-theory correction to eleven-dimensional supergravity,
namely the Green-Schwarz term\footnote{The reader might wonder why
we use the form $\int G\w X_7$ for the Green-Schwarz term, since
we used $\int\tilde C\w X_8$ in chapter \ref{G2anom}, c.f. Eq.
(\ref{GSmod}). However, first of all we noted already in chapter
\ref{G2anom} that $\int\tilde G\w X_7$ would have led to the same
results. Furthermore, on $M_{10}\times S^1$ we have from
(\ref{GdtC}) that $\int G\w X_7=\int\tilde C\w
X_8-\g\sum_i\int\bar\o_i\w X_7$. But the latter term is a local
counterterm that does not contribute to $I_{12}$.}
\begin{equation}
S_{GS}:=-{1\over(4\pi\k^2_{11})^{1/3}}\int_{M_{10}\times I}
G\wedge X_7=-{1\over2(4\pi\k^2_{11})^{1/3}}\int_{M_{10}\times S^1}
G\wedge X_7\ ,\label{Green-Schwarz term}
\end{equation}
studied in \cite{VW95}, \cite{DLM95}, \cite{BM03c}. $X_8$ is given
in (\ref{X8}) and it satisfies the descent equations $X_8=\d X_7$
and $\delta X_7=\d X_6$. Its variation gives the final
contribution to our anomaly,
\begin{equation}
\delta S_{GS}=-{1\over2(4\pi\k^2_{11})^{1/3}}\int_{M_{10}\times
S^1} G\wedge \d X_6^1
=\sum_i{\g\over2(4\pi\k^2_{11})^{1/3}}\int_{M_{10}} \bar
I_{4(i)}\wedge \bar X_6\ ,
\end{equation}
where we integrated by parts and used the properties of the
$\delta_i$. The corresponding (Minkowskian) anomaly polynomial is
\begin{equation}
I_{12}^{(GS)}=\sum_i {\g\over2(4\pi\k^2_{11})^{1/3}}\ \bar
I_{4(i)}\wedge \bar X_8=:\sum_iI_{12(i)}^{(GS)}\ .\label{I12GS}
\end{equation}
which cancels the second part of our anomaly provided
\begin{equation}\label{g2}
\g=-(32\pi\k^2_{11})^{1/3}\ .
\end{equation}
Happily, the sign is consistent with our first condition
(\ref{g1}) for anomaly cancellation and it selects $b=1$. This
value for $b$ was suggested in \cite{BDSa99} from general
considerations unrelated to anomaly cancellation. Choosing $\g$
(and thus the corresponding value for $\l$) as in (\ref{g2}) leads
to a local cancellation of the anomalies. Indeed let us collect
all the contributions to the anomaly of a single fixed plane,
namely (\ref{I12i}), (\ref{I12top}) and (\ref{I12GS})
\begin{equation}
iI_{12(i)}^{(untwisted)}+iI_{12(i)}^{(twisted)}-iI_{12(i)}^{(top)}-iI_{12(i)}^{(GS)}=0\
.
\end{equation}
The prefactors of $-i$ in the last two terms come from the fact
that we calculated the variation of the Minkowskian action, which
has to be translated to Euclidean space (c.f. Eq. \ref{master}).
Note that the anomalies cancel separately on each of the two
ten-dimensional planes. In other words, we once again find {\it
local} anomaly cancellation.

\chapter{Conclusions}
We have seen that anomalies are a powerful tool to explore some of
the phenomena of M-theory. The requirement of a cancellation of
local gauge and gravitational anomalies imposes strong constraints
on the theory, and allows us to understand its structure in more
detail. In the context of higher dimensional field theories
anomalies can cancel in two different ways. For instance, in the
case of M-theory on the product of Minkowski space with a compact
manifold, anomalies can be localised at various points in the
internal space. The requirement of global anomaly cancellation
then simply means that the sum of all these anomalies has to
vanish. The much stronger concept of local anomaly cancellation,
on the other hand, requires the anomaly to be cancelled on the
very space where it is generated. We have seen that for M-theory
on singular $G_2$-manifolds and on $M_{10}\times S^1/\mathbb{Z}_2$
anomalies do indeed cancel locally via a mechanism known as
anomaly inflow. The main idea is that the classical action is not
invariant under local gauge or Lorentz transformations, because of
``defects" in the space on which it is formulated. Such a defect
might be a (conical or orbifold) singularity or a boundary.
Furthermore, we saw that the classical action had to be modified
close to these defects. Only then does the variation of the action
give the correct contributions to cancel the anomaly. These
modifications of the action in the cases studied above are
modelled after the similar methods that had been used in
\cite{FHMM98} to cancel the normal bundle anomaly of the M5-brane.
Of course these modifications are rather ad hoc. Although the same
method seems to work in many different cases the underlying
physics has not yet been understood. One might for example ask how
the smooth function $\r$ (c.f Eq. (\ref{rho})) is generated in the
context of $G_2$-compactifications. Some progress in this
direction has been made in \cite{HR00}, \cite{BHR02}. It would
certainly be quite interesting to further explore these issues.

There are no explicit examples of metrics of compact
$G_2$-manifolds with conical singularities and therefore the
discussion above might seem quite academic. However, we were able
to write down relatively simple metrics for a compact manifold
with two conical singularities and weak $G_2$-holonomy. Although
the corresponding effective theory lives on $AdS_4$ one expects
that the entire mechanism of anomaly cancellation should be
applicable to this case as well, see e.g. \cite{ADHL03} for a
discussion. Therefore, our explicit weak $G_2$-metric might serve
as a useful toy model for the full $G_2$-case.

\begin{appendix}
\part{Appendices}

\chapter{Notation}\label{notation}
Our notation is as in \cite{Me03}. However, for the reader's
convenience we list the relevant details once again.

\section{General notation}
The metric on flat space is given by
\begin{equation}
\eta :={\rm diag}(-1,1,\ldots1)\ .
\end{equation}
The anti-symmetric tensor is defined as
\begin{eqnarray}
\widetilde{\epsilon}_{012\ldots d-1}&:=&{\widetilde{\epsilon}}^{\ 012\ldots d-1}:=+1\ ,\\
\epsilon_{M_1\ldots M_{d}}&:=&\sqrt{g}\
\widetilde{\epsilon}_{M_1\ldots M_{d}}\ .
\end{eqnarray}
That is, we define $\widetilde\epsilon$ to be the tensor density
and $\epsilon$ to be the tensor. We obtain
\begin{eqnarray}
\epsilon_{012\ldots d-1}&=&\sqrt{g}=e:=|{\rm det} \ e^A_{\ \ M}|\ , \\
\epsilon^{M_1\ldots M_{d}}&=&{\rm sig}(g){1\over {\sqrt{g}}}\
\widetilde{\epsilon}^{\ M_1\ldots M_{d}}\ ,\mbox{\
\ and}\\
\widetilde{\epsilon}^{\ M_1\ldots M_rP_1\ldots
P_{d-r}}\widetilde{\epsilon}_{N_1\ldots N_rP_1\ldots
P_{d-r}}&=&r!(d-r)!\delta^{[M_1\ldots M_r]}_{N_1\ldots N_r}\ .
\end{eqnarray}
(Anti-)Symmetrisation is defined as,
\begin{eqnarray}
A_{(M_1\ldots M_l)}&:=&{1 \over l!}\sum_{\pi}A_{M_{\pi(1)}\ldots M_{\pi(l)}}\ ,\\
A_{[M_1\ldots M_l]}&:=&{1 \over l!}\sum_{\pi} {\rm
sig}(\pi)A_{M_{\pi(1)}\ldots M_{\pi(l)}}\ .
\end{eqnarray}
p-forms come with a factor of $p!$, e.g.
\begin{equation}
\omega:={1\over p!}\omega_{M_1\ldots M_p}dz^{M_1}\wedge\ldots
\wedge dz^{M_p}\ .
\end{equation}
The Hodge dual is defined as
\begin{equation}
*\omega = {1\over p! (d-p)!}\, \omega_{M_1\ldots M_p}\
\epsilon^{M_1\ldots M_p}_{\phantom{M_1\ldots M_p} M_{p+1}\ldots
M_d} \ d z^{M_{p+1}}\wedge\ldots\wedge d z^{M_d}\ .
\end{equation}

\section{Spinors}\label{spinors}

\subsection{Clifford algebras and their representation}
\begin{definition}
{\em A} Clifford algebra {\em in $d$ dimensions is defined as a
set containing $d$ elements $\Gamma^{A}$ which satisfy the
relation} {\em
\begin{equation}\label{Clifford algebra}
\lbrace\Gamma^A,\Gamma^B\rbrace=2\eta^{AB} {\rm \opone} \ .
\end{equation}}
\end{definition}

Under multiplication this set generates a finite group, denoted
$C_d$, with elements
\begin{equation}
C_d=\lbrace\pm1,\pm\Gamma^A,
\pm\Gamma^{A_1A_2},\ldots,\pm\Gamma^{A_1\ldots A_d}\rbrace\ ,
\end{equation}
where $\Gamma^{A_1...A_l}:=\Gamma^{[A_1}\ldots \Gamma^{A_l]}$. The
order of this group is
\begin{equation}
{\rm ord}(C_d)=2\sum_{p=0}^d {d \choose p}=2\cdot2^d=2^{d+1}\ .
\end{equation}
\begin{definition}
{\em Let $G$ be a group. Then the} conjugacy class {\em $[a]$ of
$a\in G$ is defined as}
\begin{equation}
[a]:=\lbrace gag^{-1}|g\in G\rbrace.
\end{equation}
\end{definition}
\begin{proposition}
Let $G$ be a finite dimensional group. Then the number of its
irreducible representations equals the number of its conjugacy
classes.
\end{proposition}
\begin{definition}
{\em Let $G$ be a finite group. Then the} commutator group {\em
${\rm Com}(G)$ of $G$ is defined as}
\begin{equation}
{\rm Com}(G):=\lbrace ghg^{-1}h^{-1}|g,h\in G\rbrace\ .
\end{equation}
\end{definition}
\begin{proposition}
Let G be a finite group. Then the number of inequivalent
one-dimensional representations is equal to the order of $G$
divided by the order of the commutator group of $G$.
\end{proposition}
\begin{proposition}
Let $G$ be a finite group with inequivalent irreducible
representations of dimension $n_p$, where $p$ labels the
representation. Then we have
\begin{equation}
{\rm ord}(G)=\sum_p(n_p)^2\ .\label{order}
\end{equation}
\end{proposition}
\begin{proposition}
Every class of equivalent representations of a finite group $G$
contains a unitary representation.
\end{proposition}
For the unitary choice we get
$\Gamma^A{\Gamma^A}^{\dagger}=\opone$. From (\ref{Clifford
algebra}) we infer (in Minkowski space)
${\Gamma^0}^{\dagger}=-\Gamma^0$ and
$(\Gamma^A)^{\dagger}=\Gamma^A$ for $A\neq0$. This can be
rewritten as
\begin{equation}
{\Gamma^A}^{\dagger}=\Gamma^0\Gamma^A\Gamma^0\
.\label{Gammadagger}
\end{equation}

\subsubsection{Clifford algebras in even dimensions}
\begin{theorem}
For $d=2k+2$ even the group $C_d$ has $2^d+1$ inequivalent
representations. Of these $2^d$ are one-dimensional and the
remaining representation has (complex) dimension
$2^{d\over2}=2^{k+1}$.
\end{theorem}
This can be proved by noting that for even $d$ the conjugacy
classes of $C_d$ are given by
\begin{equation*}
\left\lbrace
[+1],[-1],[\Gamma^A],[\Gamma^{A_1A_2}],\ldots,[\Gamma^{A_1\ldots
A_d}]\right\rbrace\ ,
\end{equation*}
hence the number of inequivalent irreducible representations of
$C_d$ is $2^d+1$. The commutator of $C_d$ is ${\rm
Com}(C_d)=\lbrace \pm1\rbrace$ and we conclude that the number of
inequivalent one-dimensional representations of $C_d$ is $2^d$.
From (\ref{order}) we read off that the dimension of the remaining
representation has to be $2^{d\over2}$.

Having found irreducible representations of $C_d$ we turn to the
question whether we also found representations of the Clifford
algebra. In fact, for elements of the Clifford algebra we do not
only need the group multiplication, but the addition of two
elements must be well-defined as well, in order to make sense of
(\ref{Clifford algebra}). It turns out that the one-dimensional
representation of $C_d$ do not extend to representations of the
Clifford algebra, as they do not obey the rules for addition and
subtraction. Hence, we found that for $d$ even there is a unique
class of irreducible representations of the Clifford algebra of
dimension
$2^{d\over2}=2^{k+1}$.\\

Given an irreducible representation $\lbrace \Gamma^A\rbrace$ of a
Clifford algebra, it is clear that $\pm\lbrace
{\Gamma^{A}}^*\rbrace$ and $\pm\lbrace {\Gamma^{A}}^{\tau}\rbrace$
form irreducible representations as well. As there is a unique
class of representations in even dimensions, these have to be
related by
similarity transformations,\\
\parbox{14cm}{
\begin{eqnarray}
{\Gamma^{A}}^*&=&\pm(B_{\pm})^{-1}\Gamma^A B_{\pm}\ ,\nonumber\\
{\Gamma^A}^{\tau}&=&\pm (C_{\pm})^{-1}\Gamma^A C_{\pm}\ .\nonumber
\end{eqnarray}}\hfill\parbox{8mm}{\begin{eqnarray}\label{defBC}\end{eqnarray}}
The matrices $C_{\pm}$ are known as {\em charge conjugation
matrices}. Iterating this definition gives conditions for
$B_{\pm}, C_{\pm}$,
\begin{eqnarray}
(B_{\pm})^{-1}&=&b_{\pm}{B_{\pm}}^*\ ,\\
C_{\pm}&=&c_{\pm}{C_{\pm}}^{\tau}\ ,
\end{eqnarray}
with $b_{\pm}$ real, $c_{\pm}\in\lbrace\pm1\rbrace$ and $C_{\pm}$
symmetric or anti-symmetric.

\subsubsection{Clifford algebras in odd dimensions}
\begin{theorem}
For $d=2k+3$ odd the group $C_d$ has $2^d+2$ inequivalent
representations. Of these $2^d$ are one-dimensional\footnote{As
above these will not be considered any longer as they are
representations of $C_d$ but not of the Clifford algebra.} and the
remaining two representation have (complex) dimension
$2^{d-1\over2}=2^{k+1}$.
\end{theorem}
As above we note that for odd $d$ the conjugacy classes of $C_d$
are given by
\begin{equation*}
\left\lbrace[+1],[-1],[\Gamma^A],[\Gamma^{A_1A_2}],\ldots,[\Gamma^{A_1\ldots
A_d}],[-\Gamma^{A_1\ldots A_d}]\right\rbrace
\end{equation*}
and the number of inequivalent irreducible representations of
$C_d$ is $2^d+2$. Again we find the commutator ${\rm
Com}(C_d)=\lbrace \pm1\rbrace$, hence, the number of inequivalent
one-dimensional representations of $C_d$ is $2^d$. Now define the
matrix
\begin{equation}
\Gamma^d:=\Gamma^0\Gamma^1\ldots\Gamma^{d-1}\ ,
\end{equation}
which commutes with all elements of $C_d$. By Schur's lemma this
must be a multiple of the identity, $\Gamma^d=a^{-1}\opone$, with
some constant $a$. Multiplying by $\Gamma^{d-1}$ we find
\begin{equation}
\Gamma^{d-1}=a\Gamma^0\Gamma^1\ldots\Gamma^{d-2}\
.\label{Gammaodd}
\end{equation}
Furthermore,
$(\Gamma^{0}\Gamma^1\ldots\Gamma^{d-2})^2=-(-1)^{k+1}$. As we know
from (\ref{Clifford algebra}) that $(\Gamma^{d-1})^2=+1$ we
conclude that $a=\pm1$ for $d=3$ (mod 4) and $a=\pm i$ for $d=5$
(mod 4). The matrices
$\lbrace\Gamma^0,\Gamma^1,\ldots,\Gamma^{d-2}\rbrace$ generate an
even-dimensional Clifford algebra the dimension of which has been
determined to be $2^{k+1}$. Therefore, the two inequivalent
irreducible representations of $C_d$ for odd $d$ must coincide
with this irreducible representation when restricted to $C_{d-1}$.
We conclude that the two irreducible representations for $C_d$ and
odd $d$ are generated by the unique irreducible representation for
$\lbrace\Gamma^0,\Gamma^1,\Gamma^{d-2}\rbrace$, together with the
matrix $\Gamma^{d-1}=a\Gamma^0\Gamma^1\ldots\Gamma^{d-2}$. The two
possible choices of $a$ correspond to the two inequivalent
representations. The dimension of these representation is
$2^{k+1}$.

\subsection{Dirac, Weyl and Majorana spinors}
\subsubsection{Dirac spinors}
Let $(M,g)$ be an oriented pseudo-Riemannian manifold of dimension
$d$, which is identified with $d$-dimensional space-time, and let
$\lbrace\Gamma^A\rbrace$ be a $d$-dimensional Clifford algebra.
The metric and orientation induce a unique
$SO(d-1,1)$-structure\footnote{Let $M$ be a manifold of dimension
$d$, and $F$ the frame bundle over $M$. Then $F$ is a principle
bundle over $M$ with structure group $GL(d,\mathbb{R})$. A {\it
$G$-structure} on $M$ is a principle subbundle $P$ of $F$ with
fibre $G$.} $P$ on $M$. A {\em spin structure}
$(\widetilde{P},\pi)$ on $M$ is a principal bundle $\widetilde{P}$
over $M$ with fibre ${\rm Spin}(d-1,1)$, together with a map of
bundles $\pi:\widetilde{P}\rightarrow P$. ${\rm Spin}(d-1,1)$ is
the universal covering group of $SO(d-1,1)$. Spin structures do
not exist on every manifold. An oriented pseudo-Riemannian
manifold $M$ admits a spin structure if and only if $w_2(M)=0$,
where $w_2(M)\in H^2(M,\mathbb{Z})$ is the {\em second
Stiefel-Whitney class} of $M$. In that case we call $M$ a {\em
spin manifold}.

Define the anti-Hermitian generators
\begin{equation}
\Sigma^{AB}:={1\over2}\Gamma^{AB}={1\over4}[\Gamma^A,\Gamma^B]\ .
\end{equation}
Then the $\Sigma^{AB}$ form a representation of
\verb"so"$(d-1,1)$, the Lie algebra of $SO(d-1,1)$,
\begin{equation}\label{SigmaAB}
[\Sigma^{AB},\Sigma^{CD}]=-\Sigma^{AC}g^{BD}+\Sigma^{AD}g^{BC}+\Sigma^{BC}g^{AD}-\Sigma^{BD}g^{AC}\
.
\end{equation}
In fact, $\Sigma^{AB}$ are generators of ${\rm Spin}(d-1,1)$. Take
$\Delta^d$ to be the natural representation of ${\rm
Spin}(d-1,1)$. We define the {\em(complex) spin bundle}
$S\rightarrow M$ to be\footnote{$\widetilde{P}\times_{{\rm
Spin}(d-1,1)}\Delta^d$ is the fibre bundle which is associated to
the principal bundle $\widetilde{P}$ in a natural way. Details of
this construction can be found in any textbook on differential
geometry, see for example \cite{Na90}.}
$S:=\widetilde{P}\times_{{\rm Spin}(d-1,1)}\Delta^d$. Then $S$ is
a complex vector bundle over $M$, with fibre $\Delta^d$ of
dimension $2^{[d/2]}$. A {\em Dirac spinor} $\psi$ is defined as a
section of the spin bundle $S$. Under a local Lorentz
transformation with infinitesimal parameter $\a_{AB}=-\a_{BA}$ a
Dirac spinor transforms as
\begin{equation}
\psi'=\psi+\delta\psi=\psi-{1\over2}\a_{AB}\Sigma^{AB}\psi\
.\label{spinortrafo}
\end{equation}
The {\em Dirac conjugate} $\bar\psi$ of the spinor $\psi$ is
defined as
\begin{equation}
\bar\psi:=i\psi^{\dagger}\Gamma^0\ .
\end{equation}
With this definition we have $\delta(\bar\psi\eta)=0$ and
$\bar\psi\psi$ is Hermitian,
$(\bar\psi\psi)^{\dagger}=\bar\psi\psi$.

\subsubsection{Weyl spinors}
In $d=2k+2$-dimensional space-time we can construct the
matrix\footnote{This definition of the $\Gamma$-matrix in
Minkowski space agrees with the one of \cite{Pol}. Sometimes it is
useful to define a Minkowskian $\Gamma$-matrix as
$\Gamma=i^{k}\Gamma^0\ldots\Gamma^{d-1}$ as in \cite{BM03c}.
Obviously the two conventions agree in 2, 6 and 10 dimensions and
differ by a sign in dimensions 4 and 8.}
\begin{equation}
\Gamma_{d+1}:=(-i)^k\Gamma^0\Gamma^1\ldots\Gamma^{d-1}\
,\label{Gamma}
\end{equation}
which satisfies
\begin{eqnarray}
(\Gamma_{d+1})^2&=&\opone\ ,\\
\lbrace\Gamma_{d+1},\Gamma^A\rbrace&=&0\ ,\\
\ [\Gamma_{d+1},\Sigma^{AB}]&=&0\ .
\end{eqnarray}
Then, we can define the chirality projectors
\begin{equation}
P_L\equiv P_-:={1\over2}(\opone-\Gamma_{d+1})\ \ \ ,\ \ \
P_R\equiv P_+:={1\over2}(\opone+\Gamma_{d+1})\ ,
\end{equation}
satisfying\\
\parbox{14cm}{
\begin{eqnarray}
P_L+P_R&=&\opone\ ,\nonumber\\
P_{L,R}^2&=&P_{L,R}\ ,\nonumber\\
P_LP_R&=&P_RP_L=0\ ,\nonumber\\
\ [P_{L,R},\Sigma^{AB}]&=&0\ .\nonumber
\end{eqnarray}}\hfill\parbox{8mm}{\begin{eqnarray}\end{eqnarray}}
A {\em Weyl spinor} in even-dimensional spaces is defined as a
spinor satisfying the Weyl condition,
\begin{equation}
P_{L,R}\psi=\psi\ .
\end{equation}
Note that this condition is Lorentz invariant, as the projection
operators commute with $\Sigma^{AB}$. Spinors satisfying
$P_L\psi_L=\psi_L$ are called {\em left-handed} Weyl spinors and
those satisfying $P_R\psi_R=\psi_R$ are called {\em right-handed}.
The Weyl condition reduces the number of complex components of a
spinor to $2^{k}$.

Obviously, under the projections $P_{L,R}$ the space $\Delta^d$
splits into a direct sum $\Delta^d=\Delta^d_+\oplus\Delta^d_-$ and
the spin bundle is given by the Whitney sum $S=S_+\oplus S_-$.
Left- and right-handed Weyl spinors are sections of $S_-$ and
$S_+$, respectively.

\subsubsection{Majorana spinors}
In (\ref{defBC}) we defined the matrices $B_{\pm}$ and $C_{\pm}$.
We now want to explore these matrices in more detail. For $d=2k+2$
we define {\em Majorana spinors} as those spinors that satisfy
\begin{equation}
\psi=B_+\psi^*\ ,\label{Majorana}
\end{equation}
and pseudo-Majorana spinors as those satisfying
\begin{equation}
\psi=B_-\psi^*\ .\label{pseudoMajorana}
\end{equation}
As in the case of the Weyl conditions, these conditions reduce the
number of components of a spinor by one half. The definitions
imply\\
\parbox{14cm}{
\begin{eqnarray}
B_+^*B_+&=&\opone\ ,\nonumber\\
B_-^*B_-&=&\opone\ ,\nonumber
\end{eqnarray}}\hfill\parbox{8mm}{\begin{eqnarray}\label{Majimpl}\end{eqnarray}}
which in turn would give $b_+=1$ and $b_-=1$. These are
non-trivial conditions since $B_\pm$ is fixed by its definition
(\ref{defBC}). The existence of (pseudo-) Majorana spinors relies
on the possibility to construct matrices $B_+$ or $B_-$ which
satisfy (\ref{Majimpl}). It turns out that Majorana conditions can
be imposed in 2 and 4 (mod 8) dimensions. Pseudo-Majorana
conditions are possible in 2 and 8 (mod 8) dimensions.

Finally, we state that in odd dimensions Majorana spinors can be
defined in dimensions 3 (mod 8) and pseudo-Majorana in dimension 1
(mod 8)

\subsubsection{Majorana-Weyl spinors}
For $d=2k+2$ dimensions one might try to impose both the Majorana
(or pseudo-Majorana) and the Weyl condition. This certainly leads
to spinors with $2^{k-1}$ components. From (\ref{Gamma}) we get
$(\Gamma_{d+1})^*=(-1)^k B_{\pm}^{-1}\Gamma_{d+1} B_{\pm}$ and
therefore for $d=2$ (mod 4)
\begin{equation}
P_{L,R}^*=B_{\pm}^{-1}P_{L,R}B_{\pm}\ ,
\end{equation}
and for $d=4$ (mod 4)
\begin{equation}
P_{L,R}^*=B_{\pm}^{-1}P_{R,L}B_{\pm}\ .
\end{equation}
But this implies that imposing both the Majorana and the Weyl
condition is consistent only in dimensions $d=2$ (mod 4), as we
get
\begin{equation*}
B_{\pm}(P_{L,R}\psi)^*=P_{L,R}B_{\pm}\psi^*\ ,
\end{equation*}
for $d=2$ (mod 4), but
\begin{equation}
B_{\pm}(P_{L,R}\psi)^*=P_{R,L}B_{\pm}\psi^*
\end{equation}
for $d=4$ (mod 4). We see that in the latter case the operator
$B_{\pm}$ is a map between states of different chirality, which is
inconsistent with the Weyl condition. As the Majorana condition
can be imposed only in dimensions 2, 4 and 8 (mod 8) we conclude
that (pseudo-) Majorana-Weyl spinors can only exist in dimensions
2 (mod 8).

\bigskip
We summarize the results on spinors in various dimensions in the
following table.

\bigskip
\begin{center}
\begin{tabular}{|c|c|c|c|c|c|}\hline
d&Dirac& Weyl& Majorana&Pseudo-Majorana&Majorana-Weyl\\\hline
2&4&2&2&2&1\\
3&4& &2& & \\
4&8&4&4&&\\
5&8&&&&\\
6&16&8&&&\\
7&16&&&&\\
8&32&16&&16&\\
9&32&&&16&\\
10&64&32&32&32&16\\
11&64&&32&&\\
12&128&64&64&&\\\hline
\end{tabular}

\bigskip
The numbers indicate the real dimension of a spinor, whenever it
exists.\\
\end{center}

\section{Gauge theory}\label{gauge}

Gauge theories are formulated on principal bundles $P\rightarrow
M$ on a base space $M$ with fibre $G$ known as the {\em gauge
group}. Any group element $g$ of the connected component of $G$
that contains the unit element can be written as $g:=e^{\Lambda}$,
with $\Lambda:=\Lambda_a T_a$ and $T_a$ basis vectors of the Lie
algebra $\verb"g":=Lie(G)$. We always take $T_a$ to be
anti-Hermitian, s.t. $T_a=:-i t_a$ with $t_a$ Hermitian. The
elements of a Lie algebra satisfy commutation relations
\begin{equation}
[T_a,T_b]=C^c_{\ ab}T_c\ \ \ ,\ \ \ [t_a,t_b]= iC^c_{\ ab} t_c\ ,
\end{equation}
with the real valued {\em structure coefficients} $C^c_{\ ab}$.

Of course, the group $G$ can come in various representations. The
{\em adjoint representation}
\begin{equation}
(T^{Ad}_{\ a})^b_{\ c}:=-C^b_{\ ca}
\end{equation}
is particularly important. Suppose a connection is given on the
principal bundle. This induces a local Lie algebra valued
connection form $A=A_aT_a$ and the corresponding local form of the
curvature, $F=F_aT_a$. These forms are related by\footnote{The
commutator of Lie algebra valued forms $A$ and $B$ is understood
to be $[A,B]:=[A_M,B_N]\ dz^M\wedge dz^N$.}
\begin{eqnarray}
F&:=&dA+{1\over2}[A, A]\ ,\nonumber\\
F_{\mu\nu}&=&\partial_{\mu}
A_{\nu}-\partial_{\nu} A_{\mu}+[A_{\mu},A_{\nu}]\ ,\\
F_{\mu\nu}^a&=&\partial_{\mu} A^a_{\nu}-\partial_{\nu}
A^a_{\mu}+C^a_{\ bc}A^b_{\mu}A^c_{\nu}\ .\nonumber
\end{eqnarray}
In going from one chart to another they transform as
\begin{equation}
A^g:=g^{-1}(A+d)g\ \ \ ,\ \ \ F^g:=dA^g+{1\over2}[A^g,
A^g]=g^{-1}Fg\ .
\end{equation}
For $g=e^{\epsilon}=e^{\epsilon_aT_a}$ with $\epsilon$
infinitesimal we get $A^g=A+D\epsilon$.

For any object on the manifold which transforms under some
representation $\widetilde{T}$ (with $\widetilde{T}$
anti-Hermitian) of the gauge group $G$ we define a gauge covariant
derivative
\begin{equation}\label{covder}
D:=d+A\ \ \ ,\ \ \ A:=A_a\widetilde{T}_a\ .
\end{equation}
When acting on Lie algebra valued fields the covariant derivative
is understood to be $D:=d+[A,\ ]$.

We have the general operator formula
\begin{equation}
DD\phi=F\phi\ ,
\end{equation}
which reads in components
\begin{equation}
[D_M,D_N]\phi=F_{MN}\phi\ .\label{curvPB}
\end{equation}
Finally, we note that the curvature satisfies the Bianchi
identity,
\begin{eqnarray}
DF=0\ .
\end{eqnarray}

\section{Curvature}
Usually general relativity on a manifold $M$ of dimension $d$ is
formulated in a way which makes the invariance under
diffeomorphisms, $Diff(M)$, manifest. Its basic objects are
tensors which transform covariantly under $GL(d,\mathbb{R})$.
However, since $GL(d,\mathbb{R})$ does not admit a spinor
representation, the theory has to be reformulated if we want to
couple spinors to a gravitational field. This is done by choosing
an orthonormal basis in the tangent space $TM$, which is different
from the one induced by the coordinate system. From that procedure
we get an additional local Lorentz invariance of the theory. As
$SO(d-1,1)$ does have a spinor representation we can couple
spinors to this reformulated theory.

At a point $x$ on a pseudo-Riemannian manifold $(M,g)$ we define
the vielbeins $e_A(x)$ as
\begin{equation}
e_A(x):=e_A^{\ \ M}(x)\partial_M\ ,
\end{equation}
with coefficients $e_A^{\ \ M}(x)$ such that the $\lbrace
e_A\rbrace$ are orthogonal,
\begin{equation}
g(e_A,e_B)=e_A^{\ \ M}e_B^{\ \ N}g_{MN}=\eta_{AB}\ .
\end{equation}
Define the inverse coefficients via $e_A^{\ \ M}e_{\ \
M}^B=\delta_A^B$ and $e^A_{\ \ M}e_A^{\ \ N}=\delta_M^N$, which
gives $g_{MN}(x)=\eta_{AB}e^A_{\ \ M}(x)e^B_{\ \ N}(x)$. The dual
basis $\lbrace e^A \rbrace$ is defined as, $e^A:=e^A_{\ \ M}dz^M$.
The commutator of two vielbeins defines the {\em anholonomy
coefficients} $\Omega_{AB}^{\ \ \ \ C}$,
\begin{equation}
[e_A,e_B]:=[e^{\ \ M}_A\partial_M,e^{\ \
N}_B\partial_N]=\Omega_{AB}^{\ \ \ \ C}e_C\ ,
\end{equation}
and from the definition of $e_A$ one can read off
\begin{equation}
\Omega_{AB}^{\ \ \ \ C}(x)=e^C_{\ \ N}[e_A^{\ \ K}(\partial_K
e_B^{\ \ N})-e_B^{\ \ K}(\partial_K e_A^{\ \ N})](x)\ .
\end{equation}

When acting on tensors expressed in the orthogonal basis, the
covariant derivative has to be rewritten using the {\em spin
connection coefficients} $\omega_{M\ \ B}^{\ \ A}$,
\begin{equation}
\nabla_M^S V^{AB\ldots}_{CD\ldots}:=\partial_M
V^{AB\ldots}_{CD\ldots}+\omega_{M\ \ E}^{\ \ A}
V^{EB\ldots}_{CD\ldots}+\ldots-\omega_{M\ \ C}^{\ \ E}
V^{AB\ldots}_{ED\ldots}\ .\label{tensortrafo}
\end{equation}
The object $\nabla^S$ is called the {\em spin
connection}\footnote{Physicists usually use the term "spin
connection" for the connection coefficients.}. Its action can be
extended to objects transforming under an arbitrary representation
of the Lorentz group. Take a field $\phi$ which transforms as
\begin{equation}
\delta\phi^i=-{1\over2}\epsilon_{AB}(T^{AB})^i_{\ j}\phi^j
\end{equation}
under the infinitesimal Lorentz transformation $\Lambda^A_{\ \
B}(x)=\delta^A_B+\epsilon^A_{\ \
B}=\delta^A_B+{1\over2}\epsilon_{CD}(T^{CD}_{vec})^A_{\ \ B}$,
with the vector representation $(T^{CD}_{vec})^A_{\ \
B}=(\eta^{CA}\delta^D_B-\eta^{DA}\delta^C_B)$. Then its covariant
derivative is defined as
\begin{equation}
\nabla^S_M\phi^i:=\partial_M\phi^i+{1\over2}\omega_{MAB}(T^{AB})^i_{\
j}\phi^j\ .\label{covder2}
\end{equation}
We see that the spin connection coefficients can be interpreted as
the gauge field corresponding to local Lorentz invariance.
Commuting two covariant derivatives gives the general formula
\begin{equation}
[\nabla^S_M,\nabla^S_N]\phi={1\over2}R_{MNAB}T^{AB}\phi\
.\label{curvgeneral}
\end{equation}
In particular we can construct a connection on the spin bundle $S$
of $M$. As we know that for $\psi\in C^{\infty}(S)$ the
transformation law reads (with $\S^{AB}$ as defined in
(\ref{SigmaAB}))
\begin{equation}
\delta\psi=-{1\over2}\epsilon_{AB}\Sigma^{AB}\psi\ ,
\end{equation}
we find
\begin{equation}
\nabla^S_M\psi=\partial_M\psi+{1\over2}\omega_{MAB}\Sigma^{AB}\psi=\partial_M\psi+{1\over4}\omega_{MAB}\Gamma^{AB}\psi\
.
\end{equation}
If we commute two spin connections acting on spin bundles we get
\begin{equation}
[\nabla^S_M,\nabla^S_N]\psi={1\over4}R_{MNAB}\Gamma^{AB}\psi,\label{curv}
\end{equation}
where $R$ is the curvature corresponding to $\omega$, i.e. $R^A_{\
\ B}=D(\omega_{M\ \ B}^{\ \ A}dz^M)$.

\bigskip
In the vielbein formalism the property $\nabla g_{MN}=0$
translates to
\begin{equation}
\nabla^S_N e^A_{\ M}=0\ .
\end{equation}
In the absence of torsion this gives the dependence of
$\omega_{MAB}$ on the vielbeins. It can be expressed most
conveniently using the anholonomy coefficients
\begin{equation}
\omega_{MAB}(e)={1\over2}(-\Omega_{MAB}+\Omega_{ABM}-\Omega_{BMA})\
.
\end{equation}
If torsion does not vanish one finds
\begin{equation}
\omega_{MAB}=\omega_{MAB}(e)+\kappa_{MAB}\ ,
\end{equation}
where $\kappa_{MAB}$ is the contorsion tensor. It is related to
the torsion tensor $\mathcal{T}$ by
\begin{equation}
\kappa_{MAB}=\mathcal{T}_{MN}^L(e_{AL}e_B^{\ \ N}-e_{BL}e_A^{\ \ N
})+g_{ML}\mathcal{T}^L_{NR}e^{\ \ N}_Ae^{\ \ R}_B\ .
\end{equation}
Defining $\omega^A_{\ \ B}:=\omega^{\ \ A}_{M\ \ B}dz^M$ one can
derive the {\em Maurer-Cartan structure equations},
\begin{eqnarray}
de^A+\omega^A_{\ \ B}\wedge e^B=\mathcal{T}^A\ \ \ ,\ \ \
d\omega^A_{\ \ B}+\omega^A_{\ \ C}\wedge \omega^C_{\ \ B}=R^A_{\ \
B}\ ,
\end{eqnarray}
where
\begin{eqnarray}
\mathcal{T}^A={1\over 2}\mathcal{T}^A_{\ \ MN}dz^M\wedge dz^N\ \ \
,\ \ \ R^A_{\ \ B}={1\over2}R^A_{\ \ BMN}dz^M\wedge dz^N\ ,
\end{eqnarray}
and
\begin{eqnarray}
\mathcal{T}^A_{\ \ MN}=e^A_{\ \ P}T^P_{\ \ MN}\ \ \ ,\ \ \ R^A_{\
\ BMN}=e^A_{\ \ Q}e_B^{\ \ P}R^Q_{\ \ PMN}\ .
\end{eqnarray}
These equations tell us that the curvature corresponding to
$\nabla$ and the one corresponding to $\nabla^S$ are basically the
same. The Maurer-Cartan structure equations can be rewritten as
\begin{eqnarray}
\mathcal{T}=De\ \ \ ,\ \ \ R=D\omega\ ,
\end{eqnarray}
where $D=d+\omega$. $\mathcal{T}$ and $R$ satisfy the Bianchi
identities
\begin{eqnarray}
D\mathcal{T}=Re\ \ \ ,\ \ \ DR=0\ .
\end{eqnarray}

The Ricci tensor $\mathcal{R}_{MN}$ and the Ricci scalar
$\mathcal{R}$ are given by
\begin{eqnarray}
\mathcal{R}_{MN}:=R_{MPNQ}g^{PQ}\ ,
\mathcal{R}:=\mathcal{R}_{MN}g^{MN}\ .
\end{eqnarray}
Finally, we note that general relativity is a gauge theory in the
sense of appendix \ref{gauge}. If we take the induced basis as a
basis for the tangent bundle the relevant group is
$GL(d,\mathbb{R})$. If on the other hand we use the vielbein
formalism the gauge group is $SO(d-1,1)$. The gauge fields are
$\Gamma$ and $\omega$, respectively. The curvature of these
one-forms is the Riemann curvature tensor and the curvature
two-form, respectively. However, general relativity is a very
special gauge theory, as its connection coefficients can be
constructed from another basic object on the manifold, namely the
metric tensor $g_{MN}$ or the vielbein $e_A^{\ \ M}$.

\chapter{Some Mathematical Background}\label{mathbackground}

\section{Useful facts from complex geometry}\label{CG}
Let $X$ be a complex manifold and define
$\L^{p,q}(X):=\L^p((T^{(1,0)}X)^*)\otimes\L^q((T^{(0,1)}X)^*)$.
Then we have the decomposition
\begin{equation}
\L^k(T^*X)\otimes\mathbb{C} =\bigoplus_{j=0}^k\L^{j,k-j}(X)\ .
\end{equation}
$\L^{p,q}(X)$ are complex vector bundles, but in general they are
not holomorphic vector bundles. One can show that the only
holomorphic vector bundles are those with $q=0$, i.e.
$\L^{0,0}(X), \L^{1,0}(X),\ldots\L^{m,0}(X)$, where $m:={\rm
dim}_{\mathbb{C}}X$, \cite{Jo00}. Let $s$ be a smooth section of
$\L^{p,0}(X)$ on $X$. $s$ is a holomorphic section if and only if
\begin{equation}
\bar\partial s=0\ .
\end{equation}
Such a holomorphic section is called a {\it holomorphic $p$-form}.
Thus, the Dolbeault group $H_{\bar\partial}^{(p,0)}(X)$ is the
vector space of holomorphic $p$-forms on $X$.\\

\begin{definition}\label{smallresolution}
{\em Let $X$ be a complex three manifold and $p\in X$. The} small
resolution of $X$ in $p$ {\em is given by the pair $(\tilde
X,\pi)$ defined s.t.
\begin{eqnarray}
\pi:\tilde X&\rightarrow& X\nonumber\\
\pi: \tilde X\backslash \pi^{-1}(p)&\rightarrow& X\backslash \{p\}
\ \ \
\mbox{is one to one.}\\
\pi^{-1}(p)&\cong& S^2\nonumber
\end{eqnarray}}
\end{definition}

\subsubsection{The moduli space of complex structures}
\begin{proposition}
The tangent space of the moduli space of complex structures
$\mathcal{M}_{cs}$ of a complex manifold $X$ is isomorphic to
$H^1_{\bar\partial}(TX)$.
\end{proposition}
Loosely speaking, the complex structure on a manifold $X$ with
coordinates $z^i,\zb^{\bar i}$ tells us which functions are
holomorphic, $\bar
\partial f(z,\bar z)=0$. Changing the complex structure therefore
amounts to changing the operator $\bar \partial:=\d\zb^{\bar
i}\partial_{\zb^{\bar i}}$. Consider
\begin{equation}
\bar\partial\rightarrow\bar\partial':=\bar\partial+A
\end{equation}
where $A:=A^i\partial_i=A^i\partial_{z^i}$ and $A^i$ is a
one-form. Then linearizing $(\bar\partial +A)^2=0$ gives $\bar
\partial A=0$. On the other hand a change of coordinates
$(z,\bar z)\rightarrow (w,\bar w)$ with $z^i=w^i+v^i(\bar w^{\bar
j}),\ \zb^{\bar i}=\bar w^{\bar i}$ leads to
\begin{equation}
\bar \partial\rightarrow\bar\partial'=\bar \partial +(\bar
\partial v^i)\partial_i\ .
\end{equation}
This means that those transformations of the operator $\bar
\partial$ that are exact, i.e. $A^i=\bar\partial v^i$, can be
undone by a coordinate transformations. For $A^i$ closed but not
exact on the other hand, one changes the complex structure of the
manifold. The corresponding $A$ lie in $H^1_{\bar \partial}(TX)$
which was to be shown.

\section{The theory of divisors}\label{div}
The concept of a divisor is a quite general and powerful tool in
algebraic geometry. We will only be interested in divisors of
forms and functions on compact Riemann surfaces. Indeed, the
theory of divisors is quite convenient to keep track of the
position and degree of zeros and poles on a Riemann surface. The
general concept is defined in \cite{GH}, we follow the exposition
of \cite{FK}.

Let then $\S$ be a compact Riemann surface of genus $\gh$.
\begin{definition}
{\em A} divisor {\em on $\S$ is a formal symbol
\begin{equation}
\mathbf{A}=P_1^{\a_1}\ldots P^{\a_k}_k\ ,
\end{equation}
with $P_j\in\S$ and $\a_j\in\mathbb{Z}$. }
\end{definition}
This can be rewritten as
\begin{equation}
\mathbf{A}=\prod_{P\in\S}P^{\a(P)}\ ,
\end{equation}
with $\a(P)\in\mathbb{Z}$ and $\a(P)\neq0$ for only finitely many
$P\in\S$. The divisors on $\S$ form a group, ${\rm Div}(\S)$, if
we define the multiplication of $\mathbf{A}$ with
\begin{equation}
\mathbf{B}=\prod_{P\in\S}P^{\b(P)}\ ,
\end{equation}
by
\begin{equation}
\mathbf{A}\mathbf{B}:=\prod_{P\in\S}P^{\a(P)+\b(P)}\ ,
\end{equation}
The inverse of $\mathbf{A}$ is given by
\begin{equation}
\mathbf{A}^{-1}=\prod_{P\in\S}P^{-\a(P)}\ .
\end{equation}

Quite interestingly, there is a map from the set of non-zero
meromorphic function $f$ on $\S$ to ${\rm Div}(\S)$, given by
\begin{equation}
f\mapsto(f):=\prod_{P\in\S}P^{{\rm ord}_Pf}\ .
\end{equation}
Furthermore, we can define the {\it divisor of poles},
\begin{equation}
f^{-1}(\infty):=\prod_{P\in\S}P^{{\rm max}\{-{\rm ord}_Pf,0\}}
\end{equation}
and the {\it divisor of zeros}
\begin{equation}
f^{-1}(0):=\prod_{P\in\S}P^{{\rm max}\{{\rm ord}_Pf,0\}}\ .
\end{equation}
Clearly,
\begin{equation}
(f)={f^{-1}(0)\over f^{-1}(\infty)}\ .
\end{equation}

Let $\o$ be a non-zero meromorphic $p$-form on $\S$. Then one
defines its divisor as
\begin{equation}
(\o):=\prod_{P\in\S}P^{{\rm ord}_P\o}\ .
\end{equation}

\bigskip
{\bf Divisors on hyperelliptic Riemann surfaces}\\
Let $\S$ be a hyperelliptic Riemann surface of genus $\gh$, and
$P_1,\ldots P_{2\gh+2}$ points in $\S$. Let further
$z:\S\rightarrow \mathbb{C}$ be a function on $\S$, s.t.
$z(P_j)\neq \infty$. Consider the following function on the
Riemann surface,\footnote{One can show that for two points
$P\neq\tilde P$ on $\S$ for which $z(P)=z(\tilde P)$ one has
$y(P)=-y(\tilde P)$. These two branches of $y$ are denoted $y_0$
and $y_1=-y_0$ in the main text.}
\begin{equation}
y=\sqrt{\prod_{j=1}^{2\gh+2}(z-z(P_j))}\ .
\end{equation}
If $Q,Q'$ are those points on $\S$ for which $z(Q)=z(Q')=\infty$,
i.e. $QQ'=z^{-1}(\infty)$ is the polar divisor of $z$, then the
divisor of $y$ is given by
\begin{equation}
(y)={P_1\ldots P_{2\gh+2}\over Q^{\gh+1}Q'^{\gh+1}}\ .
\end{equation}
To see this one has to introduce local coordinate patches around
points $P\in\S$. If $P$ does not coincide with one of the $P_i$ or
$Q,Q'$ local coordinates are simply $z-z(P)$. Around the points
$Q,Q'$ we have the local coordinate $1/z$, and finally around the
$P_i$ one has $\sqrt{z-z(P_i)}$.

Let $R,R'$ be those points on $\S$ for which $z(R)=z(R')=0$, i.e.
$RR'=z^{-1}(0)$ is the divisor of zeros of $z$. The divisors for
$z$ reads
\begin{equation}
(z)={z^{-1}(0)\over z^{-1}(\infty)}={RR'\over QQ'}\ .
\end{equation}
Furthermore, it is not hard to see that
\begin{equation}
(\d z)={P_1\ldots P_{2\gh+2}\over Q^2Q'^2}\ .
\end{equation}
By simply multiplying the corresponding divisors it is then easy
to see that the forms ${z^k\d z\over y}$ are holomorphic for
$k<\gh$.

\section{Relative homology and relative cohomology}\label{relhom}
A first introduction to algebraic topology can be found in
\cite{Na90}, for a more comprehensive treatment of this beautiful
subject see for example \cite{SZ94}. We only present the
definition of relative (co-)homology, a concept that is important
in a variety of problems in string theory. For example it appears
naturally if one studies world-sheet instantons in the presence of
D-branes, since then the world-sheet can wrap around relative
cycles ending on the branes.

\subsection{Relative homology}
Let $X$ be a triangulable manifold, $Y$ a triangulable submanifold
of $X$ and $i:Y\rightarrow X$ its embedding. Consider chain
complexes $C(X;\mathbb{Z}):=(C_j(X;\mathbb{Z}),\partial)$ and
$C(Y;\mathbb{Z}):=(C_j(Y;\mathbb{Z}),\partial)$ on $X$ and $Y$.
For the pair $(X,Y)$ we can define relative chain groups by
\begin{equation}
C_j(X,Y;\mathbb{Z}):=C_j(X;\mathbb{Z})/C_j(Y;Z)\ .
\end{equation}
This means that elements of $C_j(X,Y;\mathbb{Z})$ are equivalence
classes $\{c\}:=c+C_j(Y;\mathbb{Z})$ and two chains $c, c'$ in
$C_j(X;\mathbb{Z})$ are in the same equivalence class if they
differ only by an element $c_0$ of $C_j(Y;\mathbb{Z})$,
$c'=c+c_0$. The relative boundary operator
\begin{equation}
\partial: C_j(X,Y;\mathbb{Z})\rightarrow C_{j-1}(X,Y;\mathbb{Z})
\end{equation}
is induced by the usual boundary operator on $X$ and $Y$. Indeed,
two representatives $c$ and $c'=c+c_0$ of the same equivalence
class get mapped to $\partial c$ and $\partial c'$, which satisfy
$\partial c' =\partial c+\partial c_0$, and since $\partial c_0$
is an element of $C_{j-1}(Y;\mathbb{Z})$ their images represent
the same equivalence class in $C_{j-1}(X,Y;\mathbb{Z})$.

Very importantly, the property $\partial^2=0$ on $C(X;\mathbb{Z})$
and $C(Y;\mathbb{Z})$ implies $\partial^2=0$ for the
$C_j(X,Y;\mathbb{Z})$. All this defines the relative chain complex
$C(X,Y;\mathbb{Z}):=(C_j(X,Y;\mathbb{Z}),\partial)$. It is natural
to define the relative homology as
\begin{equation}
H_j(X,Y;\mathbb{Z}):=Z_j(X,Y;\mathbb{Z})/B_j(X,Y;\mathbb{\mathbb{Z}})\
,
\end{equation}
where $Z_j(X,Y;\mathbb{Z}):={\rm ker}(\partial):=\{\{c\}\in
C_j(X,Y;\mathbb{Z}):\partial \{c\}=0\}$ and
$B_j(X,Y;\mathbb{Z}):={\rm Im}(\partial):=\{\{c\}\in
C_j(X,Y;\mathbb{Z}):\{c\}=\partial \{\hat c\}\ \mbox{with}\ \{
\hat c\}\in C_{j+1}(X,Y;\mathbb{Z})\}=\partial
C_{j+1}(X,Y;\mathbb{Z})$. Elements of ${\rm ker}(\partial)$ are
called {\em relative cycles} and elements of ${\rm Im}(\partial)$
are called {\em relative boundaries}. Note that the requirement
that the class $\{c\}$ has no boundary, $\partial \{c\}=0$, or
more precisely $\partial\{c\}=\{0\}$, does not necessarily mean
that its representative has no boundary. It rather means that it
may have a boundary but this boundary is forced to lie in $Y$.
Note further that an element in $H_j(X,Y;\mathbb{Z})$ is an
equivalence class of equivalence classes and we will denote it by
$[\{c\}]:=\{c\}+B_j(X,Y;\mathbb{Z})$ for $c\in C_j(X;\mathbb{Z})$
s.t. $\partial c\subset Y$.

We have the short exact sequence of chain complexes
\begin{equation}\label{ses}
0\rightarrow C(Y;\mathbb{Z})\stackrel{i}{\rightarrow}
C(X;\mathbb{Z})\stackrel{p}{\rightarrow}
C(X,Y;\mathbb{Z})\rightarrow 0
\end{equation}
where $i:C_j(Y;\mathbb{Z})\rightarrow C_j(X;\mathbb{Z})$ is the
obvious inclusion map and $p:C_j(X;\mathbb{Z})\rightarrow
C_j(X,Y;\mathbb{Z})$ is the projection onto the equivalence class,
$p(c)=\{c\}$. Note that $p$ is surjective, $i$ is injective and
$p\circ i=\{0\}$, which proves exactness. Every short exact
sequence of chain complexes comes with a long exact sequence of
homology groups. In our case
\begin{equation}
\ldots\rightarrow
H_{j+1}(X,Y;\mathbb{Z})\stackrel{\partial_*}{\rightarrow}H_j(Y;\mathbb{Z})\stackrel{i_*}\rightarrow
H_j(X;\mathbb{Z})\stackrel{p_*}{\rightarrow}
H_j(X,Y;\mathbb{Z})\stackrel{\partial_*}{\rightarrow}H_{j-1}(Y;\mathbb{Z})\rightarrow\ldots\
,
\end{equation}
Here $i_*$ and $p_*$ are the homomorphisms induced from $i$ and
$p$ in the obvious way, for example let $[c]\in H_j(X;\mathbb{Z})$
with $c\in Z_j(X;\mathbb{Z})\subset C_j(X;\mathbb{Z})$ then
$p_*([c]):=[p(c)]$. Note that $p(c)\in Z_j(X,Y;\mathbb{Z})$ and
$[p(c)]=p(c)+B_j(X,Y;\mathbb{Z})$. The operator
$\partial_*:H_j(X,Y;\mathbb{Z})\rightarrow H_j(Y;\mathbb{Z})$ is
defined as $[\{c\}]\mapsto[i^{-1}(\partial(
p^{-1}(\{c\})))]=[\partial c]$. Here we used the fact that
$\partial c$ has to lie in $Y$. The symbol $[\cdot]$ denotes both
equivalence classes in $H_j(X;\mathbb{Z})$ and
$H_j(Y;\mathbb{Z})$.

\subsection{Relative cohomology}
Define the space of relative cohomology $H^j(X,Y;\mathbb{C})$ to
be the dual space of $H_j(X,Y;\mathbb{Z})$,
$H^j(X,Y;\mathbb{C}):={\rm Hom}(H^j(X,Y;\mathbb{C}),\mathbb{C})$,
and similarly for $H^j(X;\mathbb{C}),\ H^j(Y;\mathbb{C})$. The
short exact sequence (\ref{ses}) comes with a dual exact sequence
\begin{equation}
0\rightarrow {\rm
Hom}(C(X,Y;\mathbb{Z}),\mathbb{C})\stackrel{\tilde
p}{\rightarrow}{\rm
Hom}(C(X;\mathbb{Z}),\mathbb{C})\stackrel{\tilde
i}{\rightarrow}{\rm Hom}(C(Y;\mathbb{Z}),\mathbb{C})\rightarrow0\
.
\end{equation}
Here we need the definition of the dual homomorphism for a general
chain mapping\footnote{A chain mapping $f:C(X;\mathbb{Z})
\rightarrow C(Y;\mathbb{Z})$ is a family of homomorphisms
$f_j:C_j(X;\mathbb{Z})\rightarrow C_j(Y;\mathbb{Z})$ which satisfy
$\partial\circ f_j=f_{j-1}\circ\partial$.}
$f:C(X;\mathbb{Z})\rightarrow C(Y;\mathbb{Z})$. Let $\phi\in {\rm
Hom}(C(Y;\mathbb{Z}),\mathbb{C})$, then $\tilde f(\phi):=\phi\circ
f$.\\
The corresponding long exact sequence reads
\begin{equation}\label{les}
\ldots\rightarrow H^{j-1}(X;\mathbb{C})\stackrel{i^*}{\rightarrow}
H^{j-1}(Y;\mathbb{C})\stackrel{\d^*}{\rightarrow}H^{j}(X,Y;\mathbb{C})\stackrel{p^*}{\rightarrow}
H^j(X;\mathbb{C})\stackrel{i^*}{\rightarrow}
H^j(Y;\mathbb{C})\rightarrow\ldots\ ,
\end{equation}
where $i^*:=\tilde i_*$, $p^*:=\tilde p_*$. For example let
$[\Theta]\in H^j(X;\mathbb{C})$. Then $i^*([\Theta])=\tilde
i_*([\Theta])=[\tilde i (\Theta)]$. $\d^*$ acts on $[\t]\in
H^{j-1}(Y;\mathbb{C})$ as $\d^*([\t]):=[\tilde p^{-1}(\d(\tilde
i^{-1}(\t)))]$, and $\d:=\tilde \partial$ is the {\it coboundary
operator}.

\bigskip
So far we started from simplicial complexes, introduced homology
groups on triangulable spaces and defined the cohomology groups as
their duals. On the other hand, there is a natural set of
cohomology spaces on a differential manifold, very familiar to
physicist, namely the de Rham cohomology groups. In fact, they
encode exactly the same topological information, as we have for
any triangulable differential manifold $X$ that
\begin{equation}
H^j_{de Rham}(X;\mathbb{C})\cong H^j_{simplicial}(X;\mathbb{C})\ .
\end{equation}
So we can interpret the spaces appearing in the long exact
sequence (\ref{les}) as de Rham groups, and the maps $i^*$ and
$p^*$ are the pullbacks corresponding to $i$ and $p$. Furthermore,
the coboundary operator $\d\equiv\tilde\partial$ is nothing but
the exterior derivative.

In fact, on a differentiable manifold $X$ with a closed
submanifold $Y$ the relative cohomology groups
$H^j(X,Y;\mathbb{C})$ can be defined from forms on $X$
\cite{KL87}. Let $\O^j(X,Y;\mathbb{C})$ be the $j$-forms on $X$
that vanish on $Y$, i.e.
\begin{equation}
\O^j(X,Y;\mathbb{C}):=\ker(\O^j(X;\mathbb{C})\stackrel{i^*}{\rightarrow}\O^j(Y))
\end{equation}
where $i^*$ is the pullback corresponding to the inclusion
$i:Y\rightarrow X$. Then it is natural to define
\begin{eqnarray}
Z^j(X,Y;\mathbb{C})&:=&\{\Theta\in\O^j(X,Y;\mathbb{C}):\d\Theta=0\}\ ,\nonumber\\
B^j(X,Y;\mathbb{C})&:=&\{\Theta\in\O^j(X,Y;\mathbb{C}):\Theta=\d\eta\ \mbox{for} \ \eta\in \O^{j-1}(X,Y;\mathbb{C})\}\ ,\\
H^j(X,Y;\mathbb{C})&:=&Z^j(X,Y;\mathbb{C})\slash
B^j(X,Y;\mathbb{C}) \ .\nonumber
\end{eqnarray}
As for the cohomology spaces we have
\begin{equation}
H^j_{de Rham}(X,Y;\mathbb{C})\cong
H^j_{simplicial}(X,Y;\mathbb{C})\ .
\end{equation}

There is a pairing
\begin{equation}
\langle .,.\rangle : H_j(X,Y;\mathbb{Z})\times
H^j(X,Y;\mathbb{C})\rightarrow \mathbb{C}
\end{equation}
for $[\Gamma]\in H_j(X,Y;\mathbb{Z})$ and $[\Theta]\in
H^j(X,Y;\mathbb{C})$, defined as
\begin{equation}
\langle [\G],[\Theta]\rangle:=\int_{\G}\Theta\ .
\end{equation}
Of course, this pairing is independent of the chosen
representative in the two equivalence classes. For
$\G':=\G+\partial\hat \G+\G_0$ with $\G_0\subset Y$ this follows
immediately from the fact that $\Theta$ is closed and vanishes on
$Y$. For $\Theta':=\Theta+\d \L$ this follows from
$\int_{\G}\d\L=\int_{\partial\G}\L=0$, as $\L$ vanishes on $Y$.

\bigskip
This definition can be extended to the space of forms
\begin{equation}
\tilde\O^j(X,Y;\mathbb{C}):=\{\Theta\in\O^j(X;\mathbb{C}):i^*\Theta=\d
\theta\}\ .
\end{equation}
Then one first defines the space of equivalence classes
\begin{equation}
\hat\O^j(X,Y;\mathbb{C}):=\tilde\O^j(X,Y;\mathbb{C})/\d
\O^{j-1}(X;\mathbb{C})\ .
\end{equation}
which is clearly isomorphic to $\O^j(X,Y;\mathbb{C})$. Note that
an element of this space is an equivalence class
$\{\Theta\}:=\Theta+\d\O^{j-1}(X;\mathbb{C})$. Then one continues
as before, namely one defines
\begin{eqnarray}
\hat Z^j(X,Y;\mathbb{C})&:=&\{\{\Theta\}\in\hat\O^j(X,Y;\mathbb{C}):\d\{\Theta\}=0\}\ ,\nonumber\\
\hat B^j(X,Y;\mathbb{C})&:=&\{\{\Theta\}\in\hat\O^j(X,Y;\mathbb{C}):\{\Theta\}=\d\{\eta\}\ \mbox{for} \ \{\eta\}\in \hat\O^{j-1}(X,Y;\mathbb{C})\}\ ,\nonumber\\
\hat H^j(X,Y;\mathbb{C})&:=&\hat Z^j(X,Y;\mathbb{C})\slash \hat
B^j(X,Y;\mathbb{C}) \ .
\end{eqnarray}
Obviously, $\hat H^j(X,Y;\mathbb{C})\cong H^j(X,Y;\mathbb{C})$.
Element of $\hat H^j(X,Y;\mathbb{C})$ are equivalence classes of
equivalence classes and we denote them by $[\{\Theta\}]$. The
natural pairing in this case is given by\\
\parbox{14cm}{
\begin{eqnarray}
\langle .,.\rangle &:& H_j(X,Y;\mathbb{Z})\times \hat
H^j(X,Y;\mathbb{C})\rightarrow \mathbb{C}\nonumber\\
\left([\Gamma],[\{\Theta\}]\right)&\mapsto&\langle
[\Gamma],[\{\Theta\}]\rangle:=\int_{\G}\Theta-\int_{\G}\d\theta \
\ \ \ \mbox{if} \ \ i^*\Theta=\d\theta\ ,\nonumber
\end{eqnarray}}\hfill\parbox{8mm}{\begin{eqnarray}\end{eqnarray}}
which again is independent of the representative.

The group $\hat H^j(X,Y;\mathbb{C})$ can be characterised in yet
another way. Note that a representative $\Theta$ of $\hat
Z(X,Y;\mathbb{C})$ has to be closed $\d\Theta=0$ and it must pull
back to an exact form on $Y$, $i^*\Theta=\d \theta$. This motives
us to define an exterior derivative\\
\parbox{14cm}{
\begin{eqnarray}
\d:\O^j(X;\mathbb{C})\times
\O^{j-1}(Y;\mathbb{C})&\rightarrow&\O^{j+1}(X;\mathbb{C})\times
\O^{j}(Y;\mathbb{C})\nonumber\\
(\Theta,\theta)&\mapsto&
\d(\Theta,\theta):=(\d\Theta,i^*\Theta-\d\theta)\ .\nonumber
\end{eqnarray}}\hfill\parbox{8mm}{\begin{eqnarray}\end{eqnarray}}\\
It is easy to check that $\d^2=0$. If we define
\begin{equation}
\mathcal{Z}(X,Y;\mathbb{C}):=\{(\Theta,\theta)\in\O^j(X;\mathbb{C})\times
\O^{j-1}(Y;\mathbb{C}):\d(\Theta,\theta)=0\}
\end{equation}
we find
\begin{equation}
\mathcal{Z}(X,Y;\mathbb{C})\cong \hat Z(X,Y;\mathbb{C})\ .
\end{equation}
Note further that $\hat B^j(X,Y;\mathbb{C})\cong
B^j(X;\mathbb{C})$, so we require that the representative $\Theta$
and $\Theta+\d\L$ are equivalent. But
$i^*(\Theta+\d\L)=\d(\t+i^*\L)$ so $\t$ has to be equivalent to
$\t+i^*\L$. This can be captured by
\begin{equation}
(\Theta,\t)\sim(\Theta,\t)+\d(\L,\l)\ ,
\end{equation}
and with
\begin{equation}
\mathcal{B}(X,Y;\mathbb{C}):=\{\d(\L,\l):(\L,\l)\in\O^j(X;\mathbb{C})\times
\O^{j-1}(Y;\mathbb{C})\}
\end{equation}
we have
\begin{equation}
\hat
H(X,Y;\mathbb{C})\cong\mathcal{Z}(X,Y;\mathbb{C})/\mathcal{B}(X,Y;\mathbb{C})\
.
\end{equation}

\section{Index theorems}\label{IT}
It turns out that anomalies are closely related to the index of
differential operators. A famous theorem found by Atiyah and
Singer tells us how to determine the index of these operators from
topological quantities. In this section we collect important index
theory results which are needed to calculate the anomalies.
\cite{Na90} gives a rather elementary introduction to index
theorems. Their relation to anomalies is explained in \cite{AG85}
and \cite{AGG84}.

\begin{theorem}({\rm Atiyah-Singer index theorem})\\
Let $M$ be a manifold of even dimension, $d=2n$, $G$ a Lie group,
$P(M,G)$ the principal bundle of $G$ over $M$ and let $E$ the
associated vector bundle with $k:={\rm dim}(E)$. Let $A$ be the
gauge potential corresponding to a connection on $E$ and let
$S_{\pm}$ be the positive and negative chirality part of the spin
bundle. Define the Dirac operators $D_{\pm}:S_{\pm}\otimes
E\rightarrow S_{\mp}\otimes E$ by
\begin{equation}
D_{\pm}:=i\Gamma^M\left(\partial_M+{1\over4}\omega_{MAB}\Gamma^{AB}+A_M\right)P_{\pm}\label{Diracoperator}\
.
\end{equation}
Then ${\rm ind}(D_+)$ with
\begin{equation}
{\rm ind}(D_+):={\rm dim}({\rm ker}D_+)-{\rm dim}({\rm ker}D_-)
\end{equation}
is given by
\begin{eqnarray}
{\rm ind}(D_+)&=&\int_M[{\rm ch}(F)\hat A(M)]_{{\rm vol}}\label{ASI}\ ,\\
\hat A(M)&:=&\prod_{j=1}^n{x_j/2\over \sinh(x_j/2)}\label{Ahat}\ ,\\
{\rm ch}(F)&:=&{\rm tr}\ \exp\left({iF\over2\pi}\right)\ .
\end{eqnarray}
\end{theorem}
The $x_j$ are defined as
\begin{equation}
p(E):={\rm det}\left(1+{R\over
2\pi}\right)=\prod_{j=1}^{[n/2]}(1+x_j^2)=1+p_1+p_2+\ldots\ ,
\end{equation}
where $p(E)$ is the {\em total Pontrjagin class} of the bundle E.
The $x_j$ are nothing but the skew eigenvalues of $R/2\pi$,
\begin{equation}
{R\over2\pi}=\left(\begin{matrix}
                    0&x_1&0&0&\ldots&\\
                    -x_1&0&0&0&\ldots&\\
                    0&0&0&x_2&\ldots\\
                    0&0&-x_2&0&\ldots\\
                    \vdots&\vdots&\vdots&\vdots&
                   \end{matrix}\right)\ .
\end{equation}
$\hat A(M)$ is known as the {\em Dirac genus} and ${\rm ch}(F)$ is
the {\em total Chern character}. The subscript ${\rm vol}$ means
that one has to extract the form whose
degree equals the dimension of $M$.\\
\\
To read off the volume form both $\hat A(M)$ and ${\rm ch}(F)$
need to be expanded. We get \cite{AG85}, \cite{AGG84}
\begin{eqnarray}
\hat A(M)&=&1+{1\over
(4\pi)^2}{1\over12}{\rm tr}R^2+{1\over(4\pi)^4}\left[{1\over288}({\rm tr}R^2)^2+{1\over360}{\rm tr}R^4\right]\nonumber\\
&&+{1\over(4\pi)^6}\left[{1\over
10368}({\rm tr}R^2)^3+{1\over4320}{\rm tr}R^2{\rm tr}R^4+{1\over5670}{\rm tr}R^6\right]\nonumber\\
&&+{1\over(4\pi)^8}\left[{1\over
497664}({\rm tr}R^2)^4+{1\over103680}({\rm tr}R^2)^2{\rm tr}R^4+\right.\nonumber\\
&&+\left.{1\over68040}{\rm tr}R^2{\rm tr}R^6+{1\over259200}({\rm tr}R^4)^2+{1\over75600}{\rm tr}R^8\right]+\ldots\ ,\\
{\rm ch}(F)&:=&{\rm
tr}\exp\left({iF\over2\pi}\right)=k+{i\over2\pi}{\rm
tr}F+{i^2\over2(2\pi)^2}{\rm tr}F^2+\ldots+{i^s\over
s!(2\pi)^s}{\rm tr}F^s+\ldots\ .\nonumber\\ \label{ch(F)}
\end{eqnarray}
From these formulae we can determine the index of the Dirac
operator on arbitrary manifolds, e.g. for $d=4$ we get
\begin{equation}
{\rm ind}(D_+)={1\over(2\pi)^2}\int_M\left({i^2\over2}{\rm
tr}F^2+{k\over48}{\rm tr}R^2\right)\ .
\end{equation}

\bigskip
The Dirac operator (\ref{Diracoperator}) is not the only operator
we need in order to calculate anomalies. We also need the analogue
of (\ref{ASI}) for spin-$3/2$ fields which is given by
\cite{AG85}, \cite{AGG84}
\begin{eqnarray}
{\rm ind}(D_{3/2})&=&\int_M[ \hat A(M)({\rm tr}
\exp(iR/2\pi)-1){\rm
ch}(F)]_{{\rm vol}}\nonumber\\
&=&\int_M[ \hat A(M)({\rm tr}(\exp(iR/2\pi)-\opone)+d-1){\rm
ch}(F)]_{{\rm vol}}\ .
\end{eqnarray}
Explicitly,
\begin{eqnarray}
\hat
A(M){\rm tr}\,({\rm exp}(R/2\pi)-\opone)&=&-{1\over(4\pi)^2}\ 2\ {\rm tr}R^2\nonumber\\
&&+{1\over(4\pi)^4}\left[-{1\over6}({\rm tr}R^2)^2+{2\over3}{\rm tr}R^4\right]\nonumber\\
&&+{1\over(4\pi)^6}\left[-{1\over144}({\rm tr}R^2)^3+{1\over20}{\rm tr}R^2{\rm tr}R^4-{4\over45}{\rm tr}R^6\right]\nonumber\\
&&+{1\over (4\pi)^8}\left[-{1\over5184}({\rm tr}R^2)^4+{1\over540}({\rm tr}R^2)^2{\rm tr}R^4-\right.\nonumber\\
&&-\left.{22\over2835}{\rm tr}R^2{\rm tr}R^6+{1\over540}({\rm tr}R^4)^2+{2\over315}{\rm tr}R^8\right]\nonumber\\
&&+\ldots
\end{eqnarray}

\bigskip
Finally, in $4k+2$ dimensions there are anomalies related to forms
with (anti-)self-dual field strength. The relevant index is given
by \cite{AG85}, \cite{AGG84}
\begin{equation}
{\rm ind}(D_A)={1\over4}\int_M[L(M)]_{2n}\ ,
\end{equation}
where the subscript $A$ stands for anti-symmetric tensor. $L(M)$
is known as the {\em Hirzebruch L-polynomial} and is defined as
\begin{equation}
L(M):=\prod_{j=1}^n {x_j/2\over \tanh(x_j/2)}\ .
\end{equation}
For reference we present the expansion
\begin{eqnarray}
L(M)&=&1-{1\over(2\pi)^2}{1\over6}{\rm tr}R^2+{1\over(2\pi)^4}\left[{1\over72}({\rm tr}R^2)^2-{7\over180}{\rm tr}R^4\right]\nonumber\\
&&+ {1\over(2\pi)^6}\left[-{1\over1296}({\rm tr}R^2)^3+{7\over1080}{\rm tr}R^2{\rm tr}R^4-{31\over2835}{\rm tr}R^6\right]\nonumber\\
&&+ {1\over(2\pi)^8}\left[{1\over31104}({\rm tr}R^2)^4-{7\over12960}({\rm tr}R^2)^2{\rm tr}R^4+\right.\nonumber\\
&&+\left.{31\over17010}{\rm tr}R^2{\rm tr}R^6+{49\over64800}({\rm
tr}R^4)^2-{127\over37800}{\rm tr}R^8\right]+\ldots \ .
\end{eqnarray}

\chapter{Special Geometry and Picard-Fuchs Equations}\label{SGPF}
In section \ref{propCY} we saw that the moduli space of a compact
Calabi-Yau manifold carries a K\"ahler metric and that the
K\"ahler potential can be calculated from some holomorphic
function $\mathcal{F}$, which itself can be obtained from
geometric integrals. In this appendix we will show that this
moduli space is actually an example of a mathematical structure
known as special K\"ahler manifold. We also explain how a set of
differential equations, the so called Picard Fuchs equations
arise. In the case of compact Calabi-Yau manifolds these are
differential equations for the period integrals of $\O$. This is
interesting, since for general Calabi-Yau manifolds it may well be
simpler to solve the differential equation than to explicitly
calculate the period integral. During my thesis some progress on
the extension of the concept of special geometry to local
Calabi-Yau manifolds has been made \cite{BM05}. Unfortunately, the
analogue of the Picard-Fuchs equations is still unknown for these
manifolds. A first attempt to set up a rigorous, coordinate free
framework for special K\"ahler manifolds was made in \cite{St90}.
Various properties of special geometry and its relation to the
moduli spaces of Riemann surfaces and Calabi-Yau manifolds, as
well as to Picard-Fuchs equations, were studied in \cite{FL91},
\cite{CAFLL92}, \cite{CAF94}. For a detailed analysis of the
various possible definitions see \cite{CRTP97}.

\section{(Local) Special geometry}\label{SGCY}
We start with the definition of a (local) special K\"ahler
manifold, which is modelled after the moduli space of complex
structures of compact Calabi-Yau manifolds, and which should be
contrasted with rigid special K\"ahler manifolds, that occur in
the context of Riemann surfaces.

\bigskip
{\bf Hodge manifolds}\\
Consider a complex $n$-dimensional Kähler manifold with
coordinates $z^i, \zb^{\bar j}$ and metric
\begin{equation}
g_{i\bar{j}}=\partial_i\bar\partial_{\bar j}K(z,\bar{z})\ ,
\end{equation}
where the real function $K(z,\zb)$ is the Kähler potential,
$i,j,\ldots,\bar i,\bar j,\ldots\in\{1,\ldots,n\}$. The
Christoffel symbols and the Riemann tensor are calculated from the
standard formulae
\begin{equation}
\G^i_{jk}=g^{i\bar l}\partial_j g_{k\bar l}\ ,\ \ \ R^i_{\ j\bar
kl}=\partial_{\bar k}\G^i_{jl}\ ,
\end{equation}
with $g_{i\bar j}g^{k\bar j}=\delta^k_i$, $g_{i\bar k}g^{i\bar
j}=\delta^{\bar j}_{\bar k}$, and the Kähler form is defined as
\begin{equation}
\o:=ig_{i\bar{j}}\d z^i\w\d \zb^{\bar j}\ .
\end{equation}
Introduce the one-form
\begin{equation}
Q:=-{1\over4}\partial K+{1\over4}\bar\partial K\ .
\end{equation}
Then under a Kähler transformation,
\begin{equation}
\tilde K(\zt(z),\ztb(\zb))=K(z,\zb)+f(z)+\bar{f}(\zb)\ ,
\end{equation}
$g$ is invariant and
\begin{equation}
\tilde Q=Q-{1\over4}\partial f+{1\over4}\bar\partial \bar f\ .
\end{equation}
Clearly
\begin{equation}
\o=2i\d Q\ .
\end{equation}
\begin{definition}
{\em A} Hodge manifold {\em is a Kähler manifold carrying a $U(1)$
line bundle $\mathcal{L}$, s.t. $Q$ is the connection of
$\mathcal{L}$. Then the first Chern-class of $\mathcal{L}$ is
given in terms of the Kähler class, $2c_1(\mathcal{L})={1\over
2\pi}[\o]$. Such a manifold is sometimes also called a}  Kähler
manifold of restricted type.\footnote{Following \cite{CRTP97} we
define the Hodge manifold to have a Kähler form which, when
multiplied by ${1\over2\pi}$, is of even integer cohomology. In
the mathematical literature it is usually defined as a manifold
with Kähler form of integer homology, $c_1(\mathcal{L})={1\over
2\pi}[\o]$.}
\end{definition}
On a Hodge manifold a section of $\mathcal{L}$ of Kähler weight
$(q,\bar q)$ transforms as
\begin{equation}
\tilde\psi(\zt(z),\ztb(\zb))=\psi(z,\zb)e^{{q\over4}f(z)}e^{-{\bar{q}\over4}\bar{f}(\zb)}
\end{equation}
when going from one patch to another. For these we introduce the
Kähler covariant derivatives
\begin{equation}
\mD\psi:=\left(\partial-{q\over4}(\partial K)\right)\psi\ \ \ ,\ \
\ \bar{\mD}\psi:=\left(\bar\partial+{\bar q\over4}(\bar\partial
K)\right)\psi\ ,
\end{equation}
where $\partial:=\d z^i\partial_i$, $\bar\partial:=\d \zb^{\bar
i}\bar\partial_{ \bar i}:=\d \zb^{\bar i}{\partial\over{\partial
\zb^{\bar i}}}$. A Kähler covariantly holomorphic section, i.e.
one satisfying $\bar D\psi=0$ is related to a purely holomorphic
section $\psi_{hol}$ by
\begin{equation}
\psi_{hol}=e^{{{\bar q}\over4}K}\psi\ ,
\end{equation}
since then $\bar\partial\psi_{hol}=0$. The Kähler covariant
derivative can be extended to tensors $\Phi$ on the Hodge manifold
$\mathcal{M}$ as
\begin{equation}
\mD\Phi=\left(\nabla-{q\over4}(\partial K)\right)\Phi\ \ \ ,\ \ \
\bar \mD\Phi=\left(\bar{\nabla}+{\bar q\over4}(\bar
\partial K)\right)\Phi\ ,
\end{equation}
where $\nabla$ is the standard covariant derivative.

\bigskip
{\bf Special K\"ahler manifolds}\\
Next we want to define the notion of a special Kähler manifold,
following the conventions of \cite{CRTP97}. There are in fact
three different definitions that are useful and which are all
equivalent.
\begin{definition}\label{sg1}
{\em A} special Kähler manifold {\em $\mathcal{M}$ is a complex
$n$-dimensional Hodge manifold with the following properties:
\begin{itemize}
\item On every chart there are complex projective coordinate
functions $X^I(z)$, $I\in\{0,\ldots,n\}$, and a holomorphic
function $\mathcal{F}(X^I)$, which is homogeneous of second
degree, i.e.
$2\mathcal{F}=X^I\mathcal{F}_I:=X^I{\partial\over\partial
X^I}\mathcal{F}$, s.t. the Kähler potential is given as
\begin{equation}\label{Kahlerpot}
K(z,\zb)=-\log\left[i\bar X^I{\partial\over\partial
X^I}\mathcal{F}(X^I)-i X^I{\partial\over\partial {\bar X^I}}
\bar{\mathcal{F}}(\bar X^I)\right]\ .
\end{equation}
\item On overlaps of charts, $U_\a,U_\b$, the corresponding
functions are connected by
\begin{equation}
\left(\begin{array}{c} X\\\partial \mathcal{F}
\end{array}\right)_{(\a)}=e^{-f_{(\a\b)}}M_{(\a\b)}\left(\begin{array}{c} X\\\partial \mathcal{F}
\end{array}\right)_{(\b)},
\end{equation}
where $f_{(\a\b)}$ is holomorphic on $U_\a\cap U_\b$ and $M_{(\a\b)}\in Sp(2n+2,\mathbb{R})$.\\
\item On the overlap of three charts, $U_\a,U_\b, U_\g$, the
transition functions satisfy the cocycle condition,\\
\parbox{13cm}{
\begin{eqnarray}
e^{f_{(\a\b)}+f_{(\b\g)}+f_{(\g\a)}}&=&1\ ,\nonumber\\
M_{(\a\b)}M_{(\b\g)}M_{(\g\a)}&=&1\ .\nonumber
\end{eqnarray}}\hfill\parbox{8mm}{\begin{eqnarray}\end{eqnarray}}
\end{itemize}}
\end{definition}

\begin{definition}\label{sg2}
{\em A special Kähler manifold $\mathcal{M}$ is a complex
n-dimensional Hodge manifold, s.t.
\begin{itemize}
\item $\exists$ a holomorphic $Sp(2n+2,\mathbb{R})$ vector bundle
$\mathcal{H}$ over $\mathcal{M}$ and a holomorphic section $v(z)$
of $\mathcal{L}\otimes\mathcal{H}$, s.t. the Kähler form is
\begin{equation}
\o=-i\partial\bar\partial\log(i\langle\bar v,v\rangle)\
,\label{d2c1}
\end{equation}
where $\mathcal{L}$ denotes the holomorphic line bundle over
$\mathcal{M}$ that appeared in the definition of a Hodge manifold,
and $\langle v,w\rangle:=v^\tau\mho w$ is the symplectic product on $\mathcal{H}$;\\
\item furthermore, this section satisfies
\begin{equation}
\langle v,\partial_i v\rangle=0\ .\label{d2c2}
\end{equation}
\end{itemize}}
\end{definition}

\begin{definition}\label{sg3}
{\em Let $\mathcal{M}$ be a complex n-dimensional manifold, and
let $V(z,\zb)$ be a $2n+2$-component vector defined on each chart,
with transition function
\begin{equation}
V_{(\a)}=e^{-{1\over2}f_{(\a\b)}}e^{{1\over2}\bar
f_{(\a\b)}}M_{(\a\b)}V_{(\b)}\ ,
\end{equation}
where $f_{\a\b}$ is a holomorphic function on the overlap and
$M_{(\a\b)}$ a constant $Sp(2n+2,\mathbb{R})$ matrix. The
transition functions have to satisfy the cocycle condition. Take a
$U(1)$ connection of the form $\k_i\d z^i-\bar \k_{\bar i}\d \bar
z^{\bar i}$, under which $\bar V$ has opposite charge as $V$.
Define\\
\parbox{13cm}{
\begin{eqnarray}
U_i:=\mD_i V&:=&(\partial_i+\k_i)V\ ,\nonumber\\
\bar \mD_{\bar i} V&:=&(\bar\partial_{\bar i}-\bar\k_{\bar i})V\ ,\nonumber\\
\bar U_{\bar i}:=\bar \mD_{\bar i} \bar V&:=&(\bar\partial_{\bar i}+\bar\k_{\bar i})\bar V\ ,\nonumber\\
\mD_i \bar V&:=&(\partial_i-\k_i)\bar V\ ,\nonumber
\end{eqnarray}}\hfill\parbox{8mm}{\begin{eqnarray}\end{eqnarray}}\\
and impose\\
\parbox{13cm}{
\begin{eqnarray}
\langle V,\bar V\rangle&=&i\ ,\nonumber\\
\bar \mD_{\bar i} V&=&0\ ,\nonumber\\
\langle V,U_i\rangle&=&0\ ,\nonumber\\
\mD_{[i}U_{j]}&=&0\ .\nonumber
\end{eqnarray}}\hfill\parbox{8mm}{\begin{eqnarray}\label{condsg3}\end{eqnarray}}\\
Finally define
\begin{equation}
g_{i\bar j}:=i\langle U_{i},\bar U_{\bar j}\rangle\
.\label{metricsg3}
\end{equation}
If this is a positive metric on $\mathcal{M}$ then $\mathcal{M}$
is a special Kähler manifold.}
\end{definition}

\bigskip
The first definition is modelled after the structure found in the
scalar sector of four-dimensional supergravity theories. It is
useful for calculations as everything is given in terms of local
coordinates. However, it somehow blurs the coordinate independence
of the concept of special geometry. This is made explicit in the
second definition which, although somewhat abstract, captures the
structure of the space. The third definition, finally, can be used
to show that the moduli space of complex structures of a
Calabi-Yau manifold is a special Kähler manifold. The proof of the
equivalence of these three definitions can be found in
\cite{CRTP97}.

\bigskip
{\bf Consequences of special geometry}\\
In order to work out some of the implications of special geometry
we start from a special Kähler manifold that satisfies all the
properties of definition \ref{sg2}. Clearly, (\ref{d2c1}) is
equivalent to $K=-\log(i\langle\bar v,v\rangle)$ and from the
transformation property of $K$ we deduce that $v$ is a field of
weight $(-4,0)$ and $\bar v$ of weight (0,4). In other words they
transform as
\begin{equation}
\tilde v = e^{-f}Mv \ ,\ \ \ \bar{\tilde{v}}= e^{-\bar f}M\bar v \
,
\end{equation}
from one local patch to another. Here $M\in Sp(2n+2,\mathbb{R})$.
The corresponding derivatives
are\\
\parbox{14cm}{
\begin{eqnarray}
\mD_iv&=&\left(\partial_i+(\partial_iK)\right)v\ ,\nonumber\\
\bar \mD_{\bar i}v&=&\bar \partial_{\bar i}v\ ,\nonumber
\end{eqnarray}}\hfill\parbox{8mm}{\begin{eqnarray}\end{eqnarray}}
and their complex conjugates.
Let us define\\
\parbox{14cm}{
\begin{eqnarray}
u_i&:=&\mD_iv\ ,\nonumber\\
\bar u_{\bar i}&:=&\bar \mD_{\bar i}\bar v\ ,\nonumber
\end{eqnarray}}\hfill\parbox{8mm}{\begin{eqnarray}\label{defU}\end{eqnarray}}\\
i.e. $u_i$ is a section of $T\otimes\mathcal{L}\otimes
\mathcal{H}$ of weight $(-4,0)$ and $\bar u_{\bar i}$ a section of
$\bar T\otimes\bar{\mathcal{L}}\otimes \mathcal{H}$ of weight
$(0,4)$. Here we denoted $T:=T^{(1,0)}(\mathcal{M})$, $\bar
T:=T^{(0,1)}(\mathcal{M})$. Then the following relations hold,
together with their complex conjugates
\begin{eqnarray}
\bar \mD_{\bar i}v&=&0\ ,\label{Dv}\\
\mD_i\bar u_{\bar j}=\ [\mD_i,\bar{\mD}_{\bar j}]\bar v&=&g_{i\bar j}\bar v\ ,\label{Dicom}\\
\ [\mD_i,\mD_j]&=&0\ ,\label{DD}\\
\mD_{[i}u_{j]}&=&0\ ,\label{DiUj}\\
\langle v,v\rangle&=&0\ ,\label{vv}\\
\langle v,\bar v\rangle&=&ie^{-K}\ ,\label{vbarv}\\
\langle v, u_i\rangle&=&0\ ,\label{vui}\\
\langle u_i,u_j\rangle&=&0\ ,\label{uiuj}\\
\langle v,\bar u_{\bar i}\rangle&=&0\ , \label{vbarui}\\
\langle u_i,\bar u_{\bar j}\rangle&=&-ig_{i\bar j}e^{-K}\
.\label{uibaruj}
\end{eqnarray}
(\ref{Dv}) holds as $v$ is holomorphic by definition,
(\ref{Dicom}), (\ref{DD}) and (\ref{DiUj}) can be found by
spelling out the covariant derivatives, (\ref{vv}) is trivial.
(\ref{vbarv}) is (\ref{d2c1}). Note that (\ref{d2c2}) can be
written as $\langle v, \mD_iv\rangle=0$ because of the
antisymmetry of $\langle .,.\rangle$, this gives (\ref{vui}).
Taking the covariant derivative and antisymmetrising gives
$\langle \mD_iv, \mD_j v\rangle=0$ and thus (\ref{uiuj}). Taking
the covariant derivative of (\ref{vbarv}) leads to (\ref{vbarui}).
Taking another covariant
derivative leads to the last relation.\\
Next we define the important quantity
\begin{equation}
C_{ijk}:=-ie^{K}\langle \mD_i\mD_jv,\mD_kv\rangle=-ie^{K}\langle
\mD_iu_j,u_k\rangle\ ,
\end{equation}
which has weight $(-4,-4)$ and satisfies
\begin{eqnarray}
C_{ijk}&=&C_{(ijk)}\ ,\\
\bar \mD_{\bar l}C_{ijk}&=&0\ ,\\
\mD_{[i}C_{j]kl}&=&0\ ,\\
\mD_iu_j&=&C_{ijk}\bar u^k\ ,\label{DU}
\end{eqnarray}
where $\bar u^j:=g^{j\bar k}\bar u_{\bar k}$. The first two
relations can be proven readily from the equations that have been
derived so far. The third relation follows from $\langle
u_i,[\mathcal{D}_j,\mathcal{D}_k]u_l \rangle=0$. In order to prove
(\ref{DU}) one expands $\mathcal{D}_iu_j$ as\footnote{To
understand this it is useful to look at the example of the complex
structure moduli space $\mathcal{M}_{cs}$ of a Calabi-Yau
manifold, which is of course the example we have in mind during
the entire discussion. As we will see in more detail below, $v$ is
given by the period vector of $\O$ and the $u_i$ have to be
interpreted as the period vectors of a basis of the $(2,1)$-forms.
Then it is clear that $\bar u_{\bar i}$ should be understood as
period vectors of a basis of $(1,2)$-forms and finally $\bar v$ is
given by the period vector of $\bar \O$. The derivative of a
$(2,1)$-form can be expressed as a linear combination of the basis
elements, which is the expansion of $\mathcal{D}_iu_j$ in terms of
$v$, $u_i,\ \bar u_{\bar i}$, $\bar v$, which now is obvious as
these form a basis of three-forms.}
$\mathcal{D}_iu_j=a_{ij}v+b_{ij}^k u_k+c_{ij}^{\bar k}\bar u_{\bar
k}+d_{ij}\bar v$, and determines the coefficients by taking
symplectic products with the basis vectors. The result is that
$c_{ij}^{\ \ \bar k}=C_{ij}^{\ \ \bar k}$ and all other
coefficients vanish. Considering $\langle u_i,[\mD_k,
\bar{\mD}_{\bar l}]\bar u_{\bar j}\rangle$ we are then led to a
formula for the Riemann tensor,
\begin{equation}\label{RiemannSG}
R_{\bar i j\bar kl}=g_{j\bar k}g_{l\bar i}+g_{l\bar k}g_{j\bar
i}-C_{jlm}\bar C_{\bar k\bar i}^{\ \ m}.
\end{equation}
Here $\bar C_{\bar i\bar j\bar k}$ is the complex conjugate of
$C_{ijk}$ and $C_{\bar k\bar i}^{\ \ m}:=g^{m\bar m}C_{\bar k\bar
i\bar m}$.

\bigskip
{\bf Matrix formulation}\\
So far we collected these properties in a rather unsystematic and
not very illuminating way. In order to improve the situation we
define the $(2n+2)\times (2n+2)$-matrix
\begin{equation}
\mathcal{U}:=\left(\begin{array}{c}v^\tau \\ {u_j}^\tau\\
 (\bar u_{\bar k})^{\tau}\\\bar v^\tau
\end{array}\right)\ ,
\end{equation}
which satisfies
\begin{equation}
\mathcal{U}\mho\mathcal{U}^\tau=ie^{-K}\left(
\begin{array}{cccc}
0&0&0&1\\
0&0&-g_{j\bar k}&0\\
0&g_{\bar k j}&0&0\\
-1&0&0&0
\end{array}\right)\ ,
\end{equation}
as can be seen from (\ref{vv})-(\ref{uibaruj}). In other words one
defines the bundle $E:=(\mathcal{L}\oplus
(T\otimes\mathcal{L})\oplus (\bar T\otimes
\bar{\mathcal{L}})\oplus\bar{\mathcal{L}})\otimes\mathcal{H}={\rm
span}(v)\oplus{\rm span}(u_i)\oplus{\rm span}(\bar u_{\bar
i})\oplus{\rm span}(\bar v)$ and $\mathcal{U}$ is a section of
$E$. Let us first study the transformation of $\mathcal{U}$ under
a change of coordinate patches. We find
\begin{equation}
\tilde{\mathcal{U}}=S^{-1}\mathcal{U}M^\tau
\end{equation}
where $M\in Sp(2n+2,\mathbb{R})$,
\begin{equation}
S^{-1}=\left(\begin{array}{cccc}
e^{-f}&0&0&0\\
0&e^{-f}\xi^{-1}&0&0\\
0&0&e^{-\bar f}\bar\xi^{-1}&0\\
0&0&0&e^{-\bar f}
\end{array}
\right)
\end{equation}
and $\xi:=\xi^i_{\,j}:={\partial \tilde z^i\over{\partial z^j}}$,
$\bar\xi:=\bar\xi^{\bar i}_{\,\bar j}:={\partial \tilde{\bar
z}^{\bar i}\over{\partial \bar z^{\bar j}}}$. Using (\ref{defU}),
(\ref{DU}), (\ref{Dicom}) and (\ref{Dv}) one finds that on a
special Kähler manifold the following matrix equations are
satisfied\footnote{Note that the matrices $\bar{ \mathbb{C}}$ and
$\bar{\mathbb{A}}$ are {\it not} the complex conjugates of
$\mathbb{C}$
and $\mathbb{A}$.},\\
\parbox{13cm}{
\begin{eqnarray}
\mD_i\mathcal{U}&=&\mathbb{C}_i\mathcal{U}\ ,\nonumber\\
\bar \mD_{\bar i}\mathcal{U}&=&\bar{\mathbb{C}}_{\bar
i}\mathcal{U}\ ,\nonumber
\end{eqnarray}}\hfill\parbox{8mm}{\begin{eqnarray}\end{eqnarray}}\\
with
\begin{equation}
\mathbb{C}_i=\left(\begin{array}{cccc}0&\delta_i^j&0&0\\
0&0&C_{ij}^{\ \ \bar k}&0\\
0&0&0&g_{i\bar k}\\
0&0&0&0
\end{array}\right),\ \ \
\bar{\mathbb{C}}_{\bar i}=\left(\begin{array}{cccc}0&0&0&0\\
g_{\bar ij}&0&0&0\\
0&\bar C_{\bar i\bar j}^{\ \ k}&0&0\\
0&0&\delta_{\bar i}^{\bar k}&0
\end{array}\right)\ .
\end{equation}
Here $\delta_i^j$ is a row vector of dimension $n$ with a 1 at
position $i$ and zeros otherwise, $g_{i\bar k}$ is a column vector
of dimension $n$ with entry $g_{i\bar k}$ at position $\bar k$,
and $C_{ij}^{\ \ \bar k}$ is symbolic for the $n\times n$-matrix
$C_i$ with matrix elements $(C_i)_j^{\ \bar k}=C_{ij}^{\ \ \bar
k}$. The entries of $\bar{\mathbb{C}}_i$ are to be understood
similarly. These matrices satisfy $[\mathbb{C}_i,\mathbb{C}_j]=0$
and $\mathbb{C}_i\mathbb{C}_j\mathbb{C}_k\mathbb{C}_l=0$. It will
be useful to rephrase these equations in a slightly different
form. We introduce the operator matrix $\mathbb{D}:=\mathbb{D}_i\d
z^i$,\\
\parbox{13cm}{
\begin{eqnarray}
\mathbb{D}_i\mathcal{U}&:=&(\partial_i+\mathbb{A}_i)\mathcal{U}:=
(\mathcal{D}_i-\mathbb{C}_i)\mathcal{U}=(\partial_i+\mathbf{\G}_i-\mathbb{C}_i)\mathcal{U}=0\ ,\nonumber\\
\bar{\mathbb{D}}_{\bar i}\mathcal{U}&:=&(\bar\partial_{\bar
i}+\bar{\mathbb{A}}_i)\mathcal{U}:=(\bar {\mathcal{D}}_{\bar
i}-\bar{\mathbb{C}}_{\bar i})\mathcal{U}= (\bar \partial_{\bar
i}+\bar{\mathbf{\G}}_{\bar i}-\bar{\mathbb{C}}_{\bar
i})\mathcal{U}=0\ ,\nonumber
\end{eqnarray}}\hfill\parbox{8mm}{\begin{eqnarray}\end{eqnarray}}\\
where
\begin{equation}
\mathbf{\G}_i:=\left(
\begin{array}{cccc}
\partial_i K&0&0&0\\
0&\delta_j^k\partial_i
K-\G_{ij}^k&0&0\\
0&0&0&0\\
0&0&0&0
\end{array}
\right)\ \ \ \ ,\ \ \ \ \bar{\mathbf{\G}}_{\bar i}:=\left(
\begin{array}{cccc}
0&0&0&0\\
0&0&0&0\\
0&0&\delta_{\bar j}^{\bar k}\bar\partial_{\bar i} K-\bar{\G}_{\bar i\bar j}^{\bar k}&0\\
0&0&0&\bar\partial_{\bar i} K
\end{array}
\right)\ ,
\end{equation}
and therefore
\begin{equation}
\mathbb{A}_i:=\left(
\begin{array}{cccc}
\partial_i K&\delta_i^j&0&0\\
0&\delta_j^k\partial_i
K-\G_{ij}^k&C_{ij}^{\ \ \bar k}&0\\
0&0&0&g_{i\bar k}\\
0&0&0&0
\end{array}
\right)\ \ \ \ ,\ \ \ \ \bar{\mathbb{A}}_{\bar i}:=\left(
\begin{array}{cccc}
0&0&0&0\\
g_{\bar ij}&0&0&0\\
0&\bar C_{\bar i\bar j}^{\ \ k}&\delta_{\bar j}^{\bar k}\bar\partial_{\bar i} K-\bar{\G}_{\bar i\bar j}^{\bar k}&0\\
0&0&\delta_{\bar i}^{\bar j}&\bar\partial_{\bar i} K
\end{array}
\right)\ .
\end{equation}
Let us see how $\mathbb{D}$ transforms when we change patches.
From the transformation properties of $\mathcal{D}$ and
$\mathbb{C}$ we find\\
\parbox{13cm}{
\begin{eqnarray}
\tilde{\mathbb{A}}&=& S^{-1}\mathbb{A}S+S^{-1}\d S\ ,\nonumber\\
\tilde{\bar{\mathbb{A}}}&=& S^{-1}\bar{\mathbb{A}}S+S^{-1}\d S\
,\nonumber
\end{eqnarray}}\hfill\parbox{8mm}{\begin{eqnarray}\end{eqnarray}}\\
for $\mathbb{A}:=\mathbb{A}_i\d z^i$, so $\mathbb{A}$ is a
connection. One easily verifies that
\begin{equation}
[\mathbb{D}_i,\mathbb{D}_j]=[\bar{\mathbb{D}}_{\bar
i},\bar{\mathbb{D}}_{\bar j}]=0\ .
\end{equation}
Relation (\ref{RiemannSG}) for the Riemann tensor gives
\begin{equation}\label{com}
[\mathbb{D}_i,\bar{\mathbb{D}}_{\bar j}]\mathcal{U}=0\ .
\end{equation}
All this implies that $\mathbb{D}$ is flat on the bundle $E$.

\bigskip
{\bf Special coordinates and holomorphic connections}\\
We see that an interesting structure emerges once we write our
equations in matrix form. To push this further we need to apply
the equivalence of definition \ref{sg2} and \ref{sg1}. In
particular, we want to make use of the fact that locally, i.e. on
a given chart on $\mathcal{M}$, we can write
\begin{equation}\label{vector}
v=\left(\begin{array}{c}X^I\\{{\partial\mathcal{F}}\over {\partial
X^J}}\end{array}\right)\ ,
\end{equation}
where $X^I=X^I(z)$ and $\mathcal{F}=\mathcal{F}(X(z))$ are
holomorphic. Furthermore, the $X^I$ can be taken to be projective
coordinates on our chart. There is a preferred coordinate system
given by
\begin{equation}
t^a(z):={X^a\over X^0}(z)\ ,\ \ \ a\in\{1,\ldots,n\}\ ,
\end{equation}
and the $t^a$ are known as {\it special coordinates}. To see that
this coordinate system really is useful, let us reconsider the
covariant derivative $\mD_i$ on the special Kähler manifold
$\mathcal{M}$, which contains $\G(z,\zb)$,
$K_i(z,\zb):=\partial_iK(z,\zb)$. If we make use of special
coordinates these split into a holomorphic and a non-holomorphic
piece. We start from $\G_{ij}^k(z,\bar z)$, set $e_i^{\
a}:={\partial t^a(z)\over \partial z^i}$, which does depend on $z$
but not on $\zb$, and write
\begin{eqnarray}
\G_{ij}^k&=&g^{k\bar l}\partial_ig_{j\bar l}=g^{a\bar
b}(e^{-1})^{\ k}_a(e^{-1})^{\ \bar l}_{\bar b}e_i^{\
c}\partial_c(e^{\ d}_je^{\ \bar f}_{\bar l} g_{d\bar
f})\nonumber\\
&=&e_i^{\ c}(\partial_ce_j^{\ d})(e^{-1})_d^{\ k}+e_i^{\ c}e_j^{\
d}(\partial_cg_{d\bar f})g^{a\bar f}(e^{-1})_a^{\ k}\nonumber\\
&=:&\hat\G_{ij}^k(z)+T_{ij}^k(z,\zb)\ .
\end{eqnarray}
Note that $\hat\G_{ij}^k$ transforms as a connection under a
holomorphic coordinate transformation $z\rightarrow\tilde z(z)$,
whereas $T_{ij}^k$ transforms as a tensor. Similarly from
(\ref{Kahlerpot}) one finds that
\begin{eqnarray}
&&K_i(z,\bar
z)=-\partial_i\log(X^0(z))\nonumber\\
&&-\partial_i\log\left[i{\bar X^0(\bar z)\over
X^0(z)}\left({\partial\over
\partial X^0}\mathcal{F}(X^0(z),X^a(z))
+ \bar t^a(\zb){\partial\over
\partial X^a}\mathcal{F}(X^0(z),X^a(z))\right)\right.\nonumber\\
&&\left.-i{\partial\over
\partial \bar X^0}\bar{\mathcal{F}}(\bar X^0(\bar z),\bar X^a(\zb))-it^a(z){\partial\over
\partial \bar X^a}\bar{\mathcal{F}}(\bar X^0(\zb),\bar
X^a(\zb))\right]\nonumber\\
&&=:\hat K_i(z)+\mathcal{K}_i(z,\zb)\ .
\end{eqnarray}
Clearly, $\mathcal{K}_i(z,\zb)$ is invariant under K\"ahler
transformations and $\hat K_i(z)$ transforms as $\hat
K_i(z)\rightarrow\hat K_i(z)+\partial_i f(z)$, which is precisely
the transformation law of $K_i(z,\bar z)$. Thus, the
transformation properties of $\G$ and $K_i$ are carried entirely
by the holomorphic parts and one can define {\it holomorphic}
covariant derivatives for any tensor $\Phi$ on $\mathcal{M}$ of
weight $(q,\bar q)$,\\
\parbox{13cm}{
\begin{eqnarray}
\hat \mD_i\Phi&:=&\left(\hat\nabla_i-{q\over4}(\partial_i\hat K)\right)\Phi\ ,\nonumber\\
\bar{\hat \mD}_{\bar i}\Phi&:=&\left(\bar{\hat\nabla}_{\bar
i}+{\bar q\over4}(\bar \partial_{\bar i}\hat K)\right)\Phi\
,\nonumber
\end{eqnarray}}\hfill\parbox{8mm}{\begin{eqnarray}\label{holcov}\end{eqnarray}}\\
where $\hat \nabla$ now only contains the holomorphic connection
$\hat \G$. A most important fact is that $\hat \G$ is flat, i.e.
the corresponding curvature tensor vanishes,
\begin{equation}
\hat R^k_{\
lij}:=\partial_l\hat\G^k_{ij}-\partial_i\hat\G^k_{lj}+\hat\G^m_{l
j}\hat\G^k_{mi}-\hat\G^m_{l j}\hat\G^k_{mi}=0\ ,
\end{equation}
which can be seen readily from the explicit form of $\hat\G$. Next
we take $\eta_{ab}$ to be a constant, invertible symmetric matrix
and define
\begin{equation}
\hat g_{ij}:=e^{\ a}_i e^{\ b}_j\eta_{ab}\ .
\end{equation}
It is not difficult to show that the Levi-Civita connection of
$\gh_{ij}$ is nothing but $\hat \G$. If we take $z^i=t^a$ one
finds that
\begin{equation}
e_i^{\ a}=\delta_i^a\ \ \ ,\ \ \ \hat\G_{ij}^k=0\ \ \ ,\ \ \ \hat
g_{ij}=\eta_{ij}\ .
\end{equation}
So the holomorphic part of the connection can be set to zero if we
work in special coordinates. Special or flat coordinates are a
preferred coordinate system on a chart of our base manifold
$\mathcal{M}$. But on such a chart $v$ is only defined up to a
transformation $\tilde v=e^{-f}Mv$. This tells us that we can
choose a gauge for $v$ such that $X^0=1$. This is, of course, the
gauge in which the holomorphic part of the Kähler connection
vanishes as well, $\hat K_i=0$.

\bigskip
{\bf Solving the Picard-Fuchs equation}\\
The fact that the connections on a special Kähler manifold can be
split into a holomorphic connection part and a non-holomorphic
tensor part is highly non-trivial. In the following we present one
very important consequence of this fact. Consider once again the
system
\begin{equation}
\mathbb{D}\mathcal{U}=0\ \ \ ,\ \ \
\bar{\mathbb{D}}\mathcal{U}=0\label{nonhollinPF}
\end{equation}
which holds on any special Kähler manifold. Now suppose we know of
a manifold that it is special Kähler, but we do not know the
holomorphic section $v$. Then we can understand
(\ref{nonhollinPF}) as a differential system for $\mathcal{U}$ and
therefore for $v$. These differential equations are known as the
{\it
Picard-Fuchs equations}. As to solve this system consider the transformation\\
\parbox{13cm}{
\begin{eqnarray}
\mathcal{U}&\rightarrow& \mathcal{W}:=R^{-1}\mathcal{U}\ ,\nonumber\\
\mathbb{A}&\rightarrow&
\hat{\mathbb{A}}:=R^{-1}(\mathbb{A}+\partial)R\
,\nonumber\\
\bar{\mathbb{A}}&\rightarrow&
\bar{\hat{\mathbb{A}}}:=R^{-1}(\bar{\mathbb{A}}+\bar\partial)R\
,\nonumber
\end{eqnarray}}\hfill\parbox{8mm}{\begin{eqnarray}\end{eqnarray}}\\
where
\begin{equation}
R^{-1}(z,\zb)=\left(\begin{array}{cccc}
1&0&0&0\\
*&\opone&0&0\\
*&*&\opone&0\\
*&*&*&1
\end{array}
\right)\ .
\end{equation}
Note that for such a matrix $R^{-1}$ the matrix $R$ will be lower
diagonal with all diagonal elements equal to one, as well.
Clearly, this transformation leaves $v$ invariant and\\
\parbox{13cm}{
\begin{eqnarray}
\mathbb{D}\mathcal{U}=0\rightarrow \hat{\mathbb{D}}\mathcal{W}=0\ ,\nonumber\\
\bar{\mathbb{D}}\mathcal{U}=0\rightarrow
\bar{\hat{\mathbb{D}}}\mathcal{W}=0\ ,\nonumber
\end{eqnarray}}\hfill\parbox{8mm}{\begin{eqnarray}\end{eqnarray}}\\
with $\hat{\mathbb{D}}:=\partial+\hat{\mathbb{A}}$ and
$\bar{\hat{\mathbb{D}}}:=\bar\partial+\bar{\hat{\mathbb{A}}}$. The
crucial point is that the solution $v$ that we are after does not
change under this transformation, i.e. we might as well study the
system $\hat{\mathbb{D}}\mathcal{W}=0$,
$\bar{\hat{\mathbb{D}}}\mathcal{W}=0$. Next we note that
$R^{-1}(z,\zb)$ does not have to be holomorphic. In fact, since
the curvature of $\bar{\mathbb{A}}$ vanishes we can go into a
system where $\bar{\hat{\mathbb{A}}}=0$ and therefore
$\bar{\mathbb{A}}=R\bar\partial R^{-1}$. In this gauge we have
\begin{equation}
\bar\partial\mathcal{W}=0\ ,
\end{equation}
and from (\ref{com}) one finds
\begin{equation}
\bar\partial\hat{\mathbb{A}}=0\ ,
\end{equation}
which tells us that the non-holomorphic parts of the connection
$\hat{\mathbb{A}}$ vanish and that the matrices $\mathbb{C}_i$ are
holomorphic. In other words we have $\hat{\mathbb{D}}_i=\hat
{\mathcal{D}}_i-\mathbb{C}_i$, where $\hat{\mathcal{D}}_i$ is the
holomorphic covariant derivative of Eq. (\ref{holcov}) including
$\hat K_i(z)$ and $\hat\G_{ij}^k$ only and we are interested in
the solutions of the {\it holomorphic} system
\begin{equation}
\hat{\mathbb{D}}\mathcal{W}=0\ .
\end{equation}
Note that in this system we still have holomorphic coordinate
transformations as a residual symmetry. Furthermore, if we plug
(\ref{vector}) in the definition of $C_{ijk}$, we find in the
holomorphic system
\begin{equation}
C_{ijk}=-ie^{\hat K}\left[(\hat{\mathcal{D}}_i\hat{\mathcal{D}}_j
X^I)\hat{\mathcal{D}}_k\mathcal{F}_I-(\hat{\mathcal{D}}_i\hat{\mathcal{D}}_j\mathcal{F}_I)\hat{\mathcal{D}}_kX^I\right]\
.
\end{equation}
In special coordinates this reduces to
\begin{equation}
C_{abc}=i\partial_a\partial_b\partial_c\mathcal{F}\ .\label{CdddF}
\end{equation}
The strategy is now to solve the system in special coordinates
first, and to find the general solution from a holomorphic
coordinate transformation afterwards. Let us then choose special
coordinates on a chart of $\mathcal{M}$, together with the choice
$X^0=1$, and let
\begin{equation}
\mathcal{W}=:\left(
\begin{array}{c}
v^\tau\\
v_a^\tau\\
\tilde v_b^\tau\\
(v^0)^\tau\end{array}\right)\ .
\end{equation}
In special coordinates the equation
$\hat{\mathbb{D}}_a\mathcal{W}=0$ reads
$(\partial_a-\mathbb{C}_a)\mathcal{W}=0$ with
\begin{equation}
\mathbb{C}_a=\left(\begin{array}{cccc}
0&\delta_a^b&0&0\\
0&0&C_{ab}^{\ \ c}&0\\
0&0&0&\eta_{ac}\\
0&0&0&0
\end{array}\right)\ .
\end{equation}
or\\
\parbox{13cm}{
\begin{eqnarray}
\partial_av&=&v_a\ ,\nonumber\\
\partial_av_b&=&C_{ab}^{\ \ c}\tilde v_c\ ,\nonumber\\
\partial_a\tilde v_c&=&\eta_{ac}v^0\ ,\nonumber\\
\partial_a{v^0}&=&0\ .\nonumber
\end{eqnarray}}\hfill\parbox{8mm}{\begin{eqnarray}\end{eqnarray}}\\
Recalling (\ref{CdddF}) the solution can be found to be
\begin{equation}
\left(\begin{array}{c}
v^\tau\\
v_b^\tau\\
\tilde v_c^\tau\\
(v^0)^\tau
\end{array}\right)=\left(\begin{array}{cccc}
1&t^d&\partial_d\mF&2\mF-t^e\,\partial_e \mF\\
0&\delta_b^d&\partial_b\partial_d\mF&\partial_b\mF-t^e\,\partial_e \partial_b\mF\\
0&0&-i\eta_{cd}&it^e\eta_{ec}\\
0&0&0&i
\end{array}\right)\ .\label{solspecial}
\end{equation}
Of course, the solution of $\hat{\mathbb{D}}\mathcal{W}=0$ in
general (holomorphic) coordinates can then be obtained from a
(holomorphic) coordinate transformation of (\ref{solspecial}).

\bigskip
{\bf Special geometry and Calabi-Yau manifolds}\\
It was shown in section \ref{propCY} that the moduli space of
complex structures of a compact Calabi-Yau manifold is K\"ahler
with K\"ahler potential $K=-\log\left(i\int\O\w\bar\O\right)$. Let
\begin{equation}
V:=e^{K\over2}\int_{\G^{(3)}_i}\O^{(3,0)}\ ,
\end{equation}
where $\{\G^{(3)}_i\}=\{\G_{\a^I},\G_{\b_J}\}$ is the set of all
three-cycles in $X$. These cycles are only defined up to a
symplectic transformation and $\O$ is defined up to a
transformation $\O\rightarrow e^{f(z)}\O$, see Eq.
(\ref{transOmega}). Therefore $V$ is defined only up to a
transformation
\begin{equation}
V\rightarrow \tilde V=e^{-{f\over2}}e^{\bar f\over2}M V\ ,
\end{equation}
with $M\in Sp(2n+2;\mathbb{Z})$ and $f(z)$ holomorphic. Next we
set
\begin{eqnarray}
U_i&:=&\mathcal{D}_iV:=\left(\partial_i+{1\over2}K_i\right)V=\left(\partial_i+{1\over2}(\partial_iK)\right)V\ ,\\
\bar{\mathcal{D}}_iV&:=&\left(\bar\partial_{\bar
i}-{1\over2}(\bar\partial_{\bar i}K)\right)V\ .
\end{eqnarray}
For vectors $V=\int_{\G^{(3)}_i}\Xi$ and $W=\int_{\G^{(3)}_i}\S$
with $\Xi,\ \S$ three-forms on $X$ there is a natural symplectic
product, $\langle V,W\rangle:=-\int_X\Xi\w\S$. Then it is easy to
verify that (\ref{condsg3}) holds and $i\langle U_i,\bar U_{\bar
j}\rangle=-{\int_X\chi_i\w\bar\chi_{\bar j}\over\int_X\O\w\bar\O}$
is indeed nothing but $G_{i\bar j}^{(CS)}$. So we find that all
the requirements of definition \ref{sg3} are satisfied and
$\mathcal{M}_{cs}$ is a special K\"ahler manifold.

In the context of Calabi-Yau manifolds the projective coordinates
are given by the integrals of the unique holomorphic three-form
$\O$ over the $\G_{\a^I}$-cycles and the holomorphic function
$\mathcal{F}$ of definition \ref{sg1} is nothing but the
prepotential constructed from integrals of $\O$ over the
$\G_{\b_J}$-cycles,
\begin{equation}
X^I=\int_{\G_{\a^I}}\O\ \ \ ,\ \ \ \mF_I=\int_{\G_{\b_I}}\O\ .
\end{equation}
In other words the vector $v$ of definition \ref{sg2} is given by
the period vector of $\O$ and the symplectic bundle is the Hodge
bundle $\mathcal{H}$. This is actually the main reason why the
structure of special geometry is so important for physicist. The
integrals of $\O$ over three cycles is interesting since it
calculates the prepotential (and therefore the physically very
interesting quantity $C_{ijk}$). On the other hand, using mirror
symmetry a period integral on one manifold can tell us something
about the instanton structure on another manifold. Unfortunately
the integrals often cannot be calculated explicitly. However, in
some cases it is possible to solve the ordinary linear
differential Picard-Fuchs equation for $\int\O$, which therefore
gives an alternative way of extracting interesting quantities.
Indeed, the differential equations that one can derive for
$\int\O$ by hand agree with the Picard-Fuchs equations of special
geometry. Finally we note that the matrix $\mathcal{U}$ is nothing
but the period matrix of the Calabi-Yau,
\begin{equation}
\mathcal{U}=\left(\begin{array}{c}
\int\O^{(3,0)}\\
\int\O_\a^{(2,1)}\\
\int\O_\a^{(1,2)}\\
\int\O^{(0,3)}
\end{array}
\right)\ ,
\end{equation}
which can be brought into holomorphic form by a (non-holomorphic)
gauge transformation.

\section{Rigid special geometry}\label{rigSG}
The quantum field theories we are interested in are generated from
local Calabi-Yau manifolds, rather than from compact Calabi-Yau
manifolds. We saw already that the integrals of $\O$ over
(relative) three-cycles maps to integrals on a Riemann surface. It
turns out that the moduli space of Riemann surfaces carries a
structure which is very similar to the special geometry of
Calabi-Yau manifolds, and which is known as rigid special
geometry.\\

\bigskip
{\bf Rigid special K\"ahler manifolds}
\begin{definition}\label{rigid1}
{\em A complex $n$-dimensional K\"ahler manifold is said to be}
rigid special Kähler {\em if it satisfies:
\begin{itemize}
\item On every chart there are $n$ independent holomorphic
functions $X^i(z)$, $i\in\{1,\ldots,n\}$ and a holomorphic
function $\mathcal{F}(X)$, s.t.
\begin{equation}
K(z,\zb)=i\left( X^i{\partial\over\partial \bar
X^i}\bar{\mathcal{F}}(\bar X^i)- \bar X^i{\partial\over\partial
{X^i}} \mathcal{F}(X^i)\right).
\end{equation}
\item On overlaps of charts there are transition functions of the
form
\begin{equation}
\left(
\begin{array}{c}X\\\partial F \end{array}
\right)_{(\a)}=e^{ic_{\a\b}}M_{\a\b}\left(
\begin{array}{c}X\\\partial F
\end{array}\right)+b_{\a\b},\label{trans}
\end{equation}
with $c_{\a\b}\in \mathbb{R}$, $M_{\a\b}\in Sp(2n,\mathbb{R})$,
$b_{\a\b}\in \mathbb{C}^{2n}$.\\
\item The transition functions satisfy the cocycle condition on
overlaps of three charts.
\end{itemize}}
\end{definition}
\begin{definition}\label{rigid2}
{\em A rigid special Kähler manifold is a Kähler manifold
$\mathcal{M}$ with the following properties:
\begin{itemize}
\item There exists a $U(1)\times ISp(2n,\mathbb{R})$ vector
bundle\footnote{This is a vector bundle with transition function
$V_{(\a)}=e^{ic_{\a\b}}M_{(\a\b)}V+b_{\a\b}$, with $c_{\a\b}\in
\mathbb{R}$, $M_{\a\b}\in Sp(2n,\mathbb{R})$, $b_{\a\b}\in
\mathbb{C}^{2n}$.} over $\mathcal{M}$ with constant transition
functions, in the sense of (\ref{trans}), i.e. with a complex
inhomogeneous piece. This bundle should have a holomorphic section
$V$, s.t. the Kähler form is
\begin{equation}
\o=-\partial\bar\partial\langle V,\bar V\rangle\ .
\end{equation}
\item We have $\langle\partial_i V,\partial_j V\rangle=0$.
\end{itemize}}
\end{definition}

\begin{definition}\label{rigid3}
{\em A rigid special manifold is a complex $n$-dimensional
K\"ahler manifold with on each chart $2n$ closed holomorphic
1-forms $U_i\d z^i$,
\begin{equation}\label{U0}
\bar\partial_i U_j=0\ ,\ \ \ \partial_{[i}U_{j]}=0\ ,
\end{equation}
with the following properties:
\begin{itemize}
\item $\langle U_i,U_j\rangle=0$ . \item The K\"ahler metric is
\begin{equation}\label{U2}
g_{ij}=i\langle U_i,\bar U_j\rangle\ .
\end{equation}
\item The transition functions read
\begin{equation}
U_{i,(\a)}\d z^i_{(\a)}=e^{ic_{\a\b}}M_{\a\b}U_{i,(\b)}\d
z^i_{(\b)}
\end{equation}
with $c_{\a\b}\in\mathbb{R}$, $M_{\a\b}\in Sp(2n,\mathbb{R})$.\\
\item The cocycle condition holds on the overlap of three
charts.\\
\end{itemize}}
\end{definition}
For the proof of the equivalence of these definitions see
\cite{CRTP97}.

\bigskip
{\bf Matrix formulation and Picard Fuchs equations}\\
As before we want to work out some of the consequences of these
definitions. Take the $U_i$ from Def. \ref{rigid3} and define
\begin{equation}
\mathcal{V}:=\left(\begin{array}{c}U_i^\tau\\\bar U_{\bar
j}^\tau\end{array}\right).
\end{equation}
Equations (\ref{U0}) and (\ref{U2}) can then be written in matrix
form
\begin{equation}
\mathcal{V}\mho\mathcal{V}^\tau=\left(\begin{array}{cc}0&-ig_{i\bar
j}\\ig_{j\bar i}&0\end{array}\right)\ .
\end{equation}
Next we define the $(2n\times 2n)$-matrix
\begin{equation}
\mathbb{A}_i:=-(\partial_i\mathcal{V})\mathcal{V}^{-1}\ ,
\end{equation}
which has the structure
\begin{equation}
\mathbb{A}_i=-\left(\begin{array}{cc}G_{ij}^{\ \ k}&C_{ij}^{\ \
\bar k}\\0&0\end{array}\right)\ ,
\end{equation}
where $G_{ij}^{\ \ k}=G_{(i,j)}^{\ \ \ \ k}$ and $C_{ij}^{\ \ \bar
k}=C_{(i,j)}^{\ \ \ \bar k}$, as can be seen from Eq. (\ref{U0}).
Then one finds
\begin{eqnarray}
\mathbb{A}_i\left(\begin{array}{cc}0&-ig_{k\bar l}\\
ig_{l\bar k}&0\end{array}\right)&=&-(\partial_i\mathcal{V})\O\mathcal{V}^\tau\nonumber\\
&=&-\partial_i\left(\begin{array}{cc}0&-ig_{j\bar l}\\
ig_{l\bar
j}&0\end{array}\right)+\mathcal{V}\O\partial_i\mathcal{V}^\tau\
,\label{eq}
\end{eqnarray}
but on the other hand
\begin{equation}
\mathbb{A}_i\left(\begin{array}{cc}0&-ig_{k\bar l}\\
ig_{l\bar k}&0\end{array}\right)=\left(\begin{array}{cc}-iC_{(i,j)l}&iG_{(i,j)\bar l}\\
0&0\end{array}\right)\ ,
\end{equation}
and therefore, taking the second line of (\ref{eq}) minus the
transpose of the first,
\begin{equation}
\left(\begin{array}{cc}iC_{(i,j)l}-iC_{(i,l)j}&-iG_{(i,j)\bar l}\\
iG_{(i,l)\bar j}&0\end{array}\right)=\partial_i\left(\begin{array}{cc}0&-ig_{j\bar l}\\
ig_{l\bar j}&0\end{array}\right)\ .
\end{equation}
We deduce that $C$ is symmetric in all its indices and that
$\partial_{[i}g_{j]\bar l}=0$, i.e. it is Kähler. Hence, we can
define the Levi-Civita connection and its covariant derivative
\begin{equation}
\G_{ij}^k=g^{k\bar l}\partial_j g_{i\bar l}\ \ \ ,\ \ \ \nabla_i
U_j:=\partial_i U_j-\G_{ij}^k U_k\ \ \ ,\ \ \ \nabla_{i}\bar
U_{\bar j}:=\partial_{i}\bar U_{\bar j}\ .
\end{equation}
Clearly, $G_{ij}^{\ \ k}=\G_{ij}^{k}$ and we find
\begin{equation}
(\partial_i+\mathbb{A}_i)\mathcal{V}=(\nabla_i-\mathbb{C}_i)\mathcal{V}=0\
\ \ ,\ \ \ (\bar\partial_{\bar i}+\bar{\mathbb{A}}_{\bar
i})\mathcal{V}=(\bar \nabla_{\bar i}-\bar{\mathbb{C}}_{\bar
i})\mathcal{V}=0\ ,
\end{equation}
with
\begin{equation}
\bar{\mathbb{A}}_{\bar i}:=\left(\begin{array}{cc}0&0\\
-\bar C_{\bar i\bar j}^{\ \ k}&-\bar\G_{\bar i\bar j}^{\bar
k}\end{array}\right)\ \ \ ,\ \ \
\mathbb{C}_i:=\left(\begin{array}{cc}0&C_{ij}^{\ \ \bar k}\\
0&0\end{array}\right)\ \ \ ,\ \ \ \bar{\mathbb{C}}_i:=\left(\begin{array}{cc}0&0\\
\bar C_{\bar i\bar j}^{\ \  k}&0\end{array}\right)
\end{equation}
and symmetric $C_{ijk}$. From $[\nabla_i,\nabla_j]\mathcal{V}=0$
we deduce that $\nabla_{[i}\mathbb{C}_{j]}=0$ or
\begin{equation}
C_{ijk}=\nabla_{i}\nabla_j\nabla_k\mathcal{S}
\end{equation}
for some function $\mathcal{S}$. Acting with
$[\nabla_i,\bar\nabla_{\bar j}]$ on $\mathcal{V}$ gives the
identity
\begin{equation}
R_{i\bar jk\bar l}=-C_{ikm}\bar C_{\bar j\bar l\bar m}g^{m\bar m}\
.
\end{equation}
Finally we define $\mathbb{D}_i:=\nabla_i-\mathbb{C}_i$ and
$\bar{\mathbb{D}}_{\bar i}:=(\bar\nabla_i-\bar{\mathbb{C}}_i)$
with the properties $[\mathbb{D}_i,\mathbb{D}_j]=0$,
$[\bar{\mathbb{D}}_{\bar i},\bar{\mathbb{D}}_{\bar j}]=0$ and
$[\mathbb{D}_i,\bar{\mathbb{D}}_{\bar j}]\mathcal{V}=0$, so
$\mathbb{D}$ is a flat connection.

Now we use the equivalence of the three definitions and define the
{\it special coordinates} to be the holomorphic functions
\begin{equation}
t^a(z):=X^i(z)\ .
\end{equation}
As above the Christoffel symbol splits into a tensor part and a
holomorphic connection part,
\begin{equation}
\G_{ij}^k(z,\zb)=\hat \G_{ij}^k(z)+T_{ij}^k(z,\zb),
\end{equation}
and once again $\hat\G$ is a flat connection, $R(\hat\G)=0$, that
vanishes if we use special coordinates. We again define a
holomorphic covariant derivative, $\hat \nabla$, that contains
only the holomorphic part of the connection. Its commutator gives
the curvature of $\hat\G$ and therefore vanishes. Next we
transform
\begin{eqnarray}
\mathcal{V}&\rightarrow& \mathcal{X}:=S^{-1}\mathcal{V}\ ,\nonumber\\
\mathbb{A}&\rightarrow&\hat{\mathbb{A}}:=S^{-1}(\mathbb{A}+\partial)S\ ,\\
\bar{\mathbb{A}}&\rightarrow&\bar{\hat{\mathbb{A}}}:=S^{-1}(\bar{\mathbb{A}}+\bar\partial)S\
,\nonumber
\end{eqnarray}
where $S$ is chosen such that $\bar{\hat{\mathbb{A}}}=0$. Then we
are left with
\begin{equation}
(\hat{\nabla}_{\a}-\hat{\mathbb{C}})\mathcal{X}=0\ \ \ ,\ \ \
\bar\partial_{\bar\a}\mathcal{X}=0\ .
\end{equation}
In special coordinates $\hat\G$ vanishes and this reads
\begin{equation}
\partial_a\mathcal{X}=\left(\begin{array}{cc}0&C_{ab}^{\ \ c}\\0&0\end{array}\right)\mathcal{X}\ .
\end{equation}
Using the equivalence of the three definitions we find that
\begin{equation}
U_i=\partial_i\left(\begin{array}{c}X^j\\\partial_j\mF\end{array}\right)\
.
\end{equation}
Then, multiplying $\partial_a U_b=C_{ab}^{\ \ d}\bar U_{ d}$ by
$U^\tau_c\mho$ we find that in special coordinates
\begin{equation}
C_{abc}=i\partial_a\partial_b\partial_c\mathcal{F}\ ,
\end{equation}
which leads to the solution
\begin{eqnarray}
\mathcal{X}=\left(\begin{array}{cc}\delta_b^d&\partial_b\partial_d\mathcal{F}\\
0&-i\eta_{cd}\end{array}\right)\ .
\end{eqnarray}

\bigskip
{\bf Rigid special geometry and Riemann surfaces}\\
For a given Riemann surface $\S$ it is natural to try to construct
a rigid special Kähler manifold using the $\hg$ holomorphic forms
$\l_i$. However, the moduli space of a Riemann surface has
dimension $3\hg-3$ for $\hg>1$, whereas we can only construct a
$\hg$-dimensional rigid special Kähler manifold from our forms.
That means that unless $\hg=1$ the special manifold will be a
submanifold of the moduli space of $\S$.\\
Let then $\mathcal{W}$ be a family of genus $\hg$ Riemann surfaces
which are parameterised by only $\hg$ complex moduli and let
$\l_i$ be the set of holomorphic one-forms on $\S$. Then we
identify
\begin{equation}
U_i=\left(\begin{array}{c}\int_{\a^j}\l_i\\\int_{\b_k}\l_i\end{array}\right).
\end{equation}
$\bar\partial_{\bar j} U_i=0$ tells us that the periods should
depend holomorphically on the moduli. Using Riemann's second
relation one can show that all requirements of definition
(\ref{rigid3}) are satisfied. (E.g. $\langle U_i, U_j\rangle=0$
follows immediately from the symmetry of the period matrix;
Riemann's second relation gives the positivity of the metric.) The
condition $\partial_{[j} U_{i]}=0$ has to be checked for the
particular example one is considering. On compact Riemann surfaces
it reduces to $\partial_{[i} \l_{j]}=(\d\eta)_{ij}$. If the
right-hand side is zero then locally there exists a meromorphic
one-form $\l$ whose derivatives give the holomorphic one-forms.
Then, using
\begin{equation}
U_i=\partial_i \left(\begin{array}{c}
X^j\\
\partial_k \mF\end{array}\right)\ ,
\end{equation}
we identify
\begin{equation}
\left(\begin{array}{c}X^i\\\partial_j
\mF\end{array}\right)=\left(\begin{array}{c}\int_{\a^i}\l\\
\int_{\b_j}\l\end{array}\right)\ .
\end{equation}
This implies that, similarly to the case of the Calabi-Yau
manifold, the holomorphic function $X^i$ and the prepotential
$\mF$ can be calculated from geometric integrals on the Riemann
surface. However, in contrast to the Calabi-Yau space, the form
$\l$ now is meromorphic and not holomorphic.

\chapter{Topological String Theory}\label{TST}
One of the central building blocks of the web of theories sketched
in Fig. \ref{bigpictureIIB} is the B-type topological string.
Indeed, although the relation between effective superpotentials
and matrix models can also be proven using field theory results
only \cite{DGLVZ02}, \cite{CDSW02}, the string theory derivation
of this relation leads to the insight that many seemingly
different theories are actually very much related, and the
topological string lies at the heart of these observations. The
reason for this central position of the topological string are a
number of its properties. It has been known for a long time
\cite{BCOV93b} that the topological string computes certain
physical amplitudes of type II string theories compactified on
Calabi-Yau manifolds. Furthermore, it turns out that the string
field theory of the open B-type topological string on the simple
manifolds $X_{res}$ is an extremely simple gauge theory, namely a
holomorphic matrix model \cite{DV02a}, as is reviewed in chapter
\ref{TSMM}. Finally, we already saw in the introduction that the
gauge theory/string theory duality can be made precise if the
string theory is topological. Here we review the definition of the
topological string, together with some of its elementary
properties. Since topological string theory is a vast and quickly
developing subject, this review will be far from complete. The
reader is referred to the book of Hori et. al. \cite{Horietal},
for more details and references. For a review of more recent
developments see for example \cite{NV04}. Here we follow the
pedagogical introduction of \cite{Vo05}.

\section{Cohomological field theories}
Before we embark on defining the topological string let us define
a {\it cohomological field theory}. It has the following
properties:
\begin{itemize}
\item It contains a fermionic symmetry operator $Q$ that squares
to zero,
\begin{equation}
Q^2=0\ .
\end{equation}
Then the {\it physical operators} of the theory are defined to be
those that are $Q$-closed,
\begin{equation}
\{Q,\mathcal{O}_i\}=0\ .\label{QO}
\end{equation}

\item The vacuum is $Q$-symmetric, i.e. $Q|0\rangle=0$. This
implies the equivalence $\mathcal{O}_i\sim\mathcal{O}_i+\{Q,\L\}$,
as can be seen from $\langle
\mathcal{O}_{i_1}\ldots\mO_{i_k}\{Q,\L\}\mO_{i_{k+2}}\ldots\rangle=0$,
where we used (\ref{QO}).

\item The energy-momentum tensor is $Q$-exact,
\begin{equation}
T_{\a\b}\equiv{\delta S\over\delta h^{\a\b}}=\{Q,G_{\a\b}\}\
.\label{Tab}
\end{equation}
This property implies that the correlation functions do not depend
on the metric,
\begin{eqnarray}
{\delta\over\delta
h^{\a\b}}\langle\mO_{i_1}\ldots\mO_{i_n}\rangle&=&{\delta\over\delta
h^{\a\b}}\int D\phi\
\mO_{i_1}\ldots\mO_{i_n}e^{iS[\phi]}\nonumber\\
&=&i\int D\phi\ \mO_{i_1}\ldots\mO_{i_n}e^{iS[\phi]}{\delta
S\over\delta
h^{\a\b}}\nonumber\\
&=&i\langle\mO_{i_1}\ldots\mO_{i_n}\{Q,G_{\a\b}\}\rangle=0\ .
\end{eqnarray}
Here we assumed that the $\mO_i$ do not depend on the metric.
\end{itemize}
The condition (\ref{Tab}) is trivially satisfied if the Lagrangian
is $Q$-exact,
\begin{equation}
L=\{Q,V\}\ ,
\end{equation}
for some operator $V$. For such a Lagrangian one can actually
calculate the correlation functions {\it exactly} in the classical
limit. To see this note that
\begin{eqnarray}
{\partial\over\partial
\hbar}\langle\mO_{i_1}\ldots\mO_{i_n}\rangle&=&{\partial\over\partial
\hbar}\int D\phi\
\mO_{i_1}\ldots\mO_{i_n}\exp\left({{i\over\hbar}\left\{Q,\int
V\right\}}\right)=0\ .\label{hbar}
\end{eqnarray}

Interestingly, from any scalar physical operator $\mO^{(0)}$, i.e.
from one that does not transform under coordinate transformation
of $M$, where $M$ is the manifold on which the theory is
formulated, one can construct a series of non-local physical
operators, which transform like forms. Integrating (\ref{Tab})
over a space-like hypersurface gives
\begin{equation}
P_\a=\{Q,G_\a\}\ .
\end{equation}
Consider
\begin{equation}
\mO_\a^{(1)}:=i\{G_\a,\mO^{(0)}\}\ ,
\end{equation}
and calculate
\begin{eqnarray}
{d\over d x^\a}\mO^{(0)}&=&i[P_\a,\mO^{(0)}]\nonumber\\
&=&i[\{Q,G_\a\},\mO^{(0)}]\nonumber\\
&=&\pm i\{\{G_\a,\mO^{(0)}\},Q\}-i\{\{\mO^{0},Q\},G_\a\}\nonumber\\
&=&\{Q,\mO^{1}\}\ .
\end{eqnarray}
Let $\mO^{1}:=\mO^{1}_\a\d x^\a$, then
\begin{equation}
\d O^{0}=\{Q,\mO^{1}\}\ .
\end{equation}
Integrating this equation over a closed curve $\g$ in $M$ gives
\begin{equation}
\left\{Q,\int_\g\mO^{(1)}\right\}=0\ .
\end{equation}
Therefore, the set of operators $\int_\g\mO^{(1)}$ are (non-local)
physical operators. Repeating this procedure gives
\begin{eqnarray}
\{Q,\mO^{(0)}\}&=&0\nonumber\\
\{Q,\mO^{(1)}\}&=&\d \mO^{(0)}\nonumber\\
\{Q,\mO^{(2)}\}&=&\d \mO^{(1)}\nonumber\\
&\ldots&\nonumber\\
\{Q,\mO^{(n)}\}&=&\d \mO^{(n-1)}\nonumber\\
0&=&\d \mO^{(n)}\ ,
\end{eqnarray}
where $n$ is the dimension of $M$. Hence, the integral of
$\mO^{(p)}$ over a $p$-dimensional submanifold of $M$ is physical.
In particular, we have
\begin{equation}
\left\{Q,\int_M\mO^{(n)}\right\}=0\ .
\end{equation}
This implies that we can add terms $t^a\mO^{(n)}_a$, with $t^a$
arbitrary coupling constants to the Lagrangian, and the
``deformed" theory then will still be cohomological.

\section{$\mathcal N=(2,2)$ supersymmetry in 1+1 dimensions}
The goal of this appendix is to construct the B-type topological
string. This can be done by twisting an $\mN=(2,2)$ supersymmetric
theory in two (real) dimensions and then coupling the twisted
theory to gravity. Let us therefore start by studying
two-dimensional $\mN=(2,2)$ theories. Since we will only be
interested in theories living on complex one dimensional
manifolds, which locally look like $\mathbb{C}$, we will
concentrate on field theories on $\mathbb{C}$ with coordinates
$z=x^1+ix^0$. Here $ix^0$ can be understood as Euclidean time.

The Lorentz group on $\mathbb{C}$ is given by $U(1)$, and Weyl
spinors have only one complex component\footnote{For a detailed
description of spinors in various dimensions see appendix
\ref{spinors}.}. On the other hand, these spinors transform under
the $U(1)$ Lorentz group and one can classify them according to
whether they have positive or negative $U(1)$-charge. A theory
with $p$ spinor supercharges of positive and $q$ spinor
supercharges of negative charge is said to be a $\mN=(p,q)$
supersymmetric field theory.

Here we study $\mN=(2,2)$ theories which are best formulated on
{\it superspace} with coordinates $z,\zb,\t^\pm,\bar\t^\pm$, where
$\t^\pm$ are Grassmann variables satisfying
$(\t^\pm)^*=\bar\t^\mp$. The superscript $\pm$ indicates how the
spinors transform under Lorentz transformation, $z\rightarrow
z'=e^{i\a}z$, namely
\begin{equation}
\t^\pm\rightarrow(\t^\pm)'=e^{\pm i\a/2}\t^\pm\ \ \ ,\ \ \
\bar\t^\pm\rightarrow(\bar\t^\pm)'=e^{\pm i\a/2}\bar\t^\pm\ .
\end{equation}
Functions that live on superspace are called {\it superfields} and
because of the Grassmannian nature of the fermionic coordinates
they can be expanded as
\begin{equation}
\Psi(z,\zb,\t^+,\t^-,\bar\t^+,\bar\t_-)=\phi(z,\zb)+\t^+\psi_+(z,\zb)+\t^-\psi_-(z,\zb)+\t^+\t^-F(z,\zb)+\ldots\
.
\end{equation}

\bigskip
{\bf Symmetries and the algebra}\\
Having established superspace, consisting of bosonic coordinates
$z,\zb$ and fermionic coordinates $\t^\pm,\bar\t^\pm$ we might ask
for the symmetry group of this space. In other words we are
interested in the linear symmetries which leave the measure $\d
z\d\zb\d\t^+\d\t^-\d\bar\t^+\d\bar\t^-$ invariant. Clearly, part
of this symmetry group is the two-dimensional Poincaré symmetry.
The translations are generated by
\begin{eqnarray}
H&=&-i{d\over d(ix^0)}=-i(\partial_z-\partial_{\zb})\ ,\nonumber\\
P&=&-i{d\over d x^1}=-i(\partial_z+\partial_{\zb})\nonumber\ .
\end{eqnarray}
We saw already how the spinors $\t^\pm,\bar\t^\pm$ transform under
Lorentz transformation, so the generator reads
\begin{equation}
M=2z\partial_z-2\zb\partial_{\zb}+\t^+{d\over d\t^+}-\t^-{d\over
d\t^-}+\bar\t^+{d\over d\bar\t^+}-\bar\t^-{d\over d\bar\t^-}\ ,
\end{equation}
where $M$ is normalised in such a way that $e^{2\pi iM}$ rotates
the Grassmann variables once and the complex variables twice.
Other linear transformations are the translation of the fermionic
coordinates $\t^\pm$ generated by $\partial\over\partial\t^\pm$
and a change of bosonic coordinates $z\rightarrow z'=z+\e\t$,
generated by the eight operators
$\t^\pm\partial_z,\t^\pm\partial_{\zb},\bar\t^\pm\partial_z,\bar\t^\pm\partial_{\zb}$.
From these various symmetry generators one defines differential
operators on superspace,
\begin{eqnarray}
\mathcal{Q}_\pm&:=&{\partial\over\partial\t^\pm}+i\bar\t^\pm\partial_\pm\ ,\\
\overline{\mathcal{Q}}_\pm&:=&-{\partial\over\partial\bar\t^\pm}-i\t^\pm\partial_\pm\ ,\\
D_\pm&:=&{\partial\over\partial\t^\pm}-i\bar\t^\pm\partial_\pm\ ,\\
\overline{D}_\pm&:=&-{\partial\over\partial\bar\t^\pm}+i\t^\pm\partial_\pm\
,
\end{eqnarray}
where  we denoted $\partial_+:=\partial_z$ and
$\partial_-:=\partial_{\zb}$.

There are two more interesting linear symmetry transformations,
known as the vector and axial R-rotations of a superfield. They
are defined as
\begin{eqnarray}
R_V(\a):&&(\t^+,\bar\t^+)\rightarrow
(e^{-i\a}\t^+,e^{i\a}\bar\t^+)\ \ \ ,\ \ \
(\t^-,\bar\t^-)\rightarrow
(e^{-i\a}\t^-,e^{i\a}\bar\t^-)\ ,\nonumber\\
R_A(\a):&&(\t^+,\bar\t^+)\rightarrow
(e^{-i\a}\t^+,e^{i\a}\bar\t^+)\ \ \ ,\ \ \
(\t^-,\bar\t^-)\rightarrow (e^{i\a}\t^-,e^{-i\a}\bar\t^-)\
,\nonumber
\end{eqnarray}
with the corresponding operators\\
\parbox{14cm}{
\begin{eqnarray}
F_V&=&-\t^+{d\over d\t^+}-\t^-{d\over d\t^-}+\bar\t^+{d\over
d\bar\t^+}+\bar\t^-{d\over d\t^-}\ ,\nonumber\\
F_A&=&-\t^+{d\over d\t^+}+\t^-{d\over d\t^-}+\bar\t^+{d\over
d\bar\t^+}-\bar\t^-{d\over d\t^-}\ .\nonumber
\end{eqnarray}}\hfill\parbox{8mm}{\begin{eqnarray}\label{Rsym}\end{eqnarray}}\\
Of course, a superfield $\Psi$ might also transform under these
transformations,
\begin{eqnarray}
e^{i\a F_V}: \Psi(x^\m,\t^\pm,\bar\t^\pm)&\mapsto& e^{i\a
q_V}\Psi(x^\m,e^{-i\a}\t^\pm,e^{i\a}\bar\t^\pm)\ ,\\
e^{i\b F_A}: \Psi(x^\m,\t^\pm,\bar\t^\pm)&\mapsto& e^{i\b
q_A}\Psi(x^\m,e^{\mp i\b}\t^\pm,e^{\pm i\b}\bar\t^\pm)\ ,
\end{eqnarray}
where $q_V$ and $q_A$ are known as the {\it vector and axial
R-charge} of $\Psi$.

From the operators constructed so far it is easy to derive the
commutation relations. One obtains the algebra of an $\mN=(2,2)$
supersymmetric theory
\begin{eqnarray}\label{22algebra}
[M,H]&=&-2P\ \ \ ,\ \ \ [M,P]=-2H\nonumber\\
\mQ_+^2=\mQ_-^2&=&\overline{\mQ}_+^2=\overline{\mQ}_-^2=0\nonumber\\
\{\mQ_\pm,\overline{\mQ}_\pm\}&=&H\pm P\nonumber\\
\ [M,\mQ_\pm]&=&\mp Q_\pm\ \ \ ,\ \ \ [M,\overline{\mQ}_\pm]=\mp \overline{\mQ}_\pm\nonumber\\
\ [F_V,\mQ_\pm]&=&-\mQ_\pm\ \ \ ,\ \ \ [F_V,\overline{\mQ}_\pm]=+\overline{\mQ}_\pm\nonumber\\
\ [F_A,\mQ_\pm]&=&\mp \mQ_\pm\ \ \ ,\ \ \
[F_A,\overline{\mQ}_\pm]=\pm \overline{\mQ}_\pm\ .
\end{eqnarray}

\bigskip
{\bf Chiral superfields}\\
A {\it chiral superfield} is a function on superspace that
satisfies
\begin{equation}
\overline{D}_\pm\Phi=0\ ,
\end{equation}
and fields satisfying
\begin{equation}
D_\pm\Upsilon=0
\end{equation}
are called {\it anti-chiral superfields}. Note that the complex
conjugate of a chiral superfield is anti-chiral and vice-versa. A
chiral superfield has the expansion
\begin{equation}
\Phi(z,\t^\pm,\bar\t_\pm)=\phi(w,\wb)+\t^+\psi_+(w,\wb)+\t^-\psi_-(w,\wb)+\t^+\t^-F(w,\wb)\
,
\end{equation}
where $w:=z-i\t^+\bar\t^+$ and $\wb:=\zb-i\t^-\bar\t^-$. Note that
a $\mQ$-transformed chiral field is still chiral, because the
$\mQ$-operators and the $D$-operators anti-commute.

\bigskip
{\bf Supersymmetric actions}\\
We are interested in actions that are invariant under the
supersymmetry transformation
\begin{equation}
\delta=\e^+\mathcal{Q}_++\e^-\mathcal{Q}_-+\bar\e^+\overline{\mathcal{Q}}_++\bar\e^-\overline{\mathcal{Q}}_-\
.\label{susytraf}
\end{equation}
Let $K(\Psi_i,\bar\Psi_i)$ be a real differentiable function of
superfields $\Psi_i$ and consider the quantity
\begin{equation}
\int\d^2z\ \d^4\t\ K(\Psi_i,\bar\Psi_i)):=\int\d z\ \d\zb\
\d\t^+\d\t^-\d\bar\t^+\d\bar\t^-K(\Psi_i,\bar\Psi_i)\
.\label{Dterm}
\end{equation}
Functionals of this form are called {\it D-terms} and it is not
hard to see that they are invariant under the supersymmetry
transformation (\ref{susytraf}). If we require that for any
$\Psi_i$ with charges $q_V^i,q_A^i$ the complex conjugate field
$\bar\Psi_i$ has opposite charges $-q_V,-q_A$, the D-term is also
invariant under
the axial and vector R-symmetry.\\
Another invariant under supersymmetry can be constructed from
chiral superfields $\Phi_i$. Let $W(\Phi_i)$ be a {\it
holomorphic} function of the $\Phi_i$, called the superpotential,
and consider
\begin{equation}
\int\d^2z\ \d^2\t\ W(\Phi_i):=\left.\int\d^2z\ \d\t^+\d\t^-\
W(\Phi_i)\right|_{\bar\t^\pm=0}\label{Fterm}
\end{equation}
This invariant is called an {\it F-term}. Clearly, this term is
invariant under the axial R-symmetry if the $\Phi_i$ have $q_A=0$.
The vector R-symmetry, however, is only conserved if we can assign
vector R-charge two to the superpotential $W(\Phi)$. For monomial
superpotentials this is always possible.

It is quite interesting to analyse the action (\ref{Dterm}) in the
case in which $K$ is a function of chiral superfields $\Phi_i$ and
their conjugates. As to do so we want to spell out the action in
terms of the component fields and keep only those fields that
contain the fields $\phi_i$. The coefficient of
$\t^+\t^-\bar\t^+\bar\t^-$ can be read off to be
\begin{equation}
{dK\over d\phi^i}\partial_+\partial_-\phi^i+{d^2K\over
d\phi^id\phi^j}\partial_+\phi^i\partial_-\phi^j =-{d^2K\over
d\phi^id\bar\phi^j}\partial_+\bar\phi^j\partial_-\phi^i+d\left({dK\over
d\phi^i}\partial_-\phi^i\right)\ .
\end{equation}
Of course, the last term vanishes under the integral. The
$\bar\phi^i$-dependent terms give the same expression with $+$ and
$-$ interchanged. This implies that the $\phi,\bar\phi$-dependent
part of the D-term action can be written as
\begin{equation}
S_\phi=-\int\d^2z\ g_{i\bar
j}\eta^{\a\b}\partial_\a\phi^i\partial_\b\bar\phi^j\ ,
\end{equation}
with $\eta^{+-}=\eta^{-+}=2$, $\eta^{++}=\eta^{--}=0$ and
\begin{equation}
g_{i\bar j}(\phi,\bar\phi)={d^2K\over d\phi^id\bar\phi^j}\ .
\end{equation}
If we interpret the $\phi^i$ as coordinates on some target space
$\mathcal{M}$, we find that this space carries as metric $g_{i\bar
j}$ which is K\"ahler with K\"ahler potential $K$. Of course, one
could also write down all the other terms appearing in the action
(\ref{Dterm}), but the expression would be rather lengthy. One the
other hand it does not contain derivatives of the $F_i$, and all
$F_i$ appear at most quadratically. Therefore, we can integrate
over $F_i$ in the path integral and the result will be to
substitute the value it has according to its equation of motion.
Then the action turns into the rather simple form
\begin{equation}\label{22action}
L=-g_{i\bar j}\partial^\a\phi^i\partial_\a\bar\phi^j-2ig_{i\bar
j}\bar\psi_-^j\Delta_+\psi_-^i-2ig_{i\bar
j}\bar\psi^j_+\Delta_-\psi_+^i-R_{i\bar jk\bar
l}\psi_+^i\psi^k_-\bar\psi_+^j\bar\psi^l_-\ ,
\end{equation}
where
\begin{eqnarray}
R_{i\bar jk\bar l}&=&g^{m\bar n}\partial_{\bar l}g_{m\bar
j}\partial_kg_{\bar
ni}-\partial_k\partial_{\bar l}g_{i\bar j}\ ,\nonumber\\
\Delta_\pm\psi^i&=&\partial_\pm\psi^i+\G_{jk}^i\partial_\pm\phi^j\psi^k\ ,\\
\G^i_{jk}&=&g^{i\bar l}\partial_kg_{\bar lj}\ .\nonumber
\end{eqnarray}

\bigskip
{\bf Quantum theory and anomalies}\\
When studying a quantum field theory it is always an interesting
question whether the symmetries of the classical actions persist
on the quantum level, or whether they are anomalous. Here we want
to check whether the quantum theory with action (\ref{Dterm}), and
with chiral superfields $\Phi_i$ instead of the general fields
$\Psi_i$, is still invariant under the vector and axial
symmetries. One possibility to analyse the anomalies of a quantum
theory is to study the path integral measure,
\begin{equation}
\prod_iD\phi^iD\psi^i_+D\psi^i_-DF^i\times c.c.\ .
\end{equation}
If we take the charges $q_V,q_A$ of all the $\Phi_i$ to zero, the
$\phi^i$ do not transform under $R$-symmetry and, as we just saw,
the $F^i$ can be integrated out, so it remains to check whether
the fermion measure is invariant. To proceed one first of all
assumes that the size of the target manifold is large compared to
the generic size of the world-sheet. In this case the Riemann
curvature will be small and one can neglect the last term in the
action (\ref{22action}). Then the path integral over $\psi_-$ has
the form
\begin{equation}
\int D\psi_-D\bar\psi_-\exp(\bar\psi_-,\Delta_+\psi_-)\ ,
\end{equation}
where the inner product is defined as
$(a,b):=\int_{\mathbb{C}}a^ib_i$. A quantity that will be
important in what follows is the index $k$ of the operator
$\Delta_+$,
\begin{equation}
k={\rm dim\ Ker}\Delta_+-{\rm dim\ Ker}\Delta_+^\dagger\ .
\end{equation}
This quantity is actually a topological invariant, and the
Atiyah-Singer index theorem tells us that in this simple case it
can be written as
\begin{equation}
k=\int_{\phi(\S)}c_1(\mathcal{M})\ .
\end{equation}
Here $\S$ is the world-sheet, i.e. it is $\mathbb{C}$ in the case
we studied so far, and $\mathcal{M}$ denotes the target space. We
will not address the problem of how to study the fermion measure
in detail, but we only list the results. First of all, because of
the Grassmannian nature of the integrals over the fermionic
variables, one can show that the quantity $\int
D\psi_-D\bar\psi_-\exp(\bar\psi_-,\Delta_+\psi_-)$ does actually
vanish, unless $k=0$. In order to obtain a non-zero result for
finite $k$ one has to insert fermions into the path integral,
\begin{equation}
\int D\psi_+D\psi_-D\bar\psi_+D\bar\psi_-\left(g_{i_1\bar
j_1}\psi_-^{i_1}(z_1)\bar\psi_+^{j_1}(z_1)\ldots g_{i_{k}\bar
j_k}\psi_-^{i_k}(z_k)\bar\psi_+^{j_k}(z_k)\right)e^{-S}\ .
\end{equation}
It is easy to see that these quantities are invariant under the
vector symmetry, but the axial symmetry is broken unless $k=0$.
This is a first indication why Calabi-Yau manifolds are
particularly useful for the topological string, since $k=0$
implies that $c_1(\mathcal{M})=0$.

\section{The topological B-model}
The starting point of this section is once again the action
\begin{equation}\label{Bmodel}
S=\int_\S\d^2z\d^4\t K(\Phi^i,\bar\Phi^i)
\end{equation}
with chiral superfields $\Phi^i$, now formulated on an arbitrary
Riemann surface $\S$.

The $\mN=(2,2)$ supersymmetric field theories constructed so far
are not yet cohomological theories. Note however, that
\begin{eqnarray}
\{\overline\mQ_++\overline\mQ_-,\mQ_+-\mQ_-\}&=&2H\ ,\nonumber\\
\{\overline\mQ_++\overline\mQ_-,\mQ_++\mQ_-\}&=&2P\ ,\nonumber
\end{eqnarray}
so if we define
\begin{equation}
Q_B:=\overline\mQ_++\overline\mQ_-
\end{equation}
then
\begin{equation}
Q_B^2=0\ ,
\end{equation}
and $H$ and $P$ are $Q_B$-exact. However, one of the central
properties of a cohomological theory is the fact that it is
independent of the world-sheet metric, i.e. one should be able to
define it for an arbitrary world-sheet metric. This can be done on
the level of the Lagrangian, by just replacing partial derivatives
with respect to the world-sheet coordinates by covariant
derivatives. However, one runs into trouble if one wants to
maintain the symmetry of the $\mN=(2,2)$ theory. In particular,
the supersymmetries should be global symmetries, not local ones.
This amounts to saying that in the transformation
$\delta\Phi^i=\e^+\mQ_+\Phi^i$ the spinor $\e^+$ has to be
covariantly constant with respect to the world-sheet metric. But
for a general metric on the world-sheet there are no covariantly
constant spinors. So, it seems to be impossible to construct a
cohomological theory from the $\mN=(2,2)$ supersymmetric one.

On the other hand, for a symmetry generated by a bosonic generator
the infinitesimal parameter is simply a number. In other words, it
lives in the trivial bundle $\S\times\mathbb{C}$ and can be chosen
to be constant. This gives us a hint on how the above problem can
be solved. If we can somehow arrange for some of the
$\mQ$-operators to live in a trivial bundle, the corresponding
symmetry can be maintained. Clearly, the type of bundle in which
an object lives is defined by its charge under the Lorentz
symmetry. We are therefore led to the requirement to modify the
Lorentz group in such a way that the $\mQ_B$-operator lives in a
trivial bundle, i.e. has spin zero. This can actually be achieved
by defining
\begin{equation}\label{twist}
M_B:=M-F_A
\end{equation}
to be the generator of the Lorentz group. One then finds the
following commutation relations
\begin{eqnarray}
[M_B,\mQ_+]=-2\mQ_+\ \ \ &,&\ \ \ [M_B,\mQ_-]=2\mQ_-\nonumber\\
\ [M_B,\overline\mQ_+]=0\ \ \ &,&\ \ \ [M_B, \overline\mQ_-]=0\
.\nonumber
\end{eqnarray}
Note that now the operator $Q_B$ indeed is a scalar, and therefore
the corresponding symmetry can be defined on an arbitrary curved
world-sheet. This construction is called {\it twisting} and we
arrive at the conclusion that the twisted theory truly is a
cohomological field theory. Of course, on the level of the
Lagrangian one has to replace partial derivatives by covariant
ones for general metrics on the world-sheet, but now the covariant
derivatives have to be covariant with respect to the modified
Lorentz group.

So far our discussion was on the level of the algebra and rather
general. Let us now come back to the action (\ref{Bmodel})
analysed in the last section. Note first of all that the B-type
twist (\ref{twist}) involves the axial vector symmetry, which
remains valid on the quantum level only if we take the target
space to have $c_1=0$. Therefore, we will take the target space
$\mathcal{M}$ to be a Calabi-Yau manifold from now on.

Let us then study in which bundles the various fields of our
theory live after twisting. It is not hard to see that
\begin{eqnarray}
\psi^i_+&\in&
\L^{1,0}(\S)\otimes\phi^*(T^{(1,0)}(\mathcal{M}))\ ,\nonumber\\
\psi^i_-&\in&\L^{0,1}(\S)\otimes\phi^*(T^{(1,0)}(\mathcal{M}))\ ,\nonumber\\
\bar\psi^i_+&\in&\phi^*(T^{(0,1)}(\mathcal{M}))\ ,\nonumber\\
\bar\psi^i_-&\in&\phi^*(T^{(0,1)}(\mathcal{M}))\ ,\nonumber
\end{eqnarray}
where $\in$ means ``is a section of". This simply says that e.g.
$\psi^i_+$ transforms as a $(1,0)$-form on the world-sheet and as
a holomorphic vector in space time, whereas e.g. $\bar\psi^i_+$
transforms as a scalar on the world-sheet, but as an
anti-holomorphic vector in space-time. It turns out that the
following reformulation is convenient,
\begin{eqnarray}
\eta^{\bar i}:=\bar\psi^i_++\bar\psi_-^{i}\ ,\nonumber\\
\theta_i:=g_{i\bar j}(\bar\psi^j_+-\bar\psi_-^{j})\ ,\nonumber\\
\r^i_z:=\psi^i_+\ ,\nonumber\\
\r^i_{\zb}:=\psi_-^i\ .\nonumber
\end{eqnarray}
The twisted $\mN=(2,2)$ supersymmetric theory with a twist
(\ref{twist}) and action (\ref{Bmodel}) is called the {\it
B-model}. Its Lagrangian can be rewritten in terms of the new
fields,
\begin{equation}
L=-t\left(g_{i\bar
j}\eta^{\a\b}\partial_\a\phi^i\partial_\b\bar\phi^j+ig_{i\bar
j}\eta^{\bar
j}(\Delta_{\zb}\r^i_z+\Delta_z\r^i_{\zb})+i\t_i(\Delta_{\zb}\r^i_z-\Delta_z\rho^i_{\zb})+{1\over2}R_{i\bar
jk}^{\ \ \ l}\r^i_z\r^k_{\zb}\eta^{\bar j}\t_l\right)\
,\label{BmodelLag}
\end{equation}
where $t$ is some coupling constant. Here we still used a flat
metric on $\S$ to write the Lagrangian, but from our discussion
above we know that we can covariantise it using an arbitrary
metric on the world-sheet, without destroying the symmetry
generated by $Q_B$. The Lagrangian can be rewritten in the form
\begin{equation}
L=-it\{Q_B,V\}-t\left(i\t_i(\Delta_{\zb}\r^i_z-\Delta_z\r^i_{\zb})+{1\over2}R_{i\bar
jk}^{\ \ \ l}\r^i_z\r^k_{\zb}\eta^{\bar j}\t_l\right)\ ,
\end{equation}
with
\begin{equation}
V=g_{i\bar
j}\left(\r^i_z\partial_{\zb}\bar\phi^j+\rho^i_{\zb}\partial_z\bar\phi^j\right)\
.
\end{equation}
Now it seems as if the B-model was not cohomological after all,
since the second term of (\ref{BmodelLag}) is not $Q_B$-exact.
However, it is anti-symmetric in $z$ and $\zb$, and therefore it
can be understood as a differential form. The integral of such a
form is independent of the metric and therefore the B-model is a
topological quantum field theory.

We mention without proof that the B-model does depend on the
complex structure of the target space, but it is actually
independent of its K\"ahler structure \cite{Wi91}. The idea of the
proof is that the variation of the action with respect to the
K\"ahler form is $Q_B$-exact.

Furthermore, the $t$-dependence of the second term in
(\ref{BmodelLag}) can be eliminated by a rescaling of $\t_i$. If
one studies only correlation function which are homogeneous in
$\t$, the path integral only changes by an overall factor of $t$
to some power. Apart from this prefactor the correlation function
is independent of $t$, as can be seen by performing a calculation
similar to the one in (\ref{hbar}). But this means that to
calculate these correlation functions one can take the limit in
which $t$ is large and the result will be exact. This fact has
interesting consequences. For example consider the equations of
motion for $\phi$ and $\bar\phi$,
\begin{equation}
\partial_z\phi^i=\partial_{\zb}\phi^i=\partial_z\bar\phi^i=\partial_{\zb}\bar\phi^i=0\
,
\end{equation}
which only have constant maps as their solutions. Since the
``classical limit" $t\rightarrow\infty$ gives the correct result
for any $t$, up to an overall power of $t$, we find that in the
path integral for $\phi$ we only have to integrate over the space
of constant maps, which is simply the target space $\mathcal{M}$
itself.

A set of metric independent local operators can be constructed
from
\begin{equation}
V:=V_{\bar i_1\ldots\bar i_p}^{j_1\ldots j_q}(\phi,\bar\phi)\d
\bar\phi^{\bar i_1}\w\ldots\d\bar\phi^{\bar
i_p}{\partial\over\partial\phi^j_1}\ldots{\partial\over\partial\phi^j_q}
\end{equation}
as
\begin{equation}
\mO_V:=V_{\bar i_1\ldots\bar i_p}^{j_1\ldots
j_q}(\phi,\bar\phi)\eta^{\bar i_1}\ldots\eta^{\bar
i_p}\t_{j_1}\ldots\t_{j_q}\ .
\end{equation}
The transformation laws
\begin{eqnarray}
\{Q_B,\phi^i\}=0\ \ \ &,&\ \ \ \{Q_B,\bar\phi^i\}=-\eta^{\bar i}\ ,\nonumber\\
\{Q_B,\t_i\}=0\ \ \ &,&\ \ \ \{Q_B,\eta^{\bar i}\}=0\ ,\nonumber
\end{eqnarray}
can then be used to show that
\begin{equation}
\{Q_B,\mO_V\}=-\mO_{\bar\partial V}\ .
\end{equation}
We see that $Q_B$ can be understood as the Dolbeault exterior
derivative $\bar\partial$ and the physical operators are in
one-to-one correspondence with the Dolbeault cohomology classes.

\section{The B-type topological string}
So far the metric on the world-sheet was taken to be a fixed
background metric. To transform the B-model to a topological
string theory one has actually to integrate over all possible
metrics on the world-sheet. If one wants to couple an ordinary
field theory to gravity the following steps have to be performed.
\begin{itemize}
\item The Lagrangian has to be rewritten in a covariant way, by
replacing the flat metrics by dynamical ones, introducing
covariant derivatives and multiplying the measure by $\sqrt{{\rm
det}h}$

\item One has to add an Einstein-Hilbert term to the action, plus
possibly other terms, to maintain the original symmetries of the
theory.

\item The path integral measure has to include a factor $Dh$,
integrating over all possible metrics.
\end{itemize}

Here we only provide a sketch of how the last step of this
procedure might be performed. We start by noting that, once we
include the metric in the Lagrangian, the theory becomes
conformal. But this means that one can use the methods of ordinary
string theory to calculate the integral over all conformally
equivalent metrics. An important issue that occurs at this point
in standard string theory is the conformal anomaly. To understand
this in our context let us review the twisting from a different
perspective. We start from the energy momentum tensor $T_{\a\b}$,
which in conformal theories is known to have the structure
$T_{z\zb}=T_{\zb z}=0$, $T_{zz}=T(z)$ and $T_{\zb\zb}=\bar
T(\zb)$. One expands $T(z)=\sum_{m=-\infty}^\infty L_mz^{-m-2}$,
and the Virasoro generators satisfy
\begin{equation}
[L_m,L_n]=(m-n)L_{m+n}+{c\over12}m(m^2-1)\delta_{m,-n}\ .
\end{equation}
$c$ is the central charge and it depends on the theory. Technical
problems occur for non-zero $c$, since the equation of motion for
the metric reads
\begin{equation}
{\delta S\over\delta h_{\a\b}}=T_{\a\b}=0\ .
\end{equation}
In conformal theories this equation is  imposed as a constraint,
i.e. one requires that a physical state $\ket{\psi}$ satisfies
\begin{equation}
L_m\ket{\psi}=0 \ \ \ \forall m\in \mathbb{Z}\ .
\end{equation}
This is compatible with the Virasoro algebra only if $c=0$. If
$c\neq0$ one speaks of a conformal anomaly. Let us then check
whether we have a central charge in the case of the twisted
theory.

Note from (\ref{Rsym}) that $F_V+F_A$ acts on objects with a +
index, i.e. on left-moving quantities, whereas $F_V-F_A$ acts on
objects with a minus index, i.e. on right-moving quantities.
Therefore we define $F_L:=F_V+F_A$ and $F_R:=F_V-F_R$. It can be
shown that these two symmetries can be identified with the two
components of a single global $U(1)$ current. To be more precise
we have
\begin{equation}
F_L=\int_{z=0}J(z)\d z\ ,
\end{equation}
and similarly for $F_R$. Expanding the current as
\begin{equation}
J(z)=\sum_{m=-\infty}^\infty J_mz^{-m-1}
\end{equation}
gives $F_L=2\pi iJ_0$. Now recall that
$M_B=M-F_A=M-{1\over2}(F_L-F_R)$ and from $M=2\pi i(L_0-\bar L_0)$
we find
\begin{equation}
L_{0,B}=L_0-{1\over2}J_0\ \ \ ,\ \ \ \bar L_{0,B}=\bar
L_0-{1\over2}\bar J_0\ .
\end{equation}
This twisted Virasoro generator can be obtained from
\begin{equation}
T_B(z)=T(z)+{1\over2}\partial J(z)\ ,
\end{equation}
with generators
\begin{equation}
L_{m,B}=L_m-{1\over2}(m+1)J_m
\end{equation}
The algebra of these modified generators can be calculated
explicitly and one finds
\begin{equation}
[L_{m,B}, L_{n,B}]=(m-n)L_{m+n,B}\ .
\end{equation}
We find that the central charge is automatically zero, and as a
consequence the topological string is actually well-defined in any
number of space-time dimensions.\\

Now we can proceed as in standard string theory, in other words we
sum over all genera $\gh$ of the world-sheet, integrate over the
moduli space of a Riemann surface of genus $\gh$ and integrate
over all conformally equivalent metrics on the surface. In close
analogy to what is done in the bosonic string the free energy at
genus $\gh$ of the B-model topological string is given by (c.f.
Eq. (5.4.19) of \cite{Pol})
\begin{equation}
F_\gh:=\int_{\mathcal{M}_\gh}\left\la\prod_{i=1}^{3\gh-3}\left(\d
m^i\w\d \bar m^{\bar i}\int_\S G_{zz}(\m_i)^z_{\zb}\int_\S
G_{\zb\zb}(\bar\m_{\bar i})^{\zb}_z\right)\right\ra\ ,
\end{equation}
where $\mathcal{M}_\gh$ is the moduli space of a Riemann surface
of genus $\gh$. As usual the $(\m_i)^z_{\zb}$ are defined from the
change of complex structure, $\d z^i\rightarrow \d
z^i+\e(\m^i)^z_{\zb}\d\zb$, and the $\d m^i$ are the dual forms of
the $\m_i$. Furthermore, the quantity $G_{zz}$ is the $Q$-partner
of the energy momentum tensor component $T_{zz}$. Interestingly,
one can show that the $F_\gh$ vanish for every $\gh>1$, unless the
target space of the topological string is of dimension three.

This elementary definition of the B-type topological string is now
the starting point for a large number of interesting applications.
However, we will have to refrain from explaining further details
and refer the interested reader to the literature \cite{Horietal},
\cite{Vo05}, \cite{NV04}.

\chapter{Anomalies}\label{anomalies}
Anomalies have played a fascinating role both in quantum field
theory and in string theory. Many of the results described in the
main text are obtained by carefully arranging a given theory to be
free of anomalies. Here we provide some background material on
anomalies and fix the notation. A more detailed discussion can be
found in \cite{Me03}. General references are \cite{AG85},
\cite{AGG84}, \cite{AGW84}, \cite{Wb00} and \cite{We86}. In this
appendix we work in Euclidean space.

\section{Elementary features of anomalies}
In order to construct a quantum field theory one usually starts
from a classical theory, which is quantized by following one of
several possible quantization schemes. Therefore, a detailed
analysis of the classical theory is a crucial prerequisite for
understanding the dynamics of the quantum theory. In particular,
the symmetries and the related conservation laws should be
mirrored on the quantum level. However, it turns out that this is
not always the case. If the classical theory possesses a symmetry
that cannot be maintained on the quantum level we speak of an {\em
anomaly}.

\bigskip
A quantum theory containing a massless gauge field $A$ is only
consistent if it is invariant under the infinitesimal local gauge
transformation
\begin{equation}
A'(x)=A(x)+D\epsilon(x)\ .\label{localgaugetrafo}
\end{equation}
The invariance of the action can be written as
\begin{equation}
D_M(x){{\delta S[A]}\over{\delta A_{aM}(x)}}=0\ ,\label{gaugeinv}
\end{equation}
where $A=A_aT_a=A_{aM}T_adx^M$. Then we can define a current
corresponding to this symmetry,
\begin{equation}
J_a^M(x):={\delta S[A]\over\delta A_{aM}(x)}\
,\label{localcurrent}
\end{equation}
and gauge invariance (\ref{gaugeinv}) of the action tells us that
this current is conserved,
\begin{equation}
D_MJ^M_a(x)=0\ .\label{classcurr}
\end{equation}
Suppose we consider a theory containing massless fermions $\psi$
in the presence of an external gauge field $A$. In such a case the
expectation value of an operator is defined as\footnote{We work in
Euclidean space after having performed a Wick rotation. Our
conventions in the Euclidean are as follows: $S_M=iS_E$,
$ix^0_M=x^1_E$, $x^1_M=x^2_E,\ldots x^{d-1}_M=x^d_E$;
$i\Gamma^0_M=\Gamma^1_E,\ \Gamma^1_M=\Gamma^2_E,\ldots
\Gamma^{d-1}_M=\Gamma^d_E$;
$\Gamma_E:=i^{d\over2}\Gamma^1_E,\ldots ,\Gamma^d_E$. For details
on conventions in Euclidean space see \cite{BM03c}.}
\begin{equation}
\langle\mathcal{O}\rangle={\int D\psi D\bar\psi \ \mathcal{O}
\exp(-S[\psi,A])\over\int D\psi D\bar\psi \ \exp(-S[\psi,A])}\ ,
\end{equation}
and we define the quantity
\begin{equation}
\exp(-X[A]):=\int D\psi D\bar\psi \ \exp(-S[\psi,A])\
.\label{generating functional}
\end{equation}
Then it is easy to see that
\begin{equation}
\langle J^M_a(x)\rangle={\delta X[A]\over\delta A_{aM}(x)}\ .
\end{equation}
An anomaly occurs if a symmetry is broken on the quantum level, or
in other words if $X[A]$ is not gauge invariant, even though
$S[\psi,A]$ is. The non-invariance of $X[A]$ can then be
understood as coming from a non-trivial transformation of the
measure. Indeed, if we have
\begin{equation}\label{trans}
D\psi D\bar\psi\rightarrow \exp\left(i\int d^dx\
\epsilon_a(x)G_a[x;A] \right)D\psi D\bar\psi,
\end{equation}
then the variation of the functional (\ref{generating functional})
gives
\begin{equation}
\exp(-X[A])\int d^dx\ D_{M}\langle
J^M_a(x)\rangle\epsilon_a(x)=\int d^dx \int D\psi
D\bar\psi[iG_a[x;A]\epsilon_a(x)]\exp(-S)\ .
\end{equation}
This means that the quantum current will no longer be conserved,
but we get a generalised version of (\ref{classcurr}),
\begin{equation}
D_M\langle J^M_a(x)\rangle=iG_a[x;A]\ .\label{localcurrentquantum}
\end{equation}
$G_a[x;A]$ is called the {\em anomaly}.

\bigskip
Not every symmetry of an action has to be a local gauge symmetry.
Sometimes there are global symmetries of the fields
\begin{equation}
\Phi'=\Phi+i\epsilon\Delta\Phi\ .\label{global symmetry}
\end{equation}
These symmetries lead to a conserved current as follows. As the
action is invariant under (\ref{global symmetry}), for
\begin{equation}
\Phi'=\Phi+i\epsilon(x)\Delta\Phi
\end{equation}
we get a transformation of the form
\begin{equation}
\delta S[\Phi]=-\int d^dx\ J^M(x)\partial_M\epsilon(x)\ .
\label{dSglobal}
\end{equation}
If the fields $\Phi$ now are taken to satisfy the field equations
then (\ref{dSglobal}) has to vanish. Integrating by parts we find
\begin{equation}
\partial_M J^M(x)=0\ ,
\end{equation}
the current is conserved on shell.\footnote{This can be
generalized to theories in curved space-time, where we get
$\nabla_MJ^M(x)=0$, with the Levi-Civita connection $\nabla$.}
Again this might no longer be true on the quantum level. An
anomaly of a global symmetry is not very problematic. It simply
states that the quantum theory is less symmetric than its
classical origin. If on the other hand a local gauge symmetry is
lost on the quantum level the theory is inconsistent. This comes
about as the gauge symmetry of a theory containing massless spin-1
fields is necessary to cancel unphysical states. In the presence
of an anomaly the quantum theory will no longer be unitary and
hence useless. This gives a strong constraint for valid quantum
theories as one has to make sure that all the local anomalies
vanish.

\bigskip
{\bf The chiral anomaly}\\
Consider the specific example of non-chiral fermions $\psi$ in
four dimensions coupled to external gauge fields $A=A_aT_a=A_{a
\mu}T_adx^{\mu}$ with Lagrangian
\begin{equation}
\mathcal{L}=\bar\psi i\gamma^{\mu}D_{\mu}\psi=\bar\psi
i\gamma^{\mu}(\partial_{\mu}+A_{\mu})\psi\ .\label{Lagr}
\end{equation}
It is invariant under the global transformation
\begin{equation}
\psi':=\exp(i\epsilon\gamma_5)\psi\ ,\label{chiralsym}
\end{equation}
with $\epsilon$ an arbitrary real parameter. This symmetry is
called the {\em chiral symmetry}. The corresponding (classical)
current is
\begin{equation}
J_{5}^{\mu }(x)=\bar{\psi}(x)\gamma^{\mu}\gamma_5\psi(x)\
,\nonumber
\end{equation}
and it is conserved $\partial_{\mu}J^{\mu}_5=0$, by means of the
equations of motion. For this theory one can now explicitly study
the transformation of the path integral measure \cite{Fu79}, see
\cite{Me03} for a review. The result is that
\begin{equation}
G[x;A]={1\over16\pi^2}{\rm
tr}[\epsilon^{\mu\nu\rho\sigma}F_{\mu\nu}(x)F_{\rho\sigma}(x)]\
.\label{G[x;A]}
\end{equation}
We conclude that the chiral symmetry is broken on the quantum
level and we are left with what is known as the {\it chiral
anomaly}
\begin{equation}
\partial_{\mu}\langle J_5^{\mu}(x)\rangle=
{i\over16\pi^2}\epsilon^{\mu\nu\rho\sigma}{\rm
tr}F_{\mu\nu}(x)F_{\rho\sigma}(x)\ .
\end{equation}

\bigskip
{\bf The non-Abelian anomaly}\\
Next we study a four-dimensional theory containing a Weyl spinor
$\chi$ coupled to an external gauge field $A=A_aT_a$. Again we
take the base manifold to be  flat and four-dimensional. The
Lagrangian of this theory is
\begin{equation}
\mathcal{L}=\bar\chi i\gamma^{\mu}D_{\mu}P_+\chi=\bar\chi
i\gamma^{\mu}(\partial_{\mu}+A_{\mu})P_+\chi\ .
\end{equation}
It is invariant under the transformations\\
\parbox{14cm}{
\begin{eqnarray}
\chi'&=&g^{-1}\chi\ ,\nonumber\\
A'&=&g^{-1}(A+d)g\ ,\nonumber
\end{eqnarray}}\hfill\parbox{8mm}{\begin{eqnarray}\label{gaugetrafo2}\end{eqnarray}}
with the corresponding current
\begin{equation}
J_a^{\mu}(x):=i\bar{\chi}(x)T_a\gamma^{\mu}P_+\chi(x)\ .
\end{equation}
Again the current is conserved on the classical level, i.e. we
have
\begin{equation}
D_{\mu}J_a^{\mu}(x)=0\ .
\end{equation}
In order to check whether the symmetry is maintained on the
quantum level, one once again has to study the transformation
properties of the measure. The result of such a calculation (see
for example \cite{Wb00}, \cite{Na90}) is\footnote{Note that this
anomaly is actually purely imaginary as it should be in Euclidean
space, since it contains three factors of $T_a=-it_a$.}
\begin{equation}
D_{\mu} \langle
J^{\mu}_a(x)\rangle={1\over24\pi^2}\epsilon^{\mu\nu\rho\sigma}{\rm
tr}[T_a\partial_{\mu}(A_{\nu}\partial_\rho
A_{\sigma}+{1\over2}A_{\nu}A_{\rho}A_{\sigma})]\
.\label{nonabelian anomaly}
\end{equation}
If the chiral fermions couple to Abelian gauge fields the anomaly
simplifies to
\begin{equation}\label{nonabelianabelian}
D_{\mu} \langle
J^{\mu}_a(x)\rangle=-{i\over24\pi^2}\epsilon^{\mu\nu\rho\sigma}\partial_{\mu}A_{\nu}^b\partial_\rho
A_{\sigma}^c\cdot (q_aq_bq_c)
=-{i\over96\pi^2}\epsilon^{\mu\nu\rho\sigma}F^b_{\mu\nu}F^c_{\rho\sigma}\cdot
(q_aq_bq_c)\ .
\end{equation}
Here we used $T_a=iq_a$ which leads to $D=d+iq_aA_a$, the correct
covariant derivative for Abelian gauge fields. The index $a$ now
runs from one to the number of Abelian gauge fields present in the
theory.

\bigskip
{\bf Consistency conditions and descent equations}\\
In this section we study anomalies related to local gauge
symmetries from a more abstract point of view. We saw above that a
theory containing massless spin-1 particles has to be invariant
under local gauge transformations to be a consistent quantum
theory. These transformations read in their infinitesimal form
$A_{\mu}(y)\rightarrow A_{\mu}(y)+D_{\mu}\epsilon(y)$. This can be
rewritten as $A_{\mu b }(y)\rightarrow A_{\mu b }(y)-i\int d^4x\
\epsilon_a(x)\mathcal{T}_a(x)A_{\mu b}(y)$, with
\begin{equation}
-i\mathcal{T}_a(x):=-{\partial\over\partial
x^{\mu}}{\delta\over\delta A_{\mu a}(x)}-C_{abc}A_{\mu
b}(x){\delta\over\delta A_{\mu c}(x)}\ .
\end{equation}
Using this operator we can rewrite the divergence of the quantum
current (\ref{localcurrentquantum}) as
\begin{equation}
\mathcal{T}_a(x)X[A]=G_a[x;A]\ .\label{anomaly}
\end{equation}
It is easy to show that the generators $\mathcal{T}_a(x)$ satisfy
the commutation relations
\begin{equation}
[\mathcal{T}_a(x),\mathcal{T}_b(y)]=iC_{abc}\mathcal{T}_c(x)\delta(x-y)\
.\label{comm}
\end{equation}
From (\ref{anomaly}) and (\ref{comm}) we derive the {\em
Wess-Zumino consistency condition} \cite{WZ71}
\begin{equation}
\mathcal{T}_a(x)G_b[y;A]-\mathcal{T}_b(y)G_a[x;A]=iC_{abc}\delta(x-y)G_c[x;A]\
.\label{WZconsistency}
\end{equation}
This condition can be conveniently reformulated using the BRST
formalism. We introduce a ghost field $c(x):=c_a(x) T_a$ and
define the BRST operator by
\begin{eqnarray}
sA&:=&-Dc\ ,\label{BRST}\\
sc&:=&-{1\over2}[c,c]\ .
\end{eqnarray}
s is nilpotent, $s^2=0$, and satisfies the Leibnitz rule
$s(AB)=s(A)B\pm As(B)$, where the minus sign occurs if $A$ is a
fermionic quantity. Furthermore, it anticommutes with the exterior
derivative, $sd+ds=0$. Next we define the anomaly functional
\begin{eqnarray}
G[c;A]:=\int d^4x\ c_a(x)G_a[x;A]\ .
\end{eqnarray}
For our example (\ref{nonabelian anomaly}) we get
\begin{equation}
G[c;A] =-{i\over24\pi^2}\int {\rm tr}\left\lbrace c\
d\left[AdA+{1\over2}A^3\right]\right\rbrace\ .\label{G example}
\end{equation}
Using the consistency condition (\ref{WZconsistency}) it is easy
to show that
\begin{equation}
sG[c;A]=0\ .\label{WZconditionBRST}
\end{equation}
Suppose $G[c;A]=sF[A]$ for some local functional $F[A]$. This
certainly satisfies (\ref{WZconditionBRST}) since $s$ is
nilpotent. However, it is possible to show that all these terms
can be cancelled by adding a local functional to the action. This
implies that anomalies of quantum field theories are characterized
by the cohomology groups of the BRST operator. They are the local
functionals $G[c;A]$ of ghost number one satisfying the
Wess-Zumino consistency condition (\ref{WZconditionBRST}), which
cannot be expressed as the BRST operator acting on some local
functional of ghost number zero.

Solutions to the consistency condition can be constructed using
the {Stora-Zumino descent equations}. To explain this formalism we
take the dimension of space-time to be $2n$. Consider the
$(2n+2)$-form
\begin{equation}
{\rm ch}_{n+1}(A):={1\over(n+1)!}{\rm
tr}\left({iF\over2\pi}\right)^{n+1}\ ,
\end{equation}
which is called the (n+1)-th {\em Chern character}\footnote{A more
precise definition of the Chern character is the following. Let
Let $E$ be a complex vector bundle over $M$ with gauge group $G$,
gauge potential $A$ and curvature $F$. Then ${\rm ch}(A):={\rm
tr}\ \exp\left({iF\over2\pi}\right)$ is called the total Chern
character. The jth Chern character is ${\rm ch}_j(A):={1\over
j!}{\rm tr}\left({iF\over2\pi}\right)^j$.}. As $F$ satisfies the
Bianchi identity we have $dF=[A,F]$, and therefore, ${\rm
tr}F^{n+1}$ is closed, $d\ {\rm tr}F^{n+1}=0$. One can show (see
\cite{Me03} for details and references) that on any coordinate
patch the Chern character can be written as
\begin{equation}
{\rm ch}_{n+1}(A)=d\Omega_{2n+1},
\end{equation}
with
\begin{equation}
\Omega_{2n+1}(A)={1\over n
!}\left({i\over2\pi}\right)^{n+1}\int_0^1dt\ {\rm tr}( AF^n_t)\ .
\end{equation}
Here $F_t:=dA_t+{1\over2}[A_t, A_t]$, and $A_t:=tA$ interpolates
between 0 and $A$, if $t$ runs from 0 to 1. $\Omega_{2n+1}(A)$ is
known as the {\em Chern-Simons form} of ${\rm ch}_{n+1}(A)$. From
the definition of the BRST operator and the gauge invariance of
${\rm tr}F^{n+1}$ we find that $s({\rm tr}F^{n+1})=0$. Hence
$d(s\Omega_{2n+1}(A))=-sd\Omega_{2n+1}(A)=-s({\rm
ch}_{n+1}(A))=0$, and, from Poincar\'{e}'s lemma,
\begin{equation}
s\Omega_{2n+1}(A)=d\Omega_{2n}^1(c,A)\ .\label{descent1}
\end{equation}
Similarly, $d(s\Omega_{2n}^1(c,A))=-s^2\Omega_{2n+1}(A)=0$, and
therefore
\begin{equation}
s\Omega_{2n}^1(c,A)=d\Omega_{2n-1}^2(c,A)\ .\label{descent2}
\end{equation}
(\ref{descent1}) and (\ref{descent2}) are known as the {\em
descent equations}. They imply that the integral of
$\Omega_{2n}^1(c,A)$ over $2n$-dimensional space-time is BRST
invariant,
\begin{equation}
s\int_{M_{2n}}\Omega_{2n}^1(c,A)=0\ .
\end{equation}
But this is a local functional of ghost number one, so it is
identified (up to possible prefactors) with the anomaly $G[c;A]$.
Thus, we found a solution of the Wess-Zumino consistency condition
by integrating the two equations $d\Omega_{2n+1}(A)={\rm
ch}_{n+1}(A)$ and $d\Omega_{2n}^1(c,A)=s\Omega_{2n+1}(A)$. As an
example let us consider the case of four dimensions. We get
\begin{eqnarray}
\Omega_5(A)&=&{1\over2}\left(i\over2\pi\right)^3\int_0^1dt\  {\rm tr}(AF_t^2)\ ,\\
\Omega_4^1(c,A)&=&{i\over48\pi^3}{\rm tr}\left\lbrace c\
d\left[AF-{1\over2}A^3\right]\right\rbrace\ .
\end{eqnarray}
Comparison with our example of the non-Abelian anomaly (\ref{G
example}) shows that indeed
\begin{equation}
G[c;A]=-2\pi \int\Omega_4^1(c,A)\ .\label{anomaly omega}
\end{equation}
Having established the relation between certain polynomials and
solutions to the Wess-Zumino consistency condition using the BRST
operators it is actually convenient to rewrite the descent
equations in terms of gauge transformations. Define
\begin{equation}
G[\epsilon;A]:=\int d^4x\ \epsilon_a(x)G_a[x;A]\ .
\end{equation}
From (\ref{BRST}) it is easy to see that we can construct an
anomaly from our polynomial by making use of the descent
\begin{equation}\label{gaugedescent}
{\rm ch}_{n+1}(A)=d\Omega_{2n+1}(A)\ \ ,\ \
\delta_{\epsilon}\Omega_{2n+1}(A)=d\Omega^1_{2n}(\epsilon,A),
\end{equation}
where $\delta_{\epsilon}A=D\epsilon$. Clearly we find for our
example
\begin{equation}
\Omega_4^1(\epsilon,A)=-{i\over48\pi^3}{\rm tr}\left\lbrace
\epsilon\  d\left[AF-{1\over2}A^3\right]\right\rbrace\ .
\end{equation}
and we have
\begin{equation}\label{GOmega}
G[\epsilon,A]=2\pi\int\Omega_4^1(\epsilon,A)\ .
\end{equation}
We close this section with two comments.\\
\\
$\bullet$ The Chern character vanishes in odd dimension and thus
we cannot get an anomaly in these cases.\\
\\
$\bullet$ The curvature and connections which have been used were
completely arbitrary. In particular all the results hold for the
curvature two-form $R$. Anomalies related to a breakdown of local
Lorentz invariance or general covariance are called {gravitational
anomalies}. Gravitational anomalies are only present in $4m+2$
dimensions.

\section{Anomalies and index theory}
Calculating an anomaly from perturbation theory is rather
cumbersome. However, it turns out that the anomaly $G[x;A]$ is
related to the index of an operator. The index in turn can be
calculated from topological invariants of a given quantum field
theory using powerful mathematical theorems, the Atiyah-Singer
index theorem and the Atiyah-Patodi-Singer index
theorem\footnote{The latter holds for manifolds with boundaries
and we will not consider it here.}. This allows us to calculate
the anomaly from the topological data of a quantum field theory,
without making use of explicit perturbation theory calculations.
We conclude, that an anomaly depends only on the field under
consideration and the dimension and topology of space, which is a
highly non-trivial result.

\bigskip
Indeed, for the operator $i\gamma^{\mu}D_{\mu}$ appearing in the
context of the chiral anomaly the Atiyah-Singer index theorem
(c.f. appendix \ref{IT} and theorem \ref{ASI}) reads
\begin{equation}
{\rm ind}(i\gamma^{\mu}D_{\mu})=\int_M[{\rm ch}(F)\hat A(M)]_{{\rm
vol}}\ .
\end{equation}
We studied the chiral anomaly on flat Minkowski space, so $\hat
A(M)=\opone$. Using (\ref{ch(F)}) we find
\begin{equation}
{\rm ind}\left(i\gamma^{\mu}D_{\mu}\right)=-{1\over8\pi^2}\int
{\rm tr}F^2\ .
\end{equation}
and
\begin{equation}
G[x;A]={1\over16\pi^2}{\rm
tr}[\epsilon^{\mu\nu\rho\sigma}F_{\mu\nu}(x)F_{\rho\sigma}(x)]\ ,
\end{equation}
which is the same result as (\ref{G[x;A]}). We see that it is
possible to determine the structure of $G[x;A]$ using the index
theorem.

\bigskip
Unfortunately, in the case of the non-Abelian or gravitational
anomaly the calculation is not so simple. The anomaly can be
calculated from the index of an operator in these cases as well.
However, the operator no longer acts on a $2n$-dimensional space,
but on a space with $2n+2$ dimensions, where $2n$ is the dimension
of space-time of the quantum field theory. Hence, non-Abelian and
gravitational anomalies in $2n$ dimensions can be calculated from
index theorems in $2n+2$ dimensions. Since we do not need the
elaborate calculations, we only present the results. They were
derived in \cite{AGW84} and \cite{AGG84} and they are reviewed in
\cite{AG85}.

We saw already that it is possible to construct solutions of the
Wess-Zumino condition, i.e. to find the structure of the anomaly
of a quantum field theory, using the descent formalism. Via
descent equations the anomaly $G[c;A]$ in dimension $2n$ is
related to a unique $2n+2$-form, known as the {\em anomaly
polynomial}. It is this $2n+2$-form which contains all the
important information of the anomaly and which can be calculated
from index theory. Furthermore, the $2n+2$-form is unique, but the
anomaly itself is not. This can be seen from the fact that if the
anomaly $G[c;A]$ is related to a $2n+2$-form $I$, then
$G[c,A]+sF[A]$, with a $2n$-form $F[A]$ of ghost number zero, is
related to the same anomaly polynomial $I$. Thus, it is very
convenient, to work with anomaly polynomials instead of anomalies.

The only fields which can lead to anomalies are spin-${1\over2}$
fermions, spin-${3\over2}$ fermions and also forms with
(anti-)self-dual field strength. Their anomalies were first
calculated in \cite{AGW84} and were related to index theorems in
\cite{AGG84}. The result is expressed most easily in terms of the
non-invariance of the Euclidean quantum effective action $X$. The
master formula for all these anomalies reads
\begin{equation}
\delta X= i \int I_{2n}^1\ ,
\end{equation}
where $d I_{2n}^1=\delta I_{2n+1}\ \ ,\ \ d I_{2n+1}= I_{2n+2}$.
The $2n+2$-forms for the three possible anomalies are
\begin{eqnarray}
I_{2n+2}^{(1/2)}&=&-2\pi\left[\hat A(M_{2n})\ {\rm
ch}(F)\right]_{2n+2}\ ,\\
I_{2n+2}^{(3/2)}&=&-2\pi\left[\hat A(M_{2n})\ \left(\tr
\exp\left({i\over2\pi}R\right)-1\right)\ {\rm
ch}(F)\right]_{2n+2}\ ,\\
I_{2n+2}^{A}&=&-2\pi\left[\left(-{1\over2}\right){1\over4}\
L(M_{2n})\right]_{2n+2}\ .
\end{eqnarray}
To be precise these are the anomalies of spin-${1\over2}$ and
spin-${3\over 2}$ particles of positive chirality and a self-dual
form in Euclidean space under the gauge transformation $\delta
A=D\epsilon$ and the local Lorentz transformations
$\delta\omega=D\epsilon$. All the objects which appear in these
formulae are explained in appendix \ref{IT}.

Let us see whether these general formula really give the correct
result for the non-Abelian anomaly. From (\ref{trans}) we have
$\delta X=-i\int \e(x)G[x;A]=-iG[\e;A]$. Next we can use
(\ref{GOmega}) to find $\delta X=-2\pi i\Omega_4^1(\epsilon,A)$.
But $-2\pi\Omega_4^1(\epsilon,A)$ is related to $-2\pi{\rm
ch}_{n+1}(A)=-2\pi[{\rm ch}(F)]_{2n+2}$ via the descent
(\ref{gaugedescent}). Finally $-2\pi[{\rm ch}(F)]_{2n+2}$ is
exactly (\ref{Ihat1/2}) as we
are working in flat space where $\hat A(M)$=1.\\

The spin-${1\over2}$ anomaly\footnote{We use the term ``anomaly"
for both $G[x;A]$ and the corresponding polynomial $I$.} is often
written as a sum
\begin{equation}
I^{(1/2)}=I^{(1/2)}_{gauge}+I^{(1/2)}_{mixed}+nI^{(1/2)}_{grav},
\end{equation}
with the pure gauge anomaly
\begin{equation}
I^{(1/2)}_{gauge}:=[{\rm ch}(A)]_{2n+2}={\rm ch}_{n+1}(A)\
,\label{Igauge}
\end{equation}
a gravitational anomaly
\begin{equation}
I^{(1/2)}_{grav}=[\hat A(M)]_{2n+2}\ ,\label{I1/2grav}
\end{equation}
and finally all the mixed terms
\begin{equation}
I^{(1/2)}_{mixed}:=I^{(1/2)}-I^{(1/2)}_{gauge}-n I^{(1/2)}_{grav}\
.\label{Imixed}
\end{equation}
$n$ is the dimension of the representation of the gauge group
under which $F$ transforms.

\bigskip
\begin{center}
{\bf Anomalies in four dimensions}
\end{center}
There are no purely gravitational anomalies in four dimensions.
The only particles which might lead to an anomaly are chiral
spin-1/2 fermions. The anomaly polynomials are six-forms and they
read for a positive chirality spinor in Euclidean
space\footnote{Note that the polynomials are real, since we have,
as usual, $A=A_aT_a$ and $T_a$ is anti-Hermitian.}
\begin{equation}
I_{gauge}^{({1/2})}(F)=-2\pi \ {\rm ch}_3(A)
={i\over{{(2\pi)}^23!}}{\rm tr}F^3\ .\label{I4gauge}
\end{equation}
The mixed anomaly polynomial of such a spinor is only present for
Abelian gauge fields as ${\rm tr}(T_a)F_a$ vanishes for all simple
Lie algebras. It reads
\begin{equation}
I_{mixed}^{({1/2})}(R,F)=-{i\over{{(2\pi)}^23!}}{1\over 8}{\rm
tr}R^2{\rm tr}F={1\over{{(2\pi)}^23!}}{1\over 8}{\rm tr}R^2
F^aq_a\label{I4mixed}\ .
\end{equation}

\bigskip
\begin{center}
{\bf Anomalies in ten dimensions}
\end{center}
In ten dimensions there are three kinds of fields which might lead
to an anomaly. These are chiral spin-3/2 fermions, chiral spin-1/2
fermions and self-dual or anti-self-dual five-forms. The
twelve-forms for gauge and gravitational anomalies are calculated
using the general formulae (\ref{Ihat1/2}) - (\ref{IhatA}),
together with the explicit expressions for $\hat{A}(M)$ and $L(M)$
given in appendix \ref{IT}. One obtains
\begin{eqnarray}
I_{gauge}^{({1/2})}(F)&=&{1\over{{(2\pi)}^56!}}{\rm Tr}F^6\nonumber\\
I_{mixed}^{({1/2})}(R,F)&=&{1\over{{(2\pi)}^56!}}\left({1\over
16}{\rm tr}R^4 {\rm Tr}F^2+{5\over 64}({\rm tr}R^2)^2 {\rm
Tr}F^2-{5\over8}{\rm tr}R^2
{\rm Tr}F^4\right)\nonumber\\
I_{grav}^{(1/2)}(R)&=&{1\over{{(2\pi)}^56!}}\left(-{1\over504}{\rm
tr}R^6-{1\over
384}{\rm tr}R^4 {\rm tr}R^2-{5\over 4608}({\rm tr}R^2)^3\right)\nonumber\\
I_{grav}^{({3/2})}(R)&=&{1\over{{(2\pi)}^56!}}\left({55\over56}{\rm
tr}R^6-{75\over
128}{\rm tr}R^4  {\rm tr}R^2+{35\over 512}({\rm tr}R^2)^3\right)\nonumber\\
I_{grav}^{(5-form)}(R)&=&{1\over{{(2\pi)}^56!}}\left(-{496\over504}{\rm
tr}R^6+{7\over 12}{\rm tr}R^4 {\rm tr}R^2-{5\over 72}({\rm
tr}R^2)^3\right)\label{anomalypolynomials10}.
\end{eqnarray}
The Riemann tensor $R$ is regarded as an $SO(9,1)$ valued
two-form, the trace ${\rm tr}$ is over $SO(1,9)$ indices. It is
important that these formulae are additive for each particular
particle type. For Majorana-Weyl spinors an extra factor of
$1\over2$ must be included, negative chirality spinors (in the
Euclidean) carry an extra minus sign.

\end{appendix}

\part{Bibliography}

\end{document}